\documentclass[a4paper,fleqn]{cas-dc}
\usepackage[authoryear]{natbib}
\usepackage{hyperref}
\usepackage{graphicx}
\usepackage{pgffor}

\usepackage{longtable}
\usepackage{tabu}
\usepackage{multirow}
\usepackage{amsmath}
\usepackage{verbatim}
\usepackage{pdflscape}
\usepackage{caption}
\usepackage{subcaption}
\usepackage{amsmath}
\usepackage[normalem]{ulem}
\usepackage{gensymb}
\usepackage{adjustbox}
\usepackage[switch]{lineno} 
\usepackage{soul}

\DeclareCaptionFormat{cont}{#1 (cont.)#2#3\par}

\begin{document}
\let\WriteBookmarks\relax
\def\floatpagepagefraction{1}
\def\textpagefraction{.001}
\shorttitle{ATLAS phase-curves}
\shortauthors{M. Colazo et~al.}

\title [mode = title]{Asteroid phase curves and phase coloring effect using the ATLAS survey data}                      

\tnotetext[1]{This document is the results of the research
   project funded by the National Center of Science, Poland.}

\author[1]{Milagros Colazo}[orcid=0000-0001-6082-2477]
\author[1]{Dagmara Oszkiewicz}[orcid=0000-0002-5356-6433]
\author[2]{Alvaro Alvarez-Candal}
\author[1]{Patrycja Poźniak}
\author[1]{Przemysław Bartczak}
\author[1]{Edyta Podlewska-Gaca}

\address[1]{Astronomical Observatory Institute, Faculty of Physics and Astronomy, A. Mickiewicz University, S{\l}oneczna 36, 60-286 Pozna{\'n}, Poland}

\address[2]{Instituto de Astrof\'isica de Andaluc\'ia, CSIC, Apt 3004, E18080 Granada, Spain}

\begin{abstract}
We determined phase curves for 301$,$272 asteroids in the orange filter and 280$,$953 in the cyan filter from the latest ATLAS Solar System Catalog V2 (SSCAT-2). Among them, 3$,$345 and 492 asteroids in the orange and cyan filters, respectively, have uncertainties below 15\%. Our simple model, which considers only the apparition effect, showed good consistency with more sophisticated methods requiring much less computational time. Database cross-matching allowed us to analyze G1 and G2 distributions according to taxonomy.
We conducted two-dimensional Kolmogorov–Smirnov tests to investigate two distinct aspects: similarities in paired G1, G2 distributions across different taxa and wavelength dependency within the same taxa. When comparing different taxa, we couldn’t reject the null hypothesis for 11$\%$ of the orange sample and 31$\%$ of the cyan sample, indicating more disparities in the orange filter. For wavelength dependency, paired distributions of G1, G2 (o) vs. G1, G2 (c) showed statistically significant differences across all complexes, except for the A class.
Our analysis suggests that while phase coloring behaviors are observed without a clear preference for reddening or bluening at phase angles below 5\textdegree, reddening predominates in the 10\textdegree–30\textdegree range. We also observed smaller uncertainties in G2 than in G1. Simulations showed that G2 is less sensitive to lack of data at small phase angles. This is related to the definition of the H,G1,G2 function, where G1 contributes more to the opposition effect and G2 the linear part of the phase curve. Our catalog-independent algorithms are adaptable to new data sets, including future LSST data.

\end{abstract}

\begin{highlights}

\item Phase Curve Analysis: Phase curves were determined for >300$\,$000 asteroids.

\item Wavelength Dependence: two-dimensional K-S tests shows differences between the phase curves in the orange and cyan filters across all taxonomic classes, except for the A class.

\item Phase Coloring: Phase reddening was observed in a greater proportion of asteroids at phase angles of 10–30 degrees, while both reddening and bluening were observed at smaller angles (below 5 degrees) without a clear preference for either.

\item Uncertainty Patterns: fitted values of G2 shows lower uncertainty than G1, with simulations confirming G2’s stability across phase angles.

\item Method Comparison: Our results align with those of \cite{alvarez2022}, \cite{wilawer2024}, and \cite{carry2024}.

\item Algorithm Adaptability: The method is adaptable to various datasets, including LSST and Data Preview 0.3.

\end{highlights}


\maketitle

\section{Introduction}
Phase curves describe changes in asteroid reduced brightness (its apparent magnitude normalized to 1 AU from both the Earth and the Sun) with respect to the phase angle (the Sun-asteroid-observer angle). This dependence is mostly linear, except at small phase angles where the opposition surge occurs. This behavior is explained by a combination of shadow-hiding in rough and porous regolith and coherent backscattering mechanisms \citep{muinonen2010three,  muinonen2010coherent}. The amplitude of the opposition effect depends on geometric albedo, with a maximum for moderate-albedo objects and decreasing for both low- and high-albedo asteroids \citep{belskaya2000opposition}. Phase curves also provide insights into the surface properties related to the texture and composition of the regolith. For example the width and height of the opposition effect relates to the compaction state of regolith and distribution of particle sizes \citep{hapke1981bidirectional}.

Several phase functions have been developed over the years to model the opposition effect and the linear behaviour. The most commonly used models are described here. The H, G model \citep{bowell1989}, adopted by the International Astronomical Union (IAU) in 1985, has limitations in explaining the opposition effect, especially for very dark or very bright objects. Consequently, in 2012, the IAU replaced it with the H, G1, G2 model \citep{muinonen2010three}. \cite{muinonen2010three} defined also a two parameter phase function the H, G12 designed for objects with a limited number of observations. Later, in 2016, \citet{penttila2016h} introduced the HG12* function to minimize potential biases when working with low-quality observations (that is having large magnitude variations, small number of observations or large photometric uncertainties). More advanced models based on these phase functions, which also account for rotational lightcurve modulation and shape and geometry effects, have recently been implemented by \cite{muinonen2020asteroid, muinonen2022asteroid, carry2024combined}.

The opposition effect was first detected for asteroid (20) Massalia in the 1950s \citep{gehrels1956photometric} and has been studied extensively based on dense ground-based lightcurve observations since then (\cite{gehrels1977minor, gehrels1979minor, harris1989phase, dovgopol1992asteroid, lagerkvist1990analysis, shevchenko2016asteroid, shevchenko2019phase, pravec2012absolute, shevchenko2003rotation, shevchenko2021photometry, oszkiewicz2021first} and many others). These observations typically require months of data for a single object, which limits the number of phase curves obtained from traditional ground-based observations.

In recent decades, phase curves have started to be derived in large quantities from sparse survey data. \cite{oszkiewicz2011online, oszkiewicz2012asteroid} derived phase curves for approximately 500$\,$000 objects based on photometry reported to the Minor Planet Center (MPC). \cite{verevs2015absolute} calculated absolute magnitudes and slope parameters for around 250$\,$000 asteroids using Pan-STARRS PS1 measurements. \cite{waszczak2015asteroid} processed data from the Palomar Transient Factory survey for about 50$\,$000 objects. \cite{mahlke2021asteroid} obtained phase curve parameters for approximately 100$\,$000 objects from filtered observations in the Asteroid Terrestrial-impact Last Alert System survey \citep[][ATLAS]{Torny2018}. \cite{martikainen2021asteroid} derived phase curves for about 500 asteroids using photometry from the Gaia mission's Data Release 2. \cite{alvarez2022phase} obtained 15$\,$000 phase curves from the Sloan Moving Object Catalogue. \cite{carry2024combined} processed about 100$\,$000 asteroids observed by the Zwicky Transient Facility (ZTF). \cite{schemel2021zwicky} explored the ZTF data with a Markov Chain Monte Carlo method accounting for the unknown rotational phase for the Jupiter Trojans and \cite{bernardinelli2023photometry} applied a similar method to the Dark Energy Survey data of trans-Neptunian objects.

While these sparse data are abundant, they are affected by rotational, shape, and geometric/aspect \& apparition effects, which can influence the derived parameters \citep{jackson2022effect, robinson2024}. These studies also vary in model complexity, ranging from models with no corrections to those that account for shape and geometry effects, providing phase curve parameters for an idealized spherical asteroid with surface characteristics matching those of the studied objects. Progress has also been made in merging spase and dense photometric datasets for phase curve fitting \cite{colazo2021determination, wilawer2022asteroid, wilawer2024}.

These advances in data availability and methodological complexity have enabled increasingly detailed studies of the dependencies of phase curve parameters on taxonomy and color effects. Early work explored these relationships, but with much smaller samples. For instance, \cite{lagerkvist1990analysis} analyzed 69 asteroids to calculate mean values for S, C, and M classes, while similar studies by \cite{goidet1995polarization} and \cite{harris1989asteroid} examined around 100 and 70 asteroids, respectively. With larger data sets and improved models, these relationships can now be investigated in far greater depth. For example, \cite{oszkiewicz2012asteroid} found that the $G_{12}$ parameter statistically varies across taxonomic complexes, while also demonstrating its homogeneity within asteroid families. \cite{shevchenko2016asteroid} examined the distribution of $G_1$ and $G_2$ parameters for approximately 200 well-observed asteroids, identifying distinct regions of parameter space occupied by different complexes. More recently, \cite{mahlke2021} extended these analyses to objects observed by the ATLAS survey, further clarifying phase curve parameter distributions across taxonomic types.

With the increasing number of observations across multiple filters, - another phenomenon - the phase reddening effect can now be studied on a large scale. This effect results in an increase in spectral slope and changes in absorption bands as the phase angle increases, and it can also be observed in multi-filter phase curves. Although phase reddening has been recognized for some time, in-depth analysis has so far focused on a limited number of asteroids, taxonomic types, and populations, with most studies concentrating on Near-Earth Asteroids \citep{reddy2012, sanchez2012, sanchez2024}. Moreover, \cite{binzel2019} and \cite{CarvanoDavalos2015} suggest that this effect may vary individually among different asteroids. \cite{mahlke2021} identified statistically distinct $G_1$ and $G_2$ parameters for cyan and orange filters (demonstrating wavelength dependency) across various populations and taxonomic types. Additionally, \cite{alvarez2024} proposed renaming the phenomenon "phase coloring," noting that some objects tend to shift to a bluer hue at higher phase angles. Similarly, \cite{wilawer2024} observed that this blue-shift can occur even at low phase angles. \cite{wilawer2024} explains this trend through a known correlation with albedo: higher albedo corresponds to flatter phase curves, while lower albedo leads to steeper curves. This correlation suggests that the effect may systematically vary across different taxonomic types. Whether the effect is individual or systematic across different taxonomic types or populations-and whether both reddening and blueing can be observed and for which objects-can now be addressed with the increased volume of data.

In this paper, we calculate phase curves for hundreds of thousands of asteroids using the latest catalog from the ATLAS survey, accounting for apparition effects. We also examine taxonomy and wavelength dependency effects for the best observed objects. Additionally, we compare different phase-curve fitting models to determine which model complexity is best suited for the upcoming data revolution in phase-curve studies, as future surveys are expected to provide data on 5-10 million objects, compared to the hundreds of thousands available today \citep{jones2015asteroid, eggl2019lsst}. In Section \ref{data}, we describe the catalog used. Section \ref{methods} presents and describes the applied methods.  In Section \ref{results} we show the results obtained along with the corresponding analysis. Finally, in Section \ref{conclusions}, we include the discussion and conclusions of this study.

\section{ATLAS Data}\label{data}

ATLAS is an asteroid impact early warning system designed primarily to detect dangerous NEAs \citep{Torny2018}. The ATLAS system comprises two telescopes situated in the Northern Hemisphere: one at Haleakala Observatory (ATLAS-HKO, Observatory code T05) and the other at Mauna Loa Observatory (ATLAS-MLO, Observatory code T08). Additionally, two telescopes are located in the Southern Hemisphere: one at Sutherland Observatory (ATLAS–SAAO, Observatory code M22) in South Africa and the other at El Sauce Observatory in Rio Hurtado, Chile (Observatory code W68). The two telescopes in the Northern Hemisphere began joint operations in 2017, while those in the Southern Hemisphere commenced joint operations in early 2022. The ATLAS Solar System Catalog (SSCAT) Version 1 was released on 2022-10-07 and has only observations from the Hawaiian telescopes. ATLAS Solar System Catalog Version 2 was uploaded on 2024-05-29 and has observations from all 4 telescopes. This second version also contains the observations of the first release, therefore covering from 2016 to early 2024; this is the version we use in this work. The ATLAS survey employs a custom filter system, including the cyan (\textit{c}, $420-650 \, \mathrm{nm}$) and orange (\textit{o}, $560-820 \, \mathrm{nm}$) filters, designed to optimize sensitivity for detecting transient events and near-Earth objects across a broad spectral range.

We extracted the modified Julian Date corrected for light travel time, the magnitudes and their uncertainties, 
the asteroid's heliocentric and topocentric distances [AU], and the phase angle from the ATLAS SSCAT\footnote{\url{http://astroportal.ifa.hawaii.edu/atlas/sscat/}}. Through a query to JPL Horizons using Astroquery \citep{astroquery2019}, we added the \texttt{ElongFlag} parameter to our database, which allows us to distinguish between different apparitions. The initial database has 596$\,$191 objects, with observations covering a range of phase angles from 0.6 to 96.32 degrees. From these objects, 587$\,$100 are Main-Belt asteroids, 4$\,$247 are Near Earth Asteroids (NEAs), 4$\,$681 are Jupiter Trojans, 163 are Transneptunian objects (TNOs) and centaurs (see Table \ref{database}). 

\begin{table}[ht]
\centering
\caption{Summary of the initial database of observed objects.}
\begin{tabular}{l r}
\hline
\textbf{Object Category} & \textbf{Count} \\
\hline
Main-Belt asteroids & 587\,100 \\
Near-Earth Asteroids (NEAs) & 4\,247 \\
Jupiter Trojans & 4\,681 \\
Transneptunian objects (TNOs) and centaurs & 163 \\
\hline
Total objects in the database & 596\,191 \\
\hline
\end{tabular}
\label{database}
\end{table}

\begin{figure}
    \centering
    \includegraphics[width=\columnwidth]{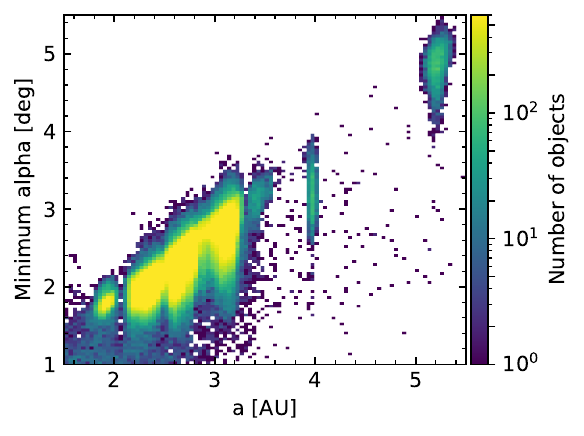}
    \caption{Observed relation between semimajor axis and minimum phase angle. Brighter colors indicate a higher number of asteroids observed in each bin.}
    \label{min_alpha}
\end{figure}

In our sample, 596$\,$191 asteroids have observations at phase angles less than 5 degrees. As shown in Figure \ref{min_alpha}, the minimum observed phase angle tends to increase with increasing semi-major axis. This implies that for the Trojan asteroid population, we lack information on the opposition effect. Conversely, NEAs have data at small phase angles, but only for a very limited number of objects. The population best sampled in the near-oppositional range is the Main Belt.

\section{Methods}\label{methods}

\subsection{Outliers treatment}
\begin{figure}
    \centering
    \includegraphics[width=\columnwidth]{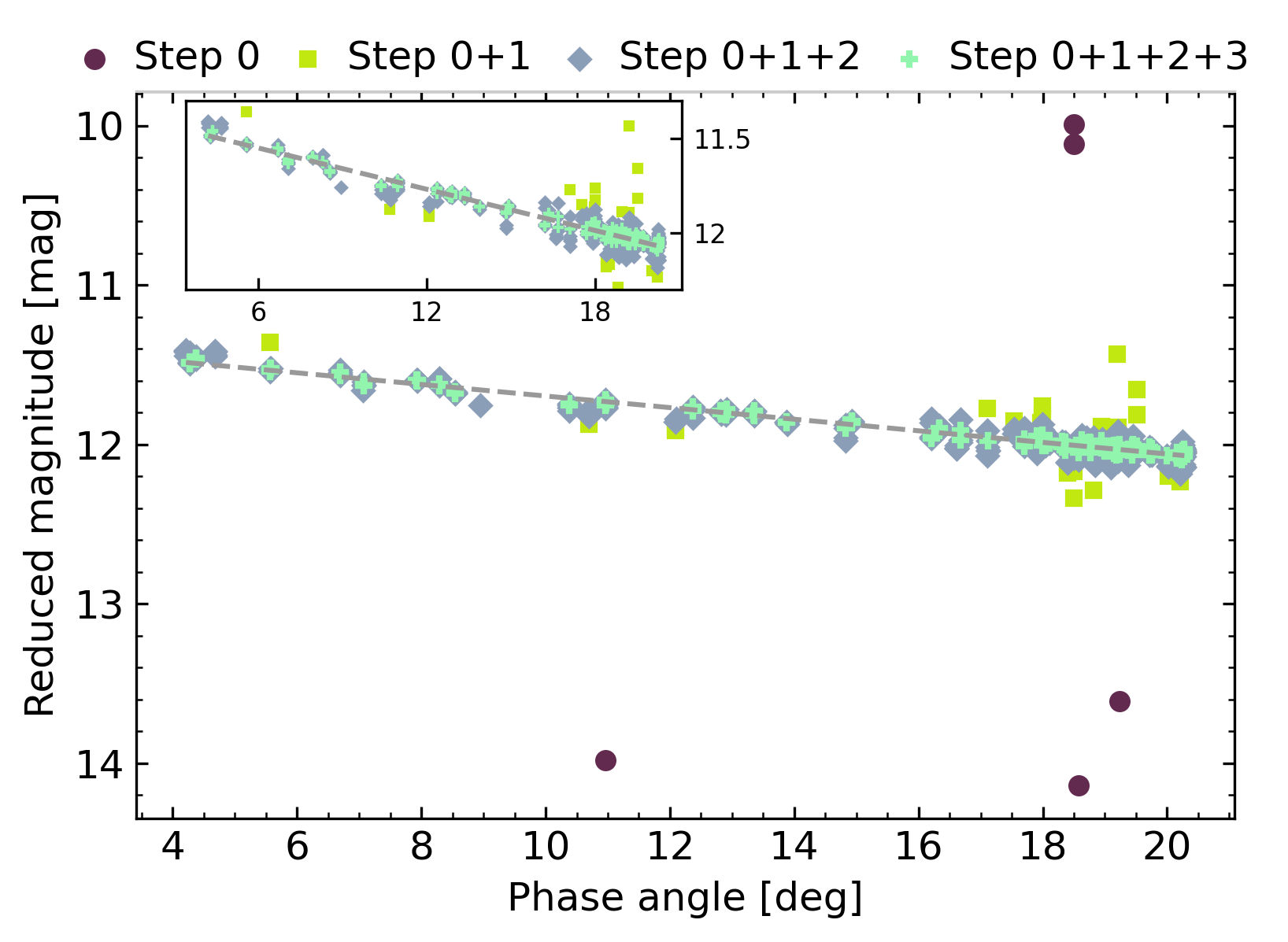}
    \caption{Outlier rejection steps for asteroid (1033) Simona. A zoomed-in view is included to highlight the dispersion amongst the data points. The purple circular points indicate observations that did not pass any steps. Green square points show those that passed the first step, light-blue diamond points represent those that passed the first two steps, and cyan plus points indicate observations that passed all steps. The gray line represents the linear-exponential function fitted during the outlier rejection process.}
    \label{outliers}
\end{figure}

Before processing we apply a three step outlier rejection procedure, an example of which is shown in Fig. \ref{outliers}. The dark purple points represent observations that did not pass any of these steps. The first step aims to remove points that lie far from the mean value of the reduced magnitude distribution, typically including observations affected by bad pixels, contamination from bright stars, and other artifacts. This step filters the data by retaining only values that do not deviate from the mean by more than three standard deviations, ensuring that only the most consistent and reliable data remain for further analysis. Points that meet this criterion are marked in green.

For the second step we used the same approach as \cite{hanus2011} and \cite{durech2020}. We fit a linear-exponential phase function to data and compute the root mean square (rms) residual. Then we removed all measurements that differed from the fitted phase function by more than 1.5 rms. After some tests and the visual inspection of several asteroids, this limit seemed to be a good threshold that removed most of the outliers. The points that passed the first two steps are light-blue in Figure \ref{outliers}.

For the third step, we apply one more rejection criterion: we consult the Light Curve Database \cite[LCDB]{warner2009} to obtain for each asteroid the reported maximum amplitude. For each phase angle, using 1-degree bins, we calculate the mean magnitude and reject those points that differ from the mean by a value greater than the maximum amplitude of the asteroid's light curve. The points that passed all the steps are cyan in Figure \ref{outliers}. In cases where the asteroid does not have a reported light curve amplitude in the LCDB, we decided to assign a value of 1 magnitude. Another option would be to use the maximum amplitude value in the database, corresponding to 3 magnitudes (as done by \citealt{robinson2024}). However, for the 31$\,$796 maximum amplitudes in the database, 95$\%$ of the objects have amplitudes less than 0.9 magnitudes. Therefore, for the sake of simplicity, we assign a value of 1. 

On average, about 8$\%$ of the data points are rejected during this process. These typically correspond to faint reduced magnitudes (around 16 mag) and occur at large phase angles ($>$20 degrees).

\subsection{Multi-apparition fitting}
\begin{figure}
    \centering
    \includegraphics[width=\columnwidth]{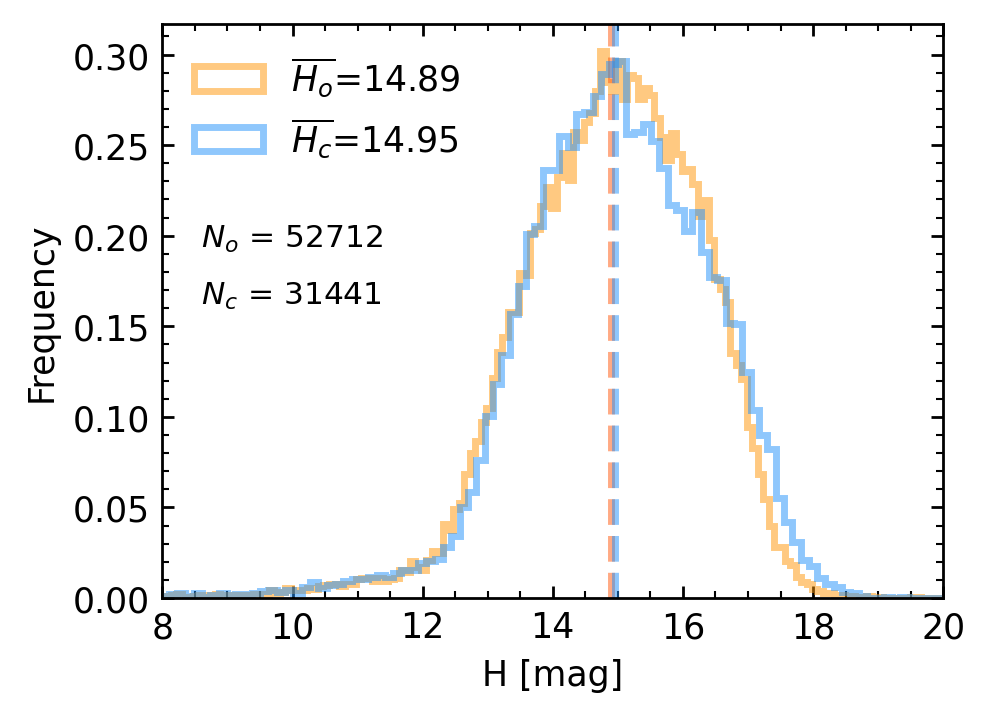}
    \caption{Distribution of absolute magnitudes in orange and cyan filters (denoted in orange and blue respectively) derived from ATLAS phase curves that meet the P16 conditions. Vertical lines represent the mean values of the distributions. The number of objects $N$ per filter is also indicated.}
    \label{Hmag}
\end{figure}

\begin{figure}
\centering
 \includegraphics[width=4cm]{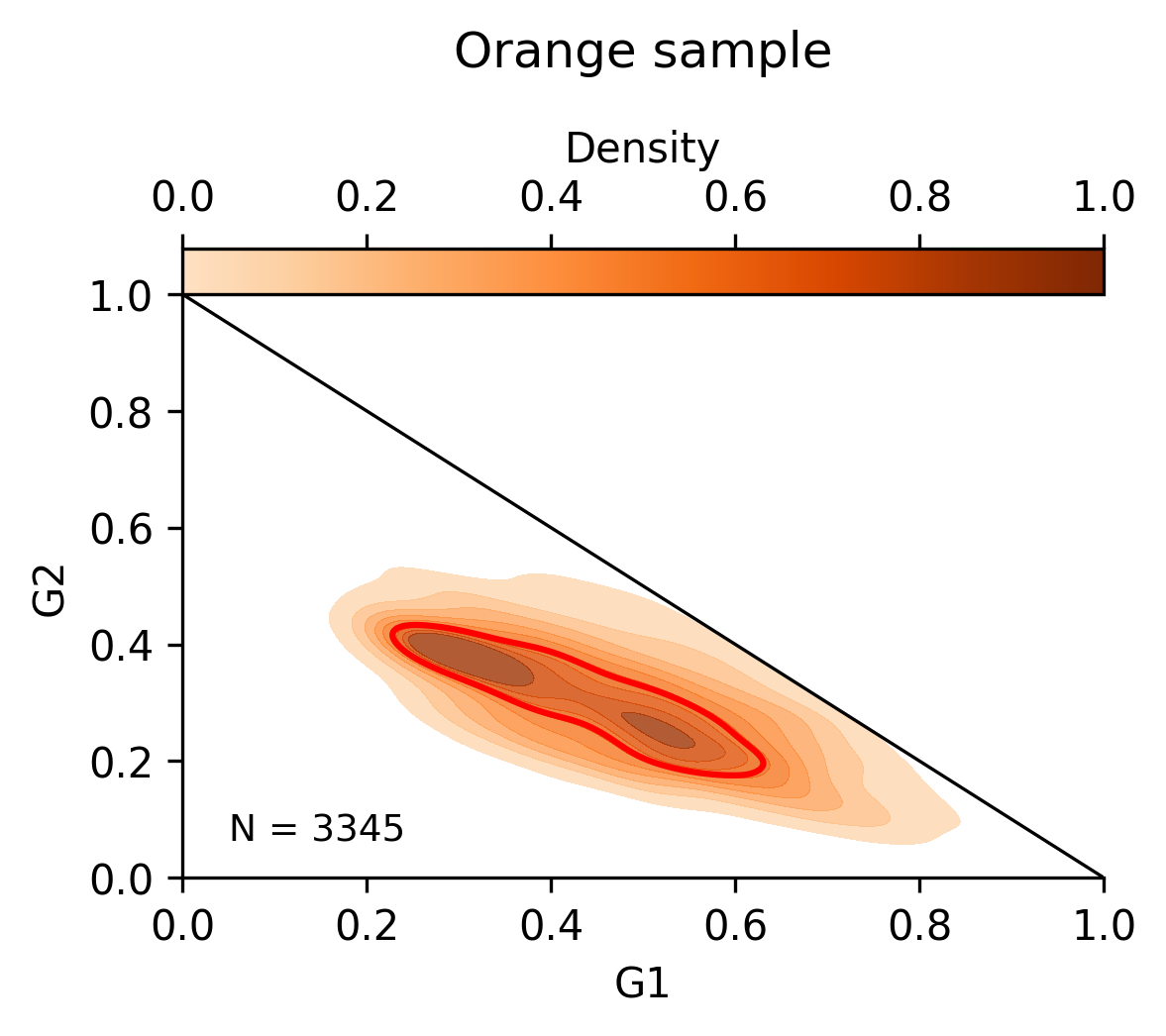}
 \includegraphics[width=4cm]{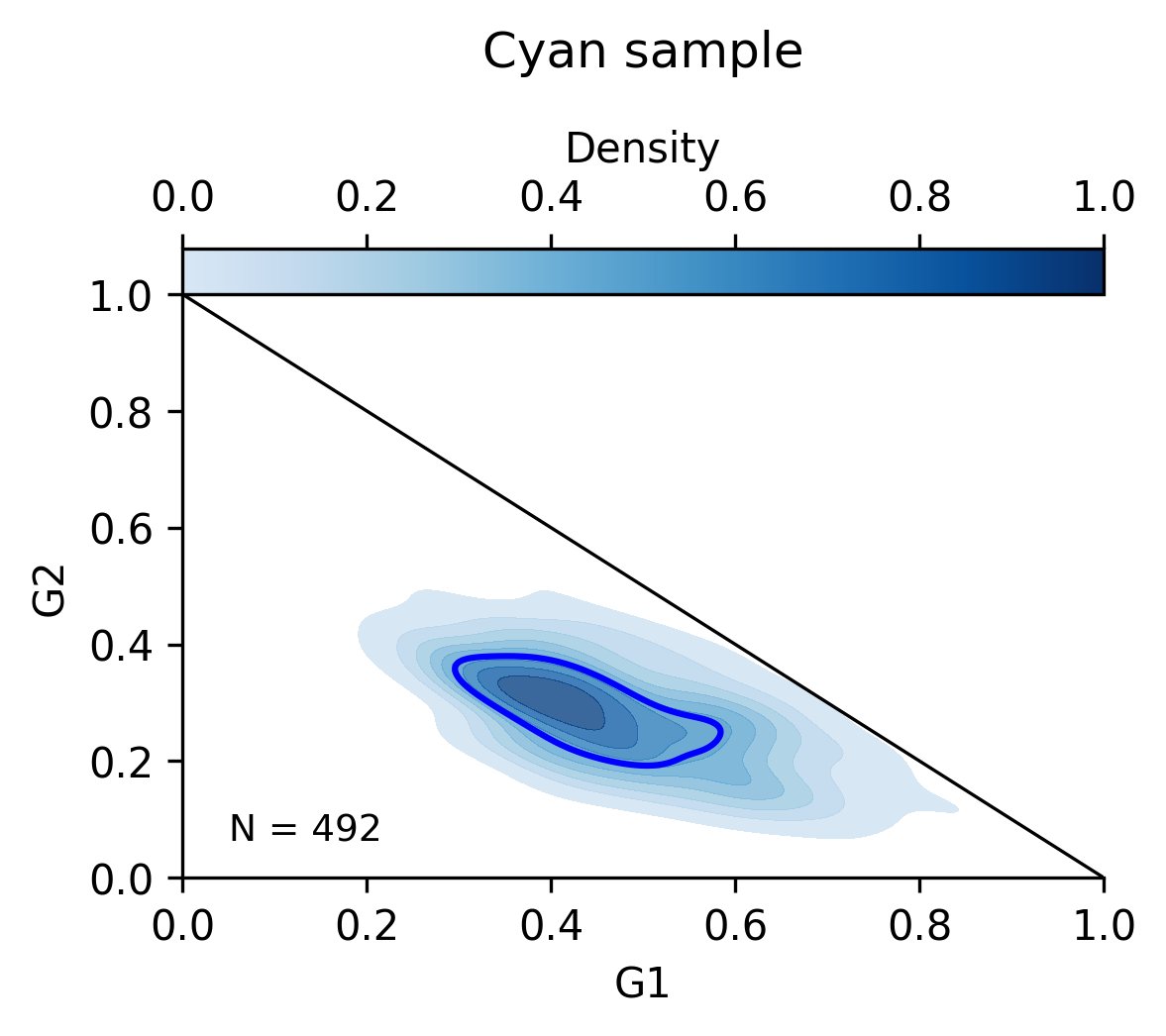}
 \includegraphics[width=4cm]{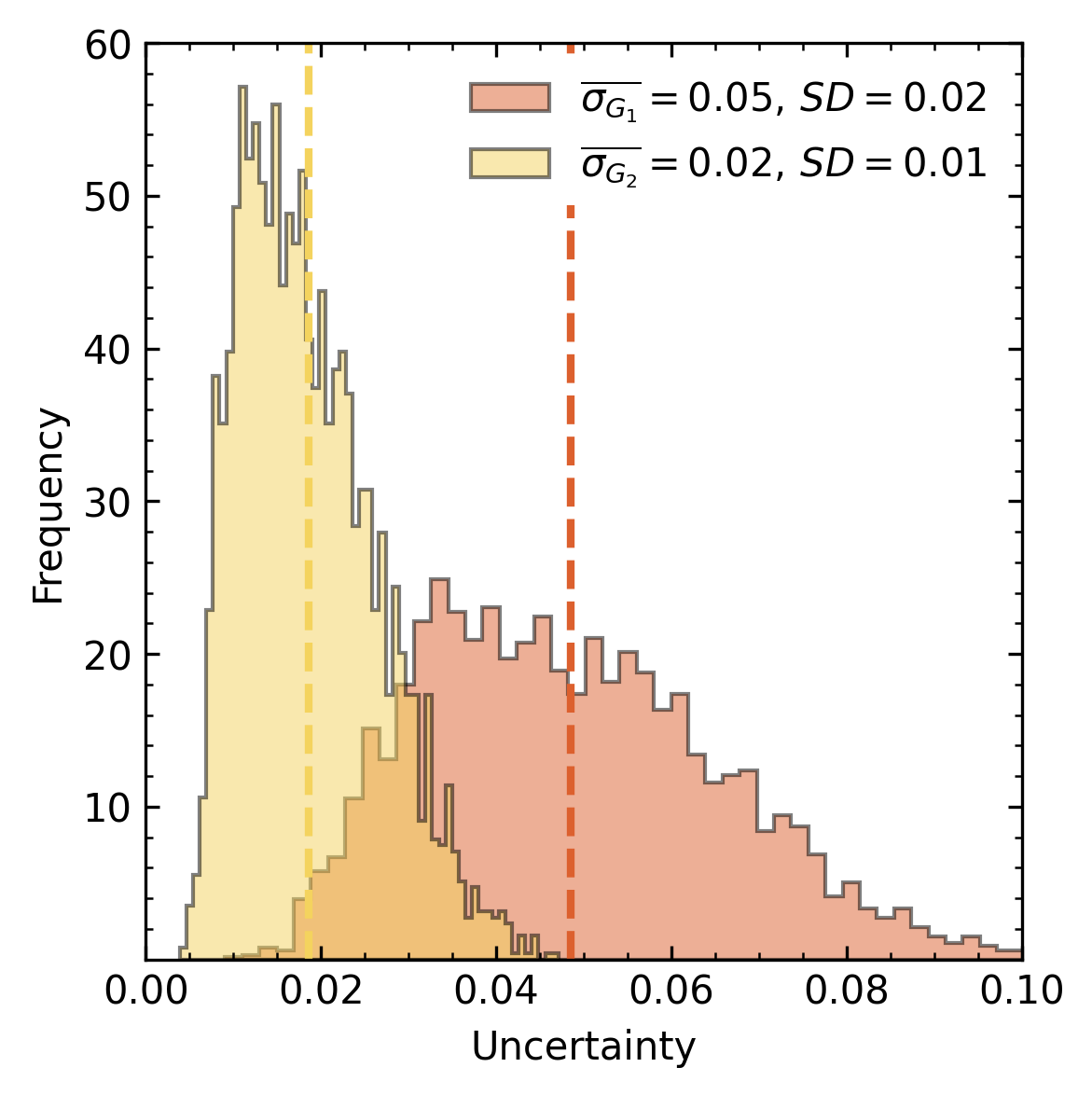}
 \includegraphics[width=4cm]{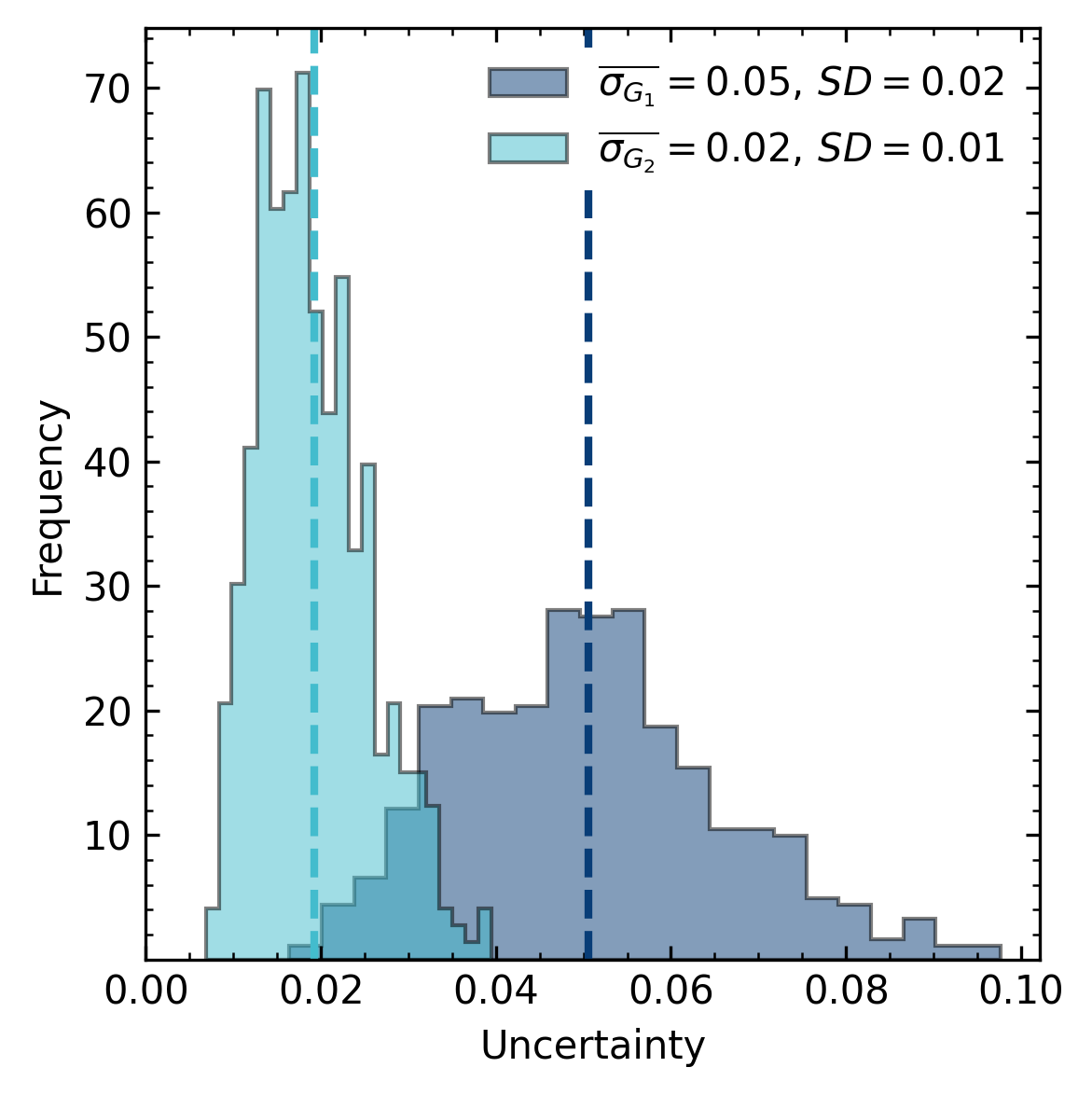}
\caption{Upper panel: G1 vs G2 parameters distributions
for the two ATLAS filters, generated using kernel density estimation (KDE), which smooths the data. The bold line represents the 1-$\sigma$ countour. Shaded regions indicate density levels, with darker regions representing higher concentrations of objects. The colorbar indicates the density values. Lower pannel: G1 and G2 uncertainty distributions. Dashed lines represent the mean values of the uncertainties in the parameters, indicated in the legend as  $\overline{\sigma_{G}}$. The legend also provides the standard deviation (SD) of the sample. The distributions are based on the most restrictive subset of data, limited to objects with an accuracy better than $15\%$ and fulfilling the criteria: $G_1, G_2 \ge 0$, and $1 - G_1 - G_2 \ge 0$. Colors in both panels correspond to the respective filters. }
\label{g1g2}
\end{figure}

\begin{figure*}
\centering
 \includegraphics[width=5cm]{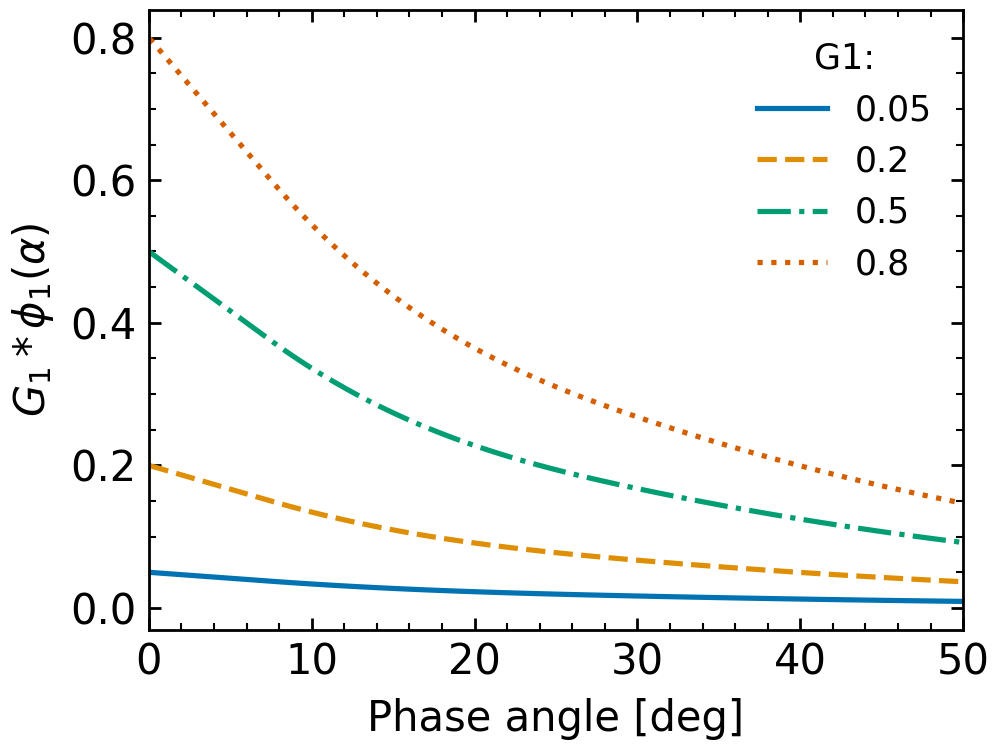}
 \includegraphics[width=5cm]{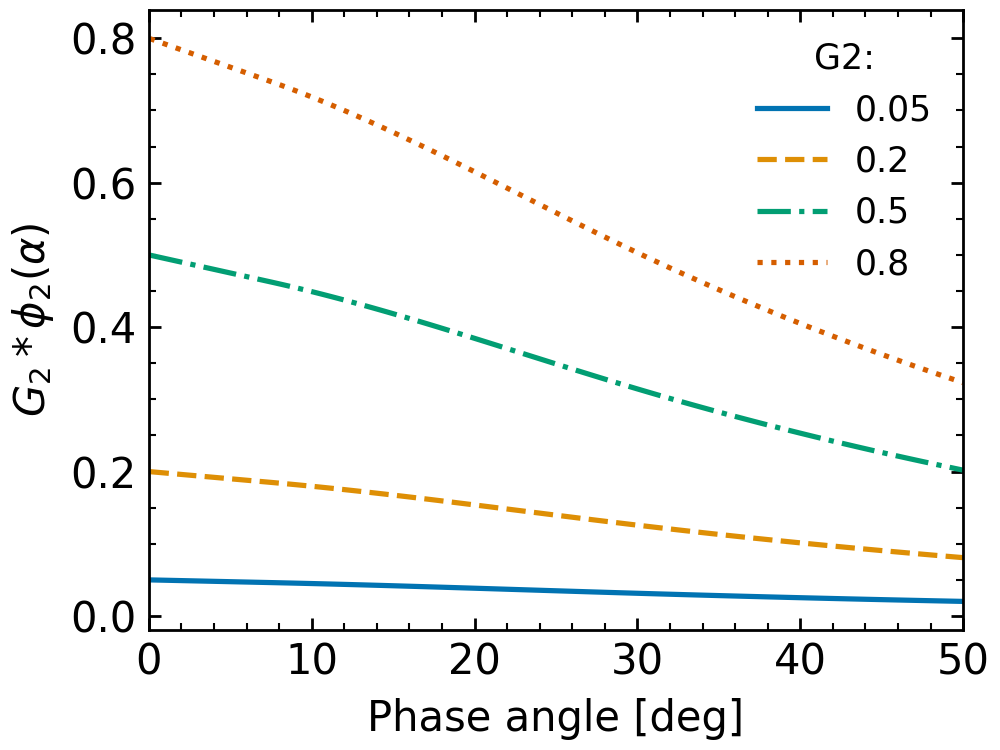}
  \includegraphics[width=5cm]{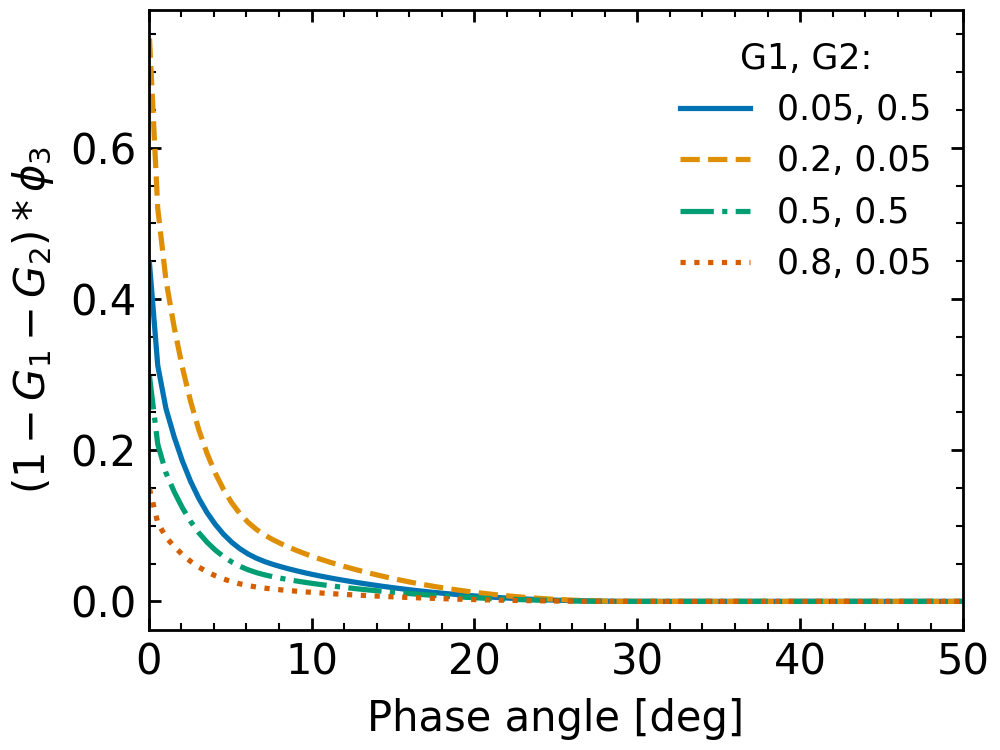}
\caption{Contribution of each term in equation \ref{HG1G2}}.
\label{basisfunc}
\end{figure*}

We utilize the H, G$_1$, G$_2$ phase function developed by \cite{muinonen2010three}, but in a multi-opposition fitting approach following the procedure of \cite{oszkiewicz2021first}. First, the magnitude data are converted to flux and divided into orange and cyan filters. Next, the data are split into different oppositions using solar elongation and phase angle information downloaded from the JPL Horizons system\footnote{\url{https://ssd.jpl.nasa.gov/horizons/app.html/}} using Python \texttt{astroquery} package. For each filter, the data from various oppositions are fitted simultaneously, using the same G$_1$ and G$_2$ parameters but different absolute magnitudes H$_i$ for each apparition. The fitting process is nonlinear and performed in the flux domain. The flux is obtained from
\begin{eqnarray}
10^{-0.4V(\alpha)}
                & = & \sum_{i=1}^{N} 10^{-0.4H_i}\big{[}G_1\phi_1(\alpha) \nonumber \\
                & + & G_2\phi_2(\alpha) \nonumber \\
                & + & (1 - G_1 - G_2)\phi_3(\alpha)\big{]} 
                \label{HG1G2},
\end{eqnarray}
\noindent
where G$_1$, G$_2$, and H$_i$ are the fitted parameters (phase function parameters and absolute magnitude for each opposition). The number of fitted parameters varies from object to object depending on number of available apparitions. {On average, each object has approximately six apparitions, with this value remaining relatively consistent across NEOs, MBAs, and Trojans. In contrast, TNOs typically exhibit fewer apparitions due to their much greater heliocentric distances. N is the number of apparitions, $\alpha$ is the phase angle, $V(\alpha)$ is the reduced magnitude and $\phi_1$, $\phi_2$, $\phi_3$ are basis functions defined in terms of cubic splines \citep{muinonen2010three}. To perform the fitting, we utilized the \texttt{least\_squares} method from the \texttt{scipy} package \citep{scipy2020}, which allows for a varied number of fitted parameters and data structured with varying lengths of points in each opposition.
We used the Levenberg-Marquardt algorithm and the \texttt{ftol} tolerance for termination set to 10$^{-8}$, which is the default value. However, since the input data are fluxes on the order of 10$^{-7}$, we must adjust the gradient tolerance parameter \texttt{gtol} to 10$^{-15}$. To determine the appropriate \texttt{gtol} value, we check at which value the algorithm converges to a solution for the parameters. Parameter uncertainties are estimated directly from the covariance matrix.

The resulting solutions may be constrained to physical solutions fulfilling the criteria: \( G_1, G_2 \ge 0 \), and \( 1 - G_1 - G_2 \ge 0 \) \citep{penttila2016h}. The \texttt{least\_squares} method allows us to set the boundary conditions of the problem beforehand. However, we observed that including these constraints before the fit generates a large number of solutions at the extremes of the range \citep[see also][]{mahlke2021,robinson2024}. These extreme values could be caused by inadequate phase angle coverage, low SNR observations, and/or large rotational variations. Therefore, we decided to perform the fit without imposing initial constraints and, for the subsequent analysis, we will use the sample that respects the conditions set by \cite{penttila2016h}. 

We did not correct for light-curve amplitude or geometry, which may introduce potential parameter biases, or misinterpretations of physical properties and erronous taxonomic classifications. To address this, we cross-validate the parameters derived using this method with those obtained from other methods in Section \ref{complex-simple}. Overall we found that that the derived parameters are consistent between the different methods. Note that additional challenges may arise for Near-Earth objects, where rapidly changing geometry, particularly the aspect angle, can lead to complex variations in brightness. Consequently, corrections for light-curve amplitude and geometry are crucial for these objects. However, most of the objects are from the main belt with slowly changing aspects and a high number of points, averaging 205 data points in orange and 80 in cyan. The resulting phase curve parameters are well-averaged over oppositions and the number of data points, as further discussed in the following sections.

\section{Results}\label{results}

\subsection{General results}

We have determined phase curves for 301$\,$272 objects in the orange pass band (o) and 280$\,$953 in cyan (c), 3$\,$345 (o, 1$\%$) and 492 (c, 0.18$\%$) with accuracy better than 15$\%$. This represents a balance between maintaining high-quality data and maximizing the number of objects available for analysis, and it is applied across all parameters. The subsequent analysis is based on objects whose parameters fall within the limits established by \cite[P16]{penttila2016h}: $G_1 > 0$, $G_2 > 0$, and $1 - G_1 - G_2 > 0$. This additional constraint reduces the number of objects to 52$\,$712 (o) and 31$\,$441 (c), ensuring that the selected parameters are physically consistent. In specific parts of the analysis, where higher precision was crucial, we used only the most restrictive subset of data —those with an accuracy better than 15$\%$. In Figure \ref{Hmag} and \ref{g1g2}, we present the results obtained for the H, G1 and G2 parameters in the orange and cyan filters. The number of objects with determined parameters in the cyan filter is smaller than in the orange filter. The cyan filter has fewer observations with \textbf{considerable dispersion}, which compromises the fitting process. The fitting procedure failed for 7$\%$ of the orange sample and 13$\%$ of the cyan sample. In Figure \ref{g1g2}, as well as in other figures throughout the paper, we use Kernel Density Estimate (KDE) plots to visualize the distribution of the G1 and G2 parameters. KDE is a non-parametric technique that estimates the probability density function of a dataset by smoothing the underlying distribution. This is achieved by placing a kernel on each data point and summing these to produce a continuous density estimate. For our plots, we used the default Gaussian kernel provided by \texttt{seaborn.kdeplot}.

As we can see in the top panels of Figure \ref{g1g2}, there is a clustering towards low G1 and medium G2 for both filters. This clustering can be explained by the predominance of S-type asteroids in the inner and middle Main Belt in terms of absolute numbers \citep{demeo2013}, which is also reflected in our sample, as shown in Figure \ref{taxas1}. This distribution was also noted by \cite{mahlke2021}. 

Figure \ref{g1g2} also includes histograms of the uncertainties in G1 and G2 for the well-determined parameters. The uncertainties in G2 are generally smaller than those in G1, as can be seen when comparing the mean values of the distributions. Additionally, the distribution of G1 uncertainties is consistently wider than that of G2. To investigate this behavior further, we analyze how the coverage of phase angle observations affects the G1 and G2 parameters of the phase function. To do so, we first examine the form taken by each term of Equation \ref{HG1G2} (see Figure \ref{basisfunc}), which depends solely on the phase angle and is thus unaffected by the fitting procedure, i.e., they are given values. Note, the highly non-linear contribution of the G1$\phi (\alpha)$ at small phase angles (particularly for the large values of G1) as compared to the G2$\phi (\alpha)$.

Next, we simulate phase curves with different phase angle coverage. We use seven different ranges: (0.1, 35), (0.1, 8), (0.1, 15), (2, 15), (2, 30), (4.3, 20), (5, 25), (10, 25). These ranges were chosen to explore how varying the span of phase angles affects the derived parameters. We generated phase angles using uniform coverage within each range. The magnitudes were simulated with random noise (standard deviation of 0.1) from two functions: a linear-exponential function and the HG function. For the HG model, we used an absolute magnitude H = 10 and a slope parameter G = 0.15. For the linear-exponential model, the coefficients were: absolute magnitude = 12, phase coefficient = 0.03, opposition amplitude = -1, and opposition slope = 0.2. Then, we performed two fits: 1) We leave G2 fixed and fit only G1 and H 2) We leave G1 fixed and fit only G2 and H. The fixed values used are: 0.05, 0.1, 0.3, 0.5, 0.6, 0.8 for both cases.

Figure \ref{simulations} shows the results obtained. The value of G1 and its uncertainty is greatly affected when there are no observations close to the opposition (especially for phase curves with large G1 values (e.g. low albedo objects). This is in agreement with what we show in Figure \ref{basisfunc}, where it is clear that the term accompanying G1 in Equation \ref{HG1G2} models the behavior for small phase angles. This finding is consistent with the results reported by \cite{mahlke2021}.

On the other hand, the value of G2 varies moderately over the different phase angles covered. Although this variation is smaller than that observed for G1, it is still noticeable given that G2 ideally spans between 0 and 1. Furthermore, the lower (and non-physical) values obtained with when fitting to data generated from} the linear-exp function suggest that this test model tends to produce flatter phase curves. Even so, the uncertainties show minimal fluctuation, highlighting its consistency despite changes in the parameter. The term associated with this parameter is more relevant to the linear behavior of the phase curve.

We also calculate the coefficient of determination (R$^2$) as a measure of the goodness of fit. The mean R$^2$ for the fit using data generated with the HG function is 0.89, while for the linear-exp data is 0.63, indicating that the HG1G2 model achieves a better fit when applied to data generated using the HG function compared to data generated with the linear-exp function.

\begin{figure*}
\centering
 \includegraphics[width=8cm]{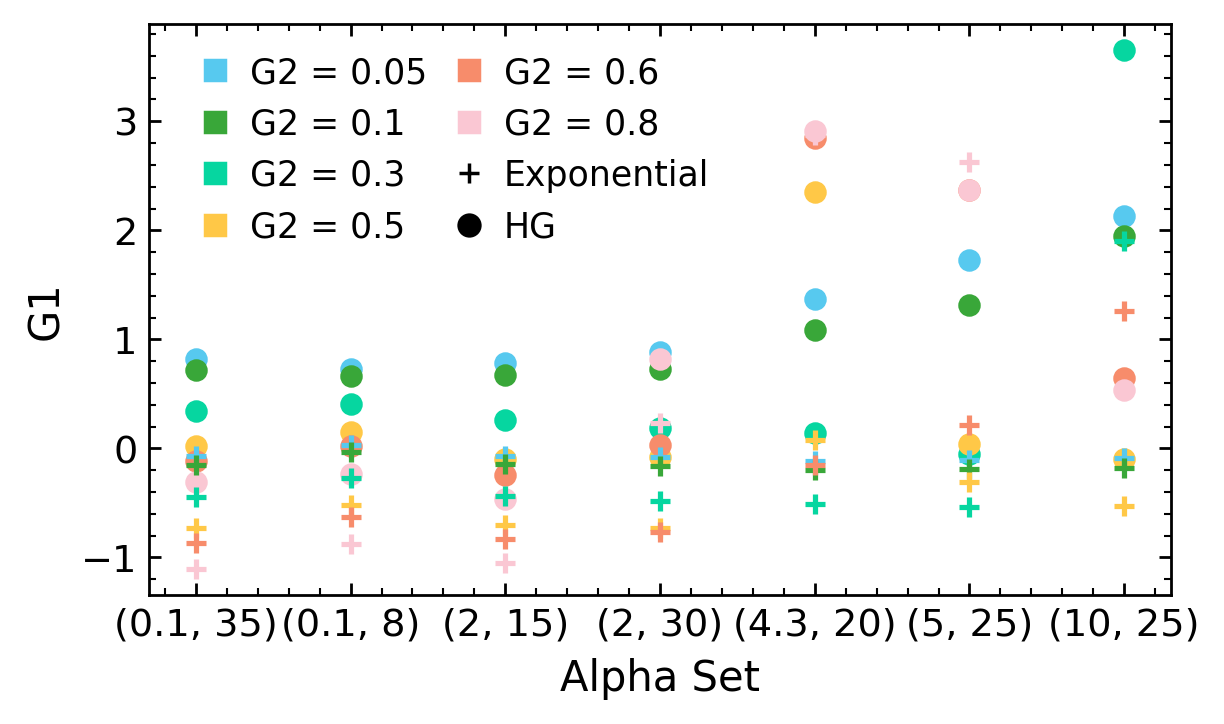}
 \includegraphics[width=8cm]{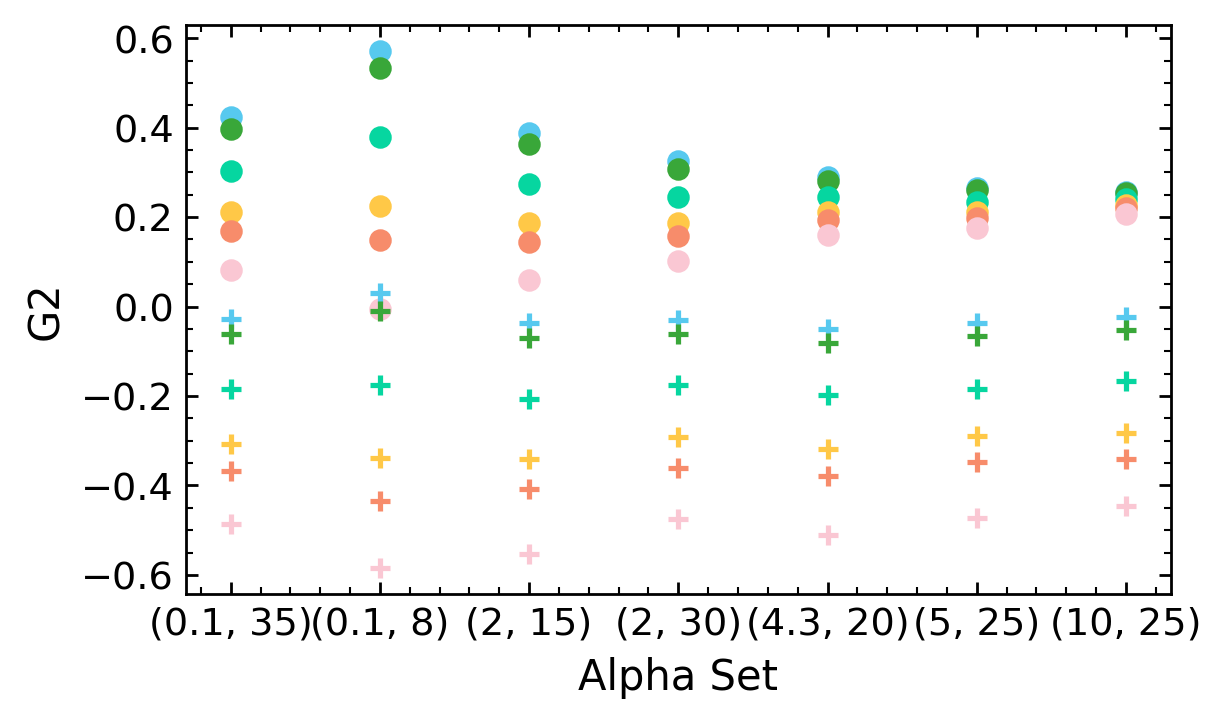}
  \includegraphics[width=8cm]{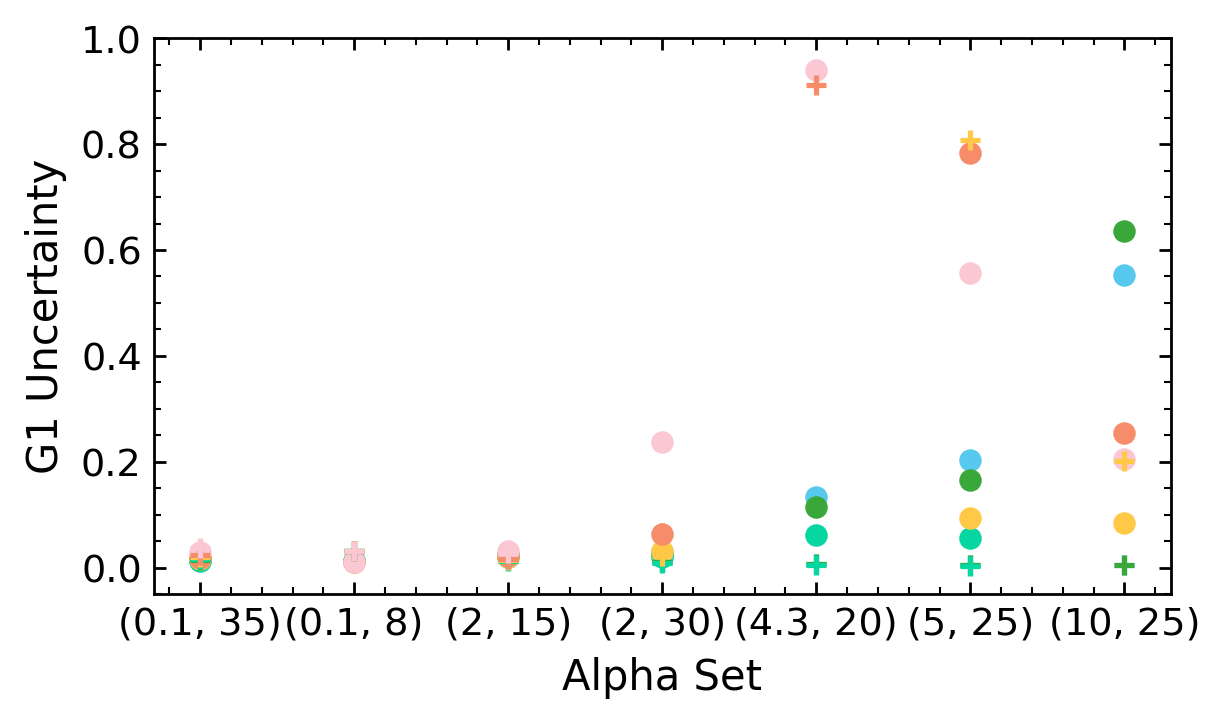}
 \includegraphics[width=8cm]{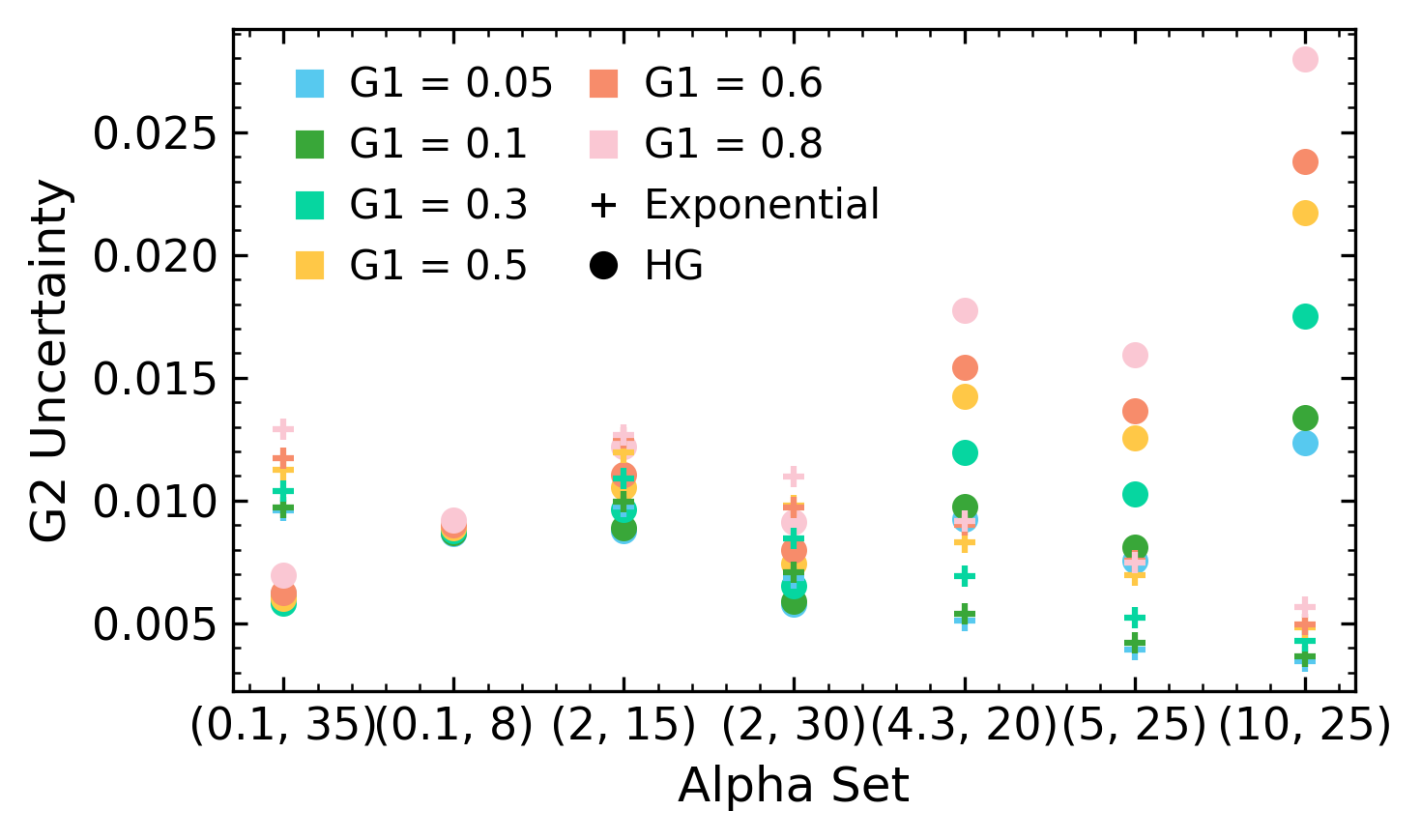}
\caption{The G1 and G2 parameters and their uncertainties derived from the simulated data.}
\label{simulations}
\end{figure*}

\subsection{ATLAS main results for taxonomic types}

\begin{figure*}[ht]
\centering
\includegraphics[width=4cm]{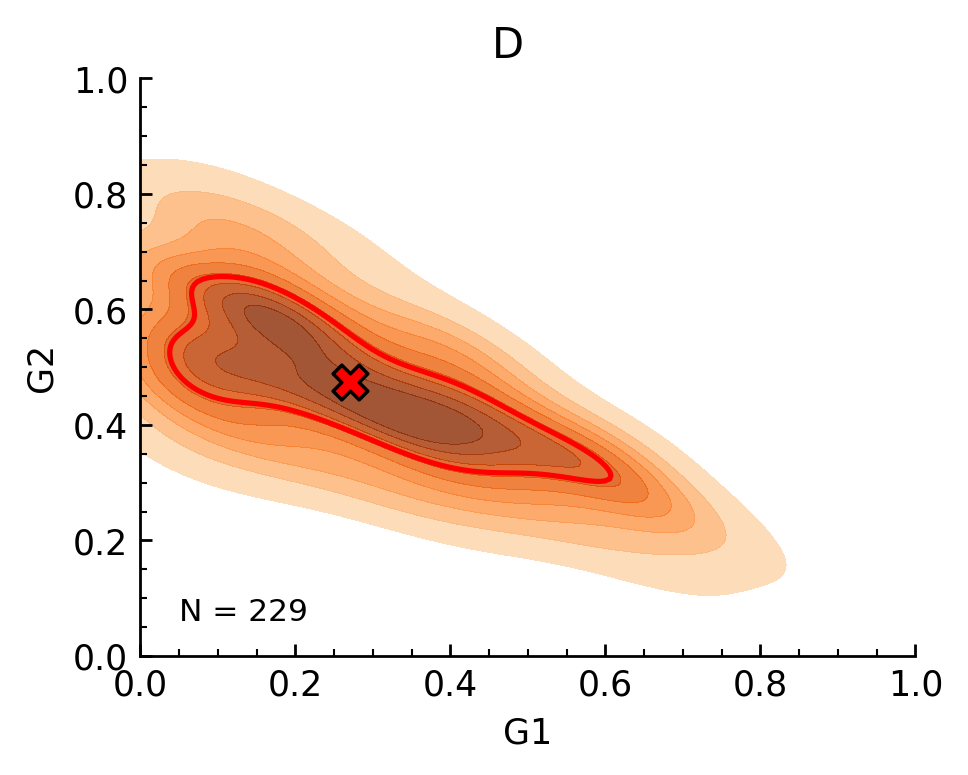}
\includegraphics[width=4cm]{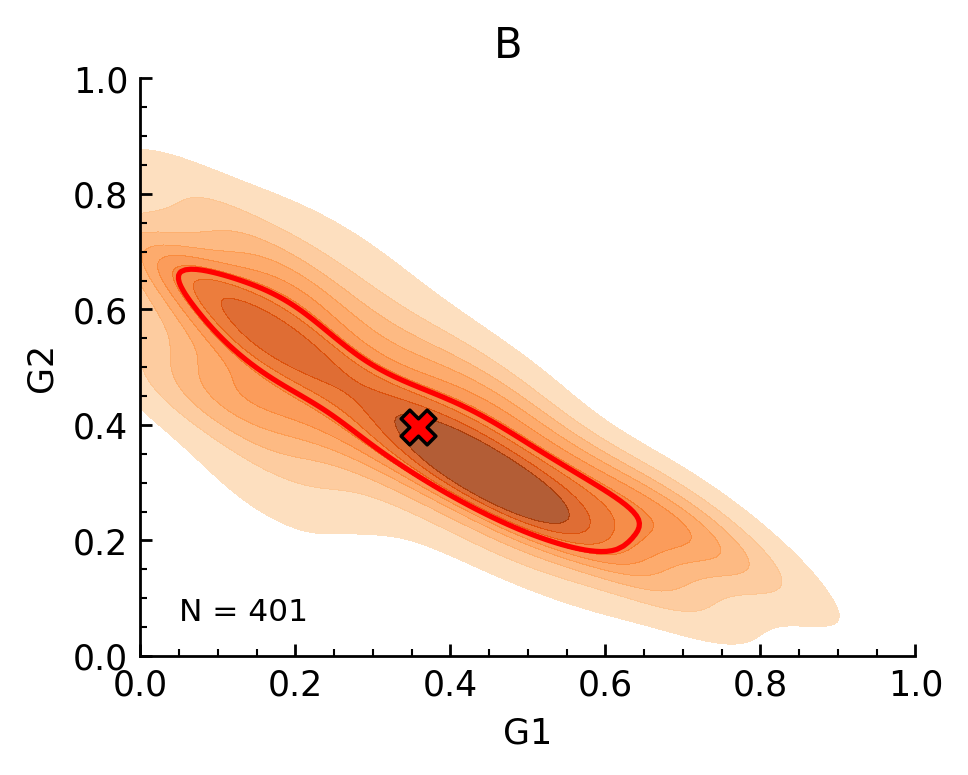}
\includegraphics[width=4cm]{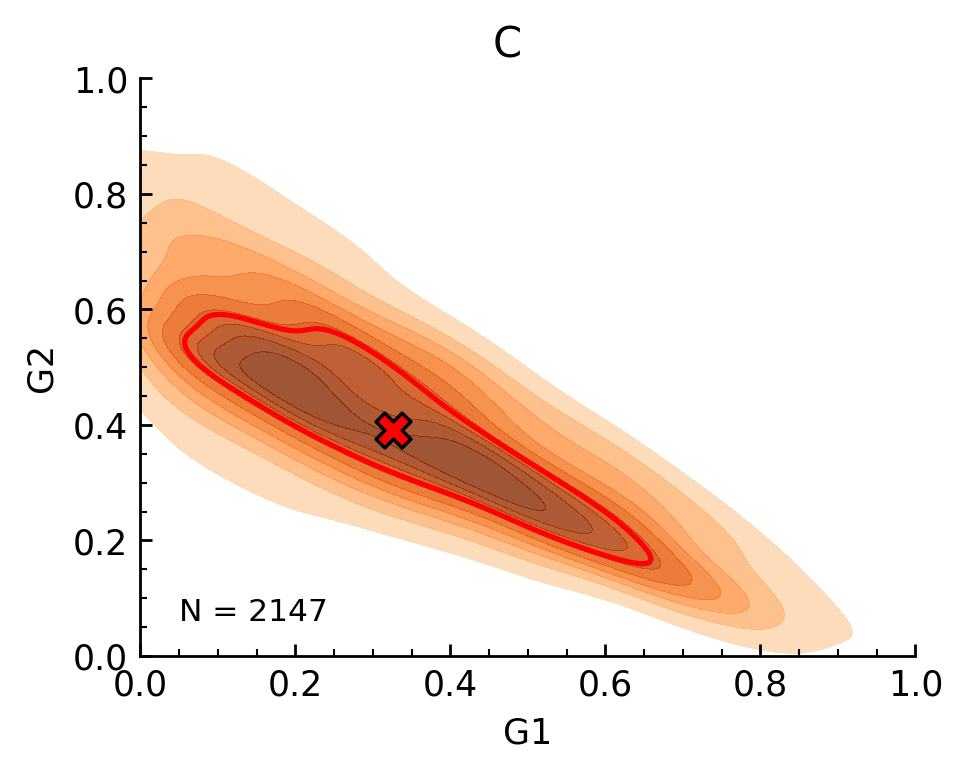}
\includegraphics[width=4cm]{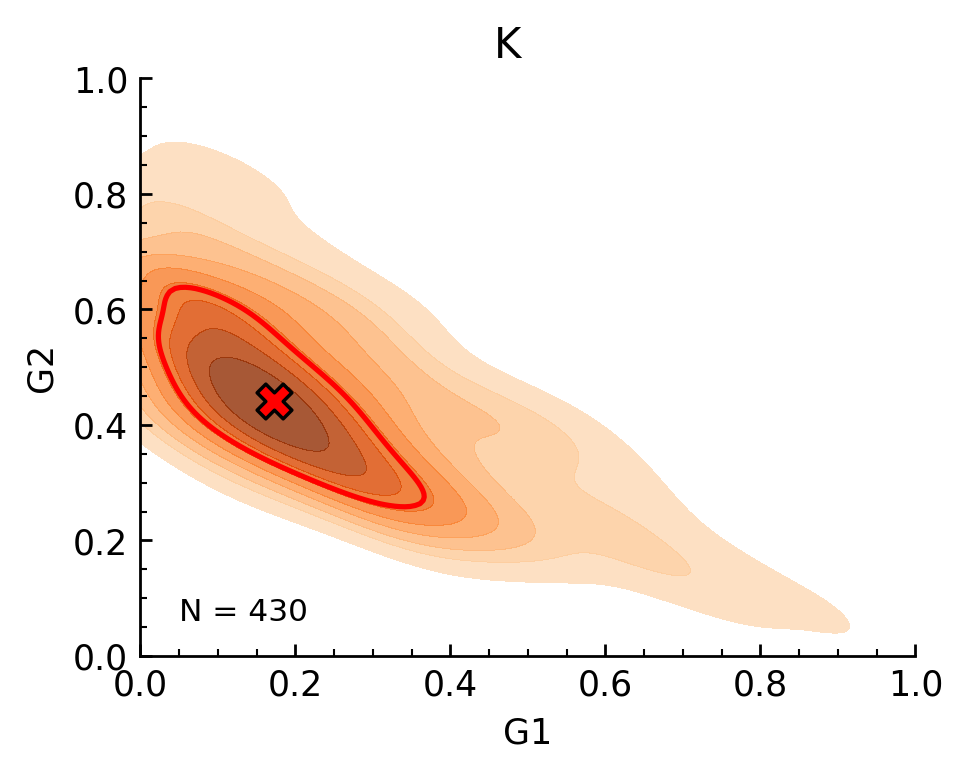}
\includegraphics[width=4cm]{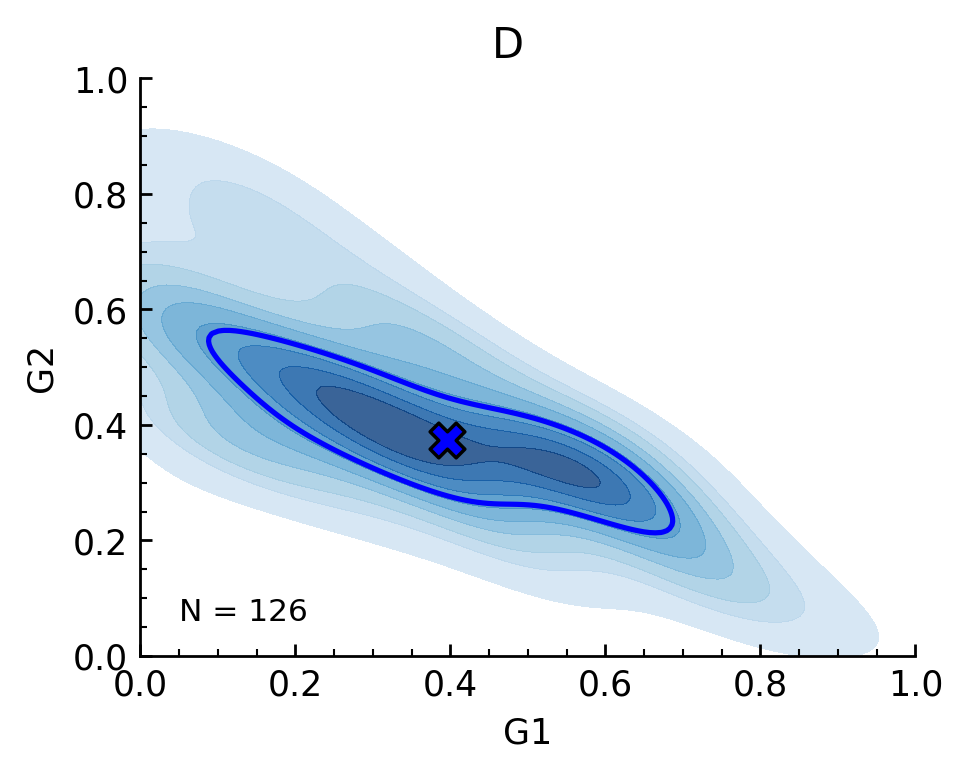}
\includegraphics[width=4cm]{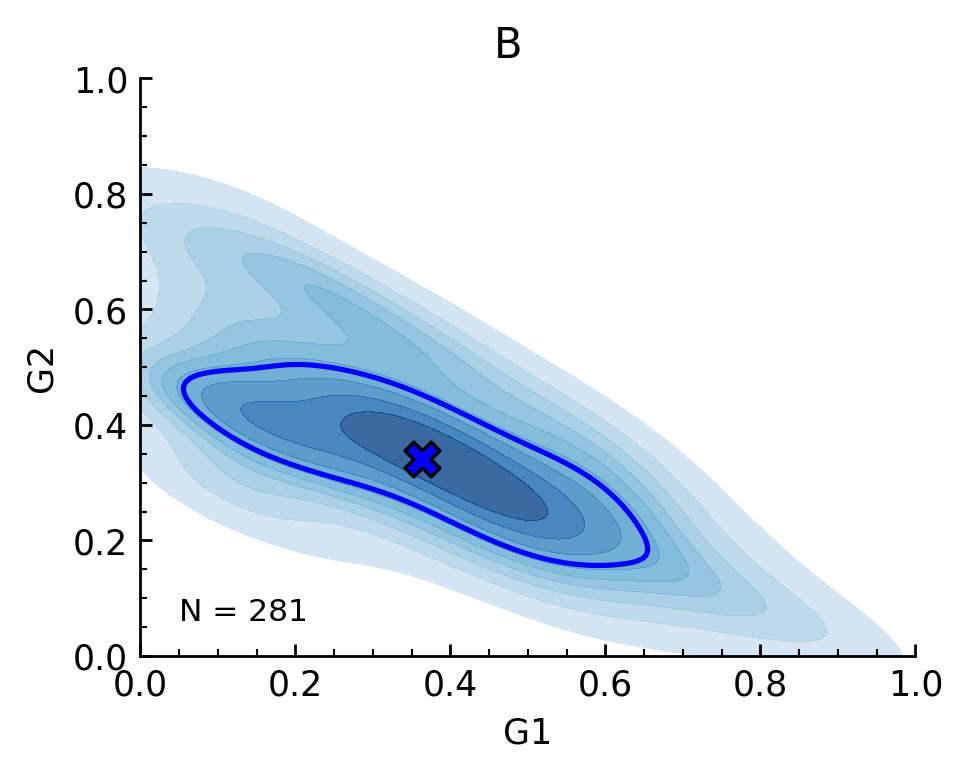}
\includegraphics[width=4cm]{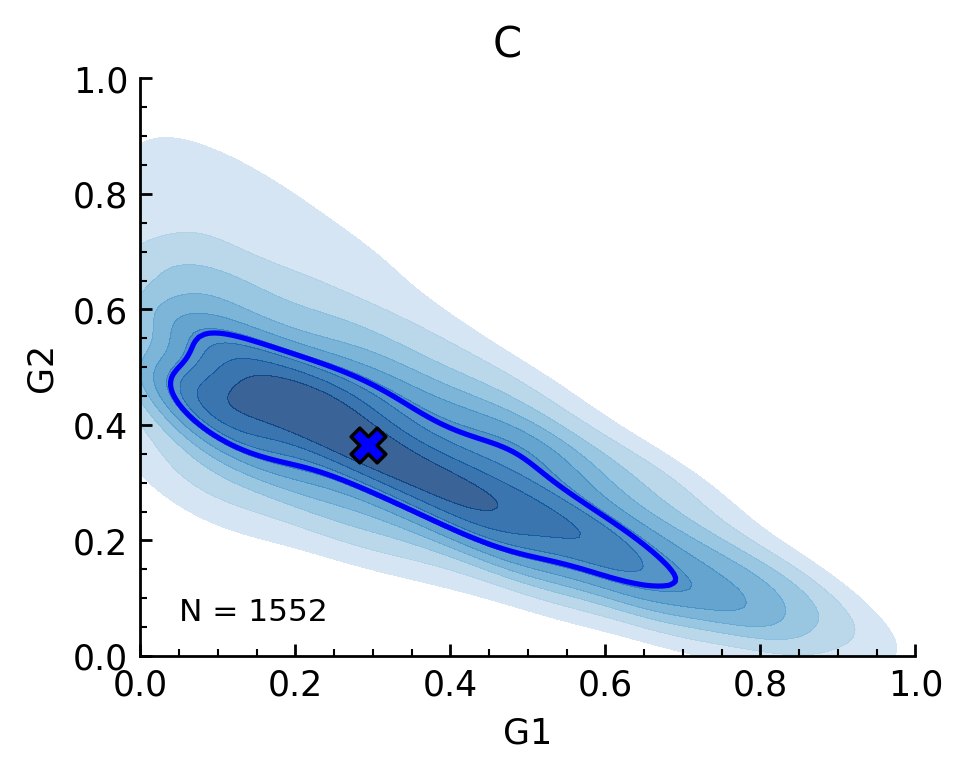}
\includegraphics[width=4cm]{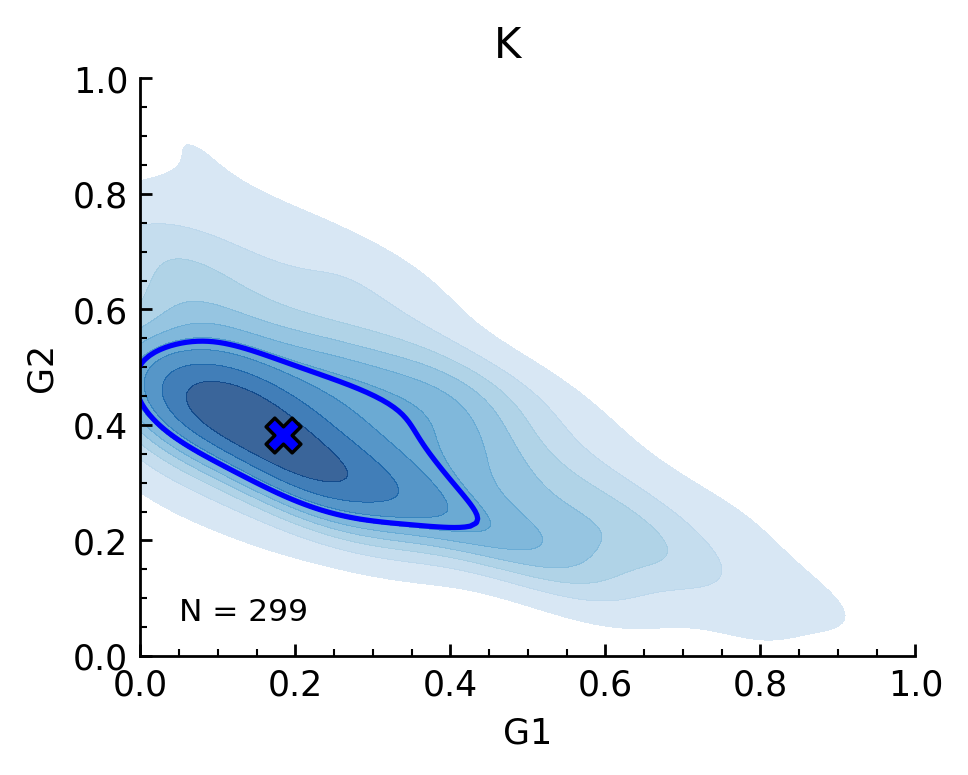}

\includegraphics[width=4cm]{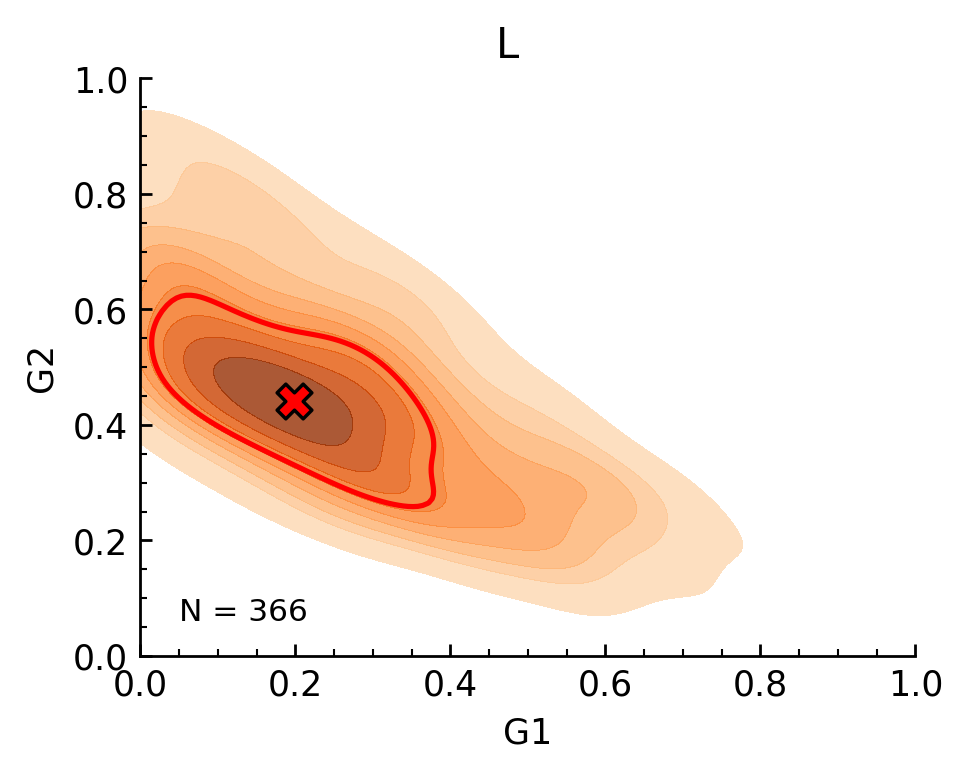}
\includegraphics[width=4cm]{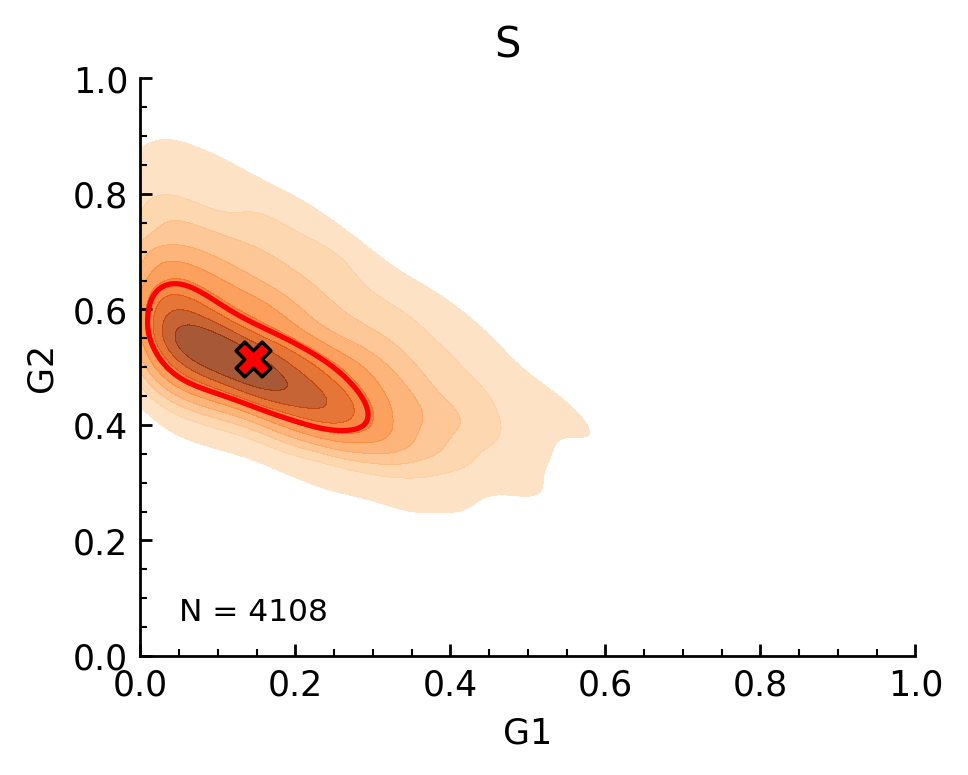}
\includegraphics[width=4cm]{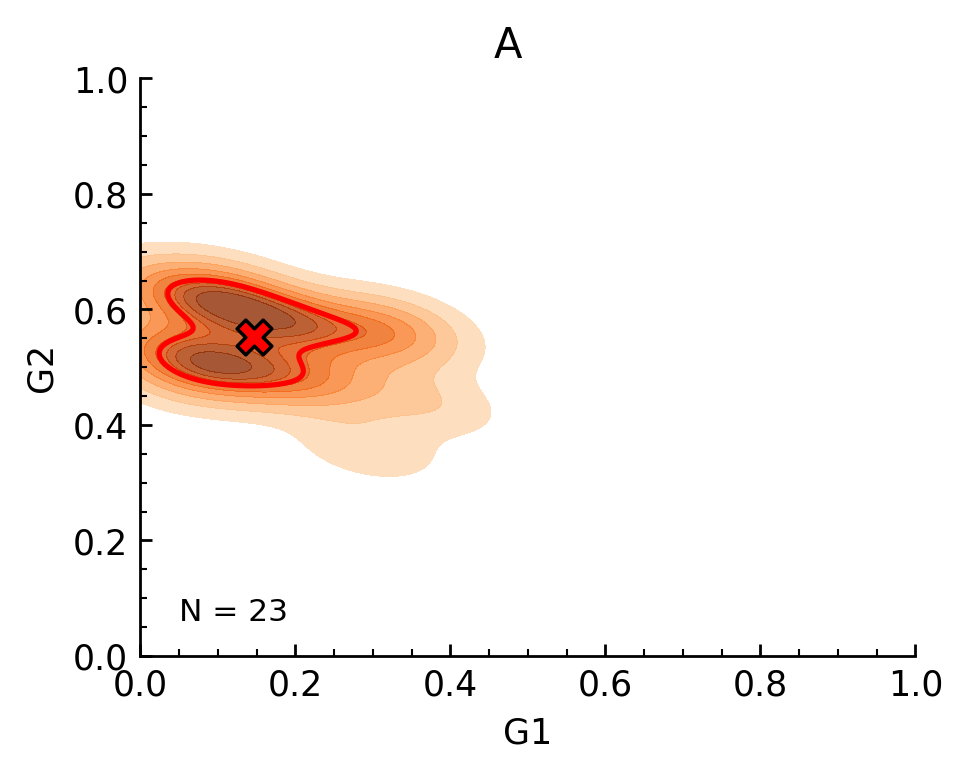}
\includegraphics[width=4cm]{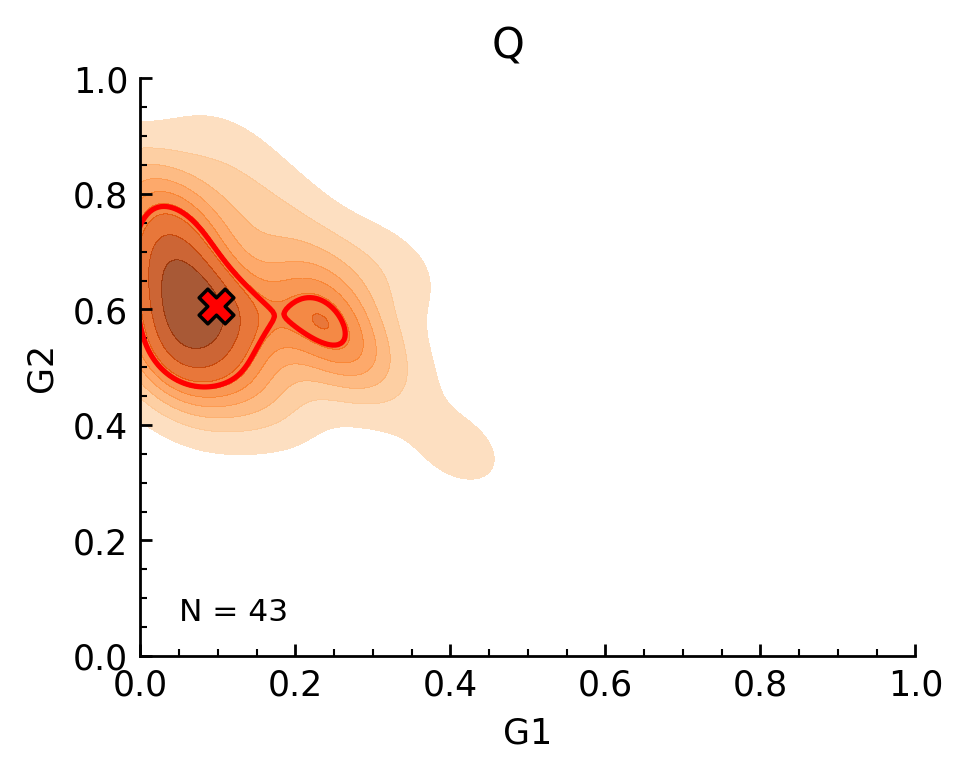}
\includegraphics[width=4cm]{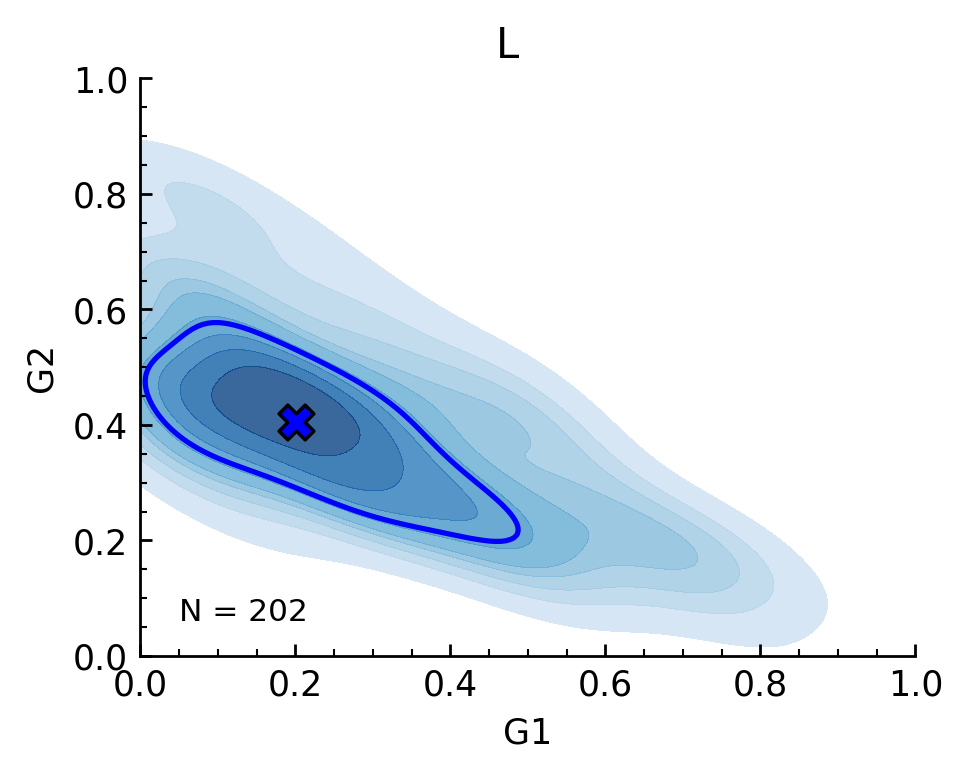}
\includegraphics[width=4cm]{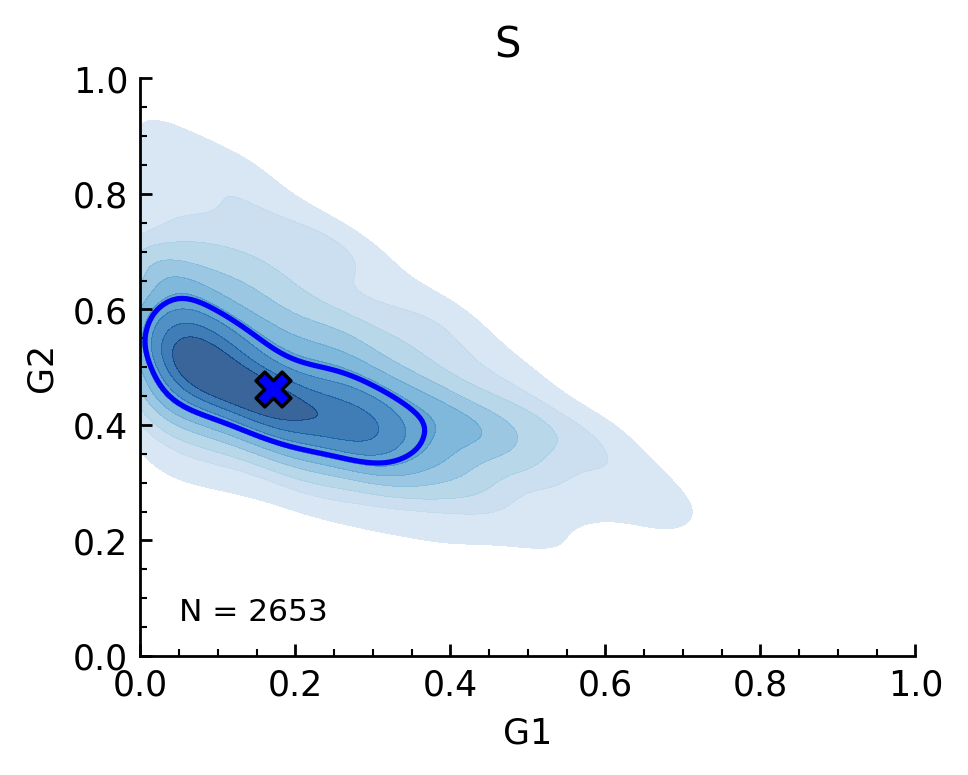}
\includegraphics[width=4cm]{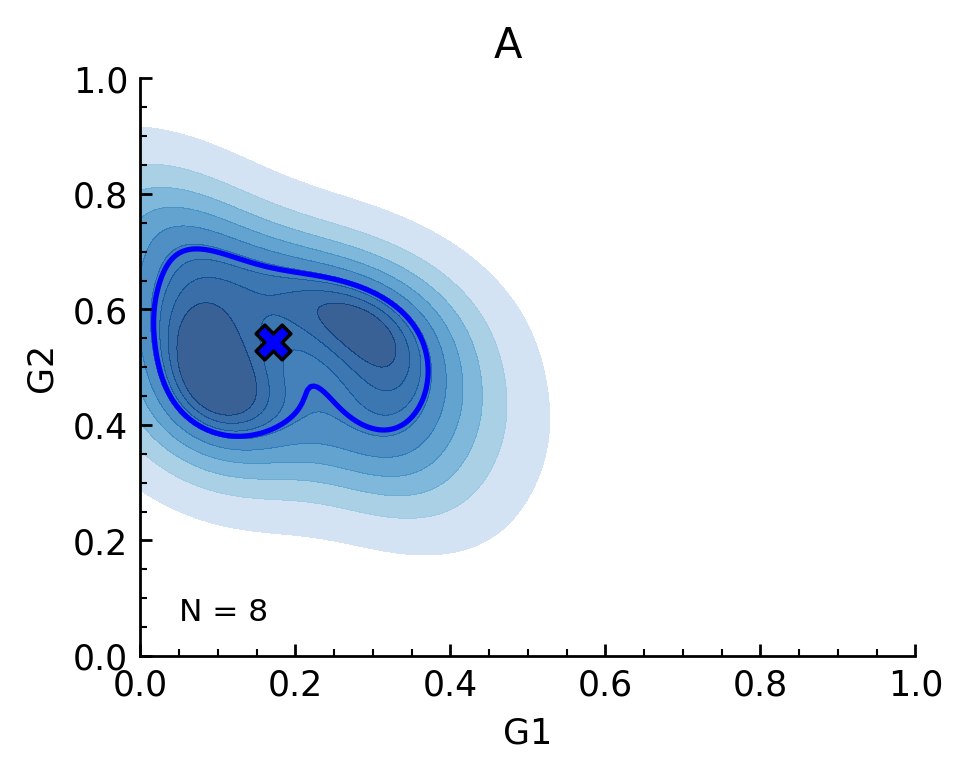}
\includegraphics[width=4cm]{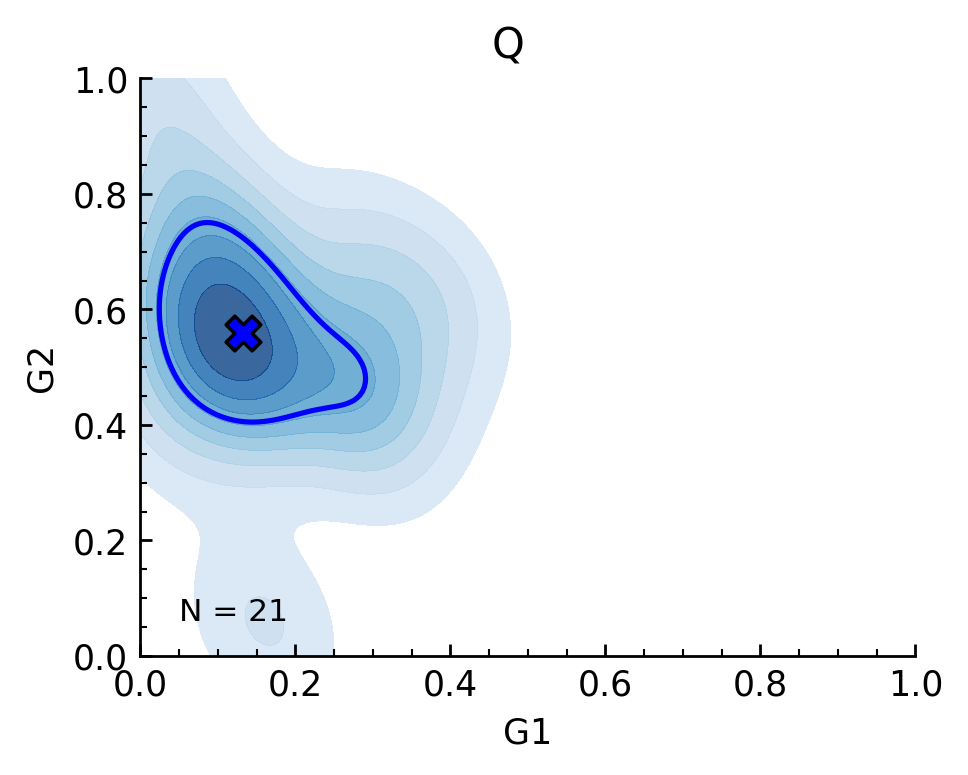}

    \begin{minipage}{0.3\textwidth}
        \centering
        \includegraphics[width=4cm]{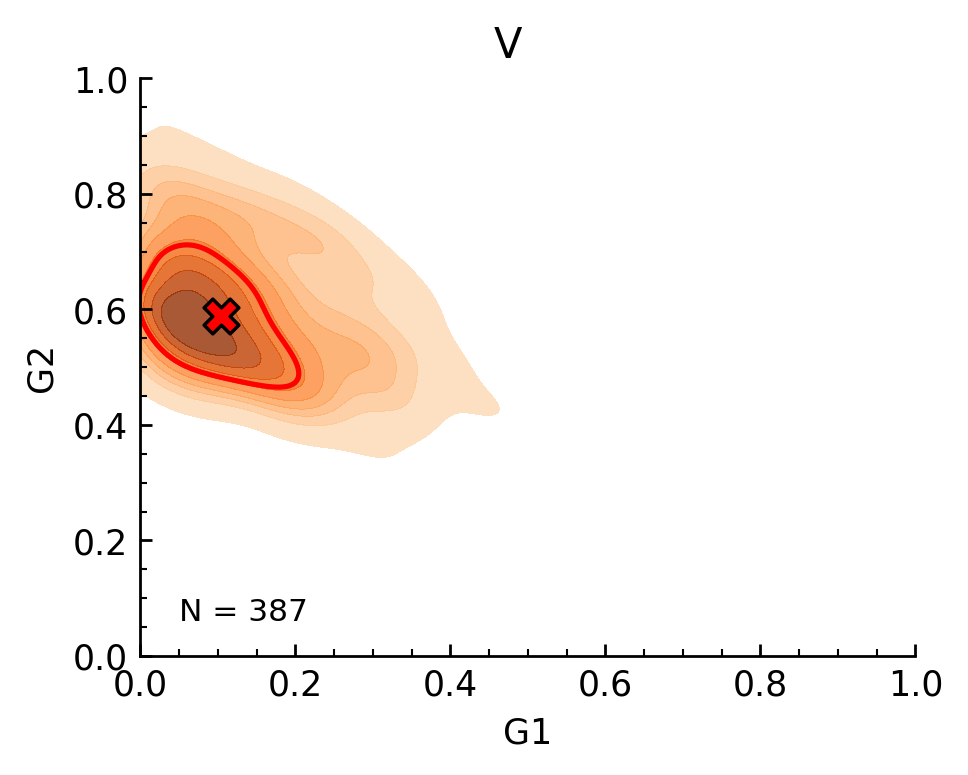}
        \includegraphics[width=4cm]{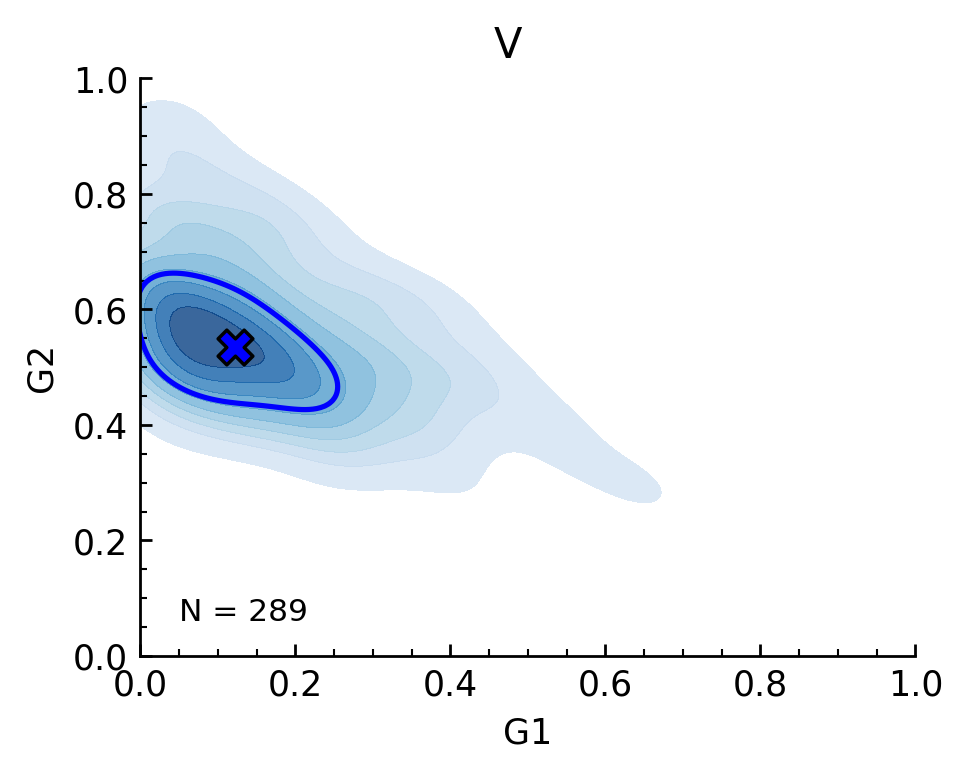}
    \end{minipage}%
    \hfill
    \begin{minipage}{0.3\textwidth}
        \centering
        \includegraphics[width=4cm]{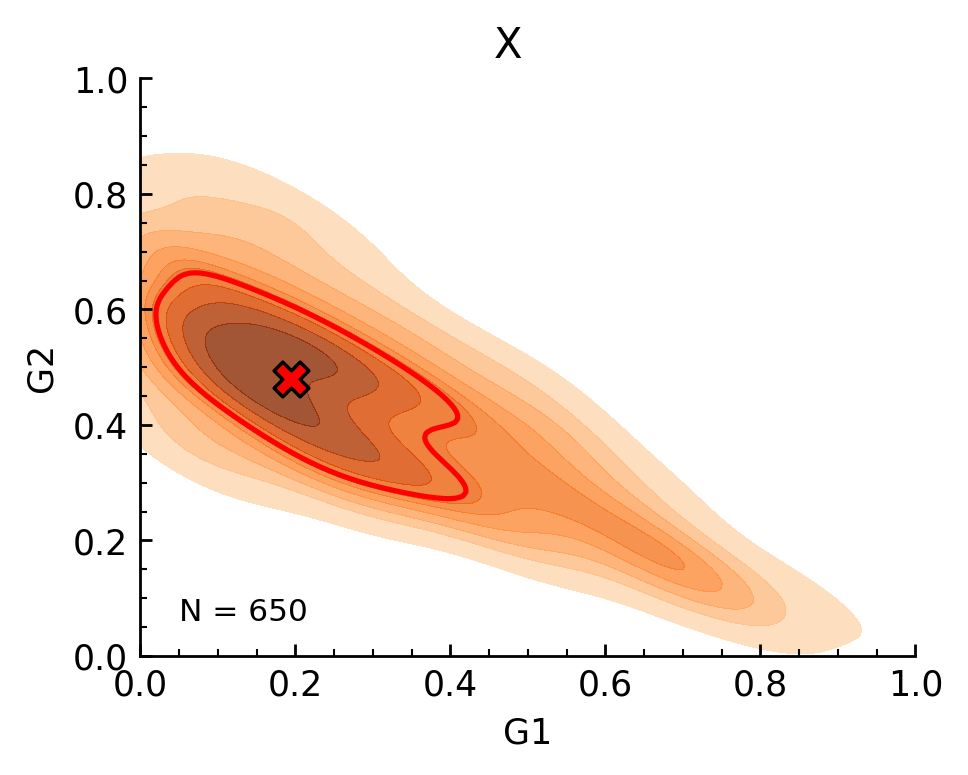}
        \includegraphics[width=4cm]{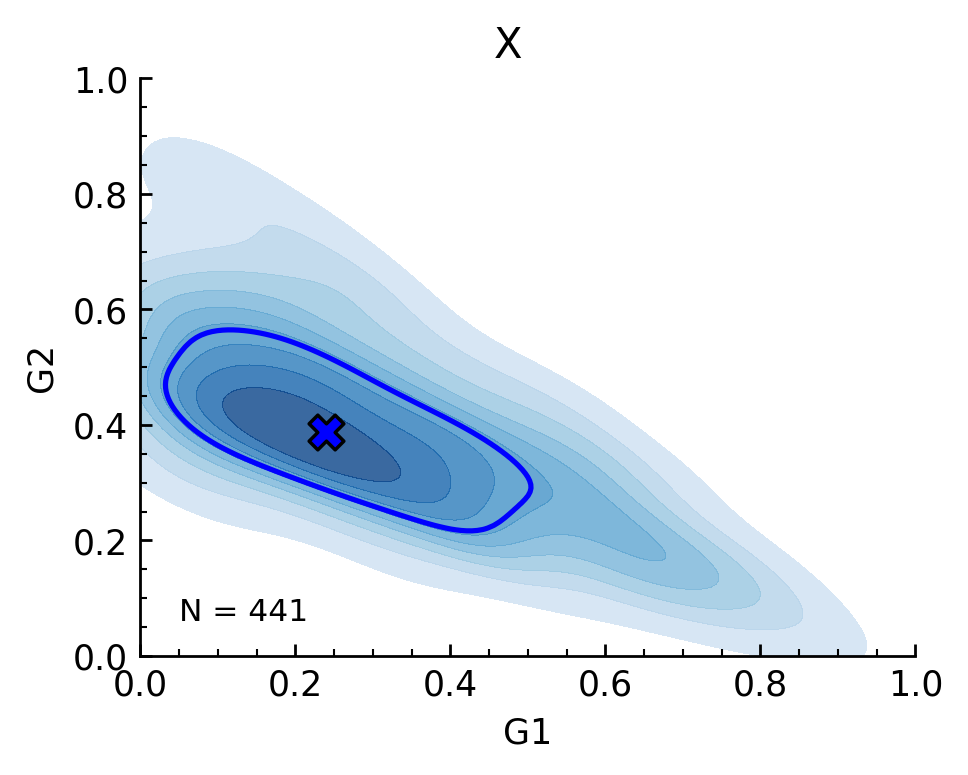}
    \end{minipage}

\caption{Density plots of the G1 vs G2 distributions for different taxonomic types. Colors represent the ATLAS filters orange and cyan. The bold lines represent the 1-$\sigma$ contours. Color scale information is provided in Figure \ref{g1g2}. The geometric center of the 95$\%$ contour is highlighted with a cross.}
\label{taxas1}
\end{figure*}

In this section, we investigate the relationship between the G1 and G2 parameters and taxonomic type. The analysis follows the approach of \cite{mahlke2021}, but applied to the newest release of the ATLAS catalog. To do this, we cross-referenced our database with the taxonomic database by \cite{demeo2013}. After merging the databases we end with 7$\,$499 asteroids in orange and 5$\,$046 asteroids in cyan. In Figure \ref{taxas1}, we show the G1-G2 distributions for several taxonomic complexes. In Table \ref{table-taxa} we present the geometric center of the 95$\%$ probability contour and the area of the 1-$\sigma$ contour. The taxonomic complexes are sorted in increasing order of their average visual albedo. Middle to high albedo complexes, such as S, A, Q and V, have distributions centered in lower G1 values and high G2 values, i.e, the left upper side of the G1-G2 parameters space. D, B, and C types asteroids, present broad distributions in both filters, centered in the middle part of the plots. 

\begin{table}[h]
\caption{Geometric center C of the 95$\%$ probability contour in orange (subscript o) and cyan (subscript c) and area A of the 1-$\sigma$ region. Also, $No$ and $Nc$ indicate the number of objects in the family for each filter. The sample used follows the conditions outlined in P16, but it is not restricted to the $15\%$ uncertainty margin. See Figure \ref{taxas1}.}
\begin{tabular}{llrrlrr}
\toprule
tax & Co & Ao & No & Cc & Ac & Nc \\
\midrule
D    & (0.32, 0.45) & 0.09 & 229 & (0.37, 0.42) & 0.09 & 126 \\
B    & (0.37, 0.41) & 0.08 & 401 & (0.39, 0.37) & 0.10 & 281 \\
C    & (0.38, 0.4) & 0.08 & 2147 & (0.4, 0.36) & 0.10 & 1552 \\
K    & (0.33, 0.42) & 0.06 & 430 & (0.33, 0.4) & 0.08 & 299 \\
L    & (0.29, 0.46) & 0.08 & 366 & (0.33, 0.4) & 0.09 & 202 \\
S    & (0.22, 0.54) & 0.04 & 4108 & (0.26, 0.5) & 0.06 & 2653 \\
A    & (0.18, 0.53) & 0.03 & 23 & (0.19, 0.54) & 0.09 & 8 \\
Q    & (0.15, 0.61) & 0.05 & 43 & (0.17, 0.54) & 0.06 & 21 \\
V    & (0.17, 0.61) & 0.03 & 387 & (0.22, 0.56) & 0.04 & 289 \\
X    & (0.38, 0.4) & 0.09 & 650 & (0.38, 0.38) & 0.09 & 441 \\
\bottomrule
\end{tabular}
\label{table-taxa}
\end{table}

\begin{figure}
\includegraphics[width=\columnwidth]{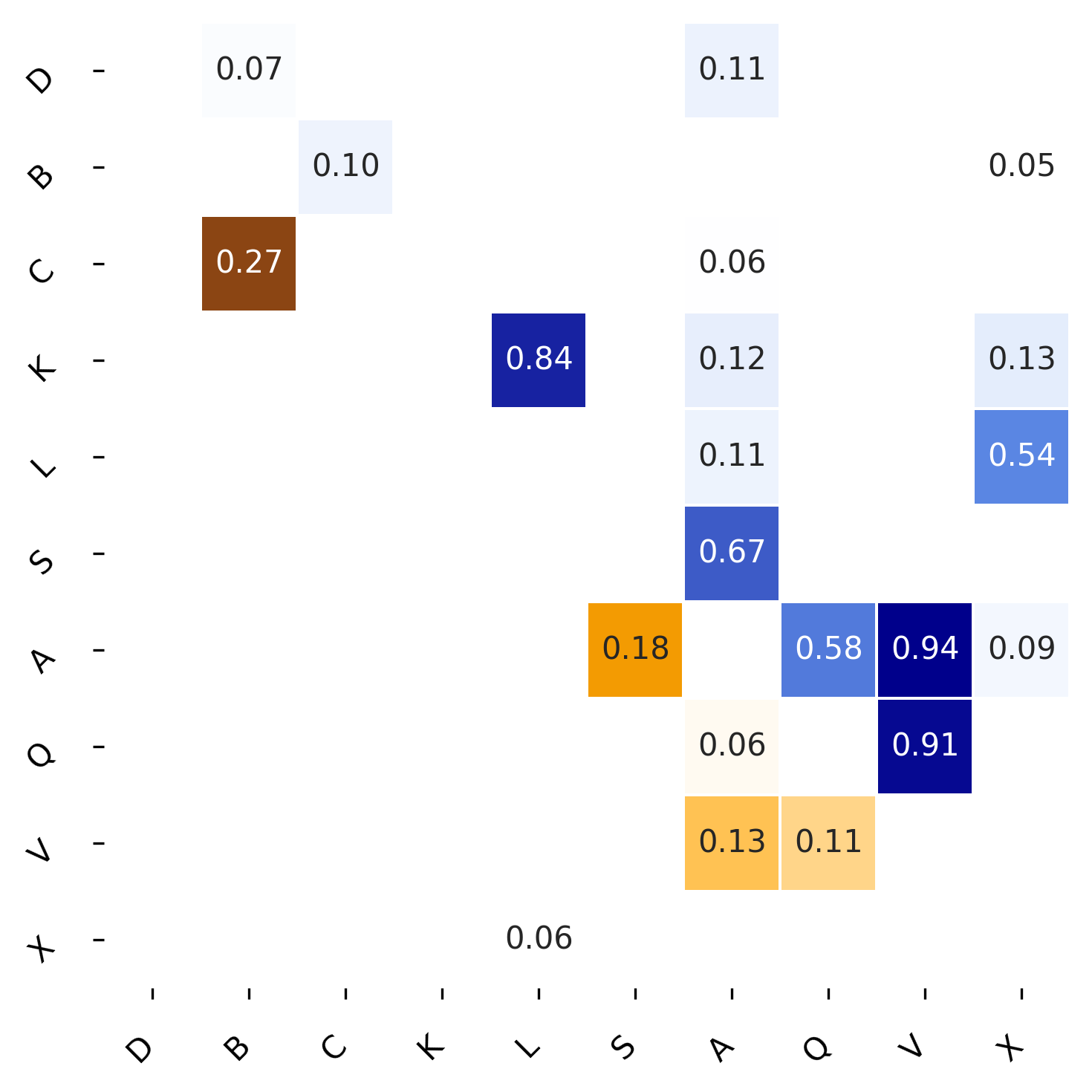}
\caption{The two-dimensional Kolmogorov–Smirnov two sample tests p-values for different taxonomic types: The upper triangle (blue tiles) shows the comparison of G1, G2 cyan distributions, while the lower triangle (orange tiles) shows the comparison of G1, G2 orange distributions. Darker colors indicate higher p-values.}
\label{ks-matrix}
\end{figure}

To analyze the similarities between paired G1, G2 distributions across different taxonomic complexes, we perform two-dimensional Kolmogorov–Smirnov two sample tests (2DKS) across all the taxa. We applied the \textsc{ndtest.ks2d2s()} code\footnote{Written by Zhaozhou Li, {\url{https://github.com/syrte/ndtest}}} with a bootstrap of 300 iterations. The bootstrap approach ensures robust and reliable statistical analysis by addressing potential biases when the model is derived from the observed dataset. Additionally, it allows the estimation of significance levels in complex or multidimensional contexts. Using a 95$\%$ confidence level, we reject the null hypothesis (that the distributions are drawn from the same population) in favor of the alternative hypothesis (that they are statistically different) if the p-value is less than 0.05 \citep{peacock1983, numerical}. The heatmap of the two-sample KS p-values is presented in Figure \ref{ks-matrix}. The upper triangle compares the paired G1, G2 cyan distributions across different taxonomic groups, while the lower triangle compares $G2$ the paired G1, G2 orange distributions. The two colors represent the two different filters, with darker colors indicating higher p-values. P-values below 0.05 are not displayed for readability. As noted by \cite{mahlke2021}, most complexes show larger p-values in the cyan filter. We cannot reject the null hypothesis in 11$\%$ of the orange sample and 31$\%$ of the cyan sample. This suggests that paired G1, G2 distributions for taxonomic complexes exhibit more disparities in orange than in cyan.

There is one pair with a p-value exactly at the 0.05 threshold: the L-X pair in the orange filter. Given that both complexes have moderate-to-high albedos, we would expect them not to exhibit disparities. As also noted by \cite{mahlke2021}, high-albedo complexes tend to show more resemblance to each other than low-albedo complexes. There are two p-values close to the maximum value, corresponding to the V-Q and V-A pairs in the cyan filter. This is consistent with expectations, as the albedos of these taxonomic types are also similar.

\subsection{ATLAS main results for asteroid families}
\begin{figure}
    \centering
    \includegraphics[width=\columnwidth]{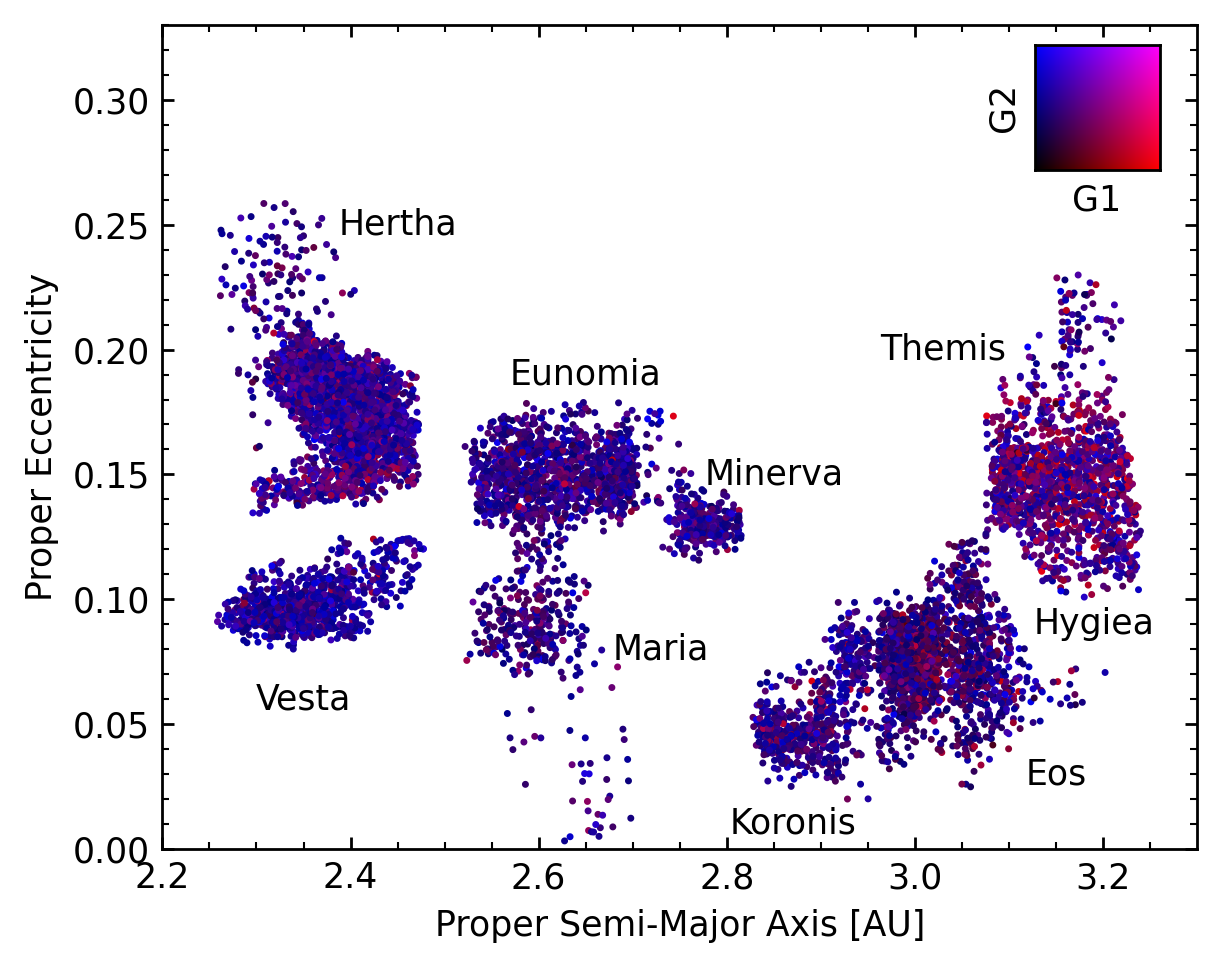}
    \caption{Proper elements distribution color-coded according to
    G1 vs G2 parameters for several asteroid families (based on orange phase curves).}
    \label{fam-g1g2}
\end{figure}

In this section, we analyze the distribution of the G1 and G2 parameters according to family membership. Family identification was performed using the Asteroids – Dynamic Site 2 (AstDyS-2), selecting families with more than 500 members. The distributions of these families in terms of proper semimajor axis and eccentricity are shown in Figure \ref{fam-g1g2}. The families exhibit good uniformity among members, which aligns with the hypothesis that asteroid families originate from the breakup of a large parent body, resulting in a shared composition. Notably, the Themis, Hygiea, and Eos families display higher G1 values, corresponding to low-albedo asteroids, as observed in the previous section. In contrast, the remaining families, which have medium to high albedos, show lower G1 values and higher G2 values, as expected. However, parameters G1 and G2 do not seem to perform as well in predicting the preponderance of taxonomic complexes compared to G12 parameter, where this distinction can be made more clearly \citep{oszkiewicz2012asteroid}. This may be due to the fact that the two parameter model was designed for poorly sampled phase curves \cite{muinonen2010three}.

\begin{figure*}[ht]
\centering
\includegraphics[width=4cm]{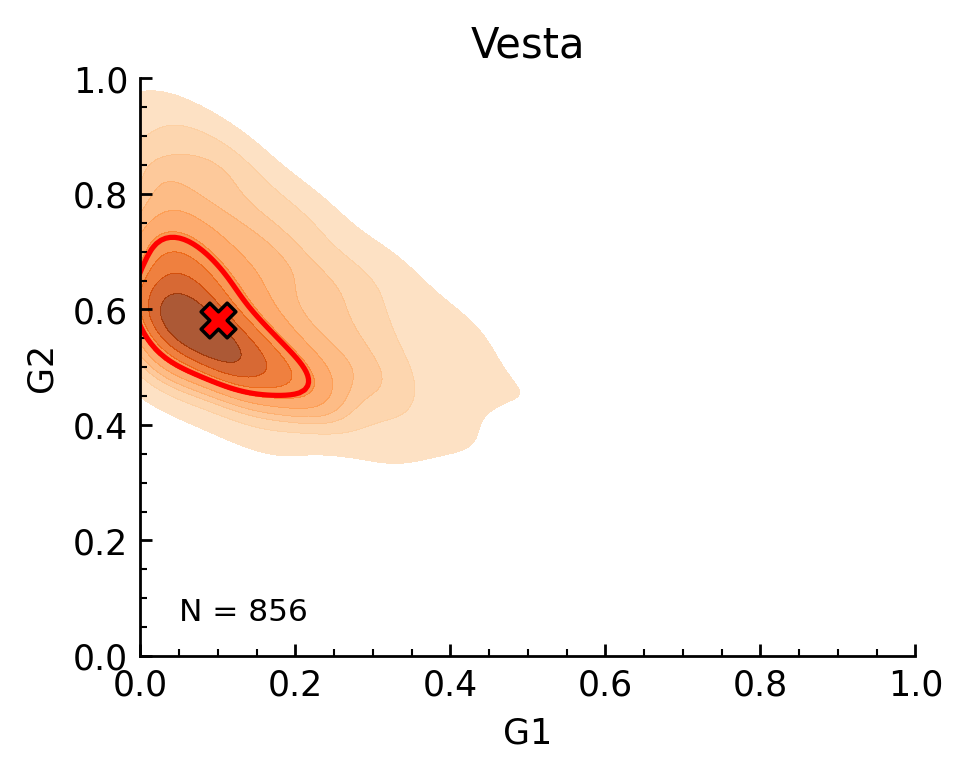}
\includegraphics[width=4cm]{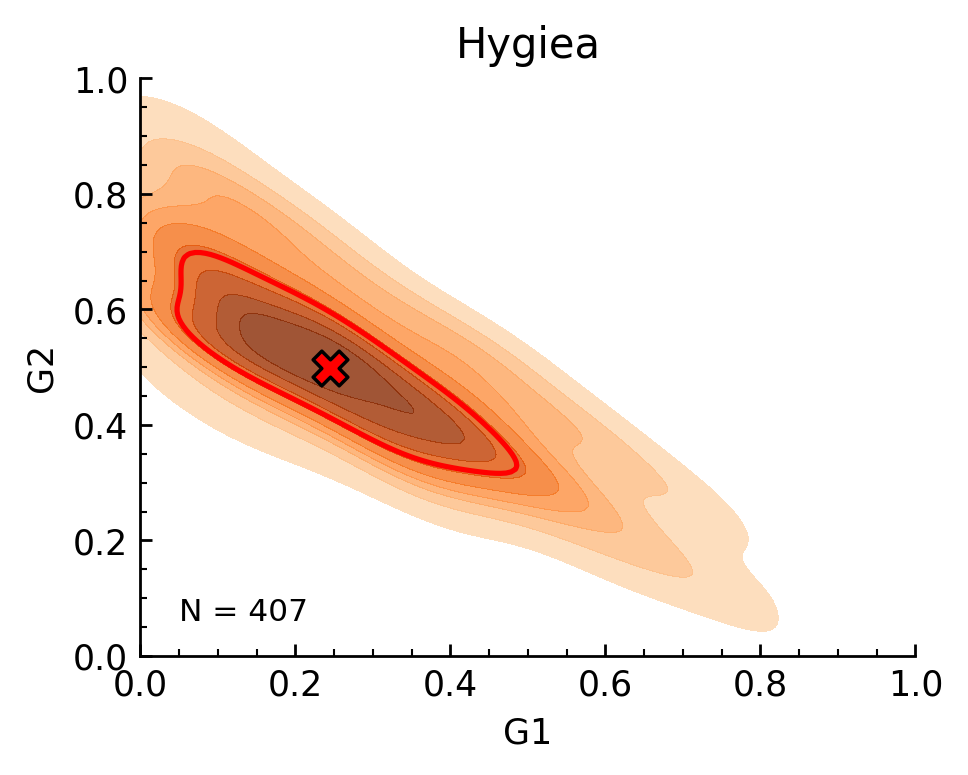}
\includegraphics[width=4cm]{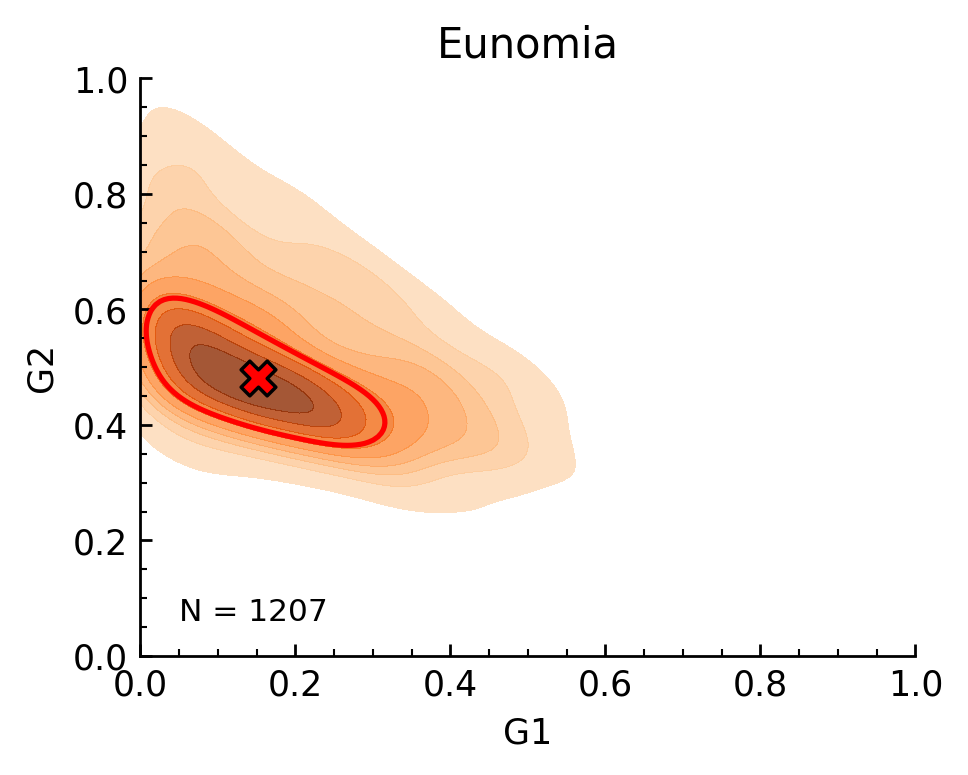}
\includegraphics[width=4cm]{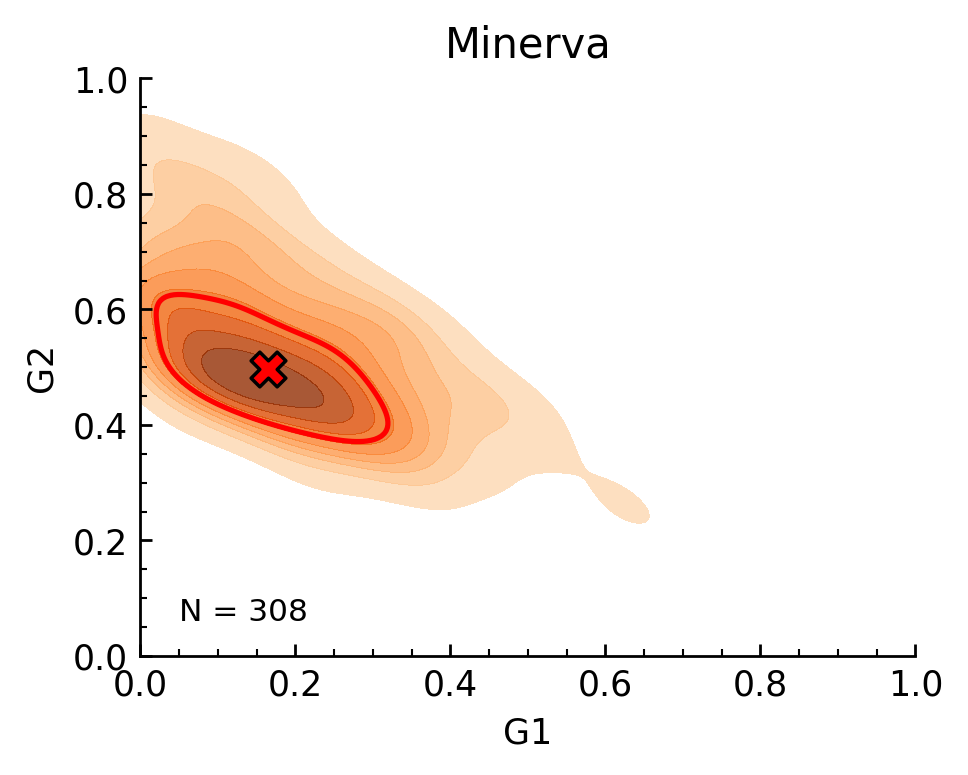}
\includegraphics[width=4cm]{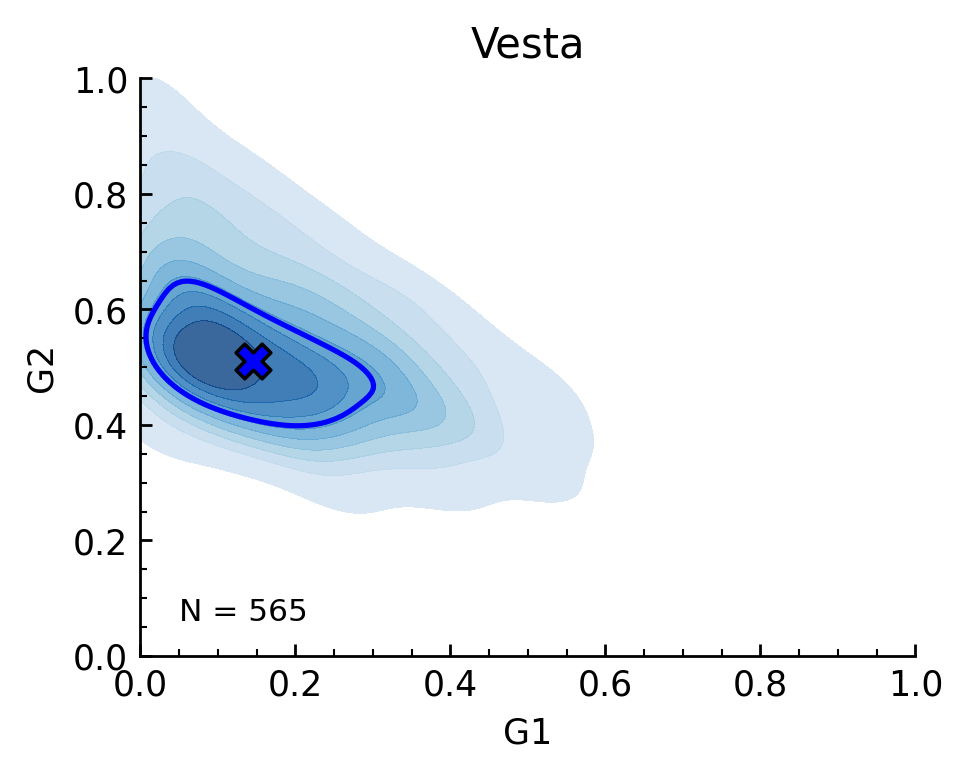}
\includegraphics[width=4cm]{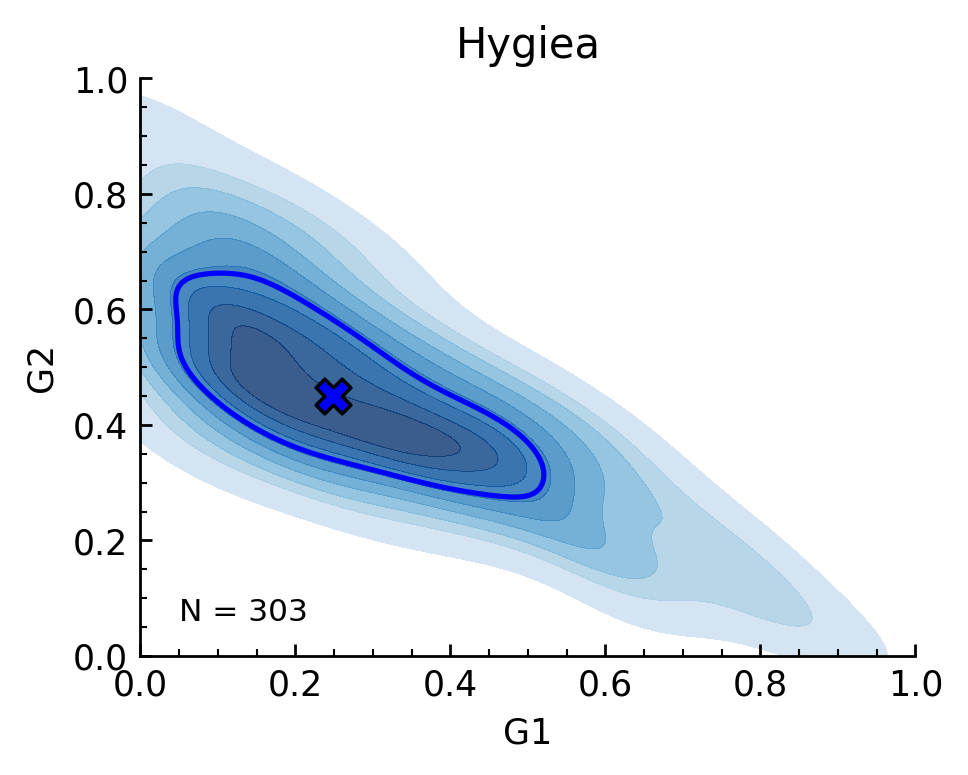}
\includegraphics[width=4cm]{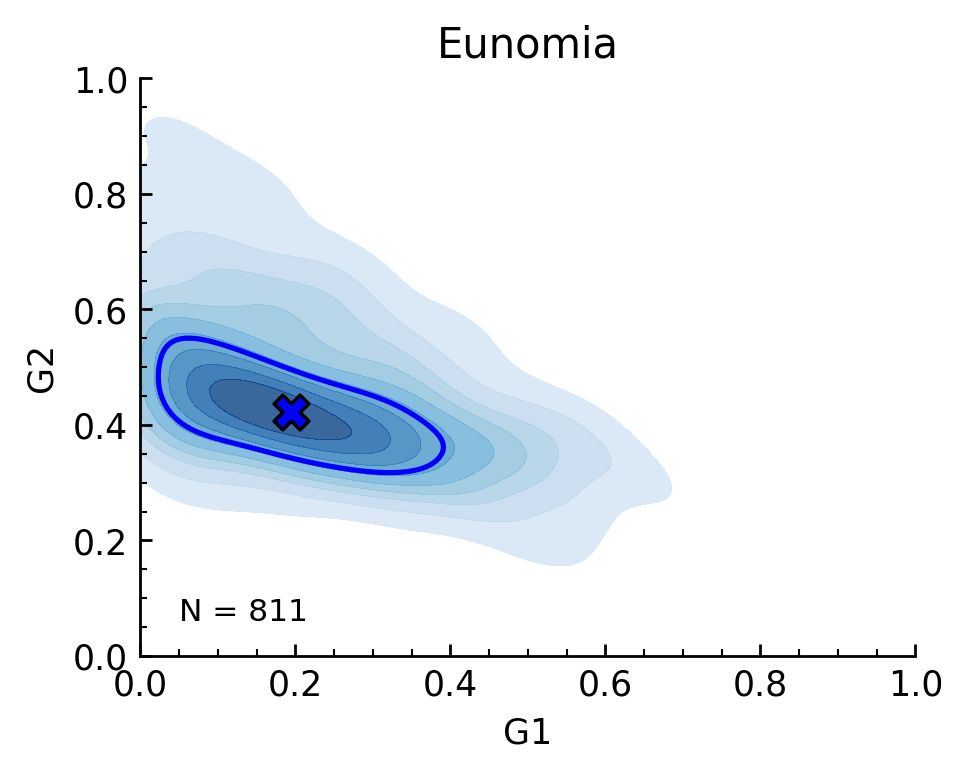}
\includegraphics[width=4cm]{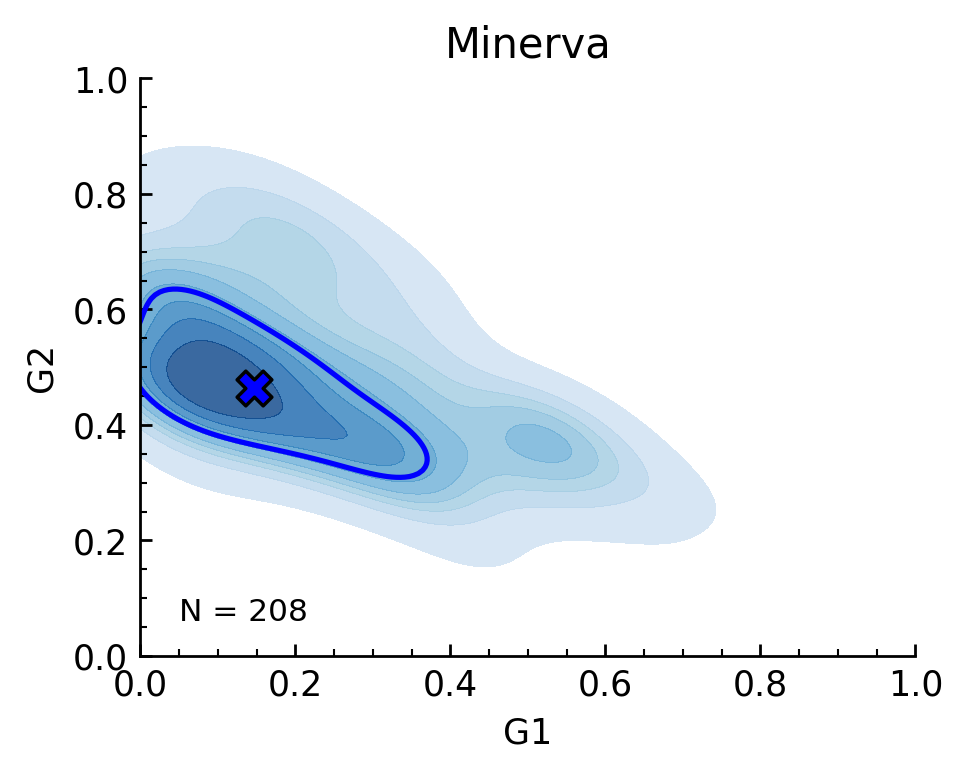}

\includegraphics[width=4cm]{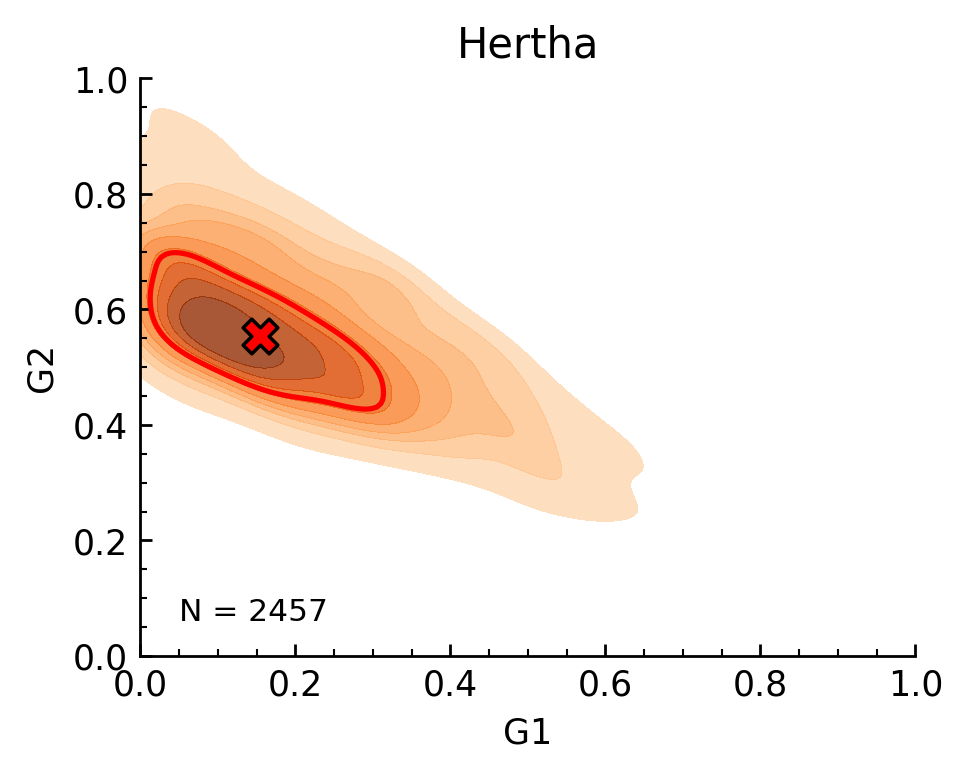}
\includegraphics[width=4cm]{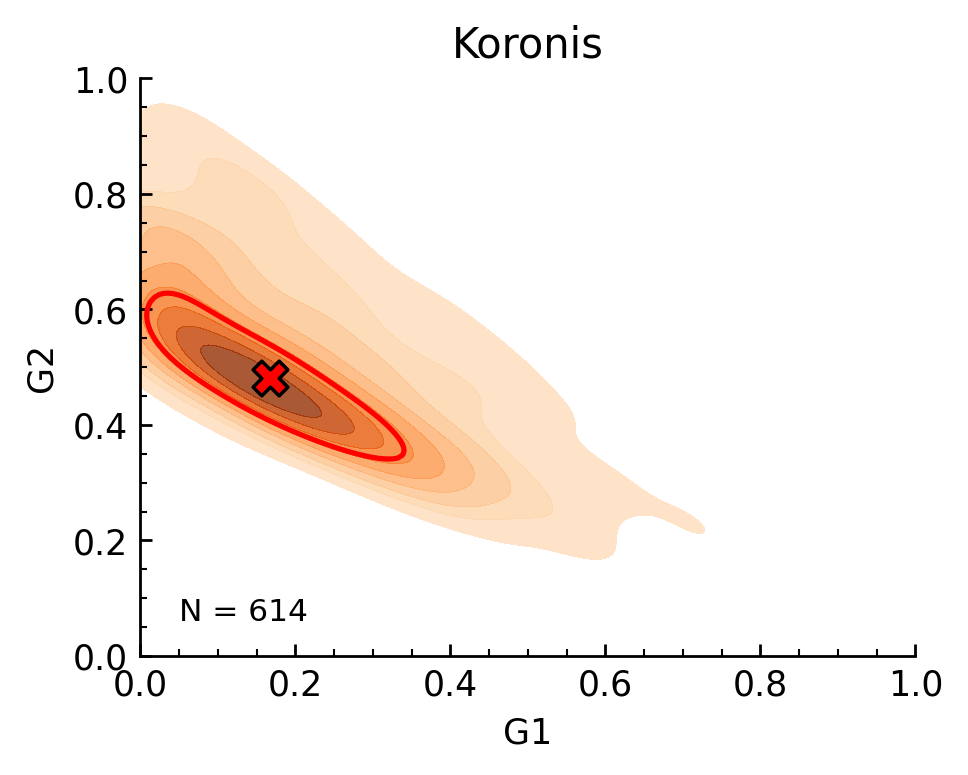}
\includegraphics[width=4cm]{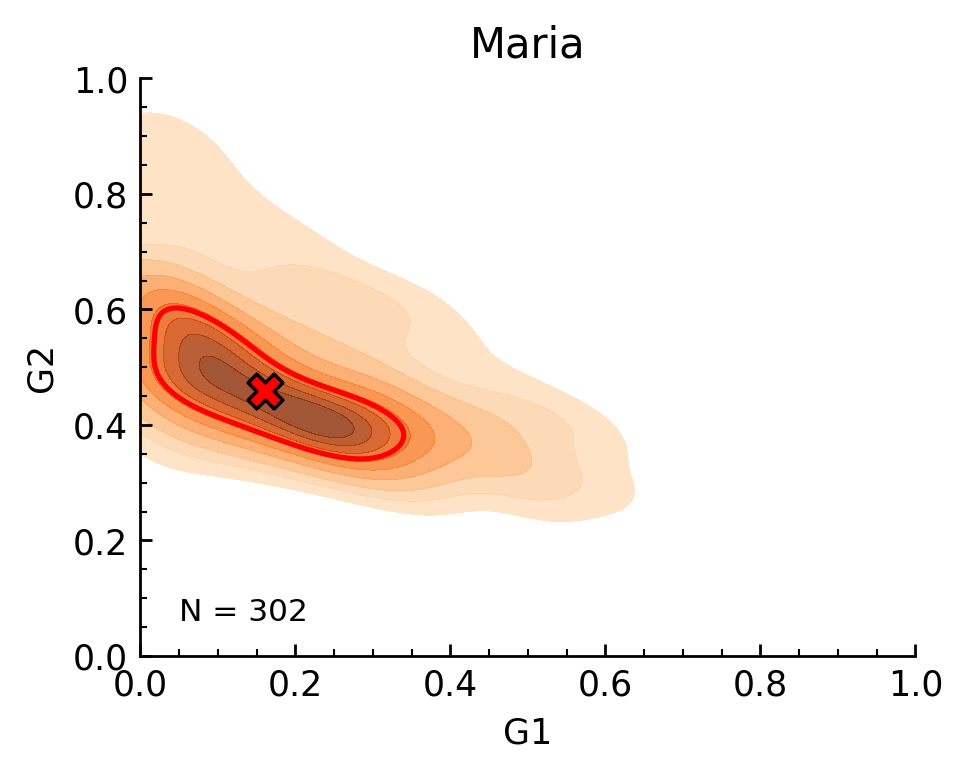}
\includegraphics[width=4cm]{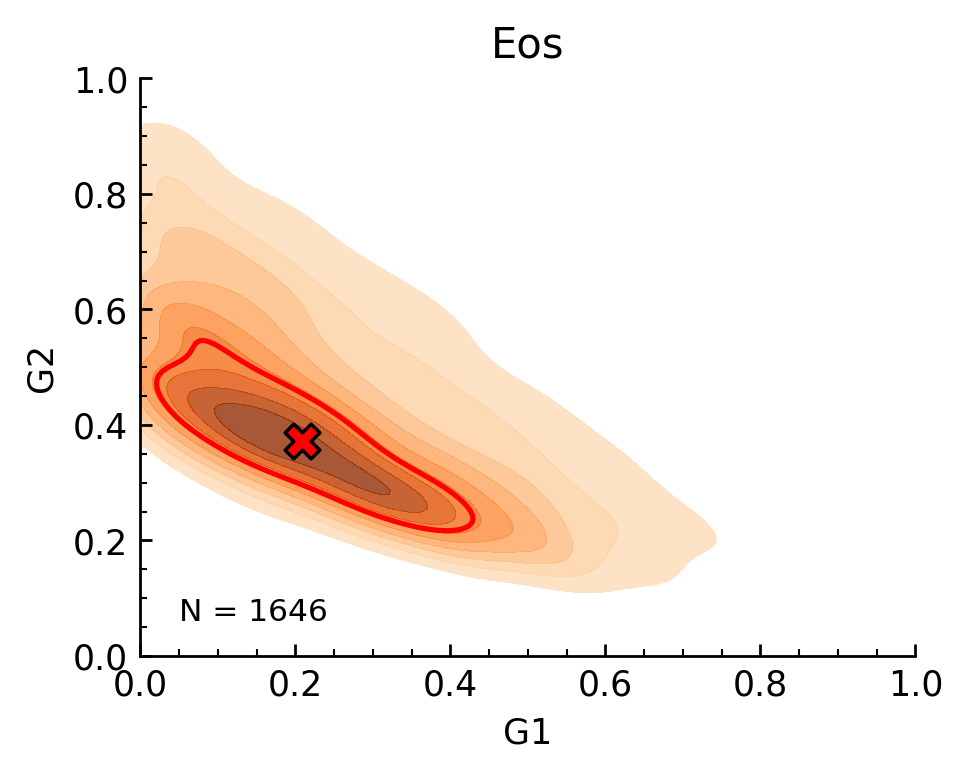}
\includegraphics[width=4cm]{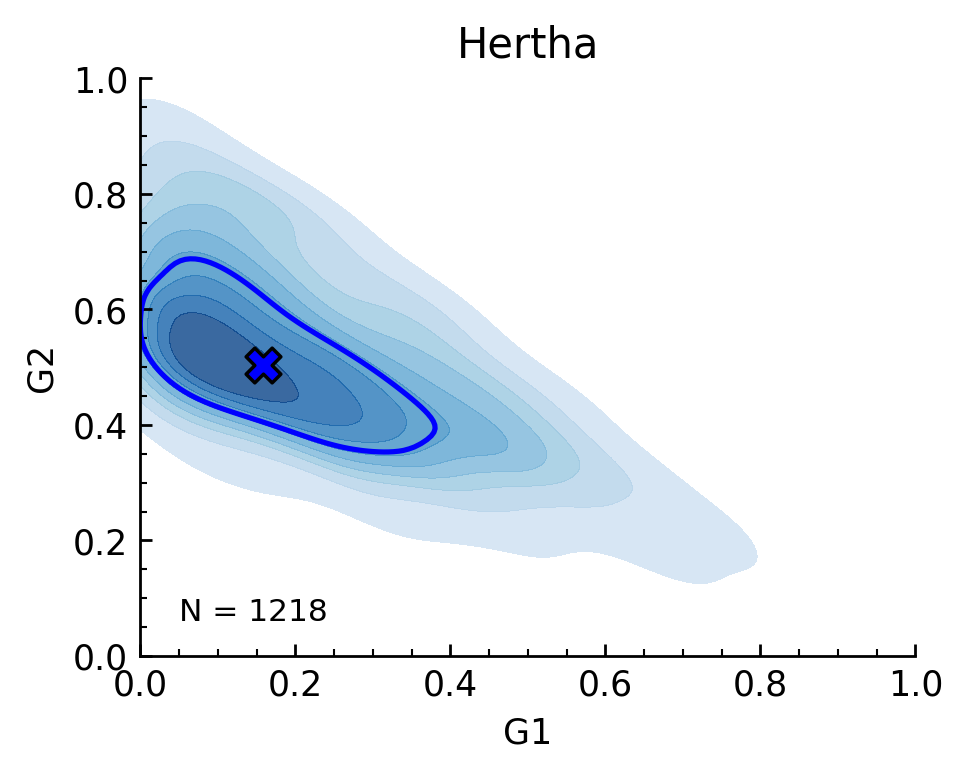}
\includegraphics[width=4cm]{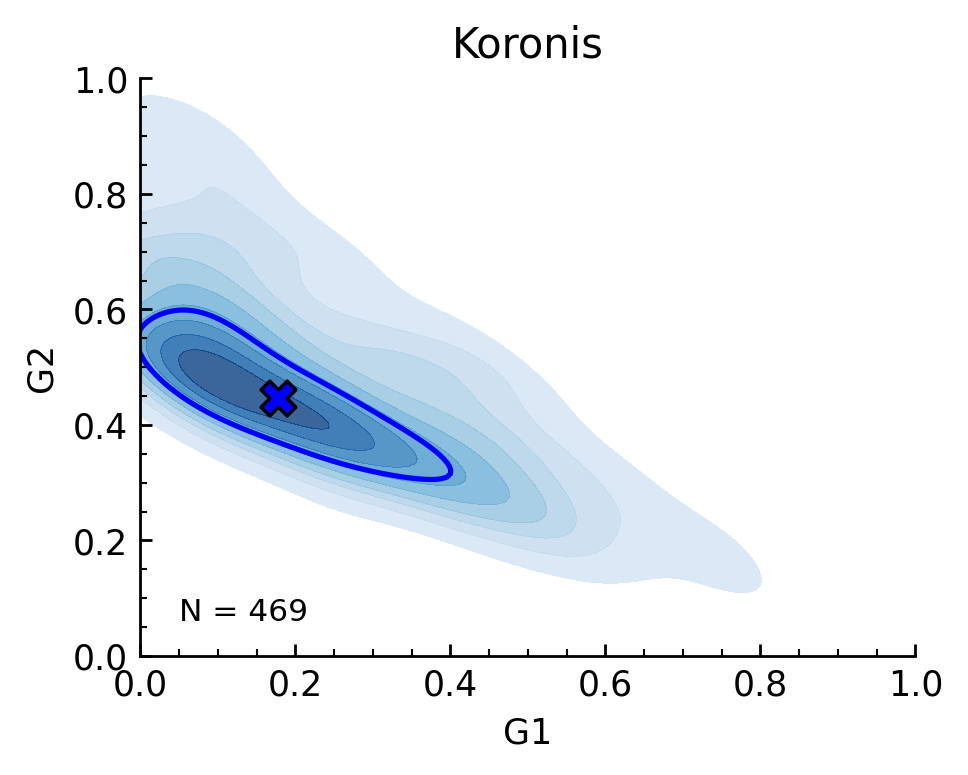}
\includegraphics[width=4cm]{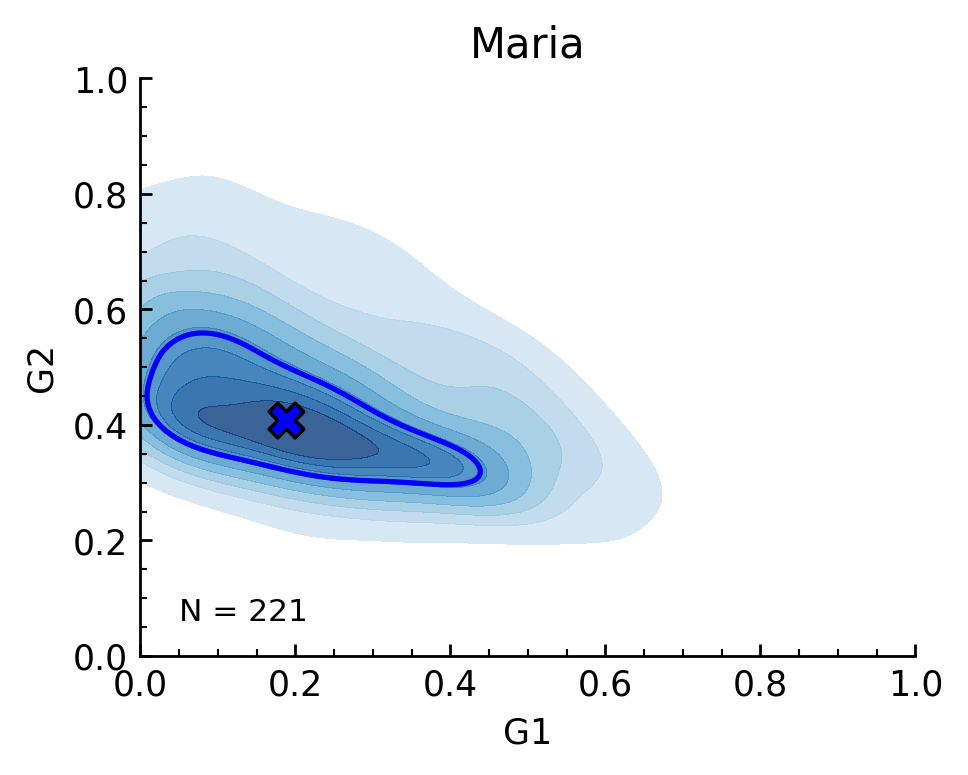}
\includegraphics[width=4cm]{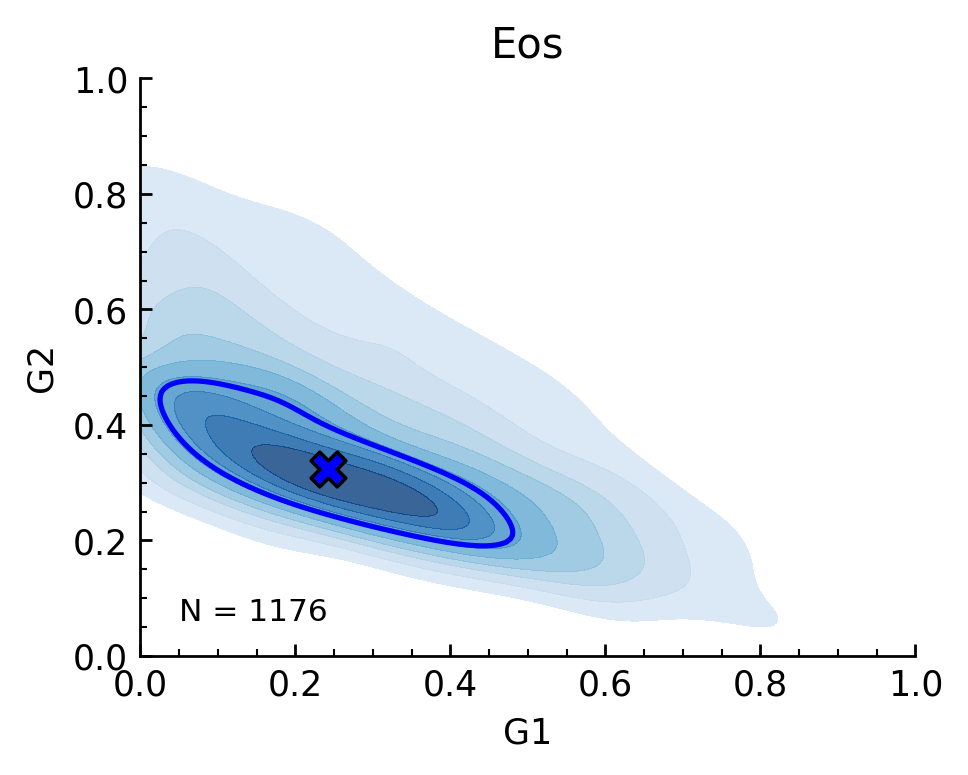}
\caption{The G1 vs G2 distributions for the most important asteroid families (more than 500 members). Colors represent the ATLAS filters orange and cyan. The bold lines represent the 1-$\sigma$ contours. Color scale information is provided in Figure \ref{g1g2}. The geometric center of the 95$\%$ contour is highlighted with a cross.}
\label{fam1}
\end{figure*}

In Table \ref{table-fam} we present the center of the distributions for most prominent asteroid families in our sample, and the corresponding plots are display in Figure \ref{fam1}. 

\begin{table*}[h]
\caption{Geometric center C of the 95$\%$ probability contour in orange (subscript o) and cyan (subscript c) and area A of the 1-$\sigma$ region. Also, $No$ and $Nc$ indicate the number of objects in the family for each filter. The sample used follows the conditions outlined in P16, but it is not restricted to the $15\%$ uncertainty margin. See Figure \ref{fam1}.}
\begin{tabular}{llrlrlr}
\toprule
fam & Co & Ao & No & Cc & Ac & Nc \\
\midrule
Vesta & (0.14, 0.6) & 0.40 & 856 & (0.18, 0.53) & 0.42 & 565 \\
Hygiea & (0.28, 0.5) & 0.30 & 407 & (0.3, 0.45) & 0.42 & 303 \\
Eunomia & (0.19, 0.51) & 0.35 & 1207 & (0.23, 0.45) & 0.41 & 811 \\
Themis & (0.39, 0.42) & 0.37 & 883 & (0.41, 0.37) & 0.44 & 622 \\
Minerva & (0.19, 0.53) & 0.23 & 308 & (0.22, 0.47) & 0.33 & 208 \\
Hertha & (0.21, 0.55) & 0.38 & 2457 & (0.23, 0.5) & 0.44 & 1218 \\
Koronis & (0.2, 0.5) & 0.35 & 614 & (0.23, 0.46) & 0.39 & 469 \\
Maria & (0.19, 0.48) & 0.25 & 302 & (0.23, 0.44) & 0.34 & 221 \\
Eos & (0.24, 0.42) & 0.43 & 1646 & (0.28, 0.36) & 0.43 & 1176 \\
\bottomrule
\end{tabular}
\label{table-fam}
\end{table*}

\subsection{ATLAS main results for asteroid populations}

\begin{figure}
\centering
 \includegraphics[width=4cm]{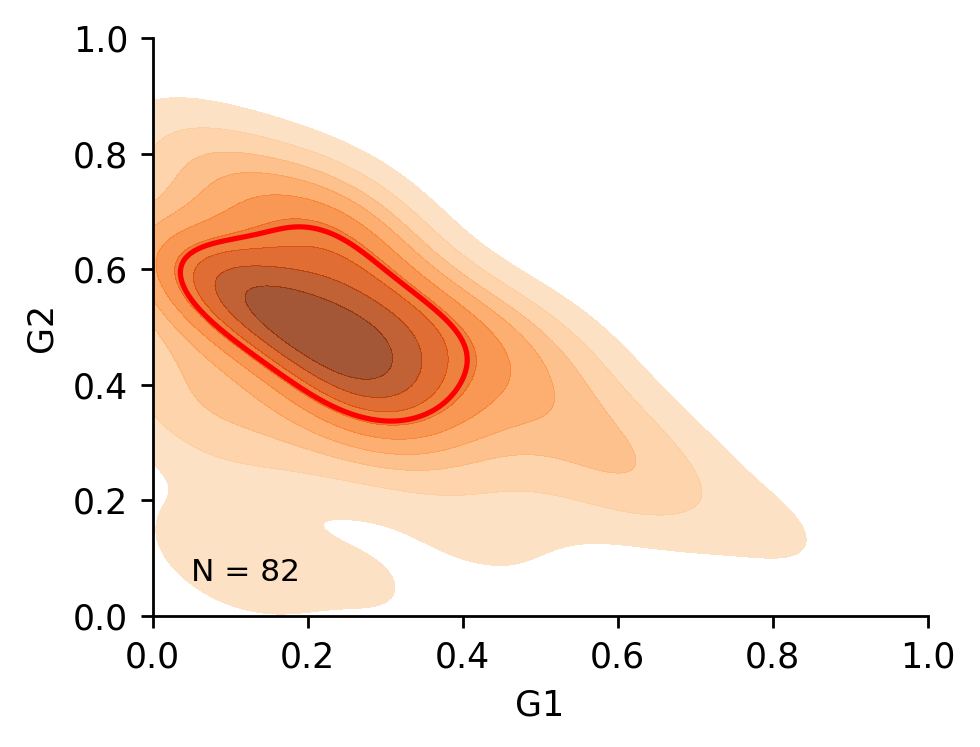}
 \includegraphics[width=4cm]{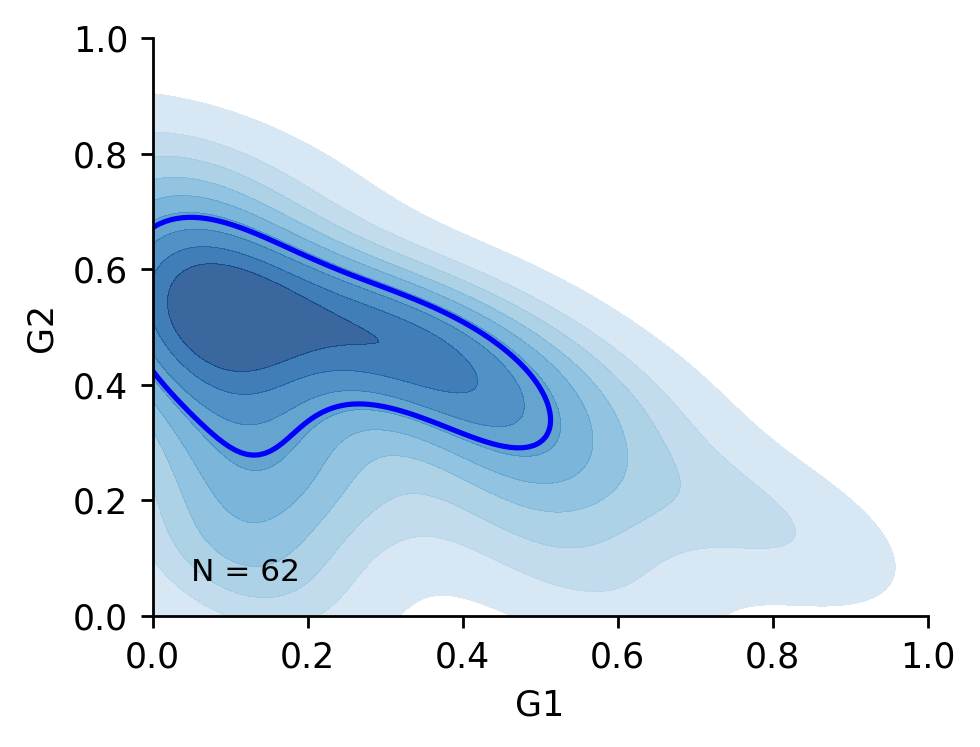}
\caption{The G1 vs G2 distributions for NEAs. The bold line represents the 1-$\sigma$ countour. The colors of the plots are according to the ATLAS filters. Color scale information is provided in Figure \ref{g1g2}.}
\label{neas}
\end{figure}

In this section, we analyze the parameter distributions of different asteroid populations. Near-Earth Asteroids (NEAs) are objects with perihelion distances $q < 1.3$ AU. These objects are believed to have evolved to their current orbital configuration from the asteroid belt, primarily due to the Yarkovsky effect and gravitational interactions with Jupiter and Saturn \citep{Bottke2006, Granvik2017}. Over 36$\,$447 Near-Earth asteroids of all sizes have been discovered\footnote{\url{http://neo.jpl.nasa.gov/stats}, (November 10th, 2024)}. These objects are of special interest as they may represent a threat to Earth \citep{Perna2016, Harris2021, Fuentes2023}. Figure \ref{neas} illustrates the distributions of G1 versus G2 for near-Earth asteroids. We obtained phase curves for 771 NEAs, of which 75 are within the P16 conditions. A notable observation is the presence of combinations with low G1 and G2 values. However, upon closer inspection, we determined that these low G1-G2 combinations correspond to NEAs with limited observations at phase angles less than 3 degrees, resulting in biased G1 values. Conversely, the peak of the KDE indicates that the NEA population is primarily composed of high-albedo asteroids. It is important to highlight that the NEA population is a weak point in the method presented in this work, as it only considers the fit of H for different apparitions. The effects of changes in aspect, which are not taken into account in this article, are significant for NEAs, as they are generally irregular objects and experience large changes in viewing aspect during observable apparitions \citep{jackson2022}. We studied the differences between the maximum and minimum H fitted at each apparition for the 75 NEAs meeting the P16 conditions, and obtained a distribution of $\overline{{\Delta H}}$ with a mean of 0.22 magnitudes. The maximum difference measured was 0.6 magnitudes.

\begin{figure}
\centering
 \includegraphics[width=4cm]{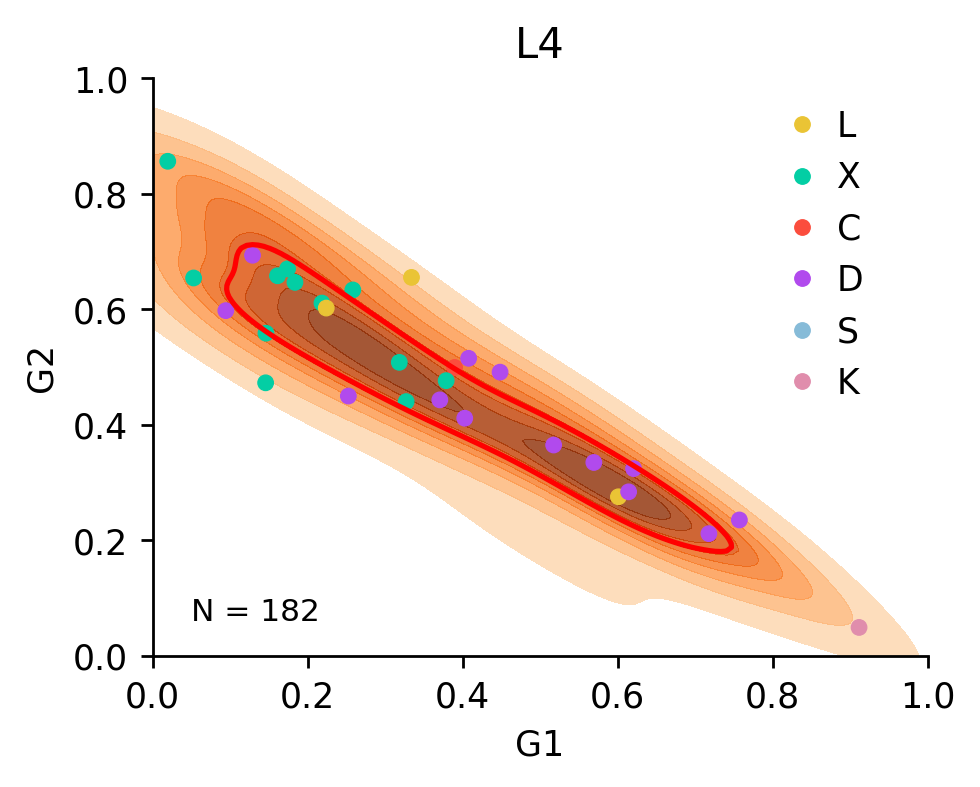}
\includegraphics[width=4cm]{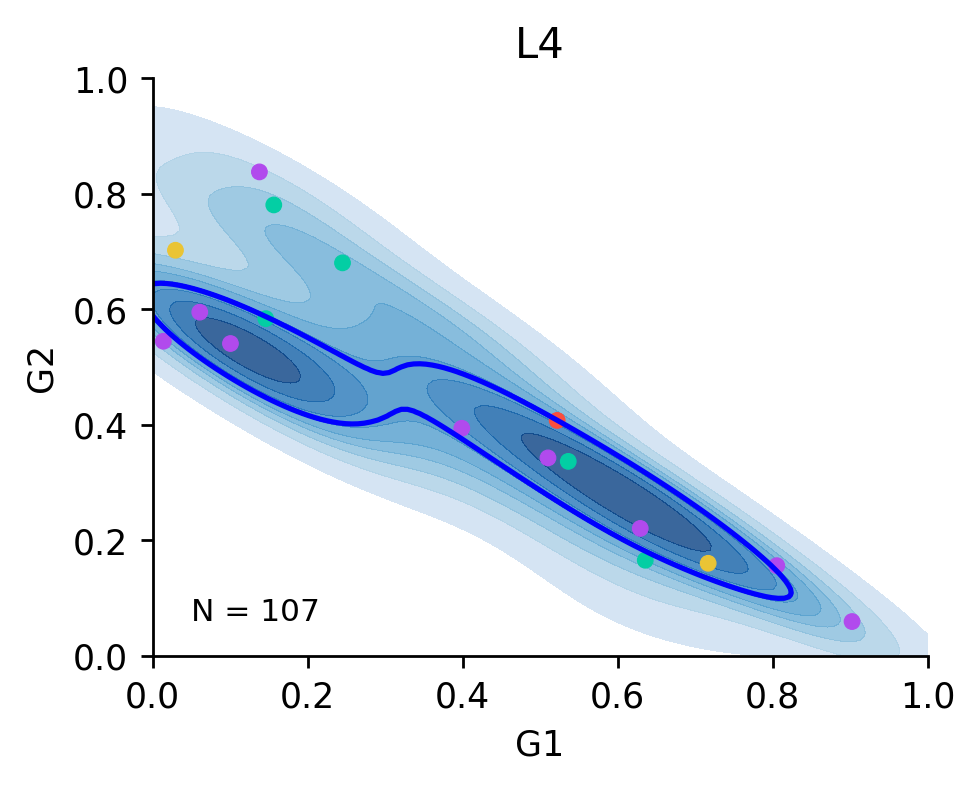}
 \includegraphics[width=4cm]{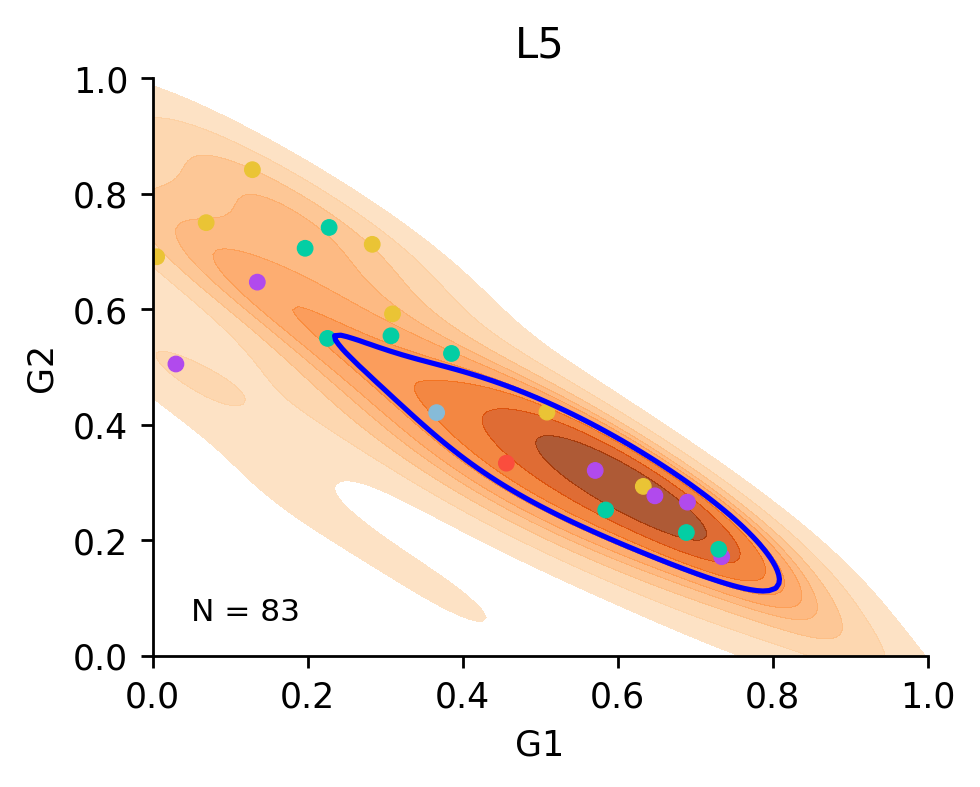}
 \includegraphics[width=4cm]{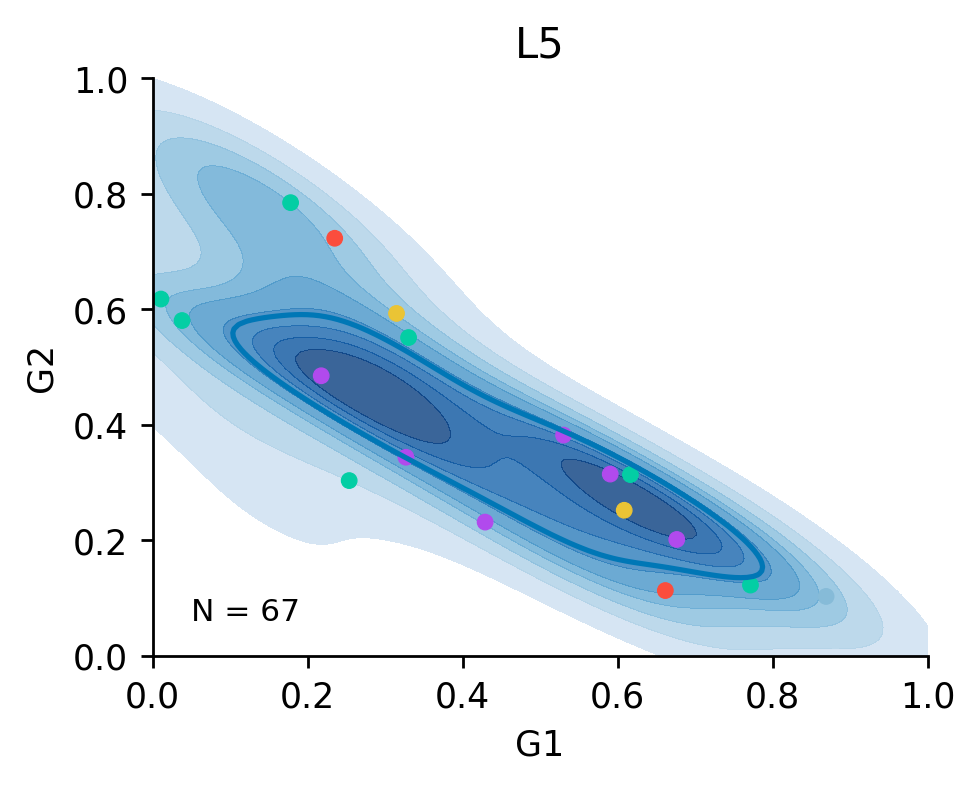}
\caption{Distribution of G1 vs G2 parameters for Jupiter Trojans. The bold line represents the 1-$\sigma$ contour. The colors of the plots correspond to the ATLAS filters, while asteroids with taxonomic classifications are represented as points, color-coded by their respective taxa. Color scale information is provided in Figure \ref{g1g2}.
}
\label{troj}
\end{figure}

Next we investigated the Jupiter Trojan population. Trojan asteroids are located at the L4 and L5 Lagrangian points with respect to the orbit of the host planet in the restricted three body problem, 60 degrees ahead of and behind the planet, respectively. The Nice model, predicts that the Trojan population originated from the outer Solar System and had been captured by Jupiter subsequently \citep{gomes2005, tsiganis2005, morbidelli2005}. We determined phase curves for 1$\,$931 objects, of which 247 are within the P16 conditions. We can also analyze the G1 vs G2 distribution for Jupiter Trojans discriminating by the L4 and L5 clouds, as can be seen in Figure \ref{troj}. We can observe that the distributions in L4 tend to be more elongated, while the distributions in L5 are more concentrated in the region of low-albedo objects. In the case of L4 in orange, where we have the largest number of objects, it is clearer that X-type asteroids (moderate to high albedo) dominate the upper left group, whereas D-type asteroids (low albedo) are more abundant in the lower right area. By applying a 2D KS test to compare the G1 and G2 distributions between the L4 and L5 clouds, we obtained a p-value of 0.01 for the orange sample and a p-value of 0.7 for the cyan sample. We can reject the null hypothesis only for the orange sample, suggesting potential differences in the physical properties between the L4 and L5 clouds for this sample. We also analyzed the H-distributions for both clouds. The results of a KS test suggest that the samples are statistically different, as we obtained a p-value of 6.6$\times 10^{-13}$ (o) and 4.4$\times 10^{-4}$ (c). These results agree with those obtained by \cite{robinson2024}, the authors attribute this difference to a bias in the ATLAS observations, as they noticed that the observations for the L4 group generally go toward fainter absolute magnitudes.

For the TNOs, only 2 of 163 phase curves were obtained that meet the parameters set by P16. The maximum mean phase angle for this population is 3.6 degrees, which makes the determination of the phase curves very challenging. Due to this limited phase angle coverage, characterizing the opposition surge is challenging. As a result, linear phase curves are commonly fitted to TNO phase curves in the literature. However, in our work, we use the HG1G2 function for TNOs to maintain consistency.

\subsection{Phase coloring}

The phase-reddening effect causes an increase in spectral slope as the phase angle increases \citep{sanchez2012}. However, \cite{alvarez2024}, proposes the term ``phase-coloring'' based on evidence suggesting that some objects tend to become bluer. In particular, intrinsically red objects become bluer in the regime of phase angles smaller than 4.5 degrees, while for larger phase angles, the behavior is as usual \citep{alvarez2022}.

\begin{figure*}
\centering
\includegraphics[width=\textwidth]{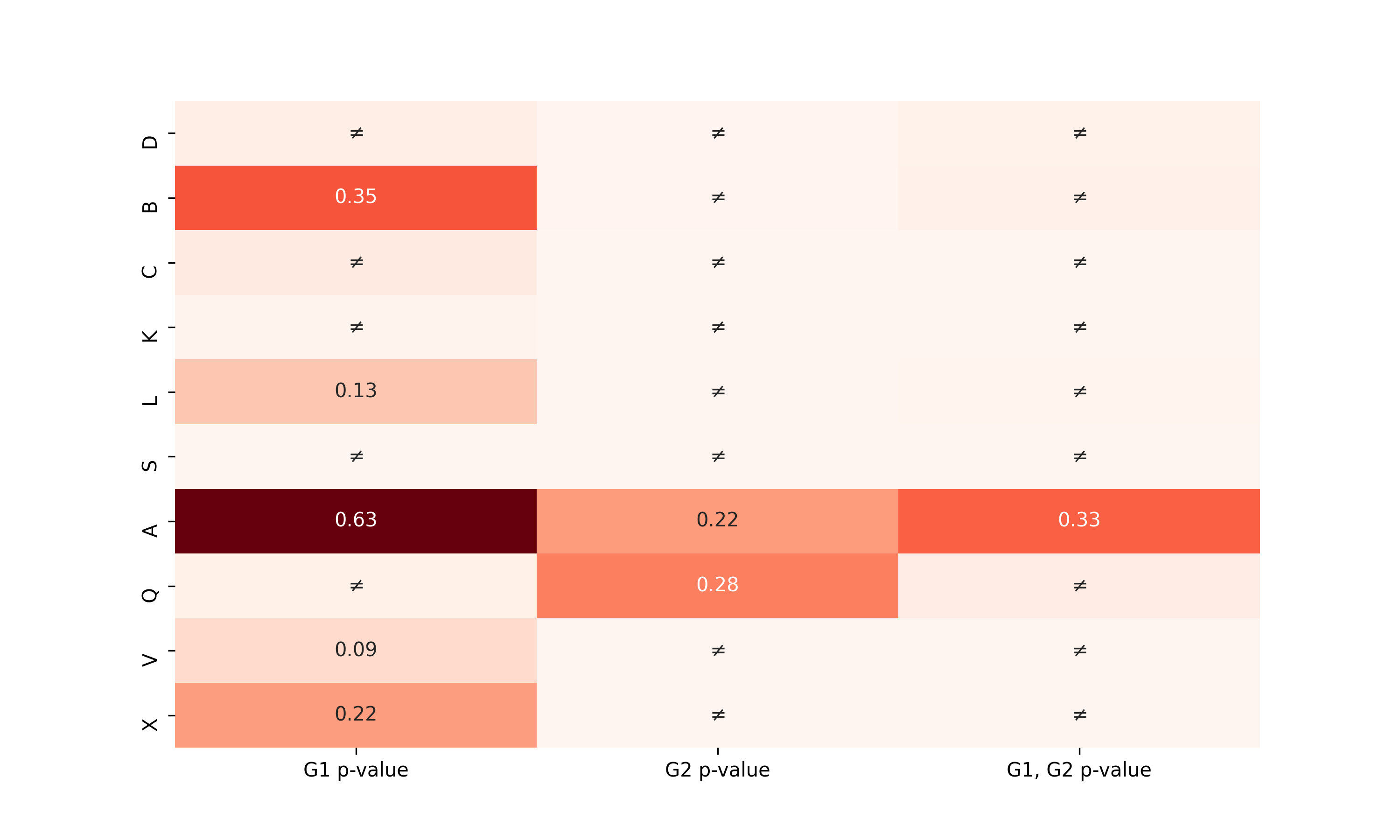}
\caption{The first two columns display the p-values for the one-dimensional K-S tests comparing G1(o) vs. G1(c) and G2(o) vs. G2(c) within the same taxonomic type. The last column shows the p-values for two-dimensional Kolmogorov–Smirnov two sample tests test between paired G1, G2 (o) vs G1, G2 (c) distributions for the same taxonomic type. Higher p-values are displayed in darker colors.}
\label{p-values_sametax}
\end{figure*}

To evaluate this wavelength dependency, we compared the paired distributions of G1, G2 (o) vs G1, G2 (c) within the same taxonomic types. Figure \ref{p-values_sametax} presents the resulting p-values. Statistically significant differences are observed in all the complexes, except for the A complex, highlighting wavelength dependency. These findings are also evident in the distribution centers shown in Table \ref{table-taxa}.

To determine which parameter primarily drives the observed differences, we conducted independent comparisons of G1 and G2 using one-dimensional KS two-sample tests. The results, illustrated in Figure \ref{p-values_sametax}, indicate that G2 exhibits the most statistically significant differences, suggesting it may dominate the conclusions drawn from the two-dimensional KS test. Notably, for taxonomic type A, the null hypothesis (H0) cannot be rejected for either parameter individually, consistent with the outcome of the two-dimensional test.

\begin{figure}
\centering
  \includegraphics[width=4cm]{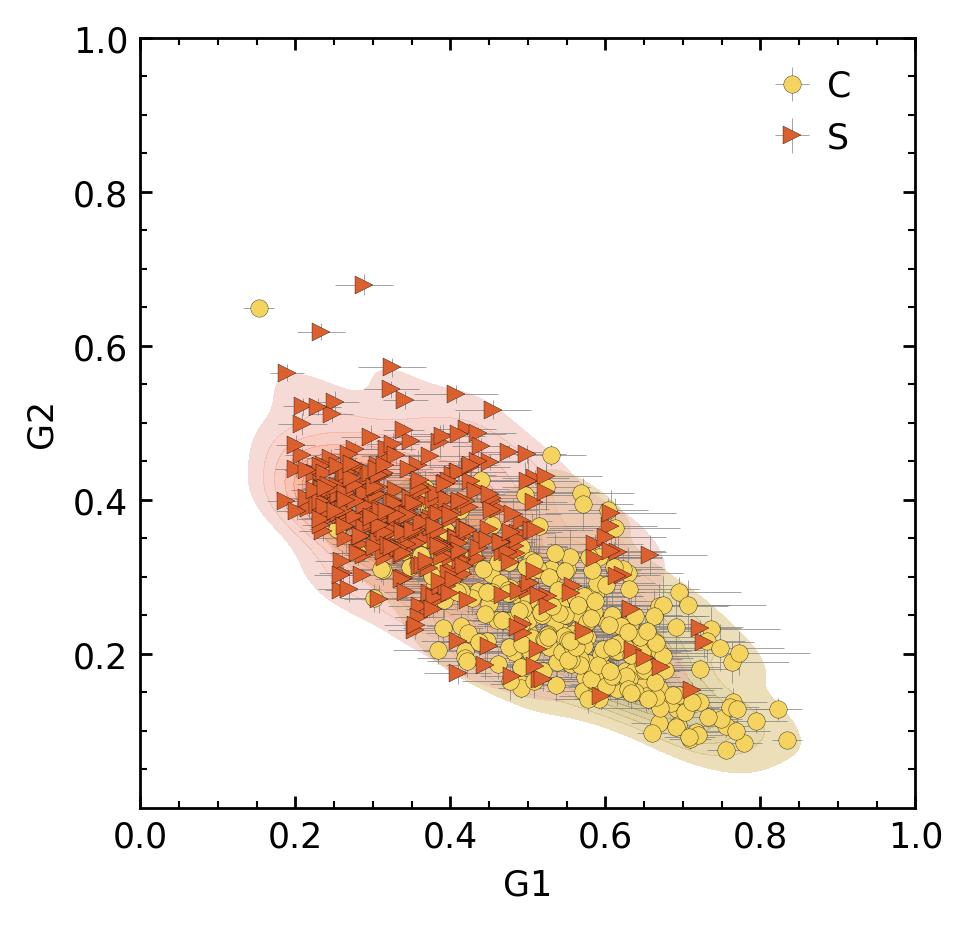}
 \includegraphics[width=4cm]{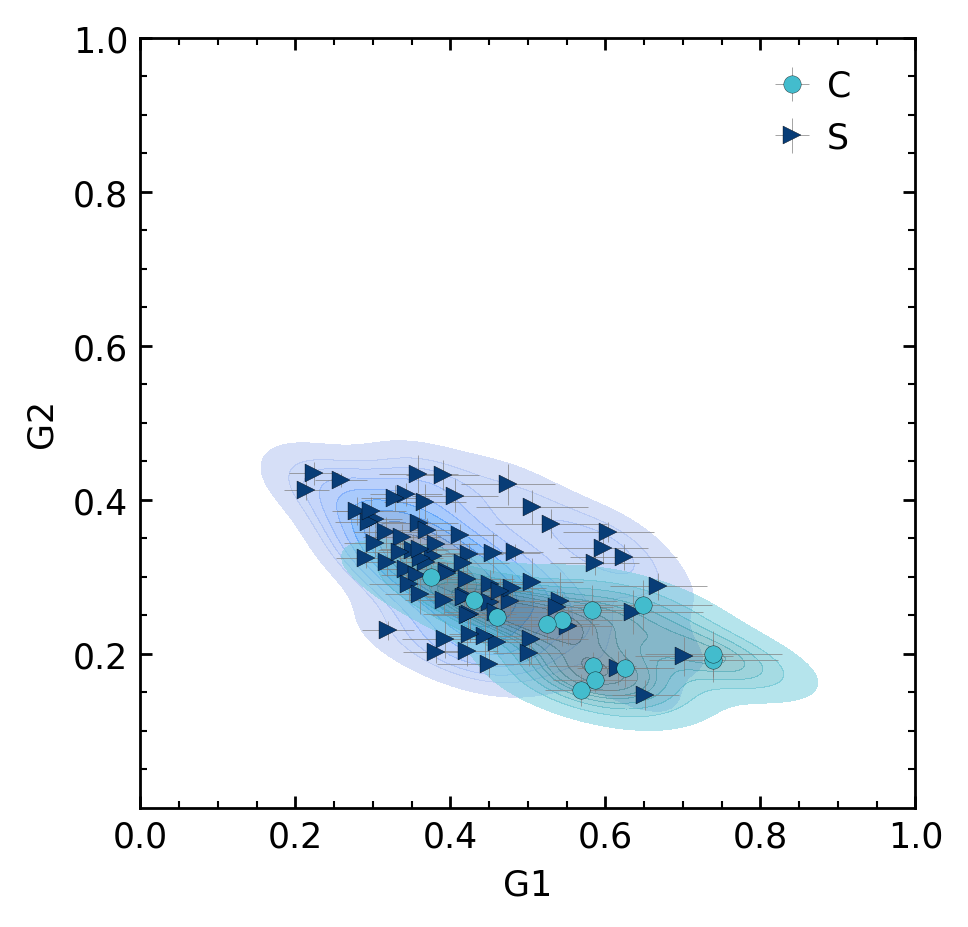}
\caption{Distribution of the G1 vs G2 parameters for C-type asteroids (circles) and for S-type asteroids (triangles). On the left, observations are in the orange filter, and on the right, in the cyan filter.}
\label{g1g2_taxas}
\end{figure}

\begin{figure}
\centering
 \includegraphics[width=4cm]{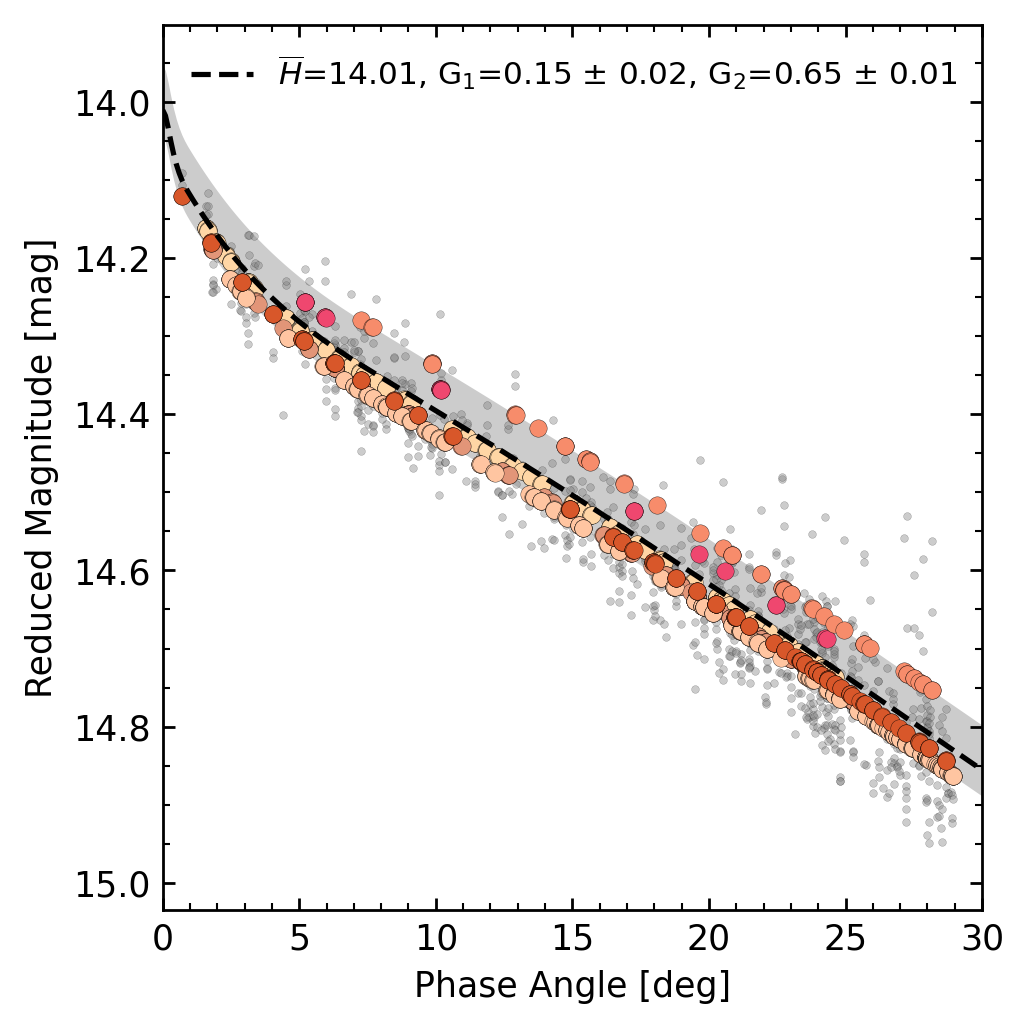}
 \includegraphics[width=4cm]{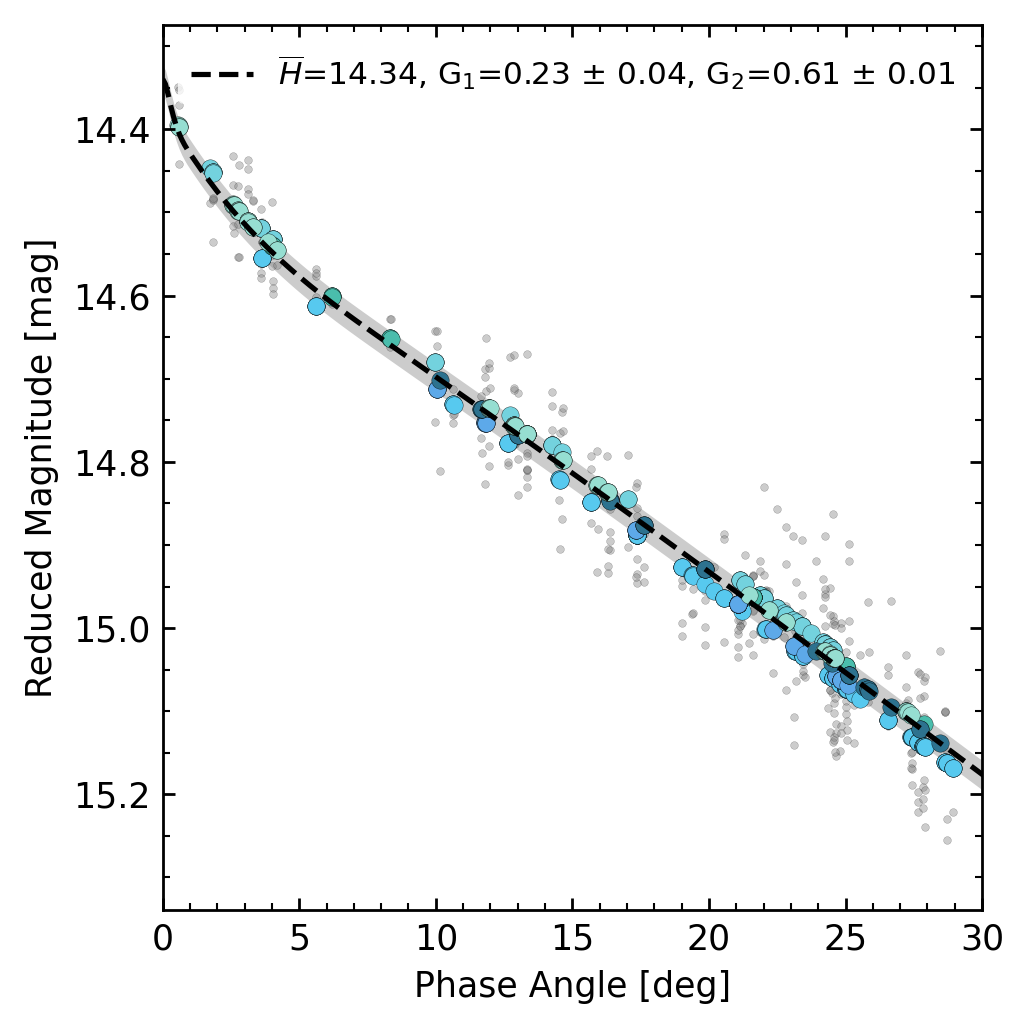}
\caption{Phase curves in orange (left) and cyan (right) for (10822) Yasunori. The dashed lines indicates our fit using the mean H value, while the gray shaded area represents the range of absolute magnitudes derived from multiple apparitions. The gray points represent the observations after outlier rejection. The colored points represent the residuals of the data from different apparitions. For each apparition, we compute a phase curve using the corresponding H value and calculate residuals by subtracting the modeled phase curve from the observed data}.
\label{yasunori}
\end{figure}

In Figure \ref{g1g2_taxas} we present the distribution of objects in the G1 vs. G2 space but restricted to the most numerous, C- and S-type asteroids, with uncertainties less than 15$\%$. Again, it is clear that S-types are located in the upper left, while C-types are in the lower right. This also aligns with the expected behavior based on the albedos of each group. However, some points deviate from this pattern. The most notable is the C-type object with G1=0.15 and G2=0.65 in the S region of the orange filter, identified as (10822) Yasunori (Fig. \ref{yasunori}). \cite{carvano2010} classified this object as X-type using SDSS data and a probabilistic approach. Both \cite{carvano2010} and \cite{demeo2013} worked with SDSS data and colors for their classifications. However, here we observe this object occupying the G1 vs G2 parameter space more typically associated with S-type objects. The best-fitting geometric albedo for this asteroid, based on the NEOWISE database V2.0 \citep{mainzer2019}, is 0.716. This value is unusually high even for S-type asteroids, suggesting an X-complex classification may be more appropriate. Furthermore, the phase curve (Figure \ref{yasunori}) does not resemble that of a C-type asteroid, indicating that the original classification may require revision.

\begin{figure}
\centering
 \includegraphics[width=4cm]{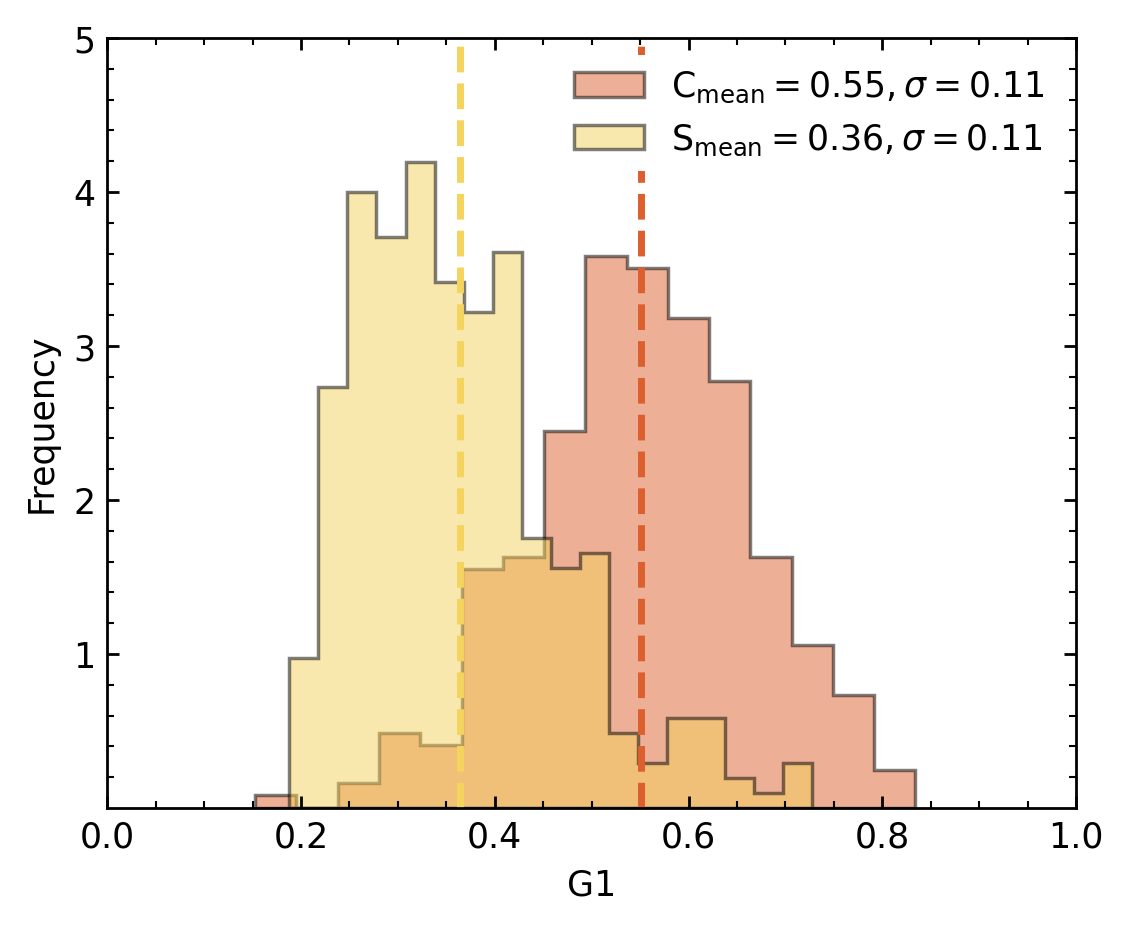}
 \includegraphics[width=4cm]{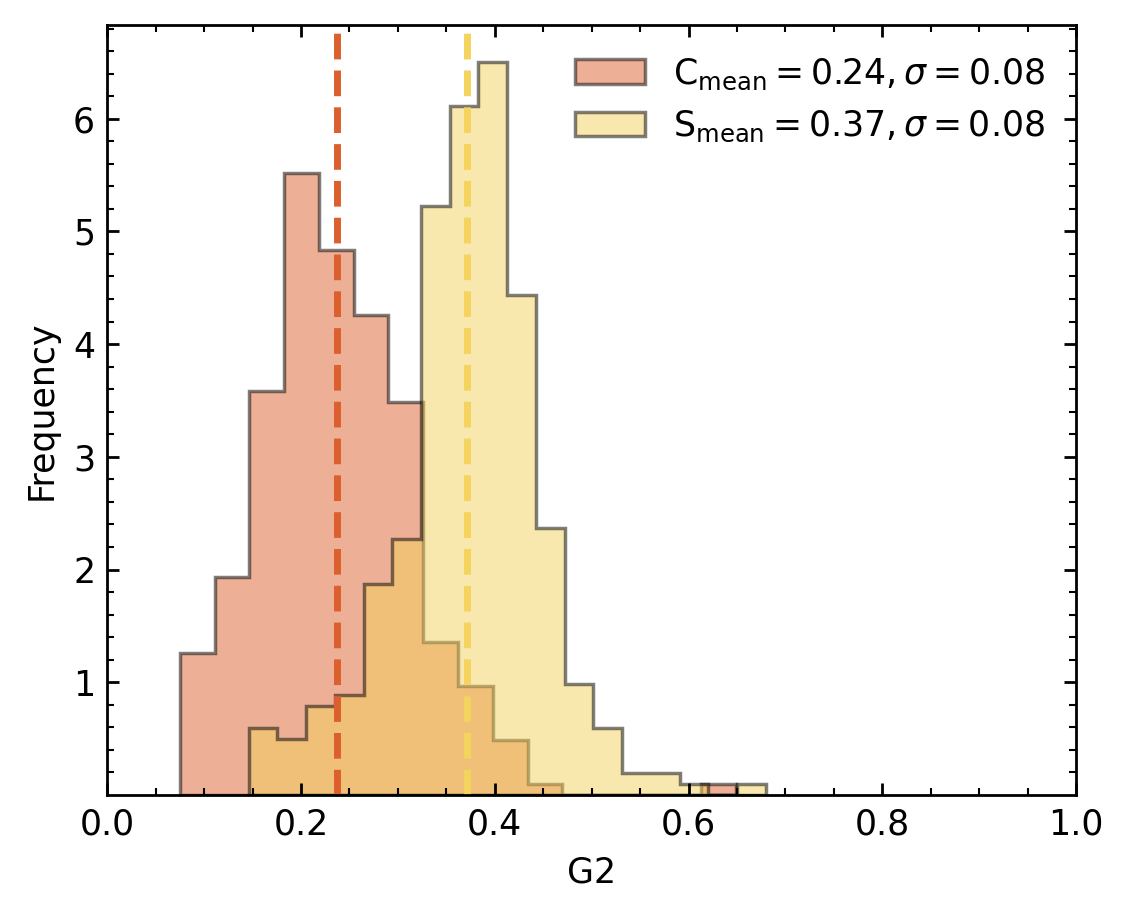}
  \includegraphics[width=4cm]{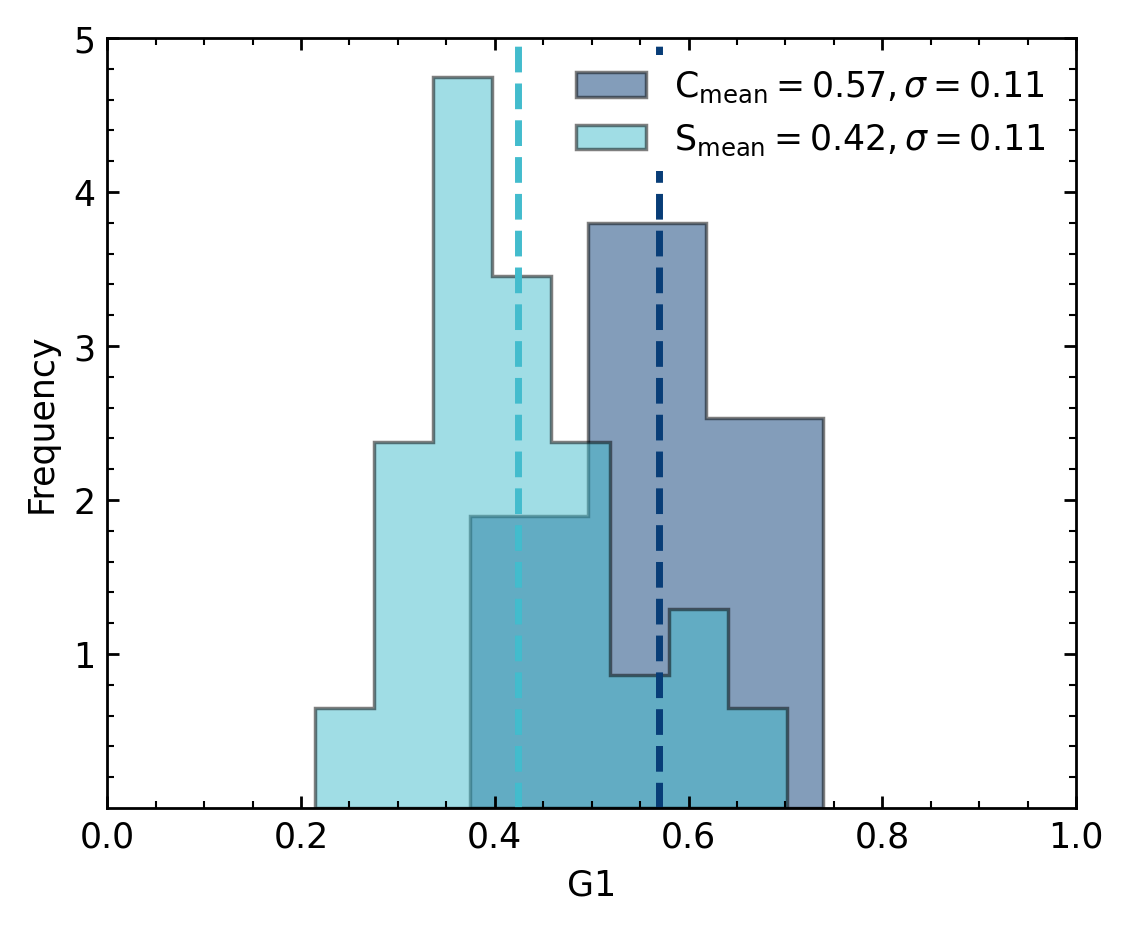}
 \includegraphics[width=4cm]{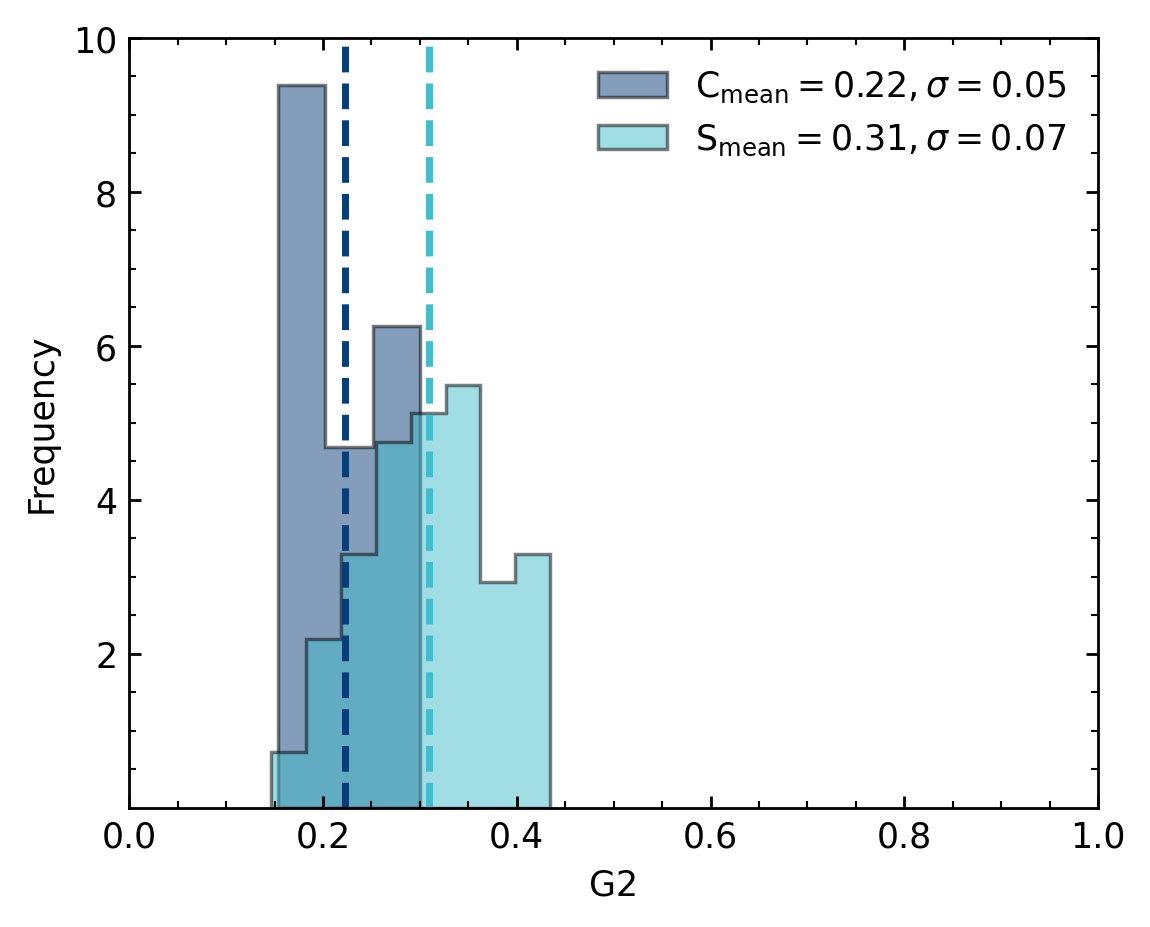}
\caption{Distributions of G1 (left) and G2 (right) parameters for C-type and S-type asteroids. The top panels represent the orange filter, and the bottom panels the cyan filter. The means of the distributions are indicated by dashed lines.}
\label{g1g2_dist_CS}
\end{figure}

Next, in Figure \ref{g1g2_dist_CS} we analyze the distributions of the parameters for C-type and S-type asteroids. One notable observation is that the mean G1 and G2 values for C-type asteroids are similar across both filters. However, for the S-complex, the mean G1 value is lower in the orange filter and the mean G2 value is lower in the cyan filter, which aligns with findings reported by \cite{wilawer2024}. According to these authors, this difference can be attributed to the well-known correlation with geometric albedo. Asteroids with higher albedos tend to exhibit flatter phase curves. Additionally, since S-type asteroids are redder, their phase curves in the cyan (blue) filter are flatter, which results in lower G2 values. To further investigate the statistical differences between the distributions of G2 values, we performed a Kolmogorov-Smirnov (KS) test. This was motivated by the fact that, although the means of the distributions were different, the high standard deviation made it insufficient to conclude that the distributions were distinct. This test assesses whether two distributions are statistically different. At a 95$\%$ confidence level, we can reject the null hypothesis—that the two distributions are identical—in favor of the alternative hypothesis, which suggests that the data come from different distributions. The p-value obtained was significantly less than 0.05, indicating a strong statistical difference between the G2 values for the cyan and orange filters in S-type asteroids (see also Figure \ref{p-values_sametax} and the accompanying discussion).

\begin{figure}
\centering
 \includegraphics[width=\columnwidth]{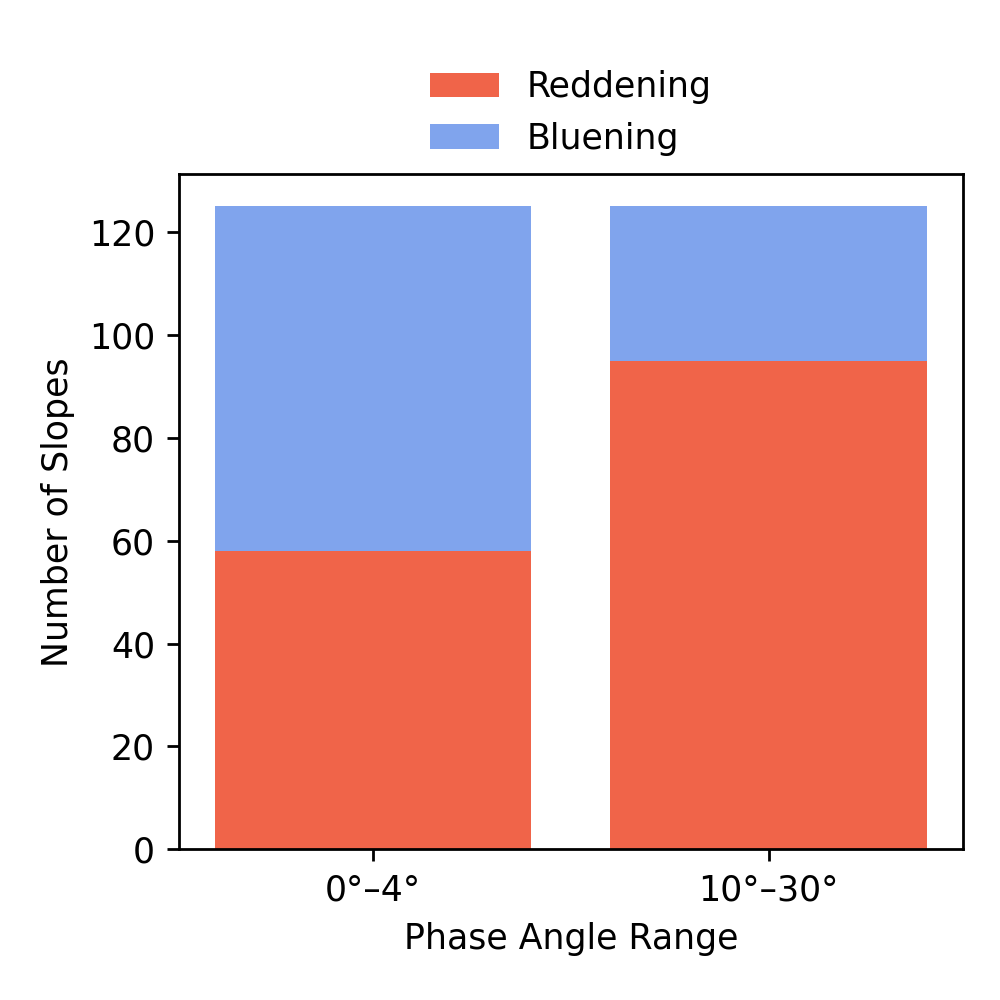}
\caption{Stacked bar chart showing the number of spectral slopes differences with reddening (positive slopes differences, in red) and bluening (negative slopes differences, in blue) across two phase angle ranges: 0\textdegree-4\textdegree and 10\textdegree–30\textdegree. The y-axis represents the count of slopes, while the x-axis categorizes the phase angle ranges.}
\label{phasered}
\end{figure}

For this work we also employ the method proposed by \cite{mahlke2021} to calculate the differences in the slopes of the phase curves observed at different effective wavelengths, using the parameters obtained by fitting them in the cyan and orange bands of ATLAS. The spectral slope, expressed in units of $\%$100 nm is: 
\begin{equation}
    S_s = \frac{10^{-0.4(m_o - m_c)}-1}{\lambda_o - \lambda_c} \times 10^4,
\end{equation}
\noindent
where $m_o$ and $m_c$ are the reduced magnitudes in the ATLAS filters, $\lambda_o$ = 518 nm and $\lambda_c$ = 663 nm are the effective wavelengths, and we are normalizing the reflectance at $\lambda_c$. The difference between the magnitudes $m_o$ and $m_c$ is obtained using the phase curves: $\Delta m = m_o(\alpha, H_o, G1_o, G2_o) - m_c(\alpha, H_c, G1_c, G2_c)$. These spectral slopes are computed at 10\textdegree-and 30\textdegree of phase angle and the difference is the slope indicates the direction of coloring (reddening [+] vs. blueing [-]). Additionally, we studied the slope difference between 0\textdegree and 4\textdegree phase angle to explore variations at lower phase angles. For this analysis, we used only the highest-quality parameters (with errors below 15$\%$). Of these 125 slopes, the nominal differences suggest bluening for 67 objects and reddening for 58 in the 0\textdegree–4\textdegree range, indicating a near balance in the direction of the coloring effect at small phase angles. However, in the 10\textdegree–30\textdegree range, reddening is more probable for 95 slopes, while bluening is more probable for 30 slopes, suggesting reddening as the dominant trend at larger phase angles (Figure \ref{phasered}). This pattern is also evident in Figures \ref{red} and \ref{blue}. It should be however noted that the uncertainties in these results are significant and should be systematically evaluated in future studies.

\begin{figure*}
    \centering
 \includegraphics[width=4cm]{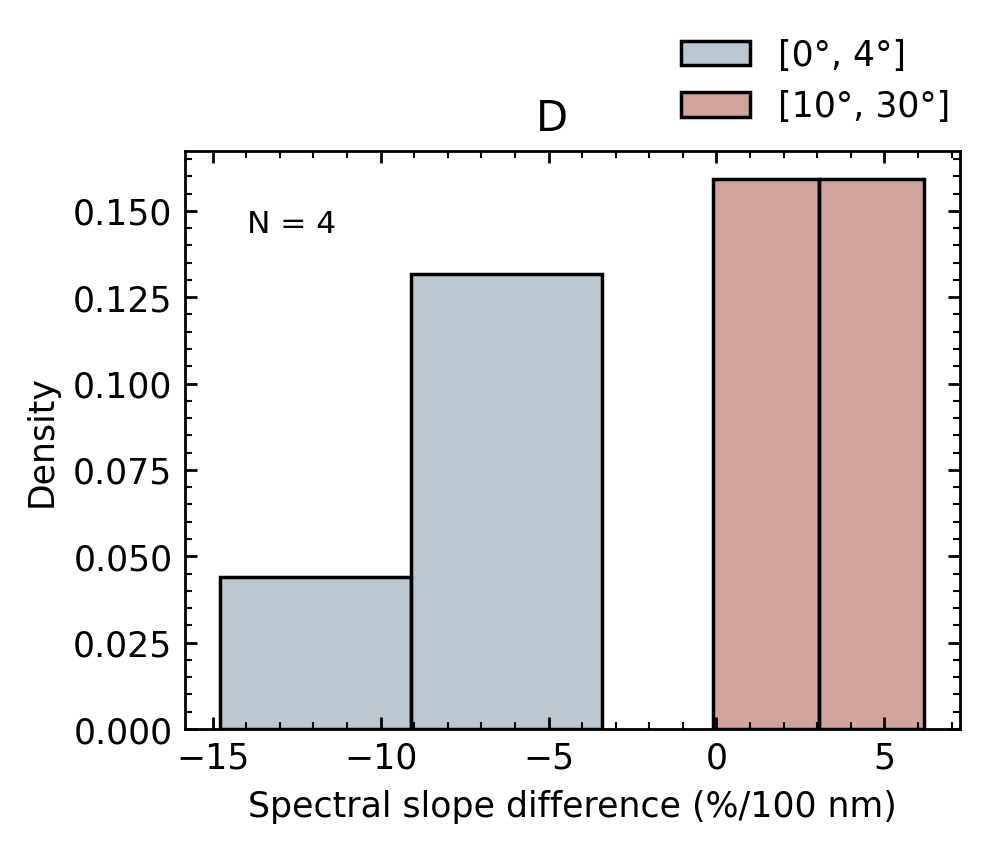}
  \includegraphics[width=4cm]{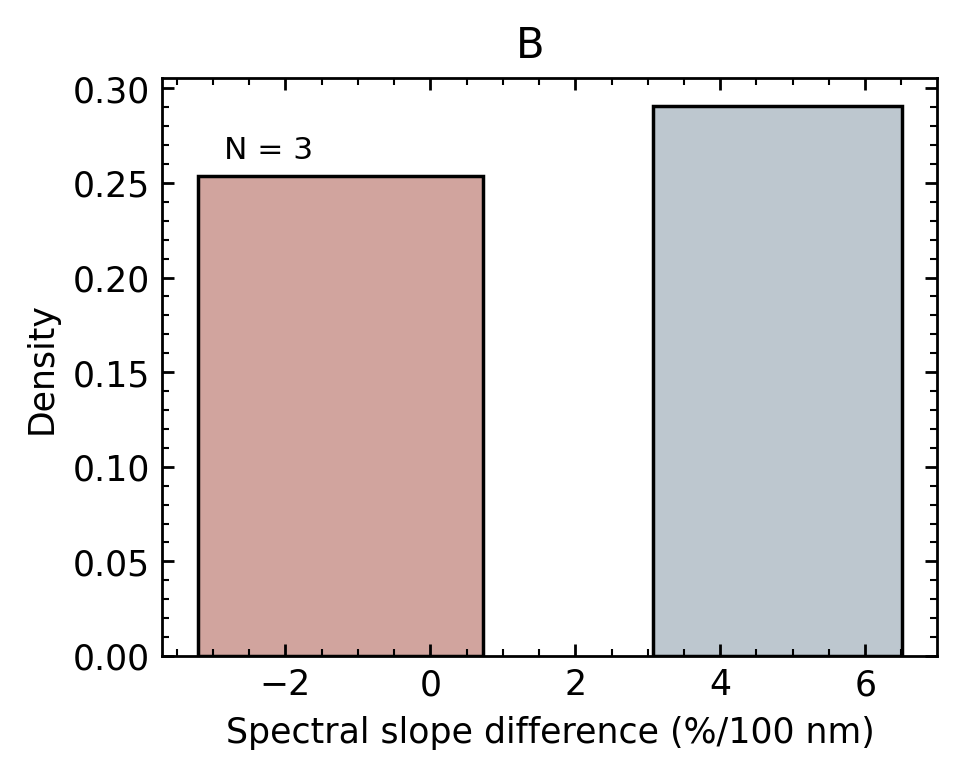}
   \includegraphics[width=4cm]{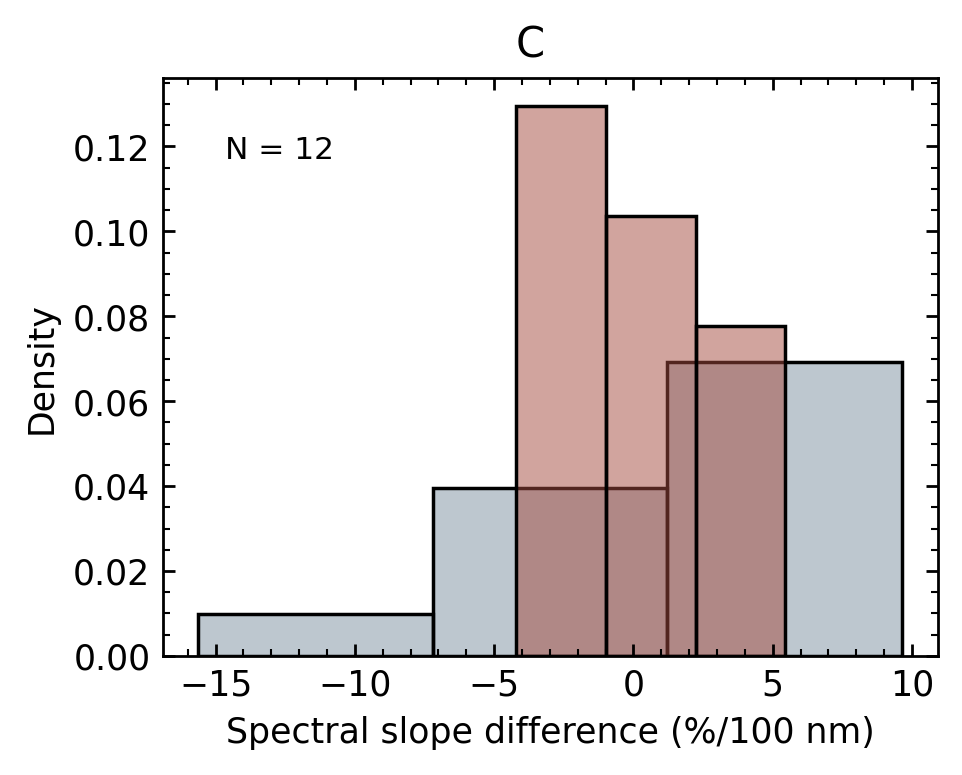}
  \includegraphics[width=4cm]{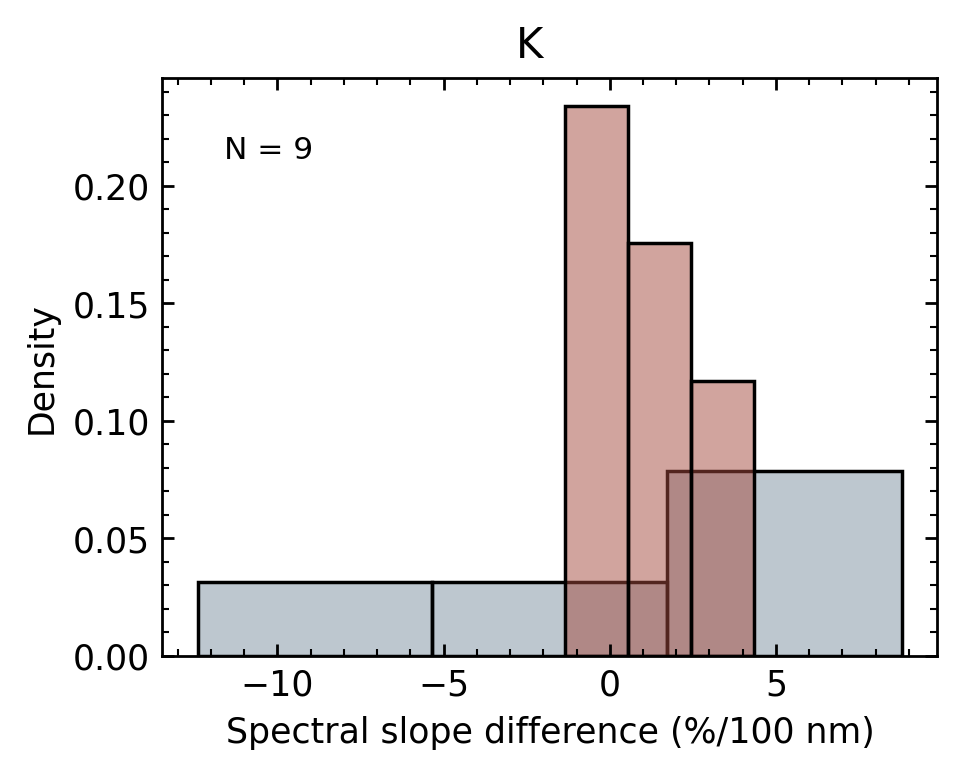}
  \includegraphics[width=4cm]{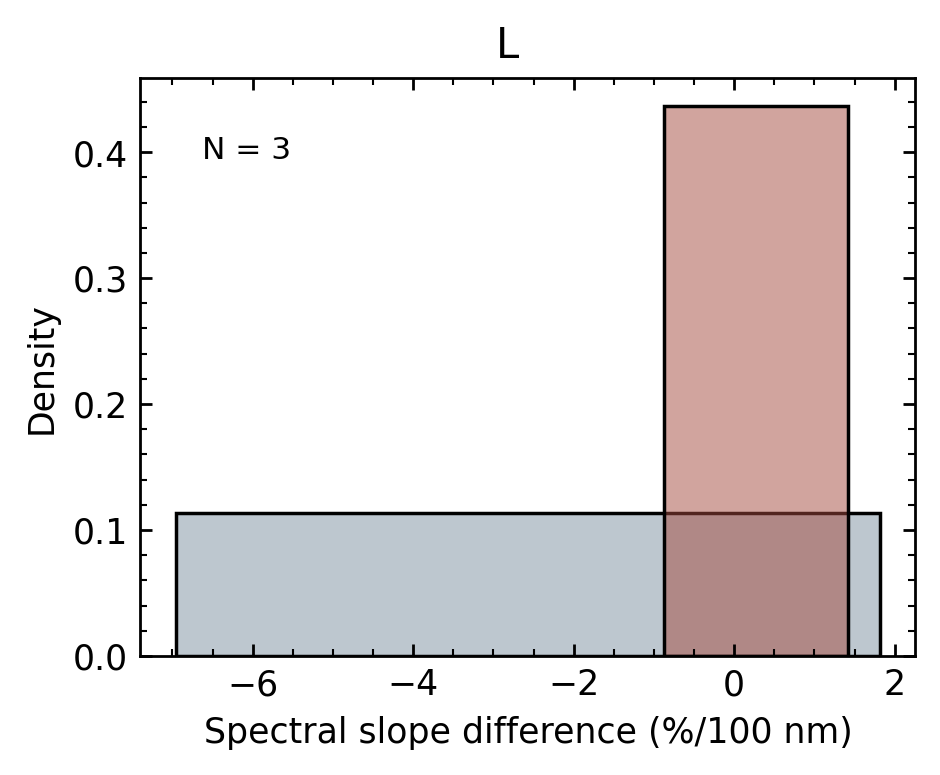}
   \includegraphics[width=4cm]{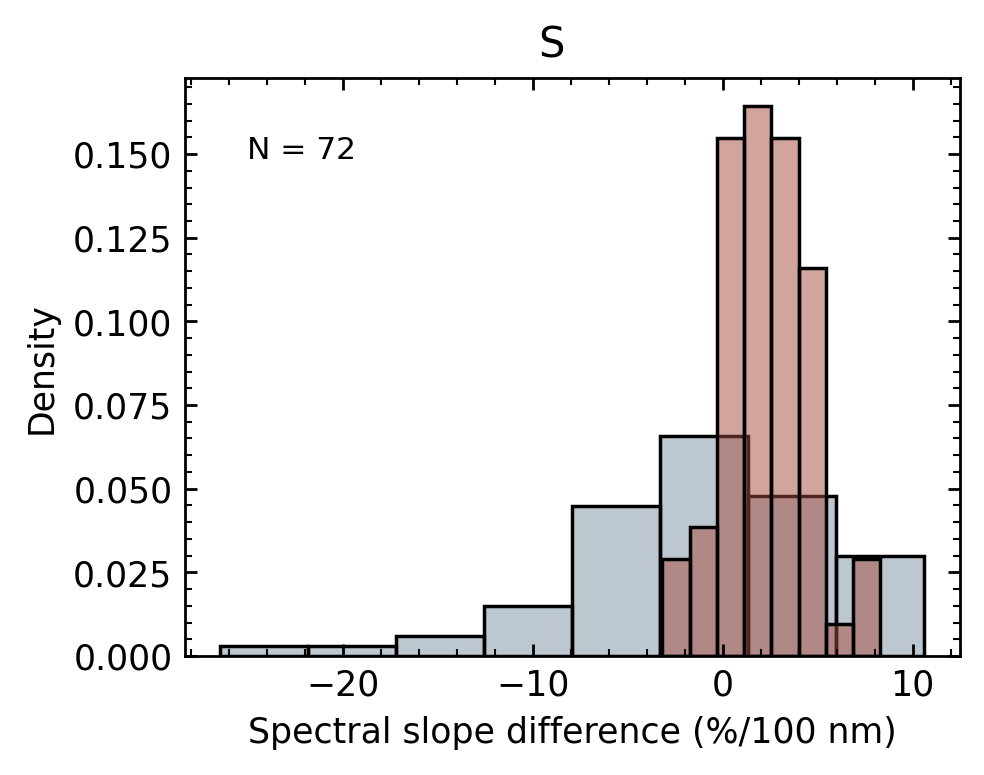}  
     \includegraphics[width=4cm]{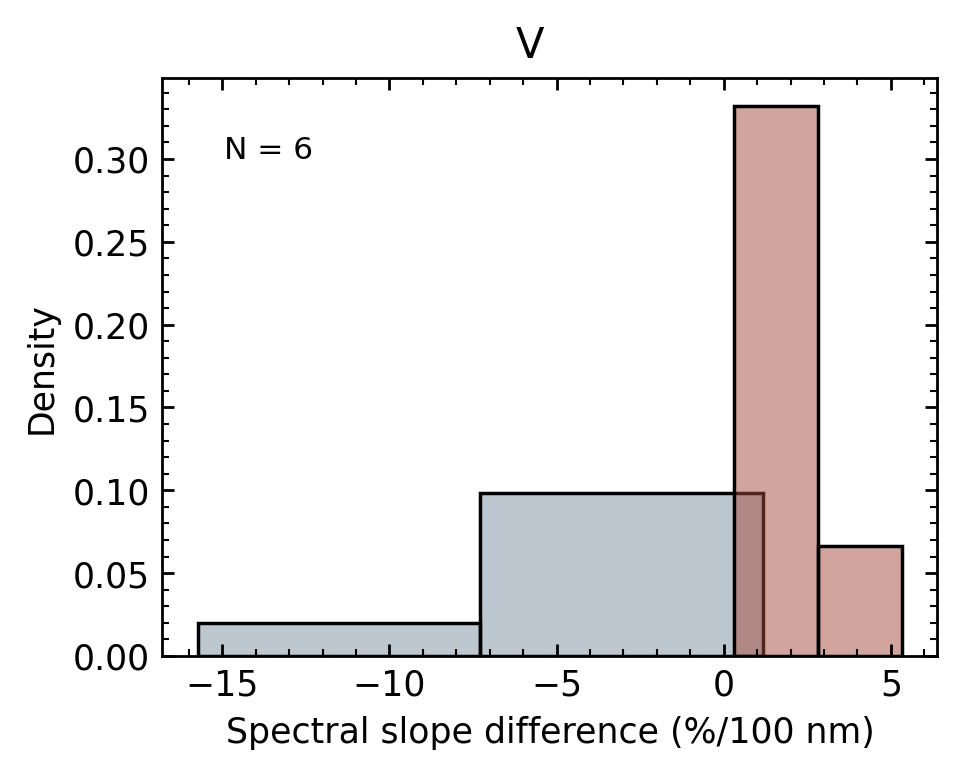}
   \includegraphics[width=4cm]{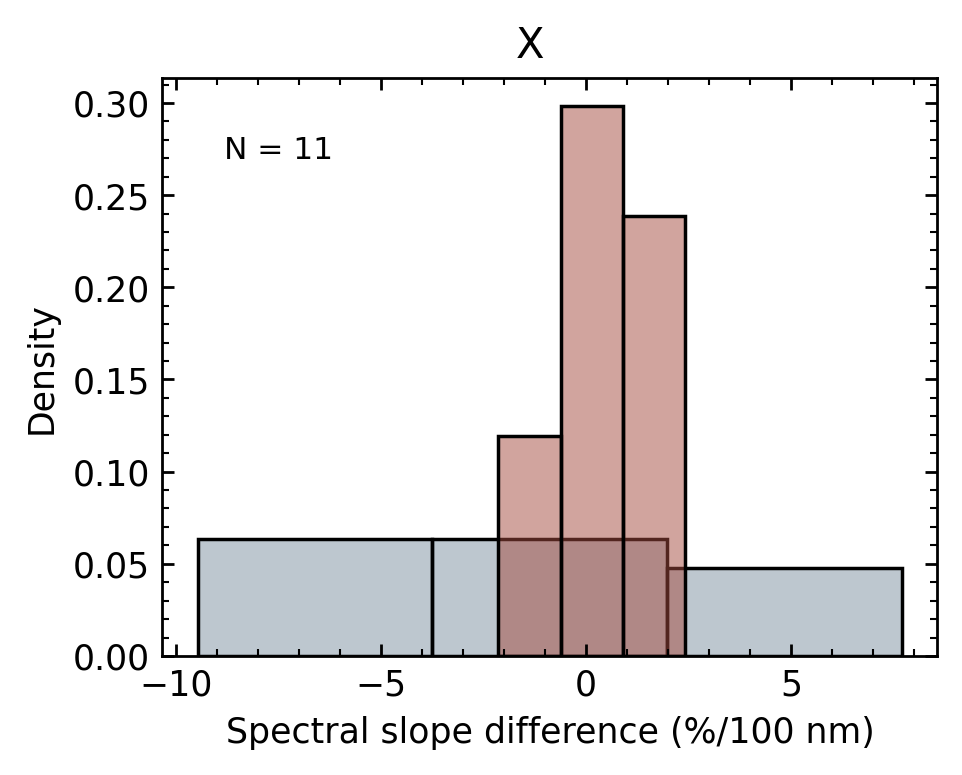}  
   \caption{Distributions of spectral slope differences between 0\textdegree–4\textdegree\ (gray) and 10\textdegree–30\textdegree\ (brown) for different taxonomic types. Positive values indicate reddening, while negative values represent bluening.}
   \label{red-blue}
\end{figure*}

In Figure \ref{red-blue}, we present the results of spectral slope differences for various taxonomic types. Although the sample sizes are relatively small, this study represents an important starting point for analyzing the phenomenon of phase coloring across different phase angle ranges and taxonomic classes. As shown in Figure \ref{red-blue}, both bluening and reddening can be detected at small phase angles. However, the distributions shift towards positive values (reddening) when analyzing spectral slope differences in the 10\textdegree–30\textdegree\ range, making reddening the more common trend. The B-complex exhibits the opposite behavior, with reddening at small phase angles and bluening at larger angles, as expected \citep{Loeffler2022}. The V-complex shows some potential bluening at small angles, but only reddening is observed in the 10\textdegree–30\textdegree\ range.

However, several caveats may be interfering with this analysis. First, as noted in \cite{wilawer2024}, ATLAS filters are too close in wavelength, making it difficult to detect significant variations across wavelenghts. Also, in our method, we are not considering the rotation effect or the shape of the asteroid, which could introduce biases in the determination of the phase curve parameters, which translate into biases in the slope determination \citep{CarvanoDavalos2015}. A detailed phase coloring analysis would require precise estimates of the slope values of the phase curves, which is a limitation for simple methods such as the one used in this work and for sparse data such as ATLAS.

We also compute the color slopes, as the difference of the first derivatives of the photometric phase function $H,G_1,G_2$ with respect to the phase angle in the orange and cyan filters, following \cite{wilawer2024} work (see appendix for the definition of color slope). In Figures \ref{pr} we present examples of the color slope as a function of phase angle, the dashed line indicates the smallest phase angle of the observations in each filter. In this analysis, we found asteroids for which, when comparing the color slopes in both filters, the reddening (red line) or bluing (blue line) effect at small phase angles falls within the 1-$\sigma$ error envelopes. The shaded gray region in the figures corresponds to the range of uncertainty in the parameters G1 and G2, calculated from their respective uncertainties. In our sample, 58 asteroids exhibit reddening effect and 55 a bluening effect, within the margin of spectral slope uncertainties. The plots of these objects can be seen in the Appendix (Figure \ref{red}). These findings are consistent with the work of \citep{alvarez2024,wilawer2024}, who observed similar effects at small phase angles. More figures can also be found in the Appendix (Figure\ref{blue}).

It is important to highlight that the critical phase angle $\alpha$ found in this work, namely the phase angle at which the color slope changes behavior, is also close to 5-10\textdegree, as previously noted by \cite{alvarez2024, wilawer2024}. This behavior remains consistent despite the different methods used to calculate the parameters. Furthermore, the data comes from different surveys, as the authors utilized SDSS, resulting in a distinct filter system as well. \cite{wilawer2024} detected this coloring effect only for S-type asteroids. This might be due to the fact that the backscattering effect dominates the opposition region for high albedo surfaces \citet{belskaya2000opposition}. This effect is strongly wavelength-dependent, as it involves the constructive interference of light scattered by surface particles. Shorter wavelengths generally exhibit a stronger backscattering enhancement due to higher scattering efficiency, especially when particle sizes are comparable to the wavelength. Consequently, the opposition surge from coherent backscattering can be more pronounced at shorter wavelengths \citep{corey1995coherent}.

We emphasize that the most pronounced changes in color slope are observed at phase angles $\alpha < 10$ degrees, consistent with the findings of \cite{wilawer2024}. While the coloring effect is more challenging to detect in the 10\textdegree–30\textdegree\ range, our results for certain taxonomic types suggest that reddening is the most prevalent trend within this phase angle interval \citep{taylor1971,millis1976,lu1990}. 

\begin{figure}
\centering
 \includegraphics[width=6cm]{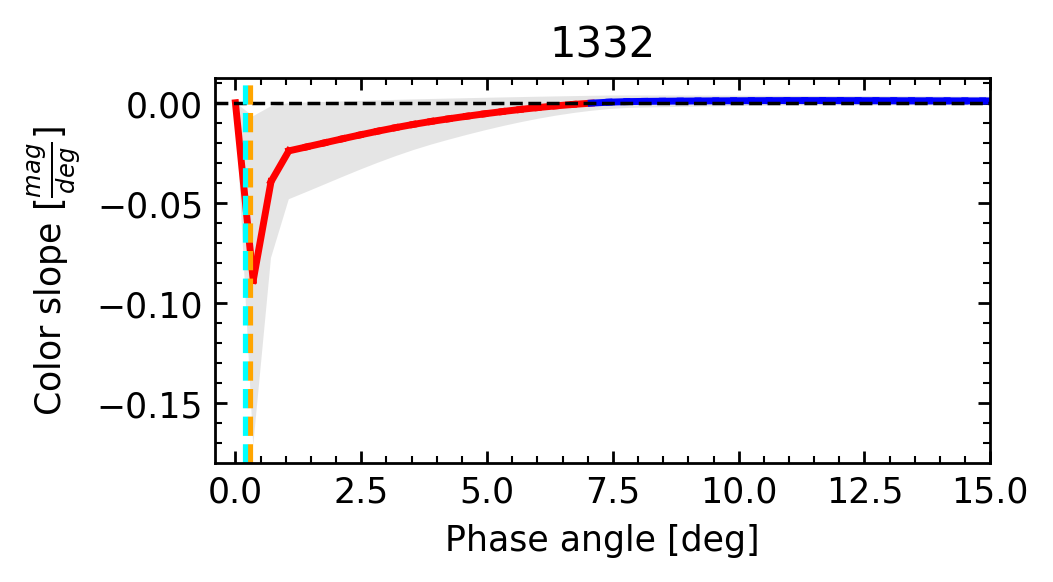}
 \includegraphics[width=6cm]{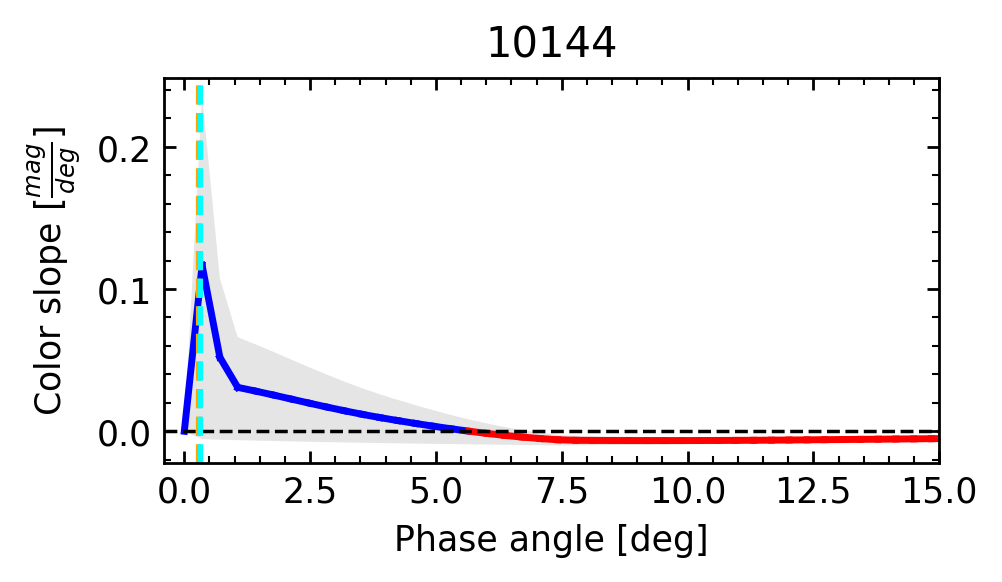}
\caption{Color slope examples, representing the difference between the first derivatives of the phase function with respect to the phase angle in the two filters for asteroids showing coloring effect (red corresponds to reddening and blue to blueing effect). Vertical lines indicate the minimum observed phase angles separately in both filters. The shaded area denotes the 1-$\sigma$ envelope.}
\label{pr}
\end{figure}

\subsection{Comparison with other methods}\label{complex-simple}

To assess the reliability of the results obtained with our algorithm, we compared the derived parameters with those published by \cite[W24]{wilawer2024}. This analysis serves as both a validation of our method and an exploration of the impact of additional data, highlighting the robustness and consistency of the results across different approaches. The authors calculated phase curves for 35 well-observed asteroids using sparsely sampled ATLAS data and other, more densely sampled, ground-based observations. The dense observations were used to derive asteroid models and their respective rotational period and rotation axis coordinates densities. Subsequently, ATLAS data were used to calculate phase curve parameters assuming a fixed shape. As shown in \citealt[Figure~2]{wilawer2024}, the procedure used is highly complex and reflects a high computational cost, including the challenge of adapting the algorithm for supercomputers. Specifically, the core-combined computation time required by the authors to derive the phase curves for 35 asteroids was 170$\,$000 hours.

As previously mentioned, our approach is much simpler and accounts for fewer effects, for example, we do not correct for rotational effect. It is important to highlight that when working with a sample of around 500$\,$000 objects, it is necessary to implement scalable algorithms with a reasonable balance between computational time and accuracy. Our algorithm takes an average of 1 core-minute to perform the outlier rejection and the phase curve fitting for the same sample of asteroids (without parallelization) using ATLAS observations.

\begin{figure}[ht]
\centering
 \includegraphics[width=8cm]{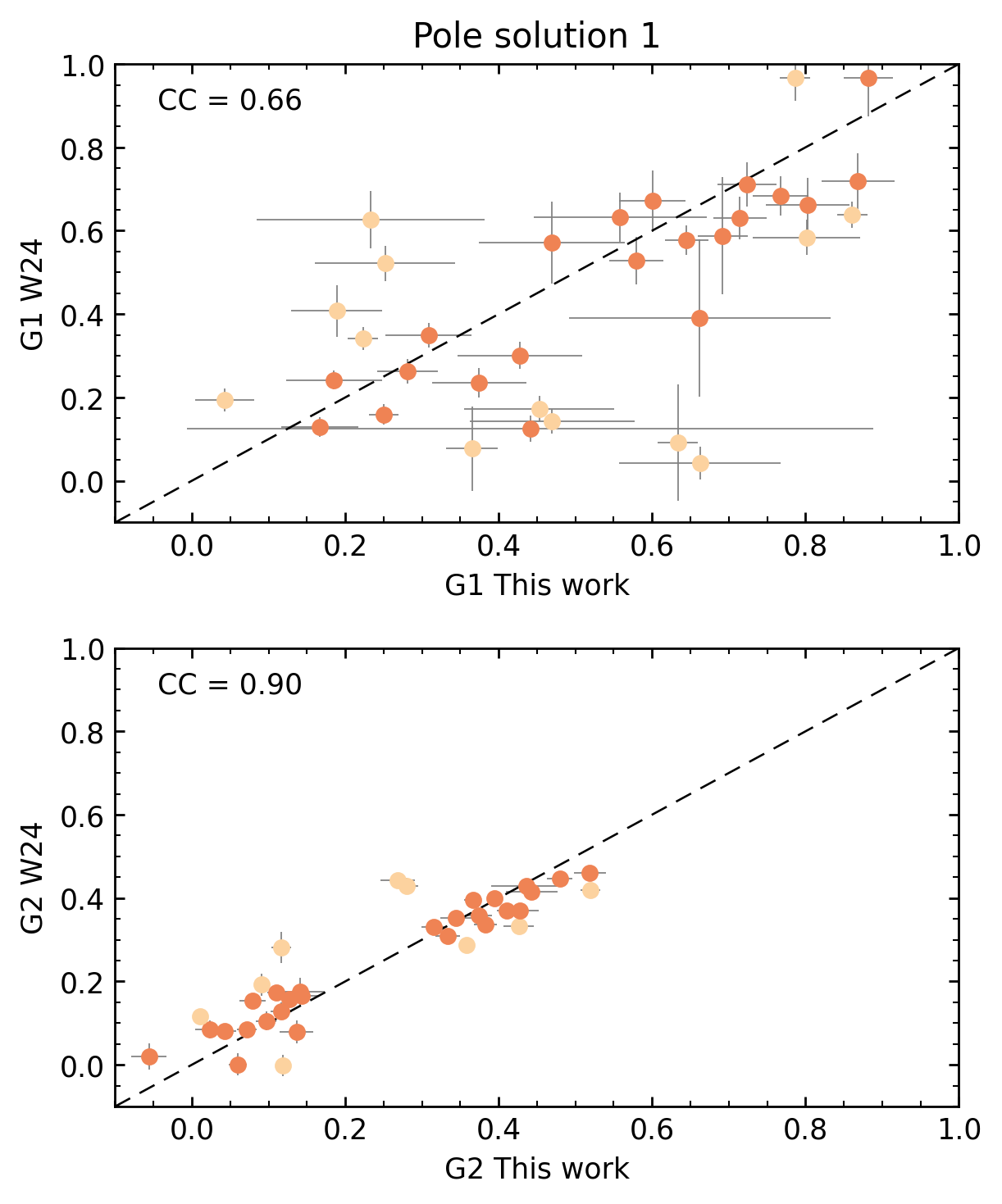}
 \includegraphics[width=8cm]{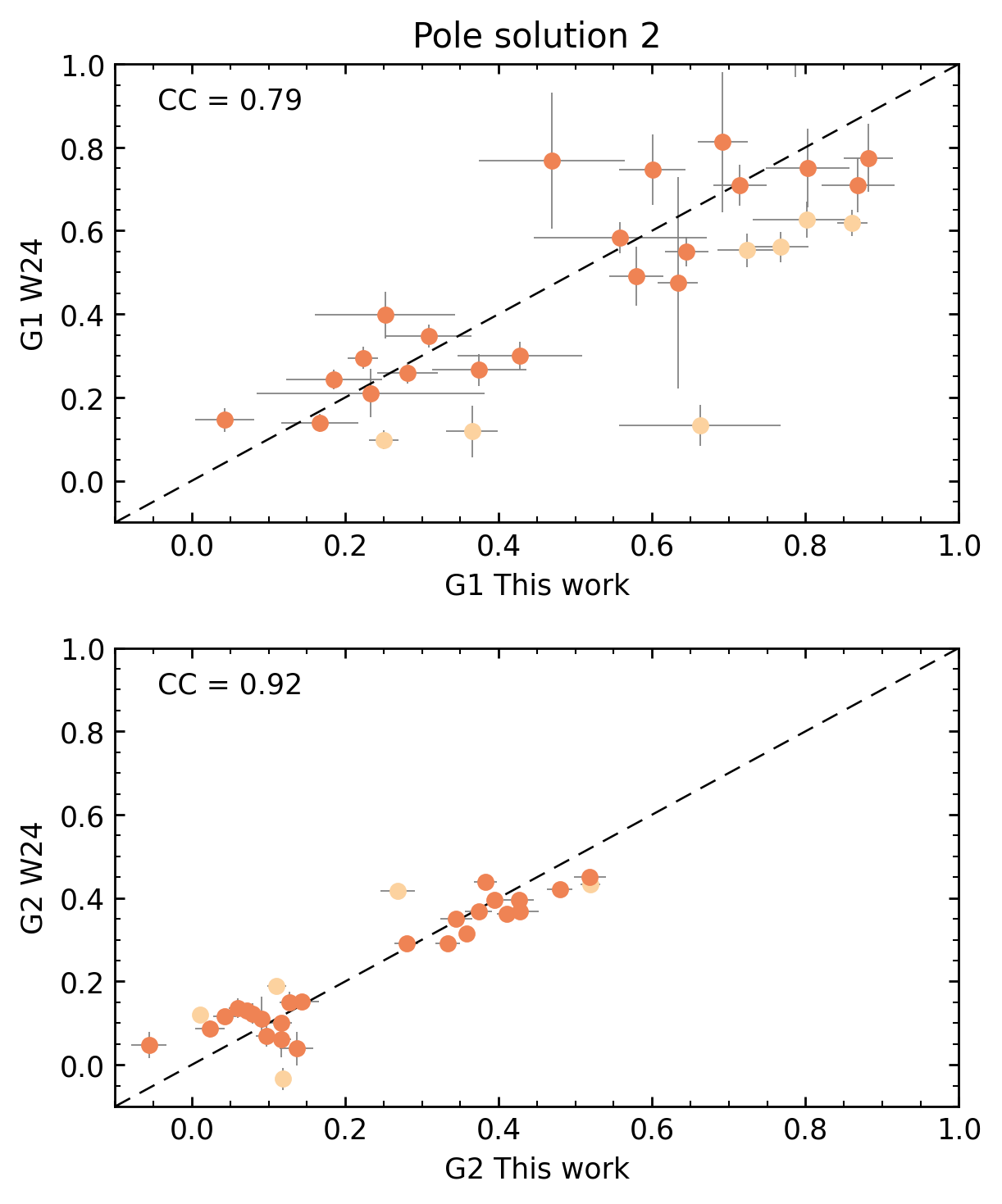}
\caption{Comparison between the G1 and G2 parameters obtained in this work and those derived by \cite{wilawer2024}, for the orange filter. The lighter colored points indicate no intersection between the parameter uncertainty intervals. The darker colored points represent regions where the uncertainty ranges do overlap. The correlation coefficient (CC) values are also indicated.}
\label{emilcomp1}
\end{figure}

\begin{figure}[ht]
\centering
 \includegraphics[width=8cm]{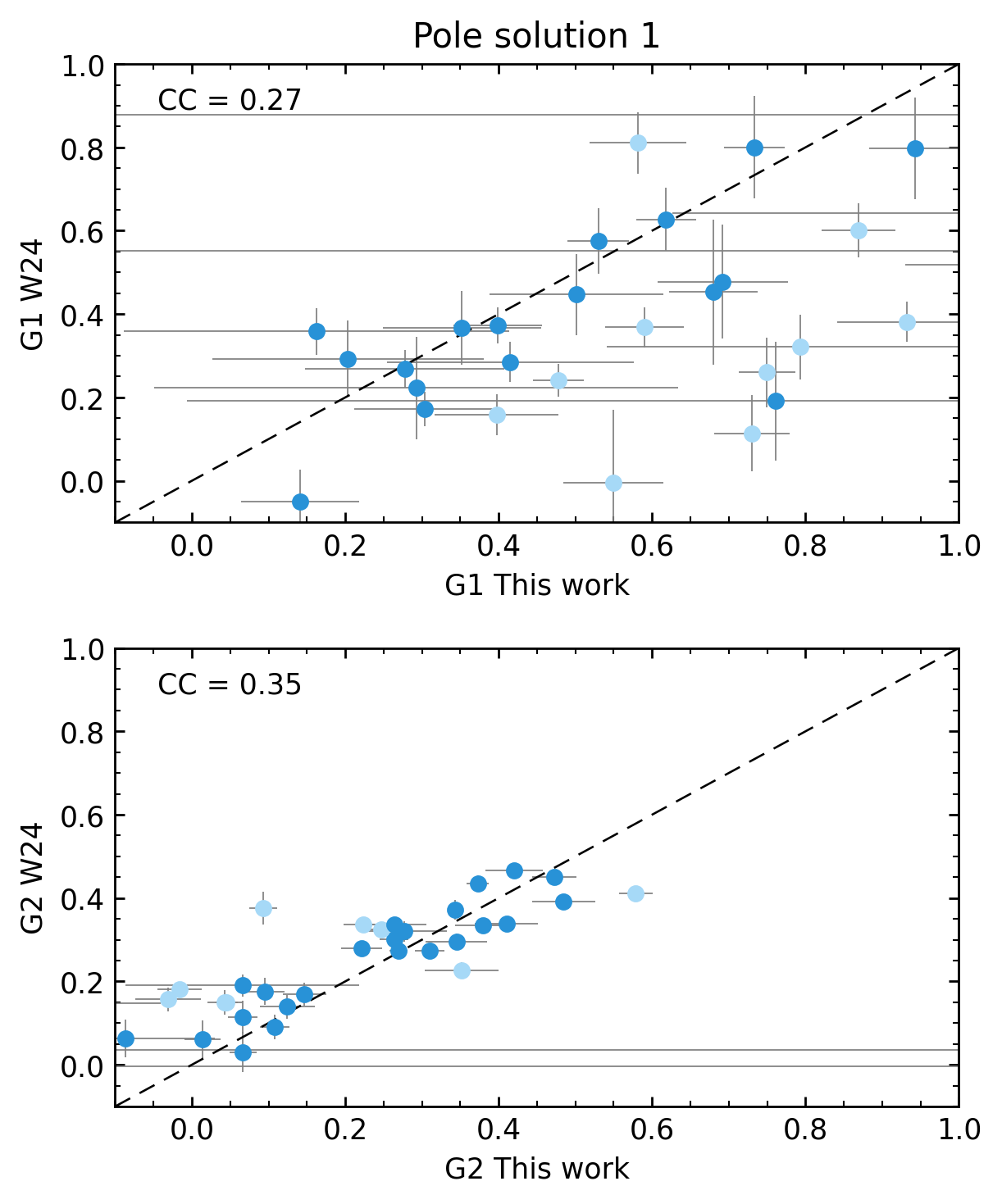}
 \includegraphics[width=8cm]{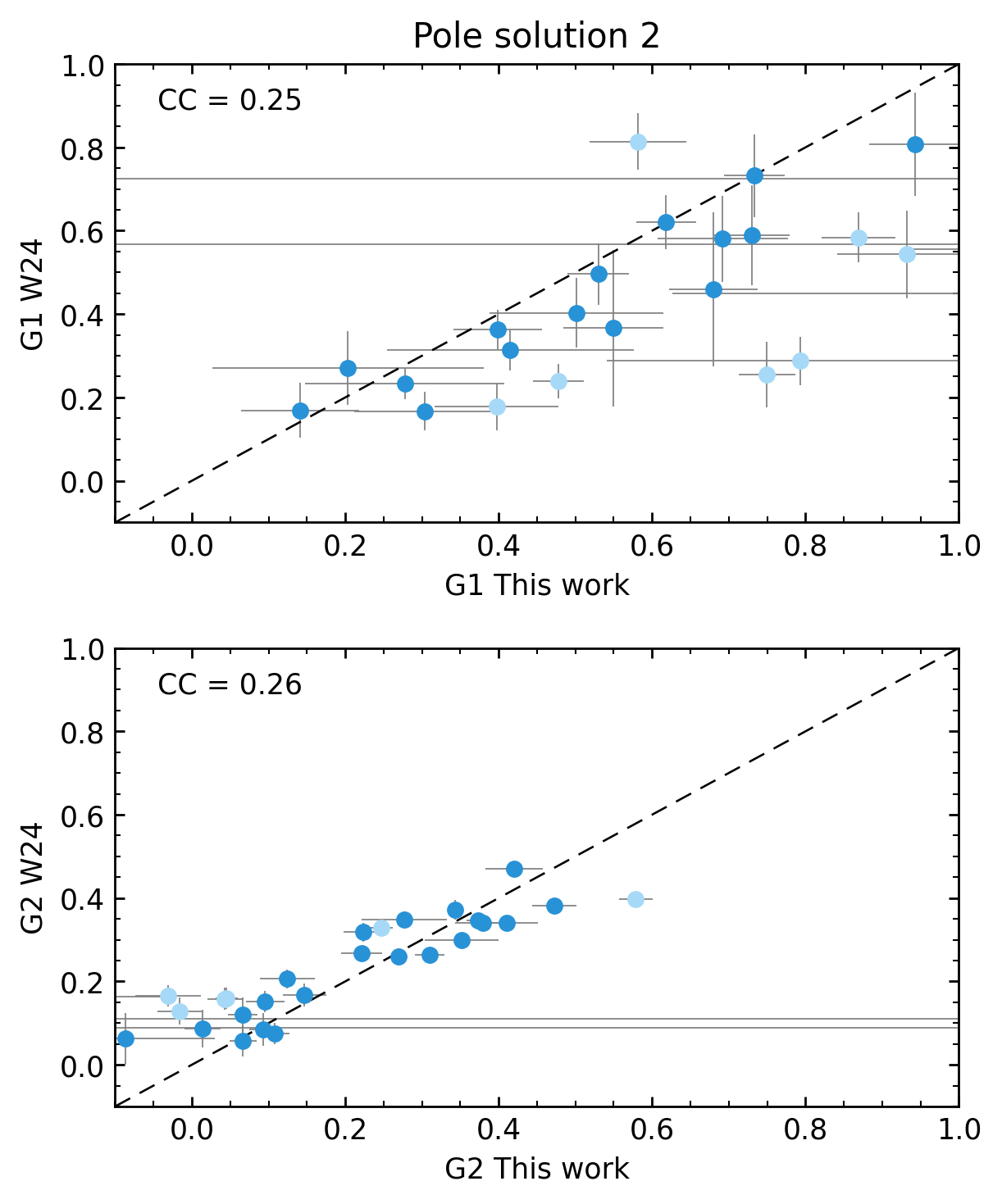}
\caption{Comparison between the G1, G2 parameters obtained in this work and those derived by \cite{wilawer2024}, for the cyan filter. The lighter colored points indicate no intersection between the parameter uncertainty intervals. The darker colored points represent regions where the uncertainty ranges do overlap. The correlation coefficient (CC) values are also indicated.}
\label{emilcomp2}
\end{figure}

As can be seen in Figures \ref{emilcomp1} and \ref{emilcomp2}, the results obtained in this work are similar to those reported by \cite{wilawer2024}. Moreover, most of them are within the uncertainties intervals. The parameters that differ significantly from the values obtained by the authors correspond to pathological cases also in their article, classified into the groups `Objects with one realistic solution and one solution with unrealistically large opposition-effect amplitudes' and `Objects for which the taxonomic type is not in agreement with the phase-curve parameters'. We also note that our algorithm differs most from their results, for values of G1 on the order of $\sim0.1$. In most cases, this is due to the low density of points in the opposition region.

\begin{figure}
\centering
 \includegraphics[width=4cm]{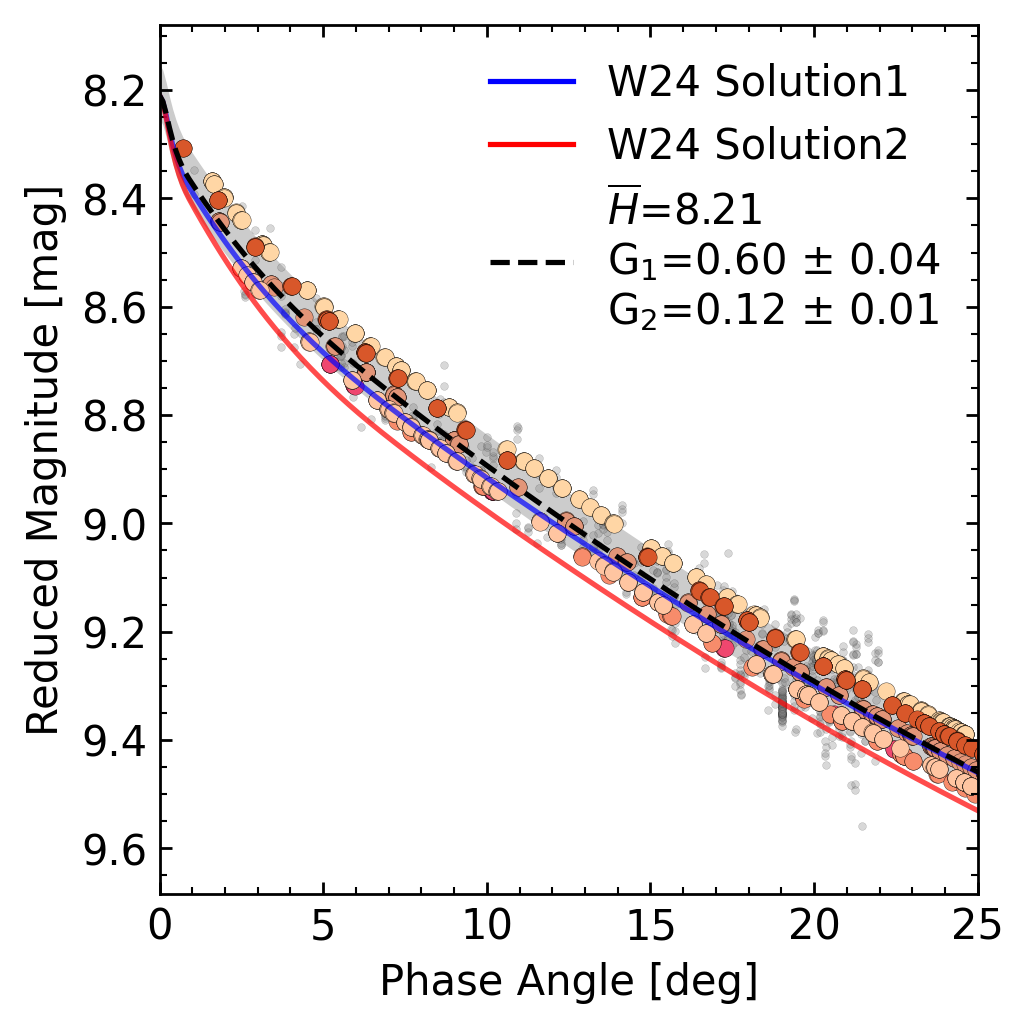}
 \includegraphics[width=4cm]{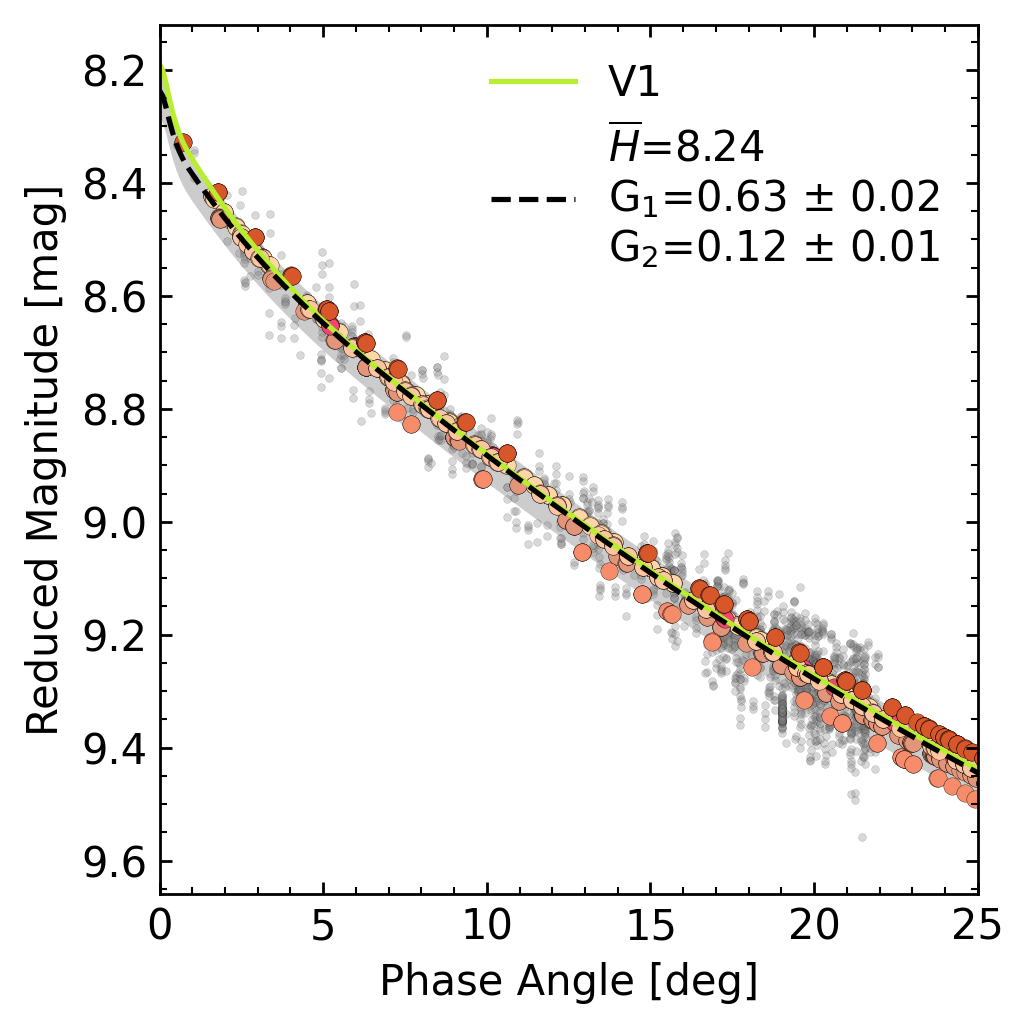}
\caption{Phase curves for (127) Johanna in orange filter for Atlas SSCAT V1 (left) and V2 (right). The dashed lines indicate our fit using the mean H value, while the gray shaded area represents the range of absolute magnitudes derived from multiple apparitions. The gray points represent the observations after outlier rejection. The colored points represent the residuals of the data from different apparitions. Left panel: The blue and red lines represent the W24 phase curves using the two pole solutions \citep{wilawer2024}. The dashed line corresponds to the fitted phase curve obtained using the algorithm presented in this work applied to Atlas SSCAT V1 observations.
Right panel: The green line indicates our Atlas SSCAT V1 fitted phase curve, while the dashed line represents the fitted phase curve for V2.}
\label{johanna}
\end{figure}

In addition, in Figure \ref{johanna}, we show an example of the phase curve of asteroid (127) Johana compared with both models of W24. In turn, we can notice that in version 2 of the ATLAS catalog (in contrast to V1 used in W24) we have more data in the opposition range. Specifically, in version 1 there are 39 observations at phase angles less than 5 degrees, while in version 2 there are 88 observations. Let us note that, by better modeling the opposition effect, the value of G1 is affected but the value of G2 remains the same. This supports the evidence that the G1 parameter is much more sensitive to the data set used for the fit (discussed in Section \ref{complex-simple}). 

\begin{figure*}
\centering
 \includegraphics[width=5cm]{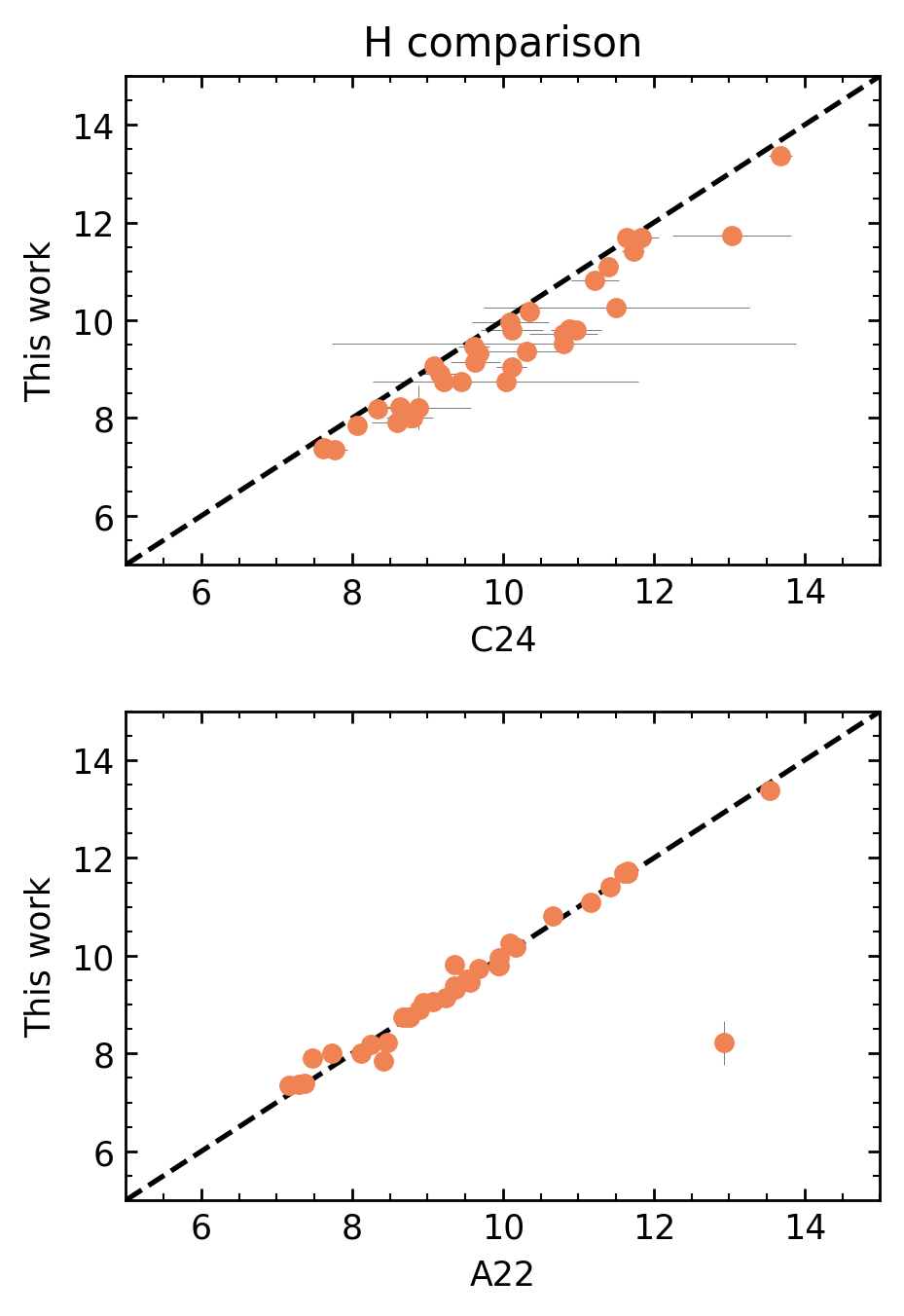}
 \includegraphics[width=5.2cm]{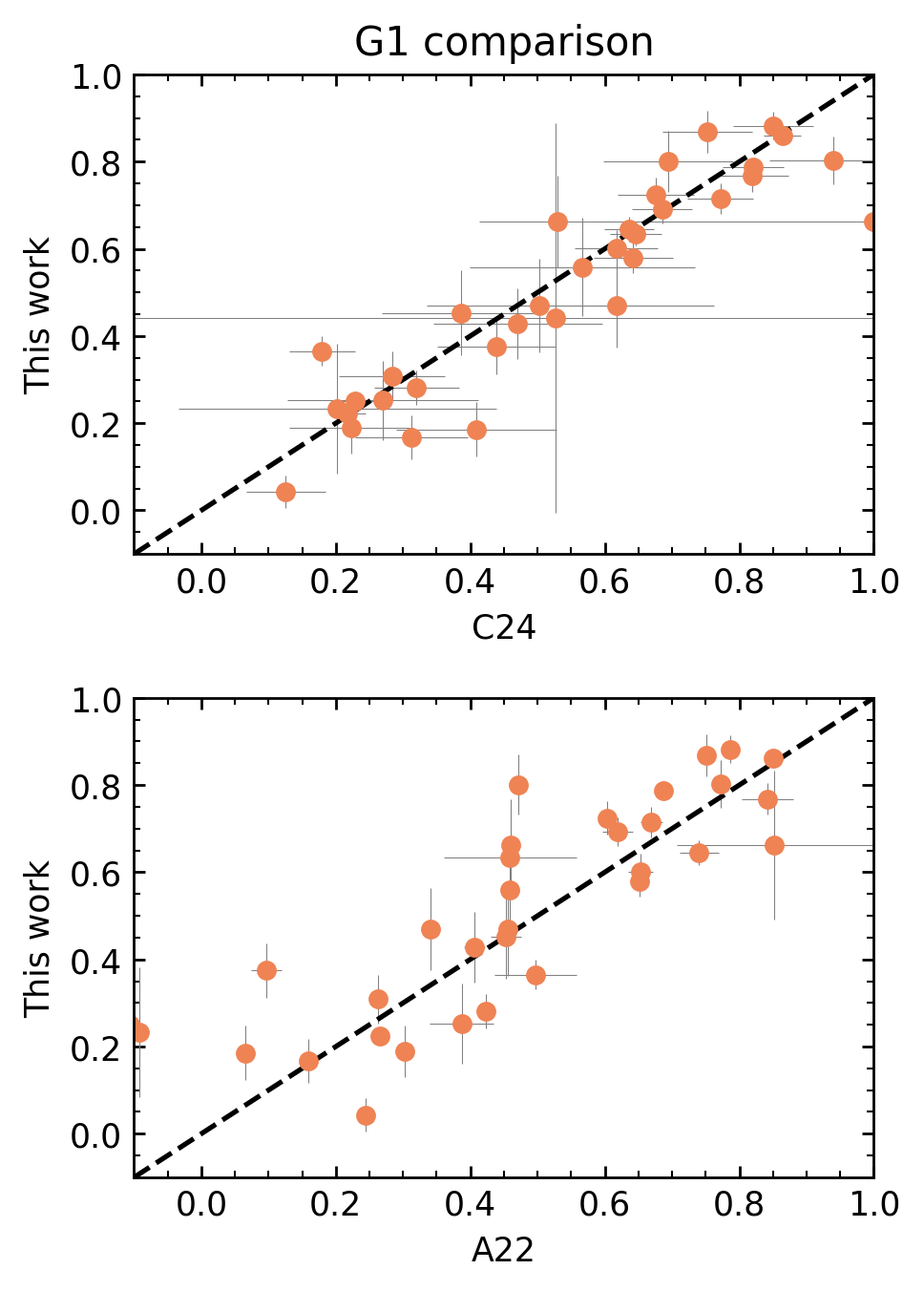}
 \includegraphics[width=5.2cm]{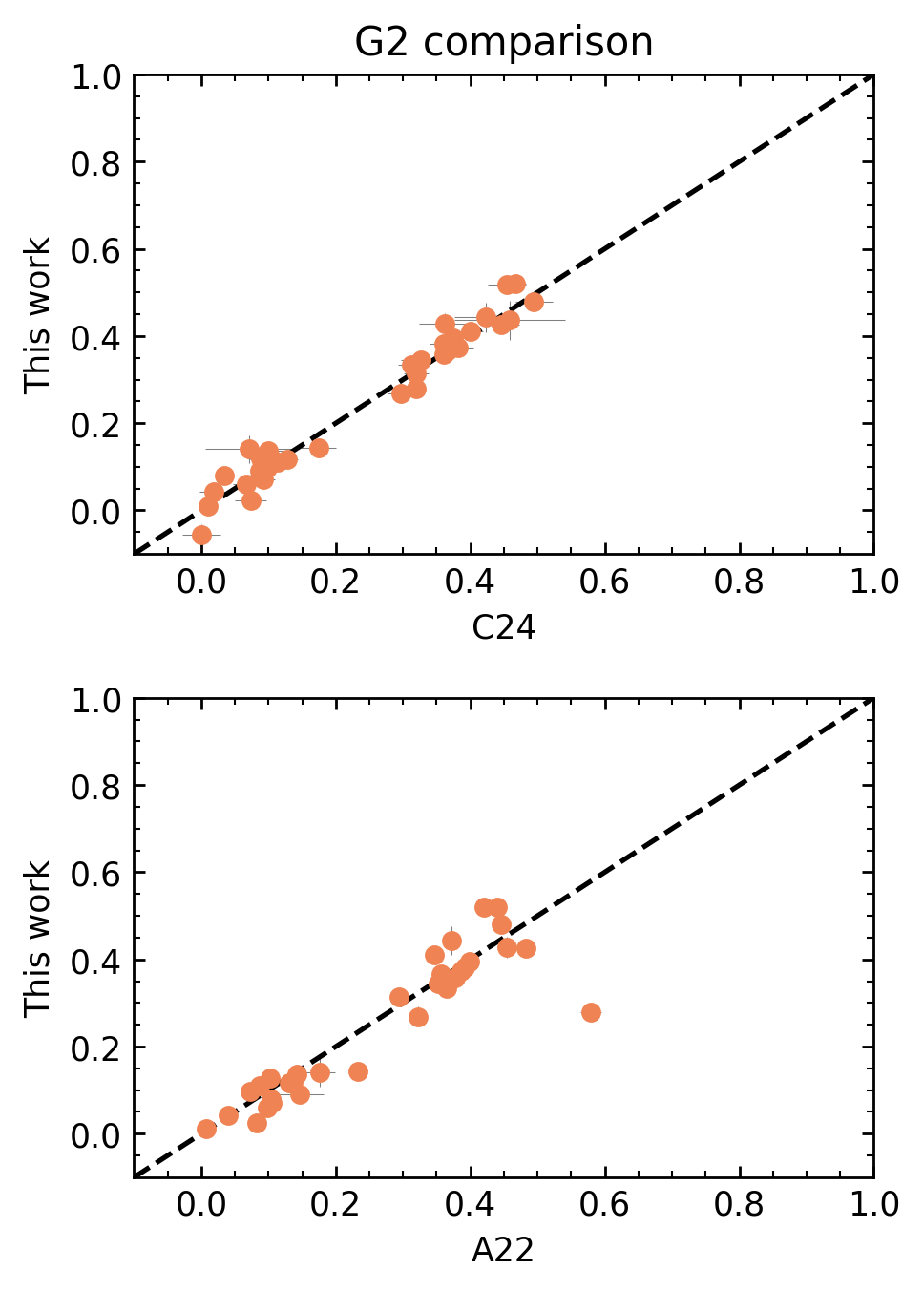}
\caption{Comparison of H, G1, G2 parameters derived with different methods using data in the orange filter.}
\label{methods-comp}
\end{figure*}

To further validate our results, we compared with two other methods in the literature: \cite[C24]{carry2024} and \cite[A22]{alvarez2022}. For this, we also use observations of the same 35 asteroids in the orange filter, that we previously used to compare with W24. \cite{carry2024} proposed a new form of the phase function H, G1, G2 to consider the variability between apparitions, adding a term dependent on the asteroid orientation, a function of its oblateness (R) and its equatorial coordinates ($\alpha$, $\delta$). For N filters, the proposed function requires determining 3 $\times$ (N + 1) parameters. The fit is performed using a least squares method, and to ensure physically plausible values it is necessary to include boundary values in the fit. As seen in Figure \ref{methods-comp}, the values obtained with our method and that of \cite{carry2024} are similar, which makes us think that fitting the traditional H, G1, G2 function separating by apparitions or adding the new term s($\alpha$, $\delta$) are comparable approaches. However, the absolute magnitudes H appear to be systematically slightly fainter than ours. First, the method of C24 corresponds to absolute magnitudes at equatorial geometry and our method to apparition-averaged geometry. Second, this discrepancy may also be due to a bias in the data, such as a higher number of observations at the lightcurve maximum for small phase angles and at the minimum for large phase angles, which influences the fit. This is a typical situation where the narrow opposition region is not well covered. Moreover, there are systematic differences when comparing with this method, as ours produces an average across multiple apparitions, while the C24 method is reduced to equatorial geometry. In this case, the computation time for both algorithms is close to one minute for 35 asteroids (per filter).

\cite{alvarez2022} proposed a method that integrates Monte Carlo simulations and Bayesian inference to consider nominal error in magnitudes and the effects of rotational variations in the calculation of phase curves, using sparse data obtained from large photometric surveys. The present code also includes an estimation of the probability of having a certain light curve amplitude (allowing amplitudes greater than 1 mag). For this method, it is not necessary to separate between apparitions, as the absolute magnitudes reflect an average value, with the changing aspect angle implicitly accounted for through the probability distribution of rotational amplitudes. As we can see in Figure \ref{methods-comp}, the results are consistent with ours, including the absolute magnitudes H. However, obtaining the parameters of 35 asteroids took 32 minutes for 1000 MC iterations, as we did not parallelize the algorithm.

As we can see, the G2 parameter shows a lower sensitivity to the data and methods used, which implies that its value remains relatively constant among all the algorithms analyzed. On the other hand, the G1 parameter shows the largest differences, indicating that it is more susceptible to fluctuations in the data and the methods used. Importantly, G1 models the opposition effect, a region that generally lacks sufficient data. This combination of the complexity of the opposition effect and the sparsity of data in that area makes accurate determination of G1 challenging.

\section{Conclusions}\label{conclusions}
We have determined the phase curves for 52$\,$712 (o), and 31$\,$441 (c) asteroids within P16 conditions using ATLAS SSCAT V2. Although the model used is simple and only considers the apparition effect without taking into account the rotation of the asteroid, the results obtained show good consistency compared to 
more sophisticated methods as the one developed by W24.
It is worth noting the advantage of our approach in terms of computational efficiency since we can calculate the parameters for 35 asteroids in only 1 minute without parallelization), while the more complex method requires up to 170 thousand total core hours to perform the same analysis. In the era of large sky surveys, it is critical to find the right balance between accuracy and computational time when developing analysis methods for parameter determination. Additionally, we compared our method with those of C24 and A22. The results among all methods turned out to be similar.

A crucial tool when working with large databases is cross-matching, which allows us to integrate additional information from different sources. For example, combining our phase curve parameter database with a taxonomy database allowed us to study the behavior of G1 and G2 parameters according to taxonomic classification. We found that S- and V-type asteroids tend to locate in the G1 < G2 region (indicative of medium-high albedos), while darker asteroids, such as C- and B-type asteroids, are located in the G1 > G2 region. Furthermore, we note that for D-, C-, K- and S-type asteroids, a dependence on G1 and G2 with the filter is observed. This behavior had already been reported by \cite{wilawer2024} for S-type asteroids, who suggest that this difference may be related to the well-known correlation with geometric albedo: asteroids with higher albedos tend to present flatter phase curves. In addition, as S-type asteroids are more reddish, their phase curves in the cyan (blue) filter tend to be flatter, which translates into lower G2 values. To analyze the similarities between G1, G2 distributions across different taxa, we performed two-dimensional Kolmogorov–Smirnov tests, using a 95$\%$ confidence level. We found that in 11$\%$ of the orange sample and 31$\%$ of the cyan sample, the null hypothesis could not be rejected, suggesting that the G1, G2 distributions differ more significantly in the orange filter. Some taxonomic pairs, such as C-X and L-X, showed expected disparities based on albedo differences, while others, like V-Q and V-A in cyan, had p-values close to the maximum, suggesting that their similar albedos result in comparable behavior.

Our analysis of the phase coloring effect suggests that at phase angles below 5 degrees, phase coloring behaviors are more evident, with both bluening and reddening observed. However, no clear preference for either reddening or bluening emerges at these smaller phase angles. In contrast, within the phase angle range of 10–30 degrees, reddening cases are more prevalent than bluening, consistent with the expectation that reddening effects become more noticeable at higher phase angles. Although the ATLAS filters are broad band and closely spaced in wavelength, which limits their ability to detect color variations, and the lack of rotational corrections adds further challenges, phase coloring remains apparent even at lower phase angles. In addition, to evaluate wavelength dependency, we compared the paired distributions of G1, G2 (o) vs. G1, G2 (c) within the same taxonomic types using a two-dimensional Kolmogorov-Smirnov test. Statistically significant differences between the orange and cyan phase curves were observed across all complexes, except for the A complex.

Another essential feature of the analysis of properties in statistically significant samples is the identification of outliers, i.e., objects that do not follow the expected pattern. In this study, we identified the asteroid Yasunori as occupying a parameter space that corresponds to a taxonomic type different from the one previously assigned to it.

During our analysis, we noticed that the uncertainties associated with the G2 parameter are consistently smaller than those of G1. We tested this phenomenon by simulating phase curve observations using two different phase functions across different phase angle ranges and found that G2 values are less affected by phase angle coverage, unlike G1. Furthermore, we compared the phase curve of an asteroid using data from the ATLAS Solar System Catalog (SSCAT) Version 1, released in 2022, with Version 2, which includes additional observations collected between 2022 and 2024, during which ATLAS incorporated two more telescopes. We observed that, due to the increased coverage in the opposition region, the value of G1 changed significantly, while G2 remained almost constant. This is due to the fact that the G1 parameter mainly models the opposition effect, while G2 describes the linear region of the phase curve. For this work, we used the more complete Version 2 dataset.

Finally, our algorithms have proven to be sufficiently general and not dependent on a specific catalog, which allows them to be easily adapted to new data sets. In fact, we have already started to prepare their adaptation for the data to be obtained from LSST, using the simulations available in Data Preview 0.3.

\section*{Data Availability}
The derived data generated in this research will be shared on reasonable request
to the corresponding author.

\section*{Author contributions}
MC wrote the majority of the paper and performed all computations and statistical tests. DO contributed parts of the code for phase curve fitting and wrote the Introduction section as well as the majority of the Multi-apparition fitting subsection. MC and DO collaborated on developing the manuscript concept. PP ran part of the scripts for data acquisition. All authors provided comments.

\section*{Acknowledgments}
\noindent
MC, DO and PB were supported by grant No. 2022/45/B/ST9/00267 from the National Science Centre, Poland. 
AAC acknowledges financial support from the Severo Ochoa grant CEX2021-001131-S funded by MCIN/AEI/10.13039/501100011033 and the Spanish project PID2023-153123NB-I00, funded by MCIN/AEI.
For language editing and translation, we utilized DeepL and ChatGPT.
MC wish to thank the "Summer School for Astrostatistics in Crete" for providing training on the statistical methods adopted in this work. 
This work uses data from the University of Hawaii's ATLAS project, funded through NASA grants NN12AR55G, 80NSSC18K0284, and 80NSSC18K1575, with contributions from the Queen's University Belfast, STScI, the South African Astronomical Observatory, and the Millennium Institute of Astrophysics, Chile.
\bibliographystyle{model5-names.bst}
\bibliography{biblio}

\onecolumn
\appendix
\section{Appendix}

\subsection{Color slope definition}
 The color slope is defined as the difference between the first derivatives of the phase functions with respect to the phase angle for orange and cyan filters. These derivatives are given by:

\begin{equation}
 \frac{dV}{d\alpha} = -\frac{2.5}{\ln(10)} \cdot \frac{G_1 \frac{d\phi_1(\alpha)}{d\alpha} + G_2 \frac{d\phi_2(\alpha)}{d\alpha} + (1 - G_1 - G_2) \frac{d\phi_3(\alpha)}{d\alpha}}{G_1 \phi_1(\alpha) + G_2 \phi_2(\alpha) + (1 - G_1 - G_2) \phi_3(\alpha)}
\end{equation}
\noindent
The derivatives are expressed in magnitudes per radian, with the cubic spline derivatives taking the form:

 \begin{eqnarray} 
 \frac{d\phi(\alpha)}{d\alpha} &=& a (\alpha -b)^2+c (\alpha-d) + e
 \end{eqnarray}  
\noindent
where the coefficents $a$, $b$ ,$c$ ,$d$ are listed in Tables \ref{coe1} - \ref{coe3}.

\begin{table}[Hb!]
    \centering
    \begin{tabular}{|c|c|c|c|c|c|} \hline
    $\alpha$ range (\textdegree) & a & b & c & d & e\\ \hline
    7.5\textdegree - 30\textdegree & -6.812748 & 0.1309 & 6.126412 & 0.1309 & -1.909859 \\
    30\textdegree - 60\textdegree &  -0.348572 & 0.5236 & 0.775692 & 0.5236 & -0.554634 \\
    60\textdegree - 90\textdegree & -0.240592 & 1.0472 & 0.410668 & 1.0472 & -0.244046 \\
    90\textdegree - 120\textdegree & -0.034786 & 1.5708 & 0.158721 & 1.5708 & -0.094980 \\
    120\textdegree -150\textdegree & -0.488589 & 2.0944 & 0.122292 & 2.0944 & -0.021411 \\
    \hline
    \end{tabular}
    \caption{Coeficients for $\frac{d\phi_1}{d\alpha}$.}
    \label{coe1}
\end{table}

\begin{table}[Hb!]
    \centering
    \begin{tabular}{|c|c|c|c|c|c|} \hline
    $\alpha$ range (\textdegree) & a & b & c & d & e\\ \hline
    7.5\textdegree - 30\textdegree & 3.274255 & 0.1309 & -1.780058 & 0.1309 & -0.572958 \\
    30\textdegree - 60\textdegree & -0.379534 & 0.5236 & 0.791536 & 0.5236 & -0.767054 \\
    60\textdegree - 90\textdegree & -0.110903 & 1.0472 & 0.394089 & 1.0472 & -0.456658 \\ 
    90\textdegree - 120\textdegree & 0.085536 & 1.5708 & 0.277952 & 1.5708 & -0.280718 \\
    120\textdegree -150\textdegree & -0.294370 & 2.0944 & 0.367525 & 2.0944 & -0.111733\\
    \hline
    \end{tabular}
    \caption{Coeficients for $\frac{d\phi_2}{d\alpha}$.}
    \label{coe2}
\end{table}

\begin{table}[Hb!]
    \centering
    \begin{tabular}{|c|c|c|c|c|c|} \hline
    $\alpha$ range (\textdegree) & a & b & c & d & e\\ \hline
    0\textdegree - 0.3 \textdegree & 2428451.278843 & 0.0000 & -20559.923775 & 0.0000 & -0.106301 \\
    0.3\textdegree - 1.0\textdegree & -192238.573083 & 0.0052 & 4870.758549 & 0.0052 & -41.180439 \\
    1.0\textdegree - 2.0\textdegree & -785.953504 & 0.0175 & 173.484079 & 0.0175 & -10.366915 \\
    2.0\textdegree - 4.0\textdegree & -997.734942 & 0.0349 & 146.049126 & 0.0349 & -7.578461 \\
    4.0\textdegree - 8.0\textdegree & -497.196388 & 0.0698 & 76.394087 & 0.0698 & -3.696095 \\
    8.0\textdegree - 12.0\textdegree & -34.054440 & 0.1396 & 6.972375 & 0.1396 & -0.786057 \\
    12.0\textdegree - 20.0\textdegree & -2.510511 & 0.2094 & 2.217478 & 0.2094 & -0.465270 \\
    20.0\textdegree - 30.0\textdegree & -1.971925 & 0.3491 & 1.516411 & 0.3491 & -0.204595 \\
    \hline
    \end{tabular}
    \caption{Coeficients for $\frac{d\phi_3}{d\alpha}$.}
    \label{coe3}
\end{table}

The color slope is then computed as:

\begin{equation}
    CS = \frac{dV_o}{d\alpha}  - \frac{dV_c}{d\alpha},
\end{equation}
\noindent
where $\frac{dV_o}{d\alpha}$ is the derivative for orange filter and $\frac{dV_c}{d\alpha}$ for cyan filter respectively.

\subsection{Color slopes}
\begin{figure}[ht]
\centering
\includegraphics[width=4cm]{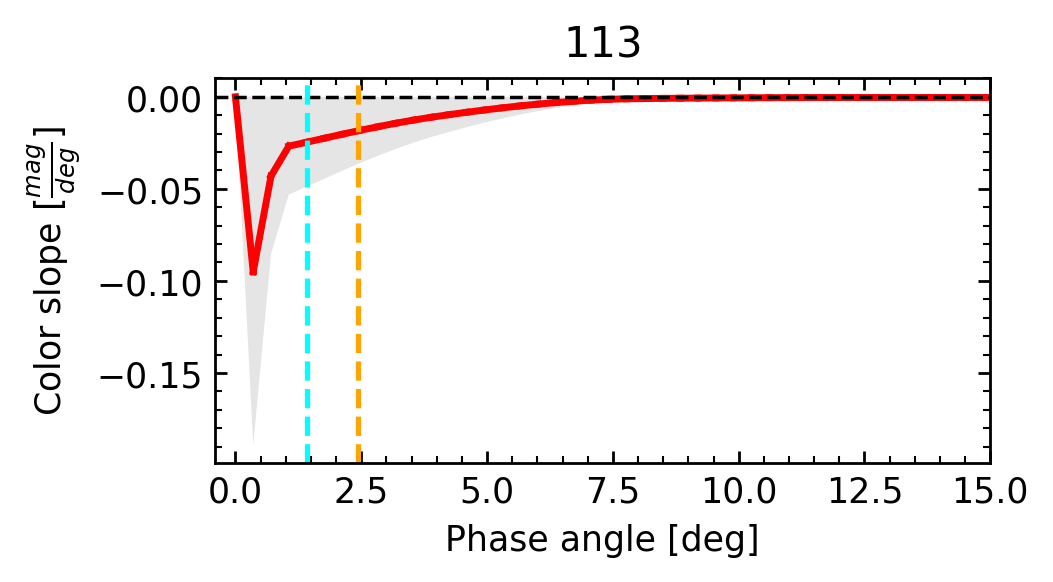}
\includegraphics[width=4cm]{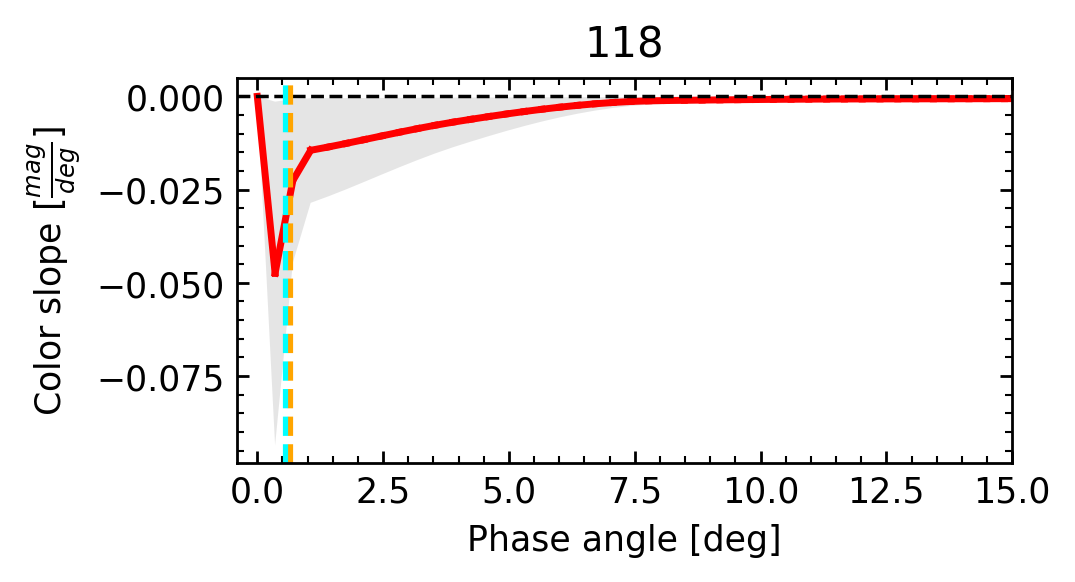}
\includegraphics[width=4cm]{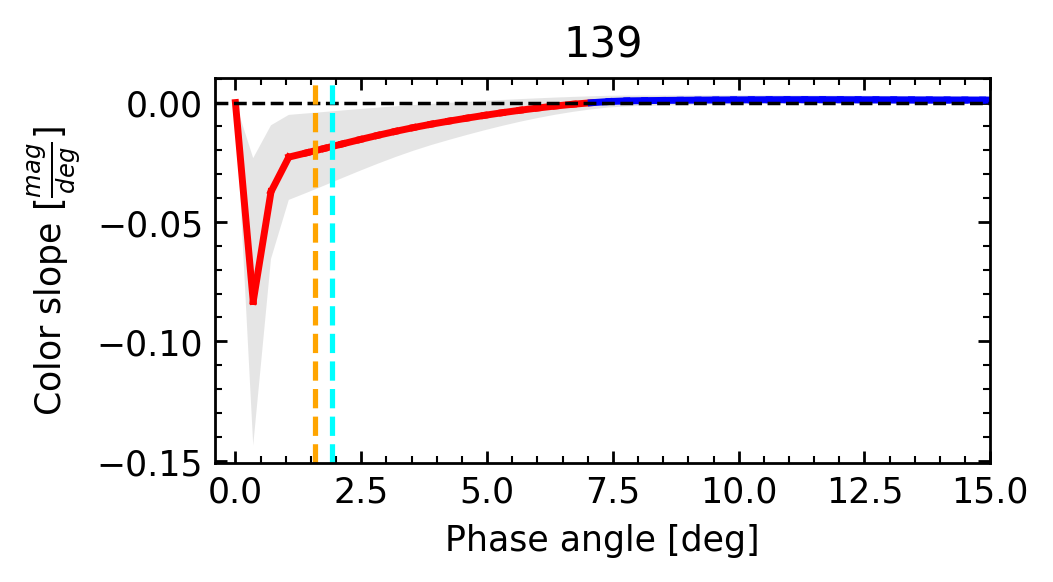}
\includegraphics[width=4cm]{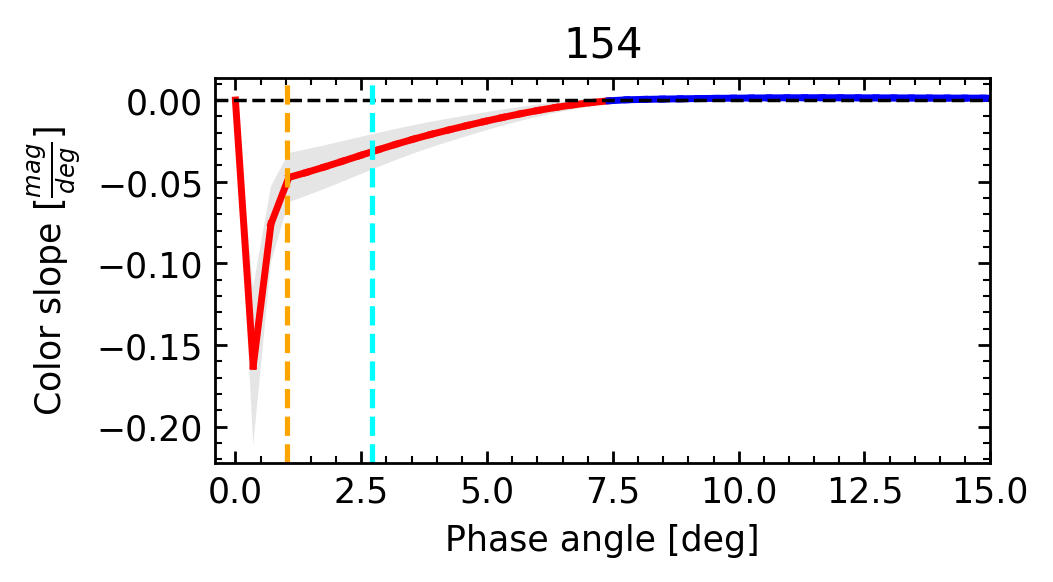}
\includegraphics[width=4cm]{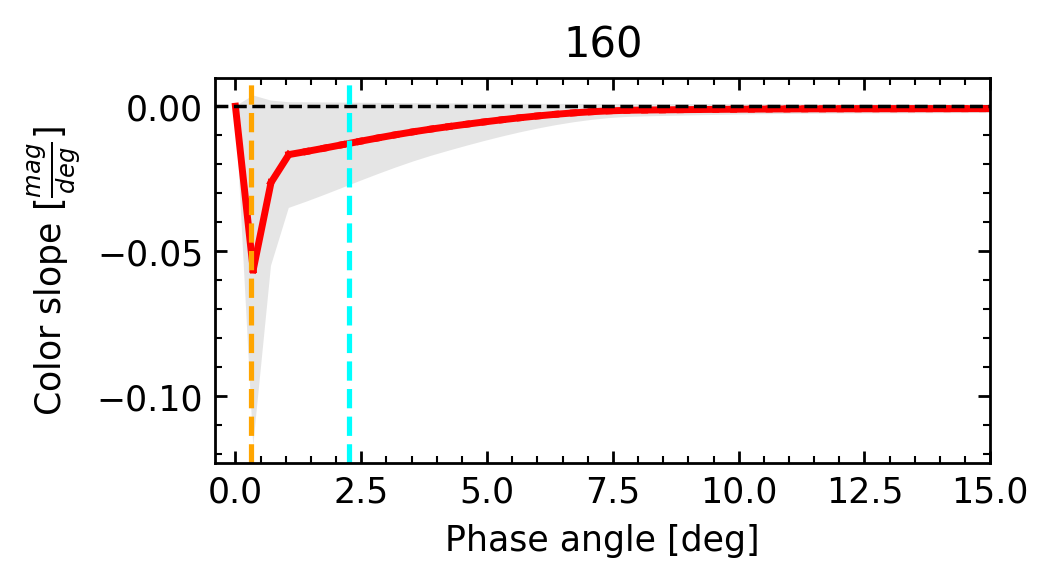}
\includegraphics[width=4cm]{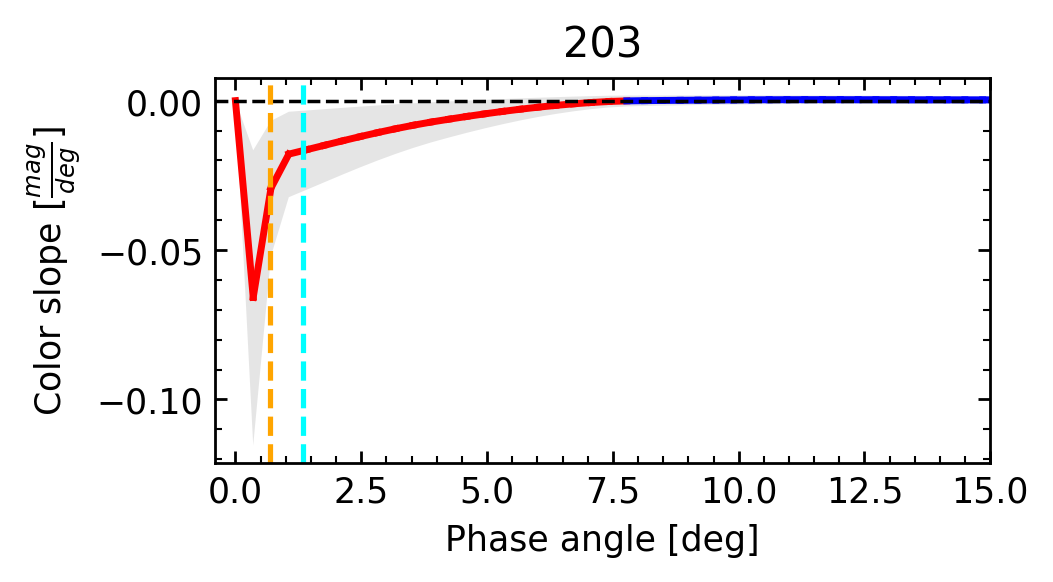}
\includegraphics[width=4cm]{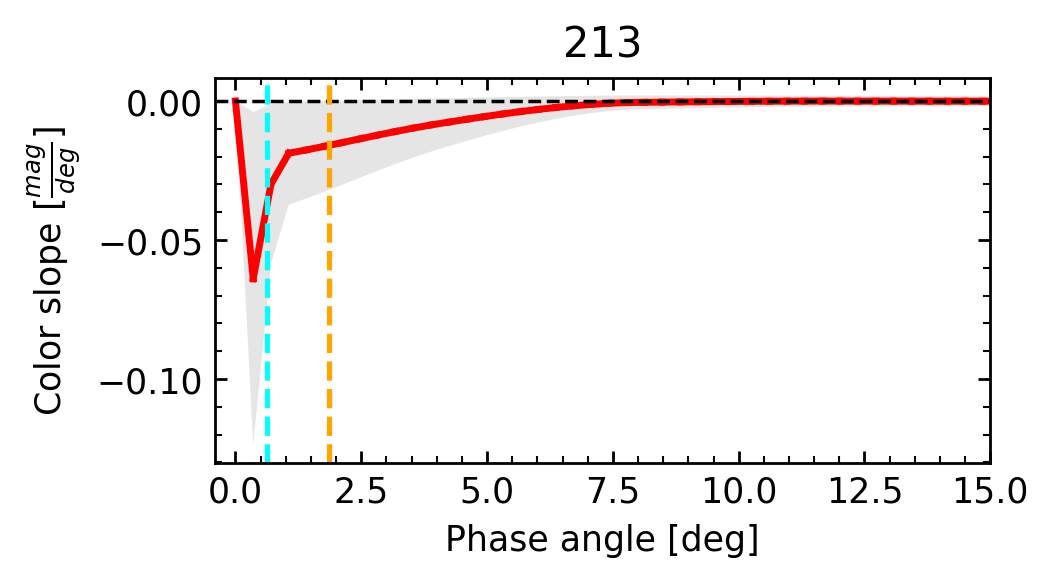}
\includegraphics[width=4cm]{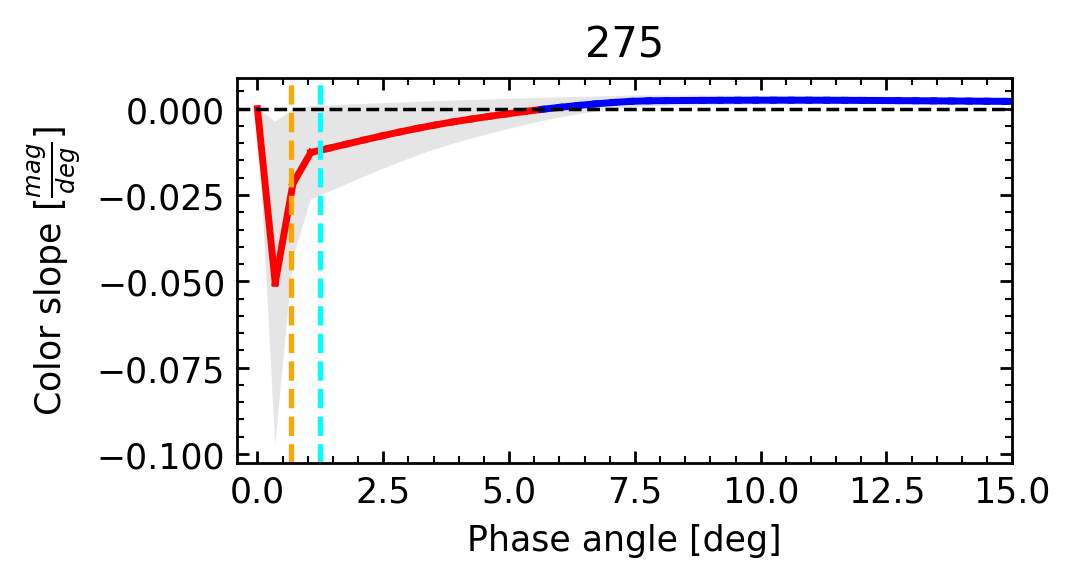}
\includegraphics[width=4cm]{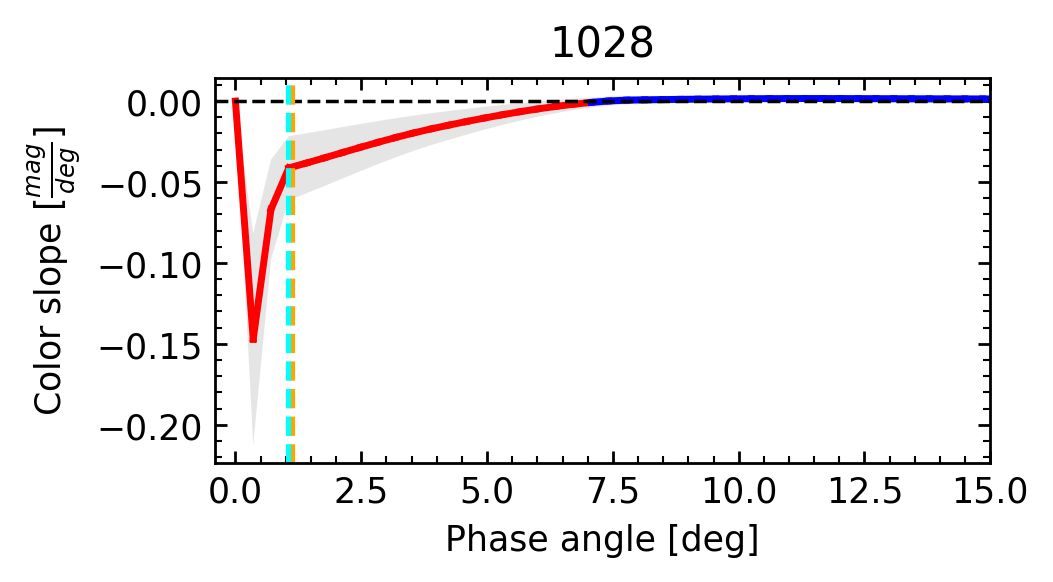}
\includegraphics[width=4cm]{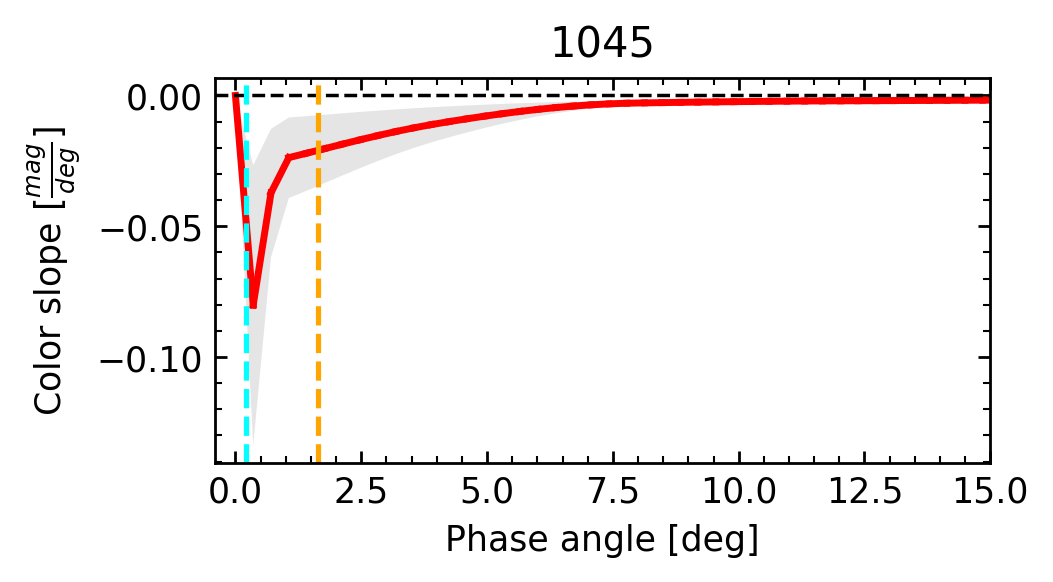}
\includegraphics[width=4cm]{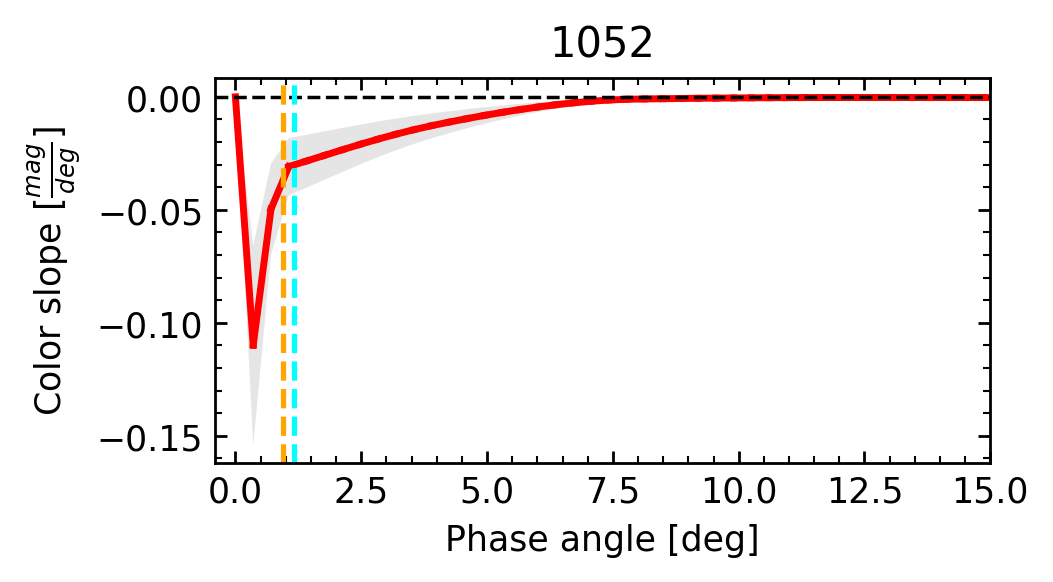}
\includegraphics[width=4cm]{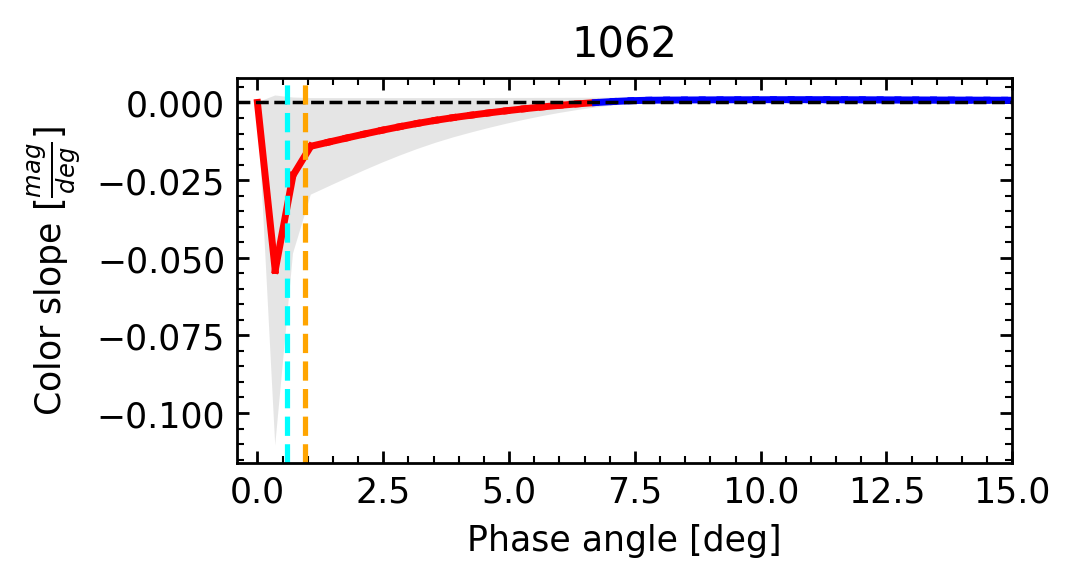}
\includegraphics[width=4cm]{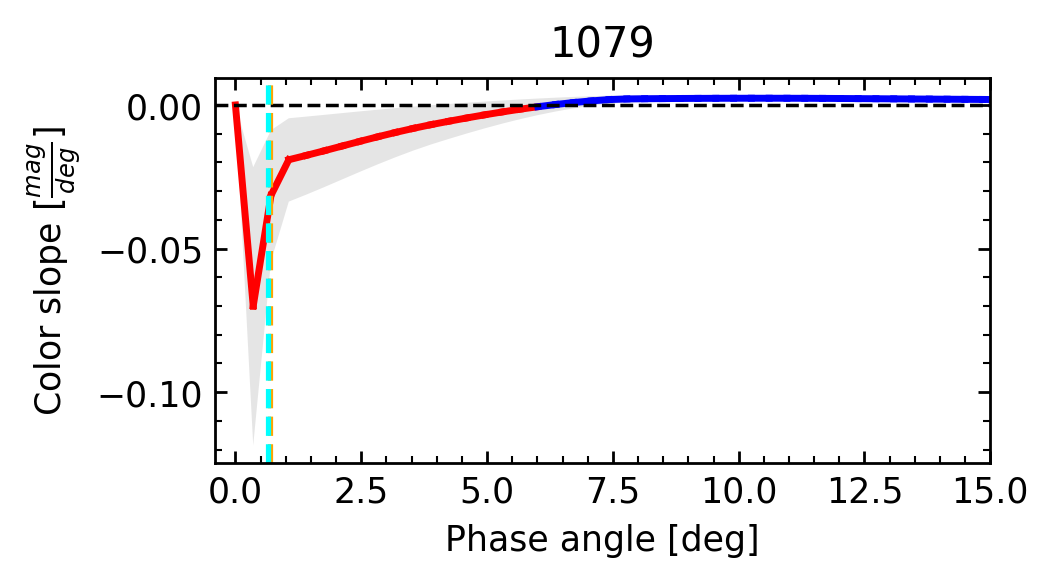}
\includegraphics[width=4cm]{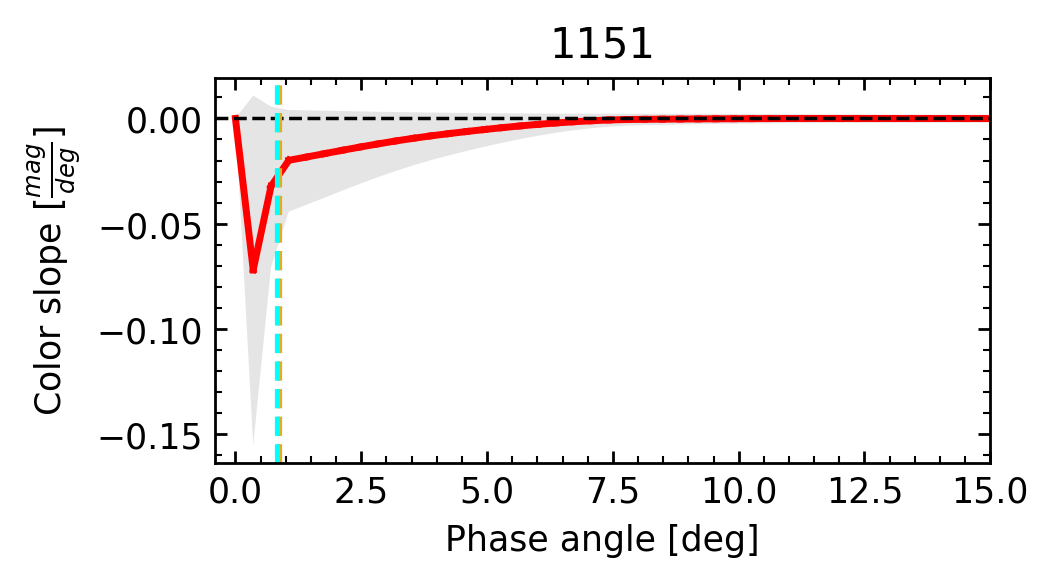}
\includegraphics[width=4cm]{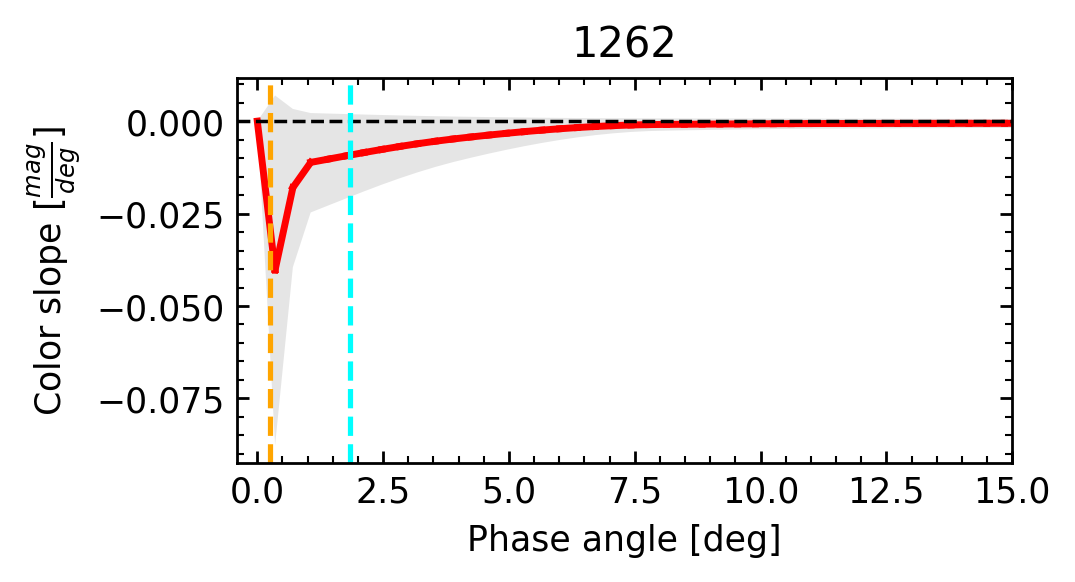}
\includegraphics[width=4cm]{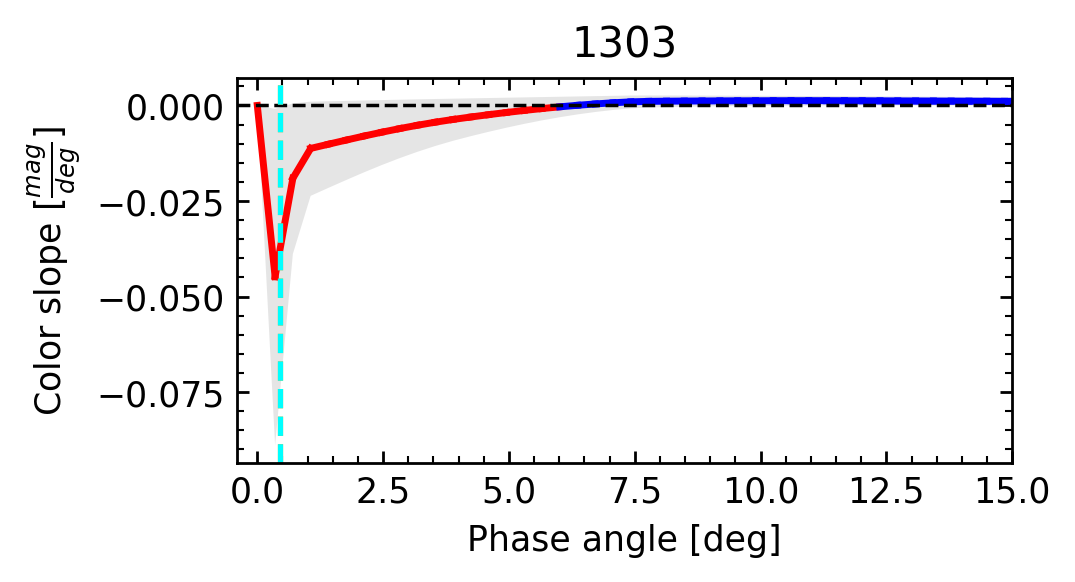}
\includegraphics[width=4cm]{color_slope_1332.png}
\includegraphics[width=4cm]{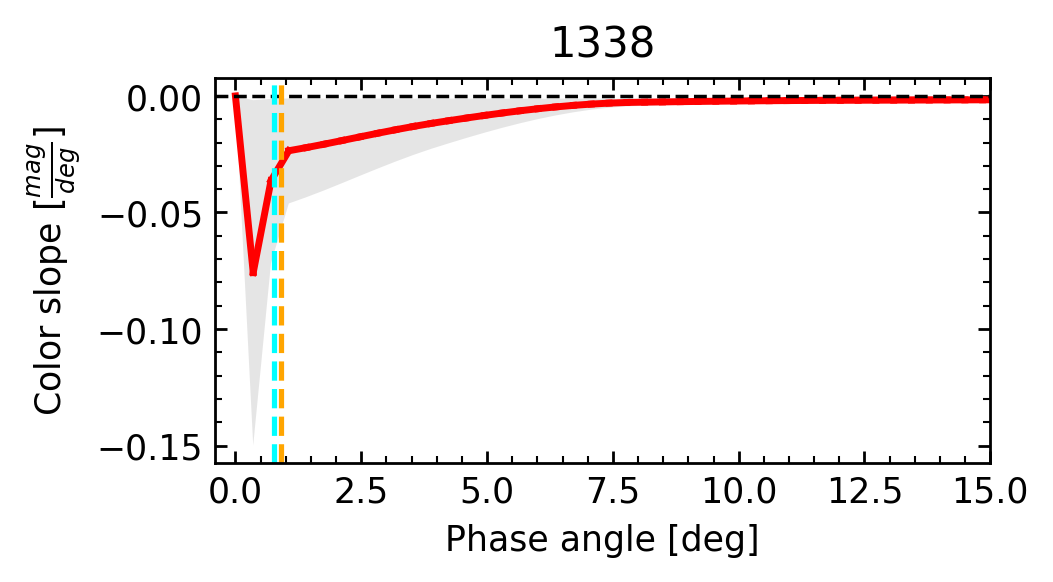}
\includegraphics[width=4cm]{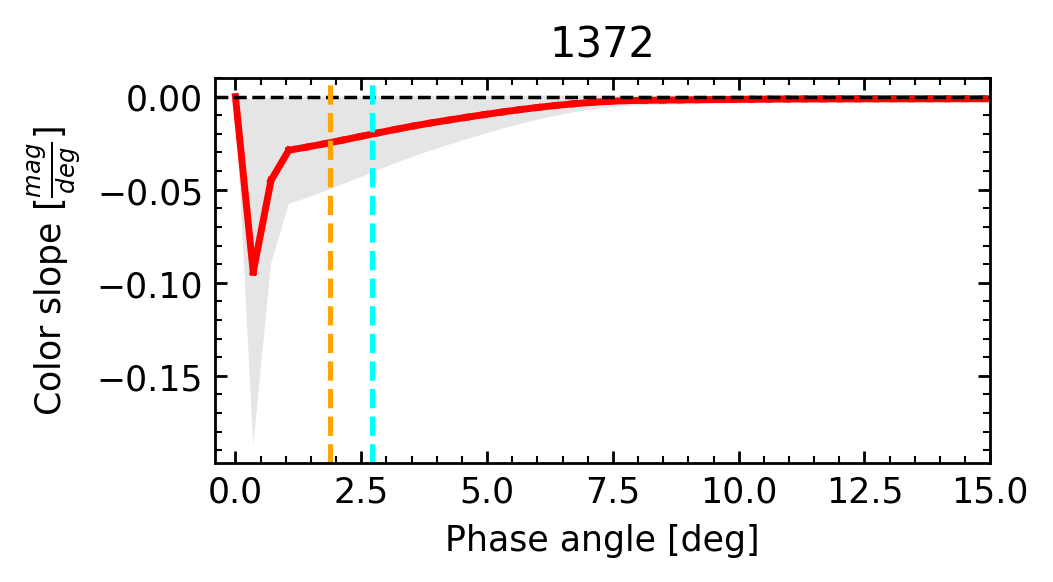}
\includegraphics[width=4cm]{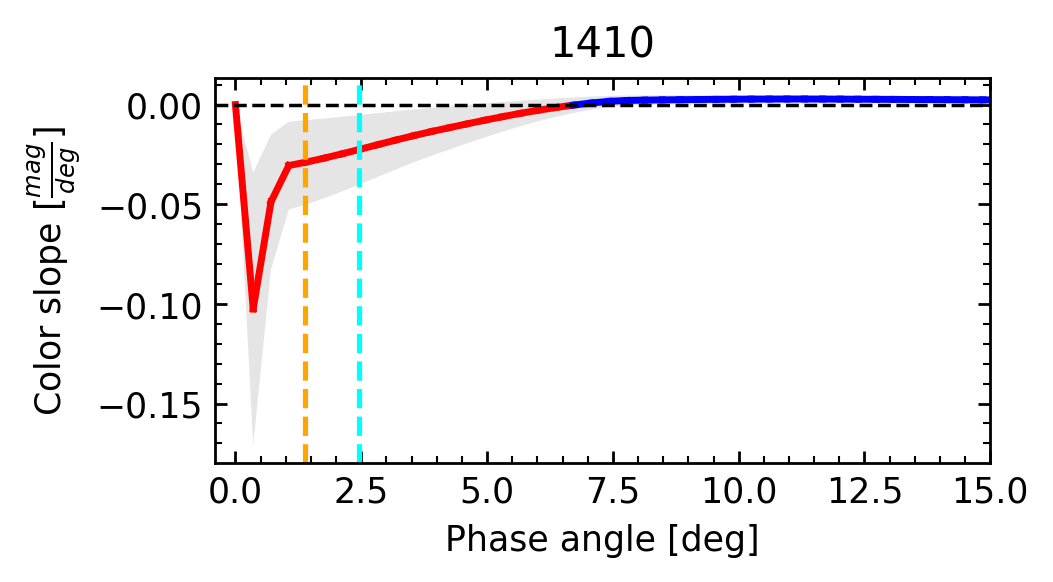}
\includegraphics[width=4cm]{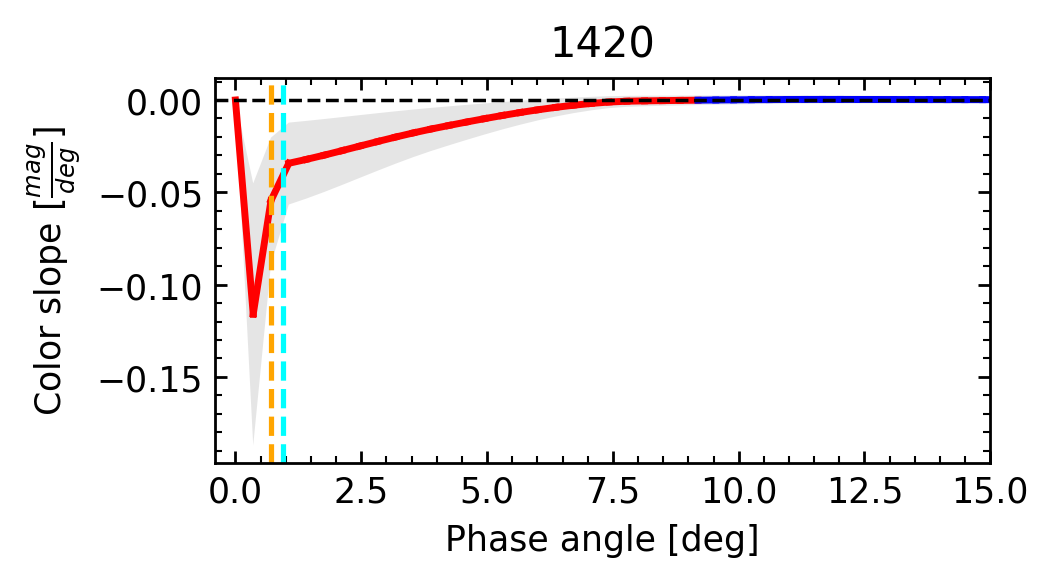}
\includegraphics[width=4cm]{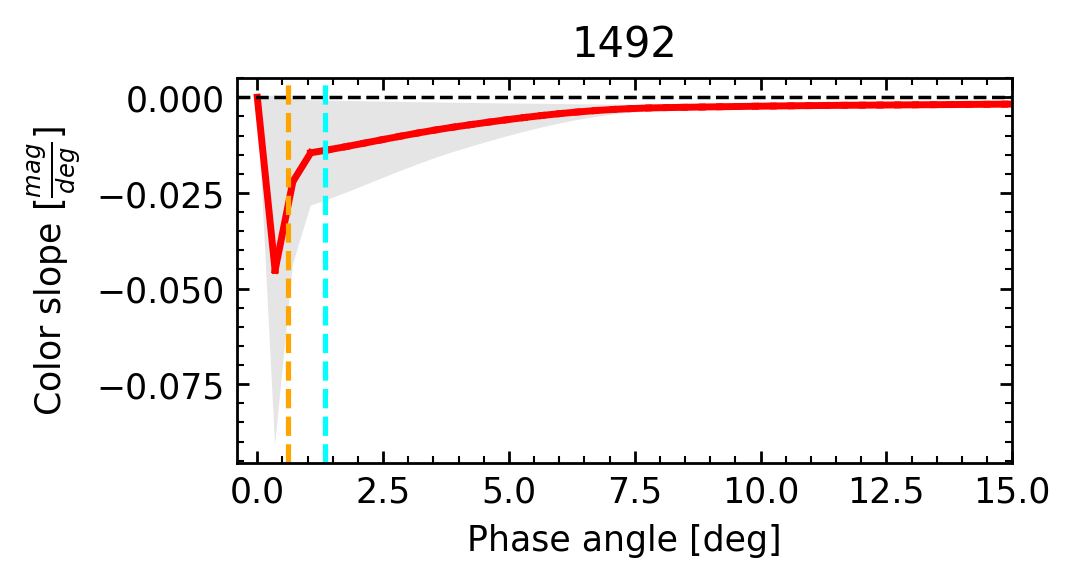}
\includegraphics[width=4cm]{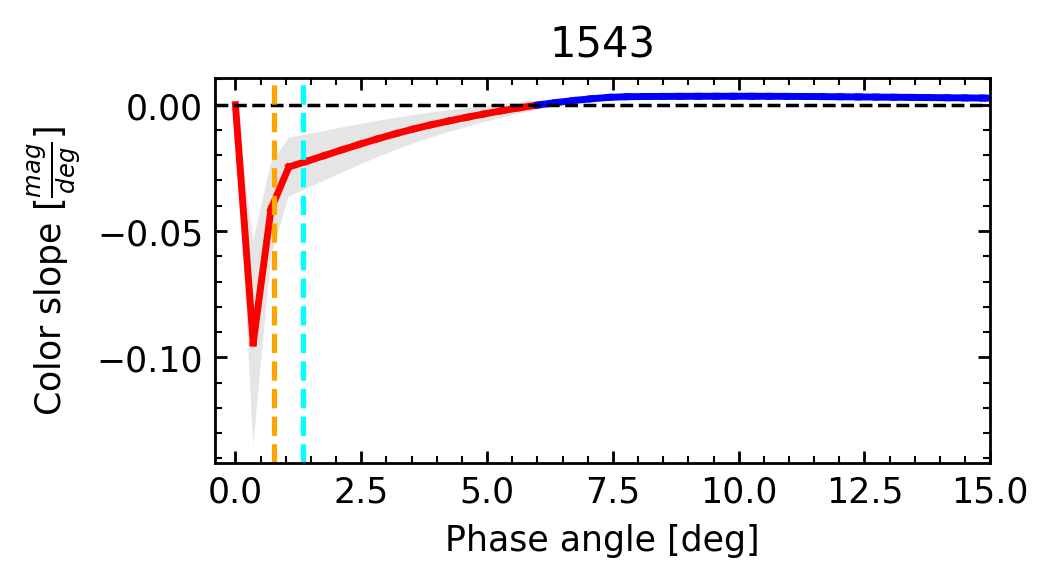}
\includegraphics[width=4cm]{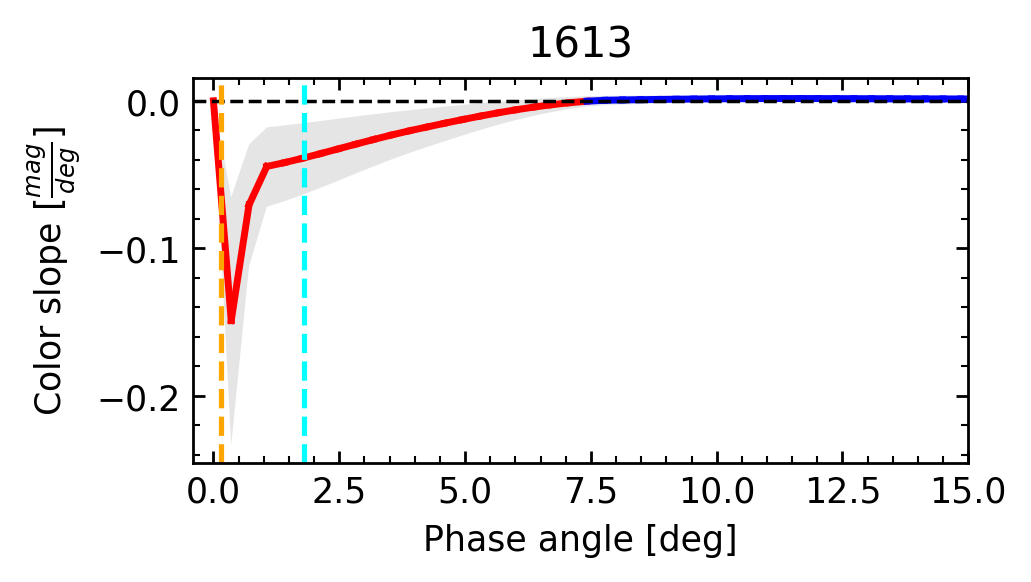}
\includegraphics[width=4cm]{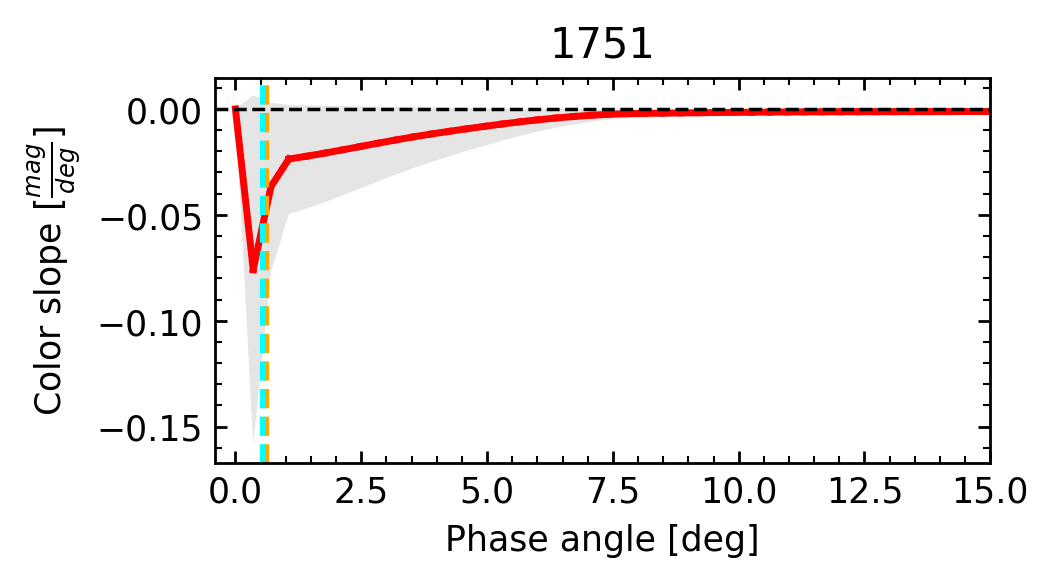}
\includegraphics[width=4cm]{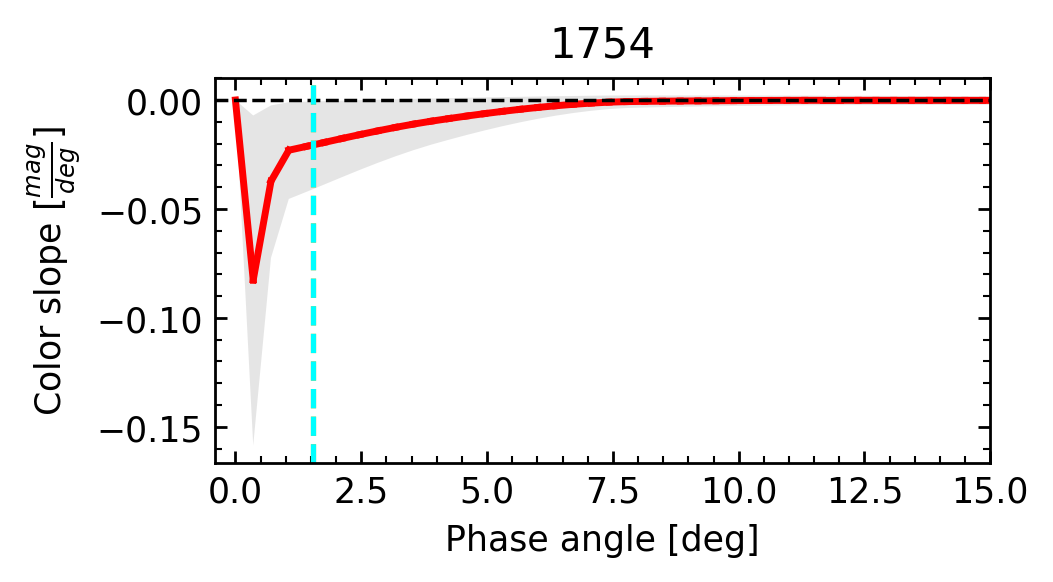}
\includegraphics[width=4cm]{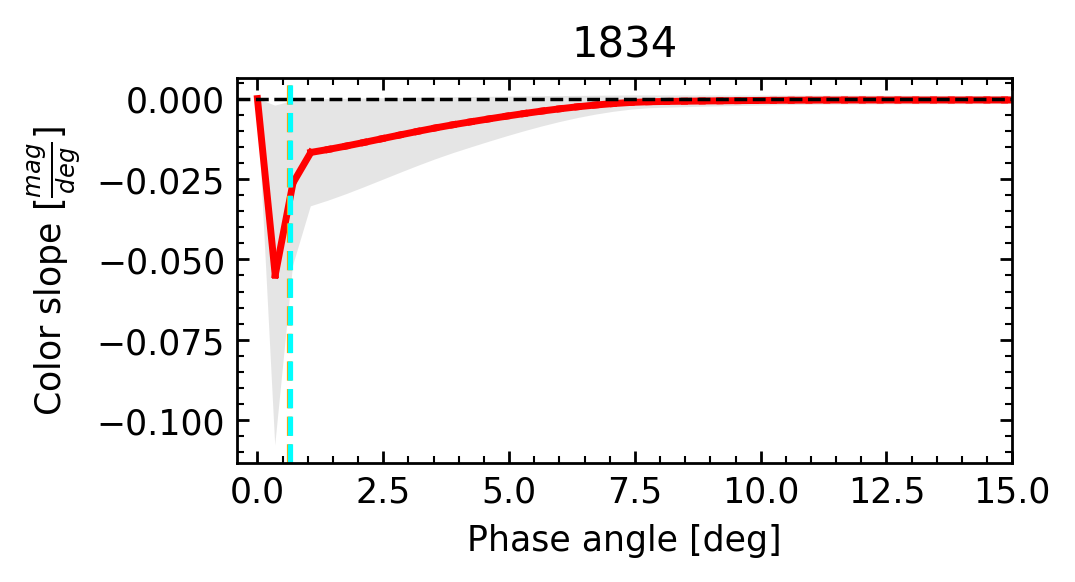}
\includegraphics[width=4cm]{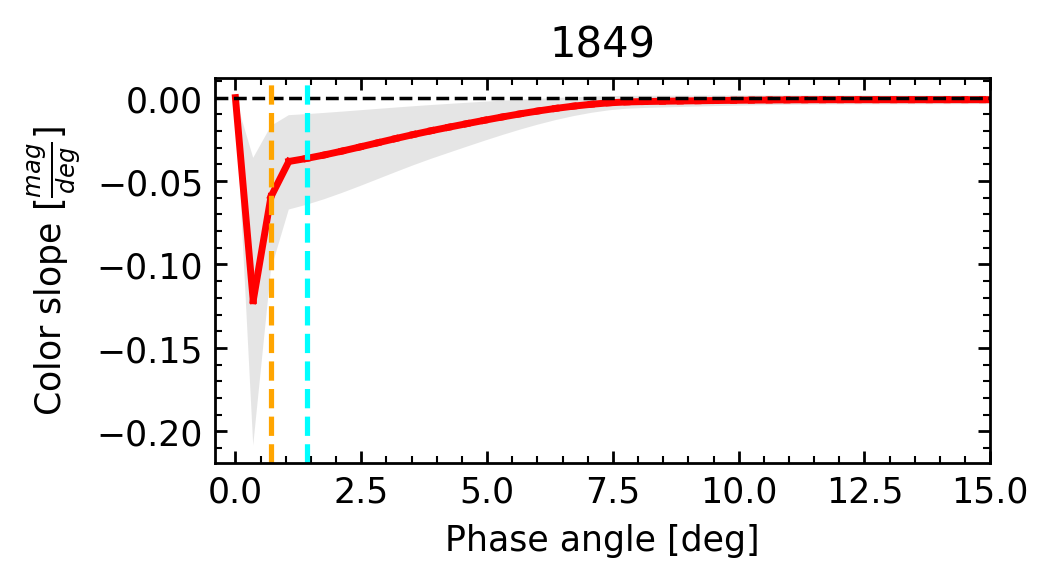}
\includegraphics[width=4cm]{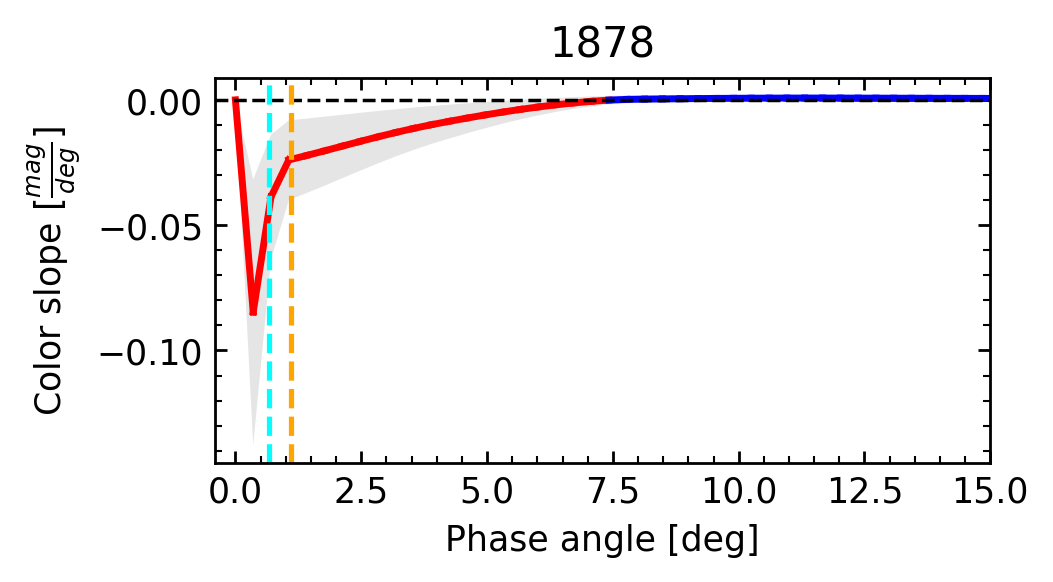}
\includegraphics[width=4cm]{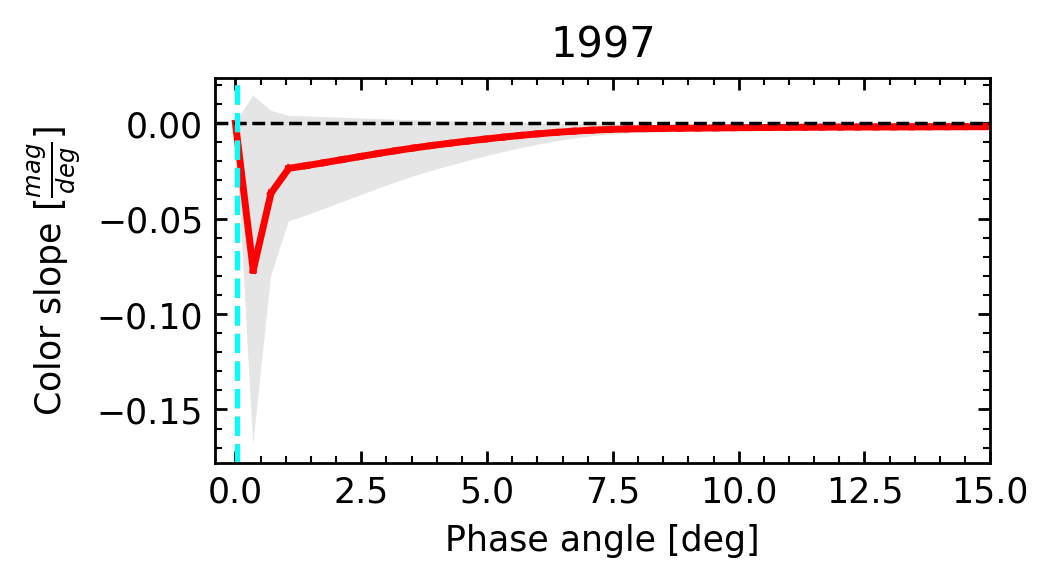}
\includegraphics[width=4cm]{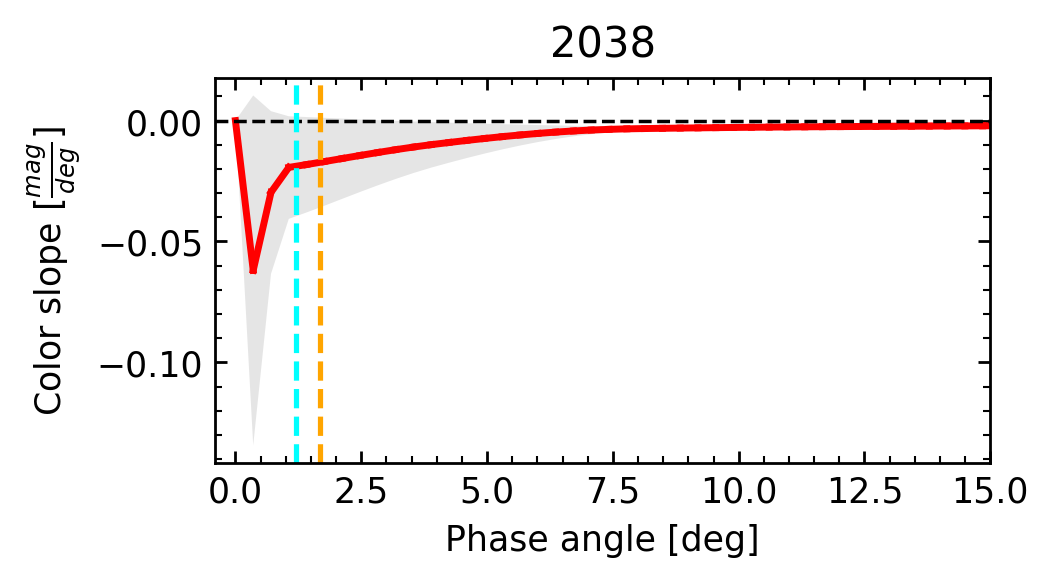}
\includegraphics[width=4cm]{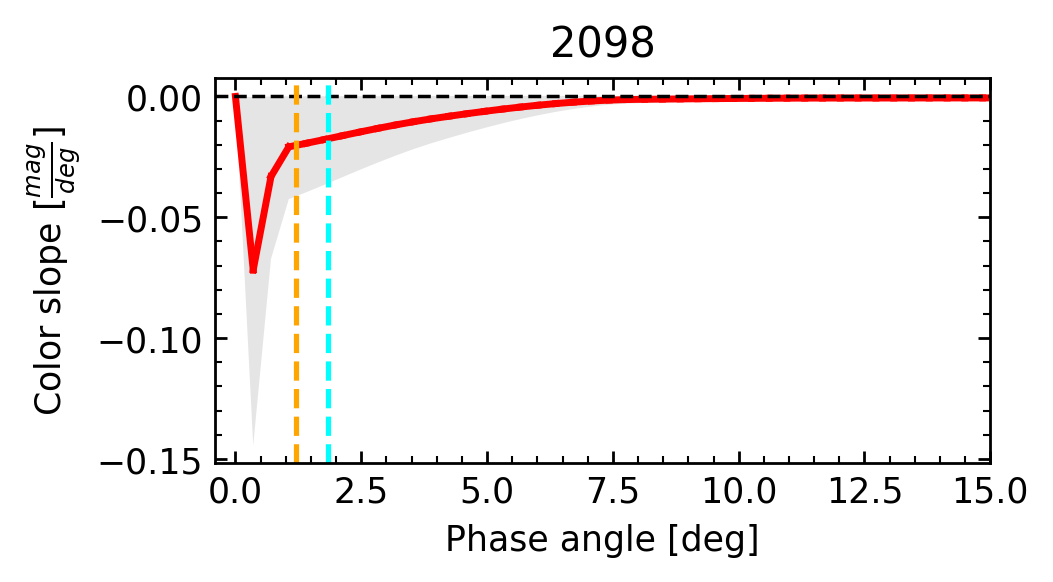}

\caption{Color slope, representing the difference between the first derivatives of the phase function with respect to the phase angle in the two filters for asteroids showing reddening effect. Vertical lines inidicates the minimum observed phase angles separately in both filters.}
\label{red}
\end{figure}

\begin{figure}[ht]
\ContinuedFloat
\captionsetup{list=off,format=cont}
\centering
\includegraphics[width=4cm]{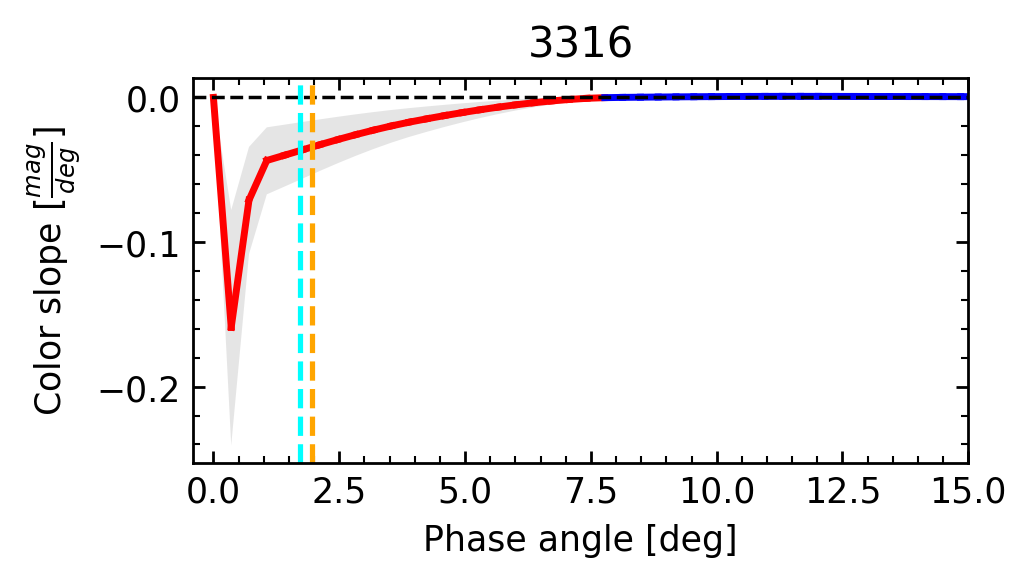}
\includegraphics[width=4cm]{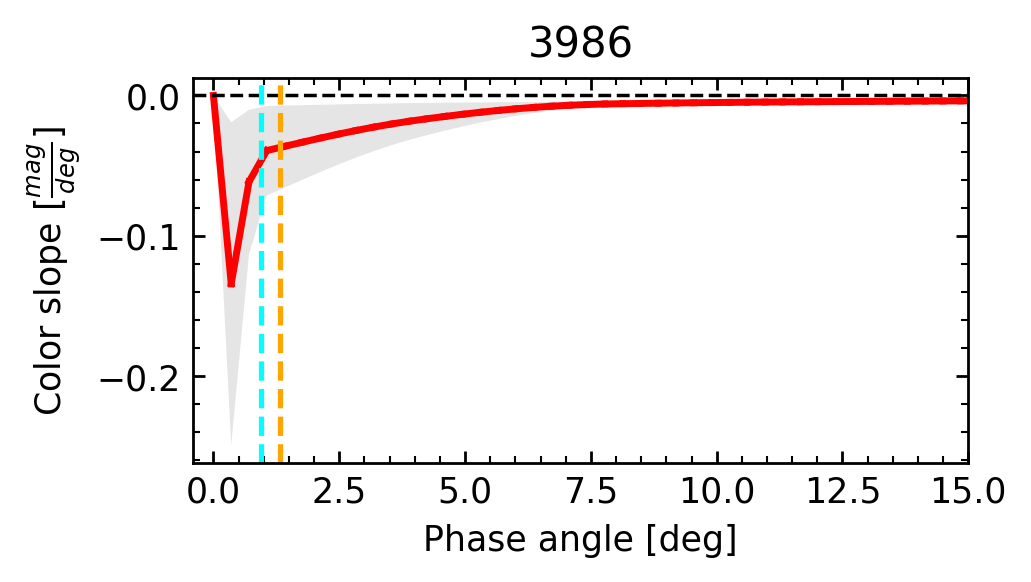}
\includegraphics[width=4cm]{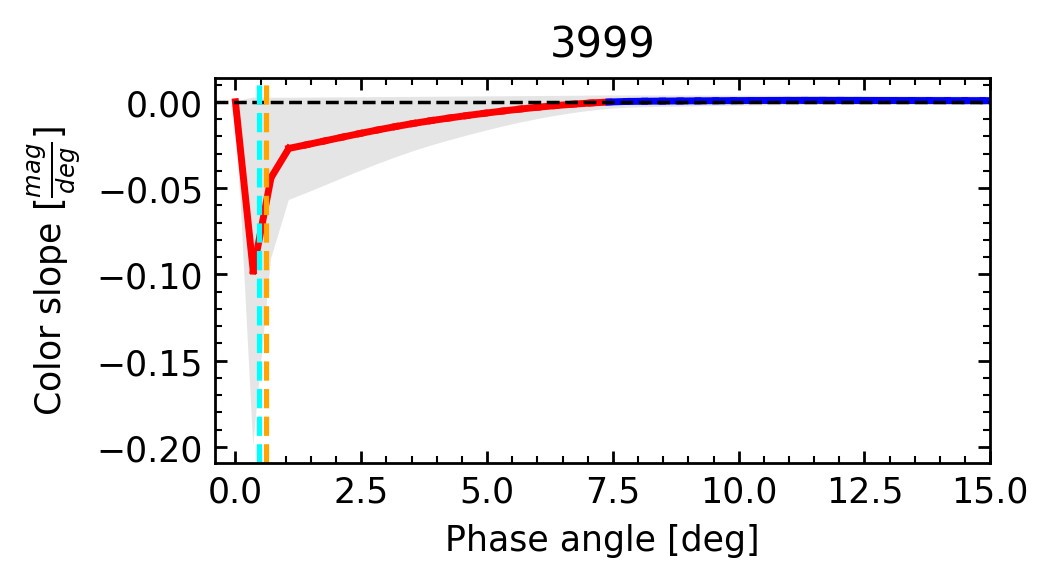}
\includegraphics[width=4cm]{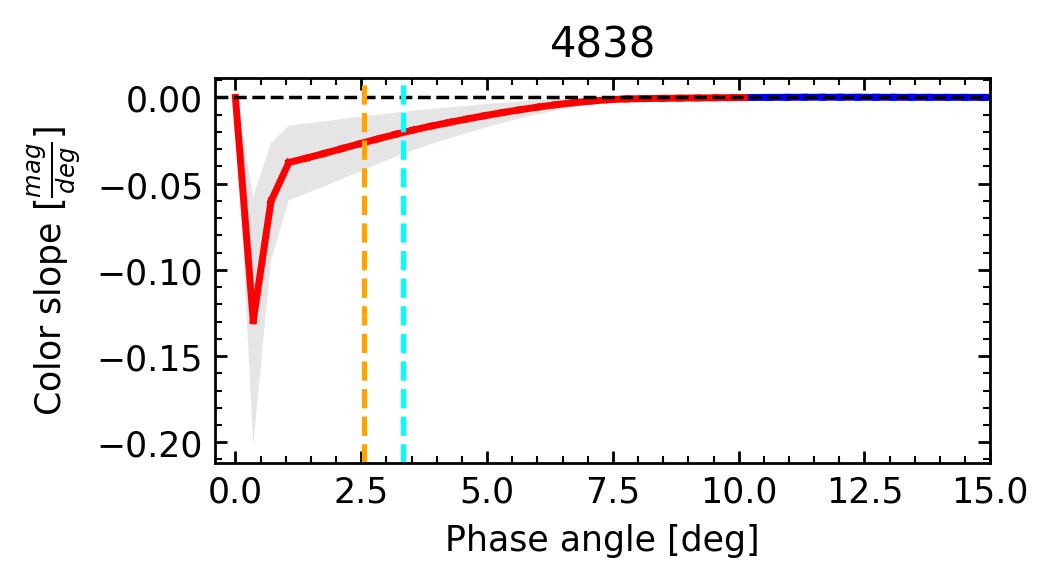}
\includegraphics[width=4cm]{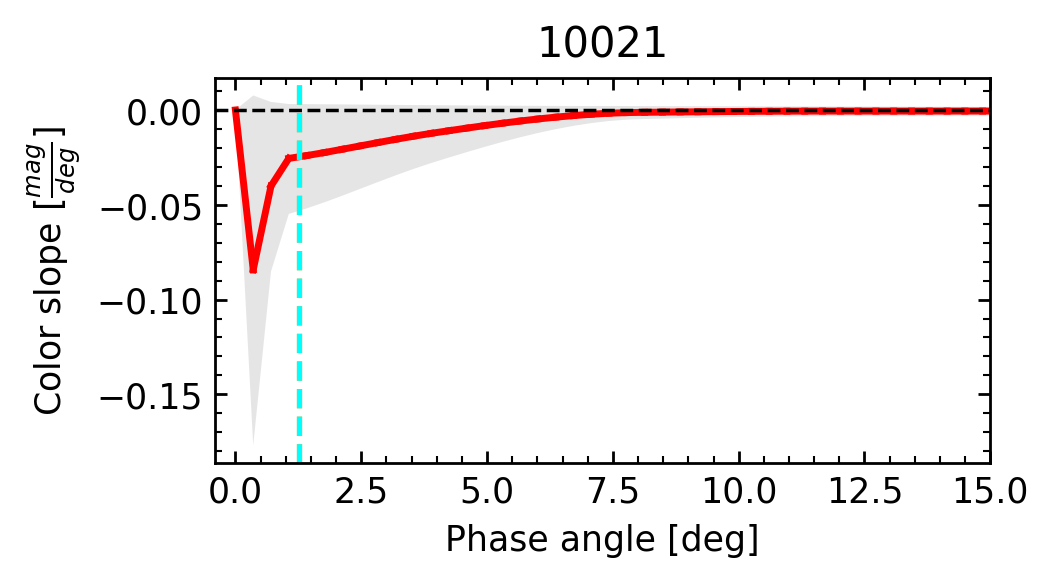}
\includegraphics[width=4cm]{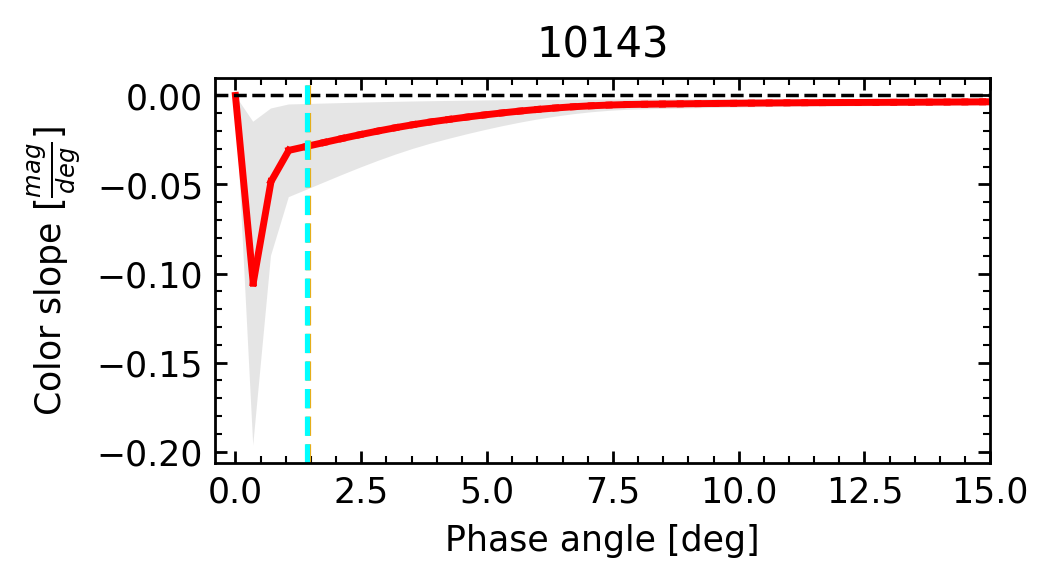}
\includegraphics[width=4cm]{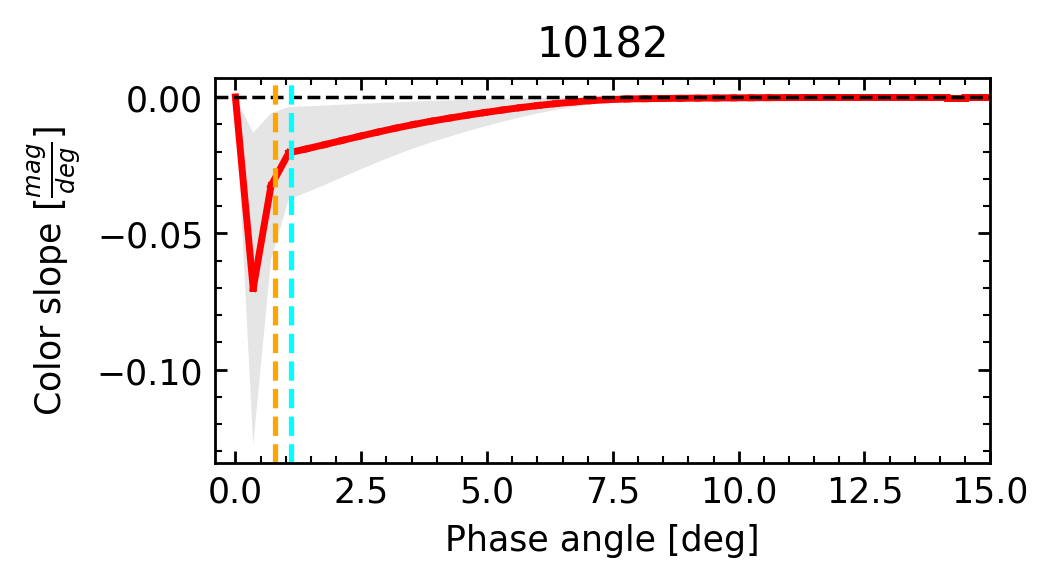}
\includegraphics[width=4cm]{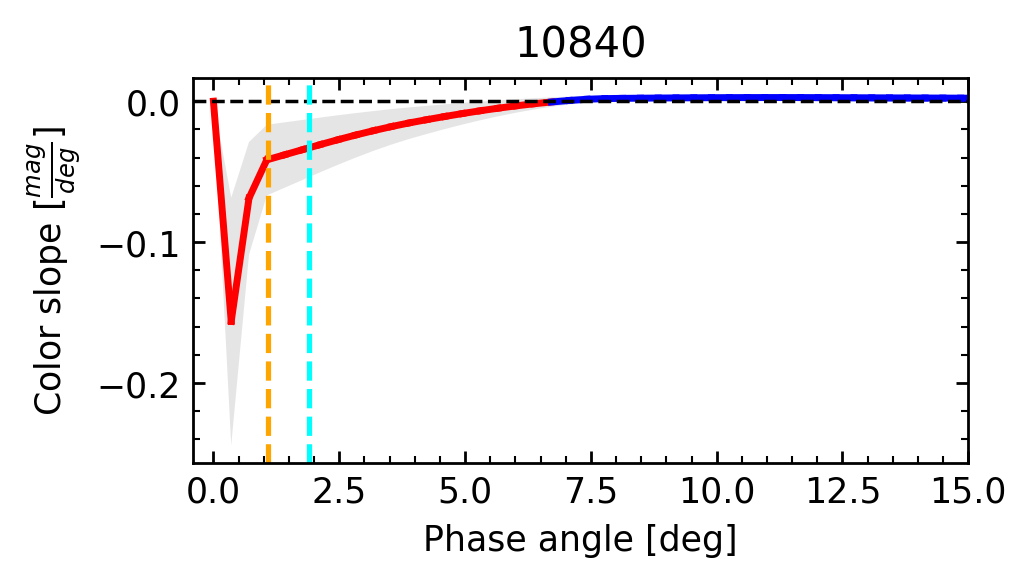}
\includegraphics[width=4cm]{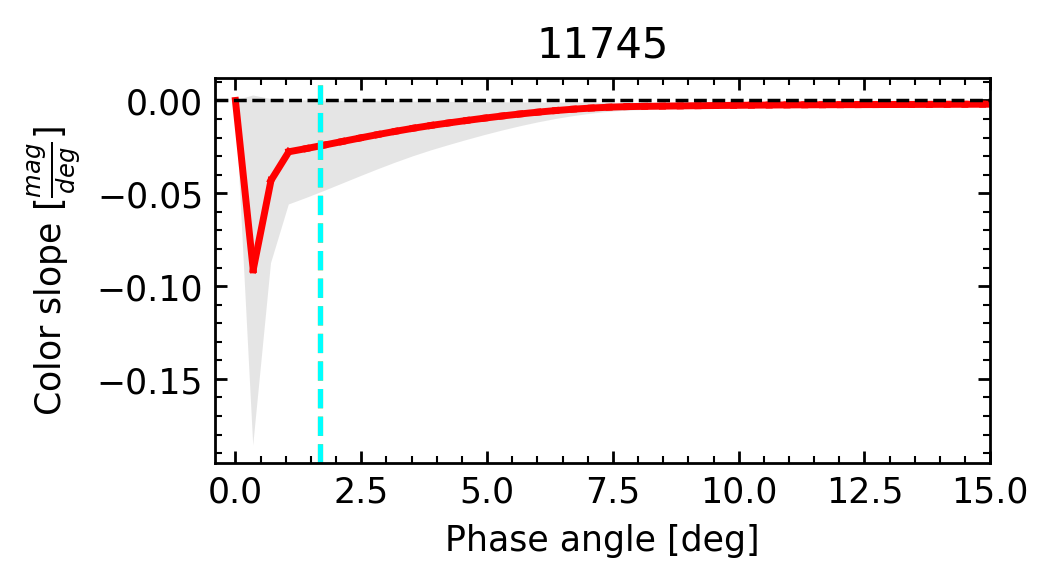}
\includegraphics[width=4cm]{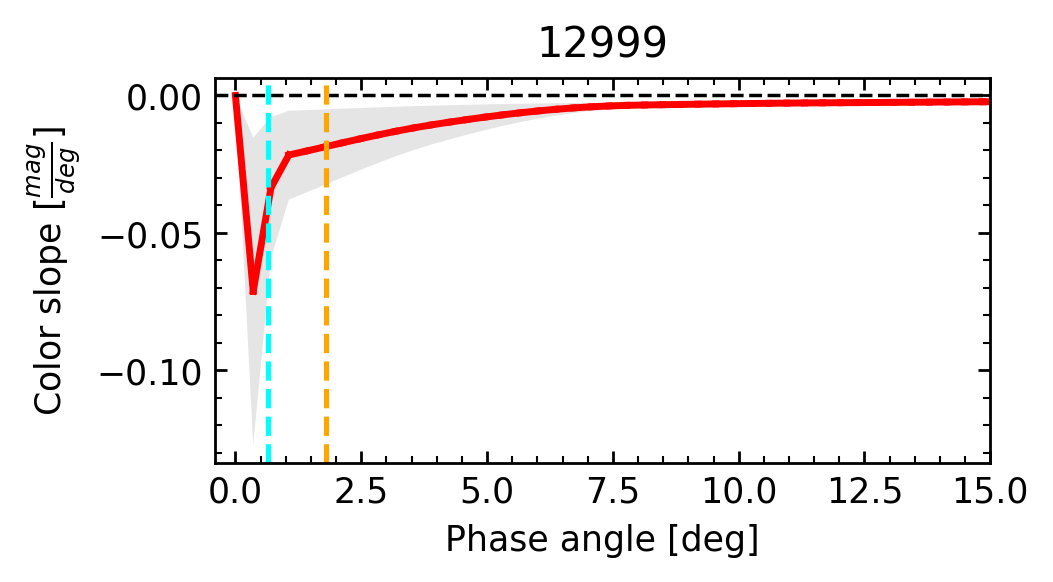}
\includegraphics[width=4cm]{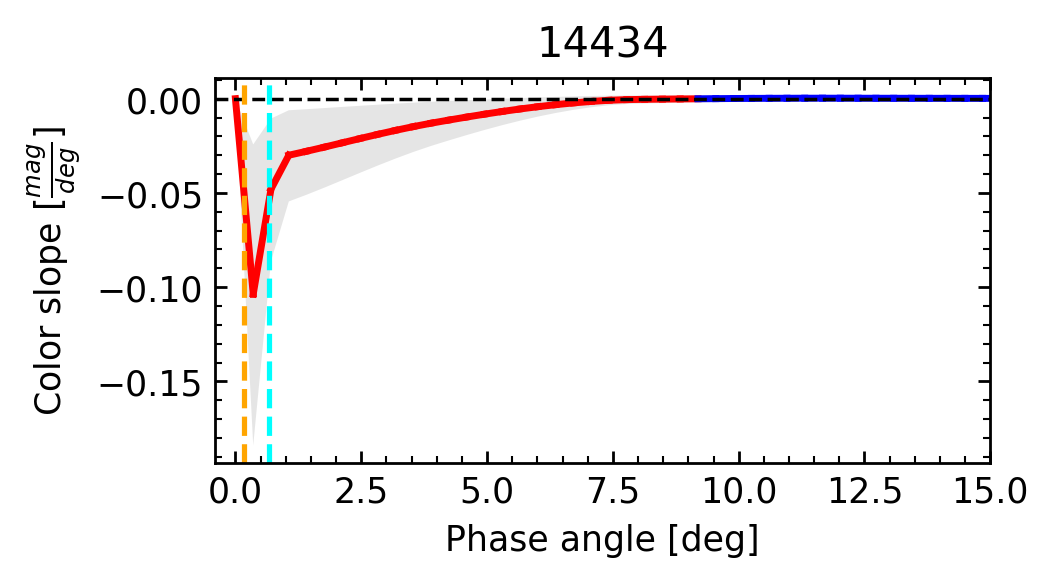}
\includegraphics[width=4cm]{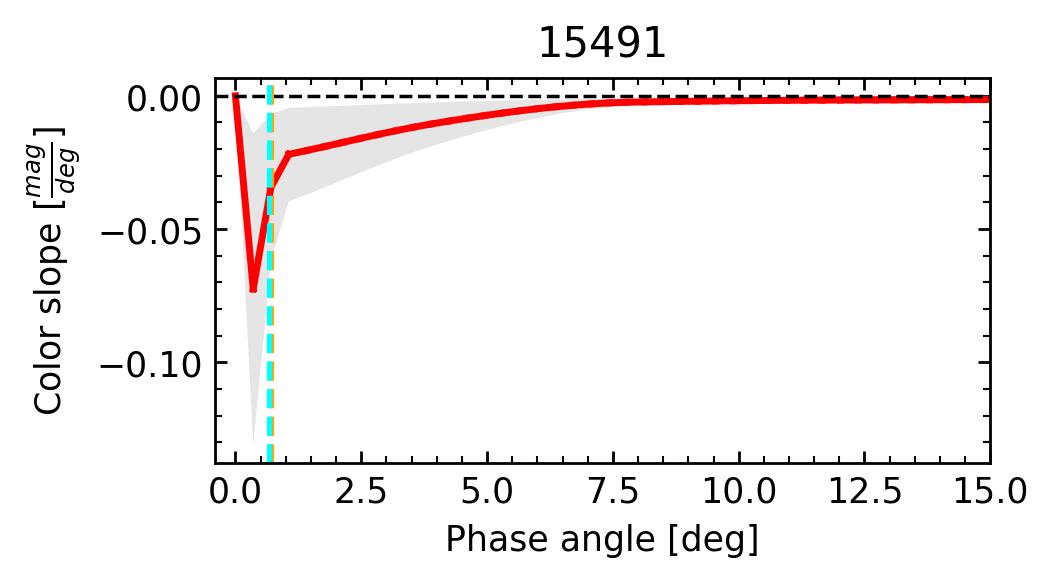}
\includegraphics[width=4cm]{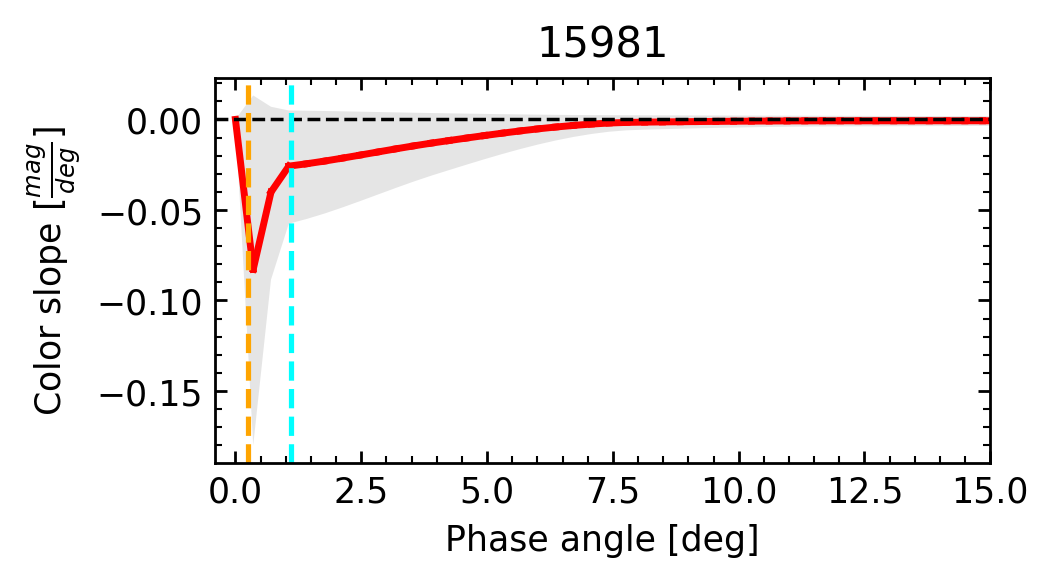}
\includegraphics[width=4cm]{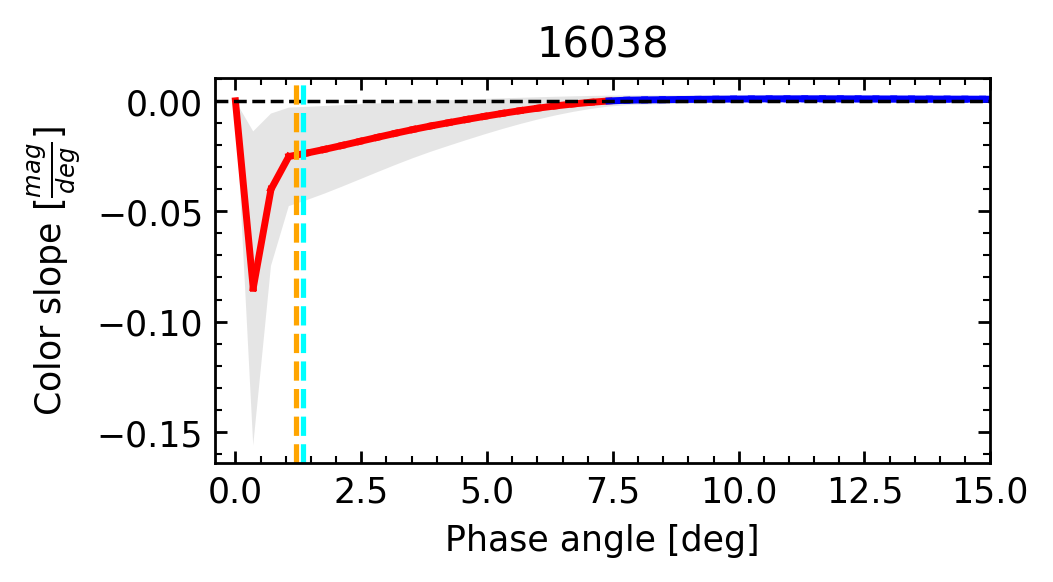}
\includegraphics[width=4cm]{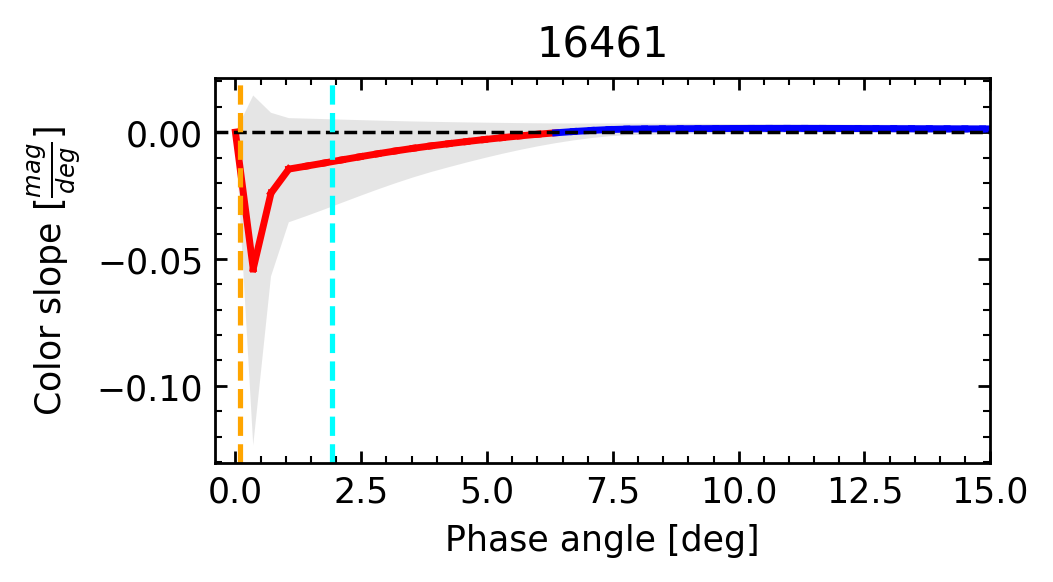}
\includegraphics[width=4cm]{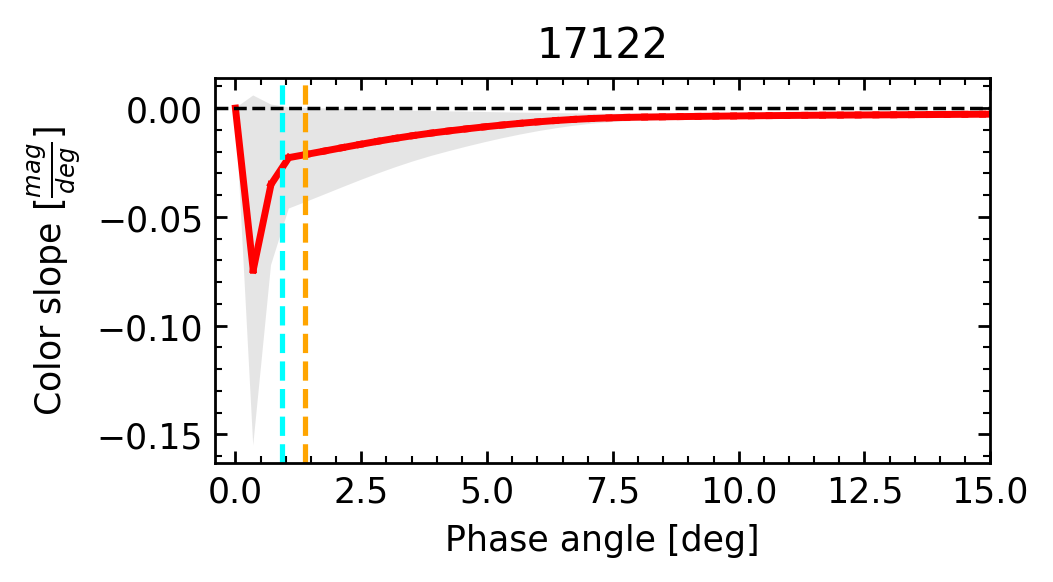}
\includegraphics[width=4cm]{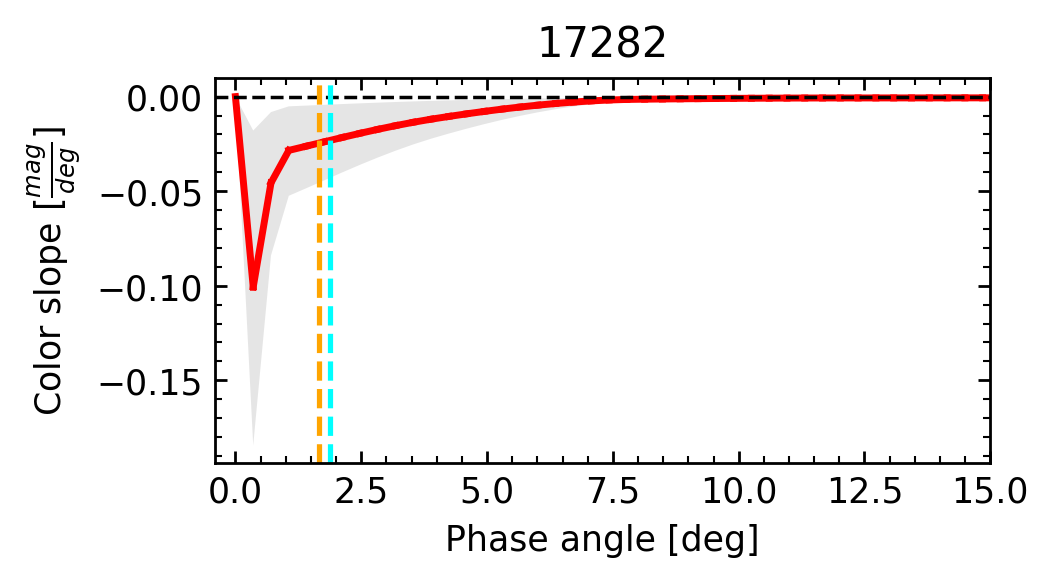}
\includegraphics[width=4cm]{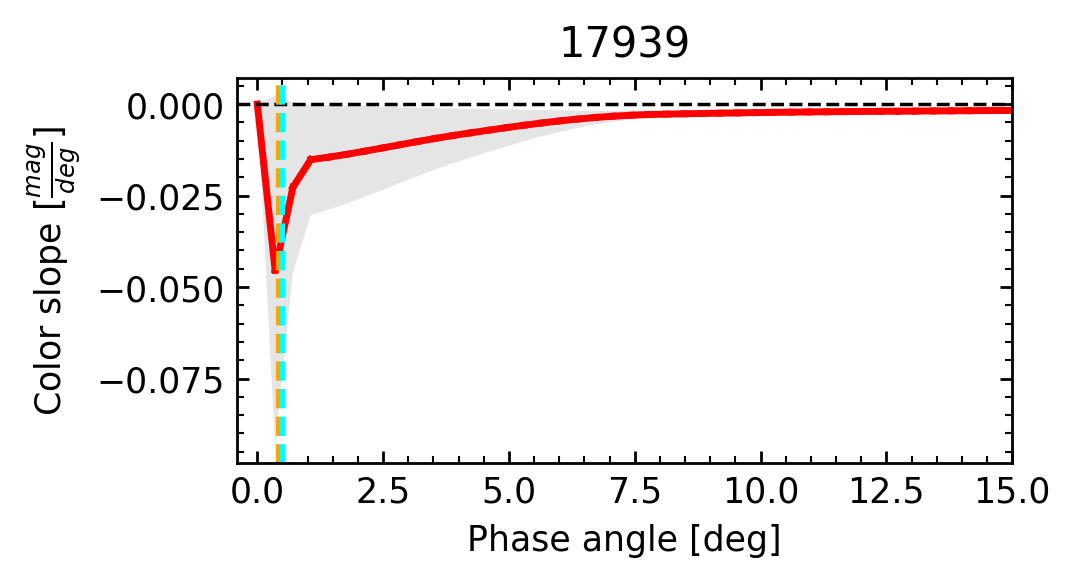}
\includegraphics[width=4cm]{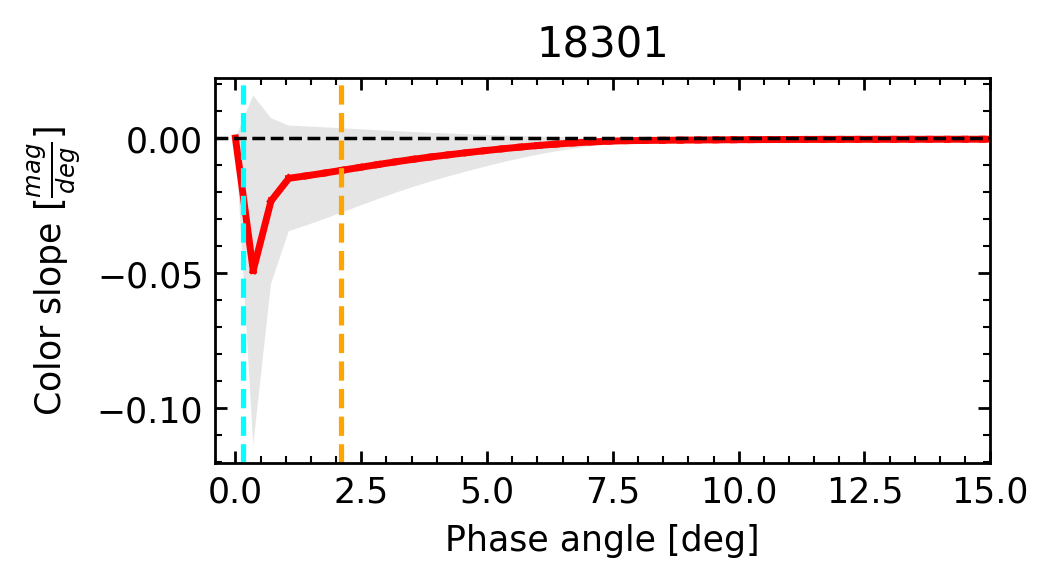}
\includegraphics[width=4cm]{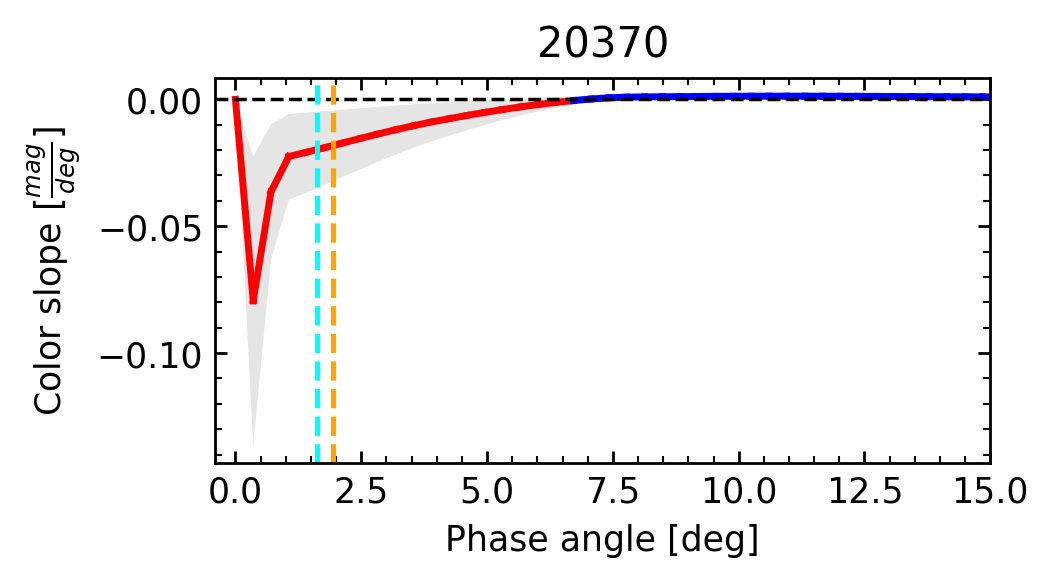}
\includegraphics[width=4cm]{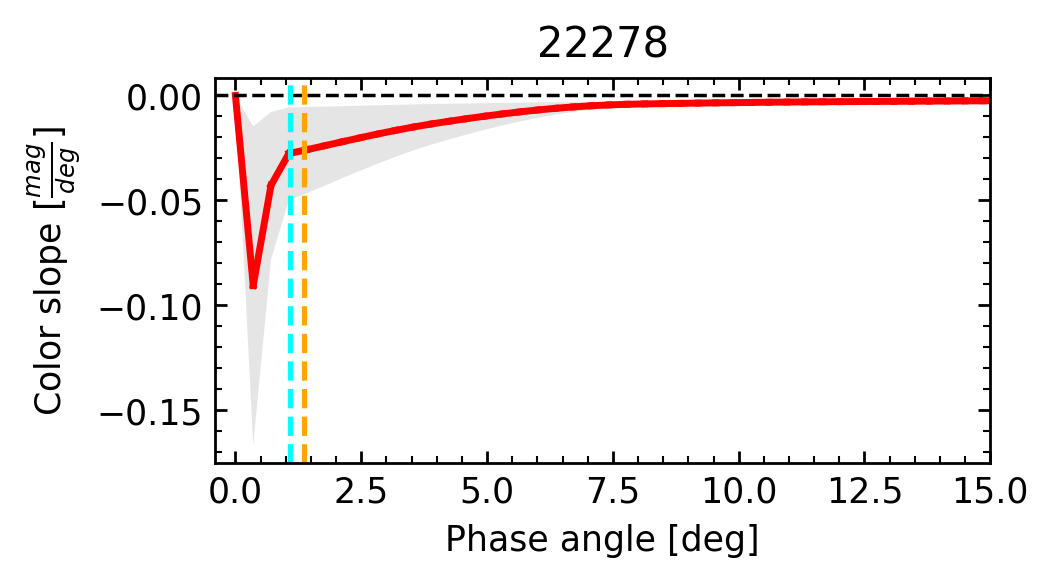}
\includegraphics[width=4cm]{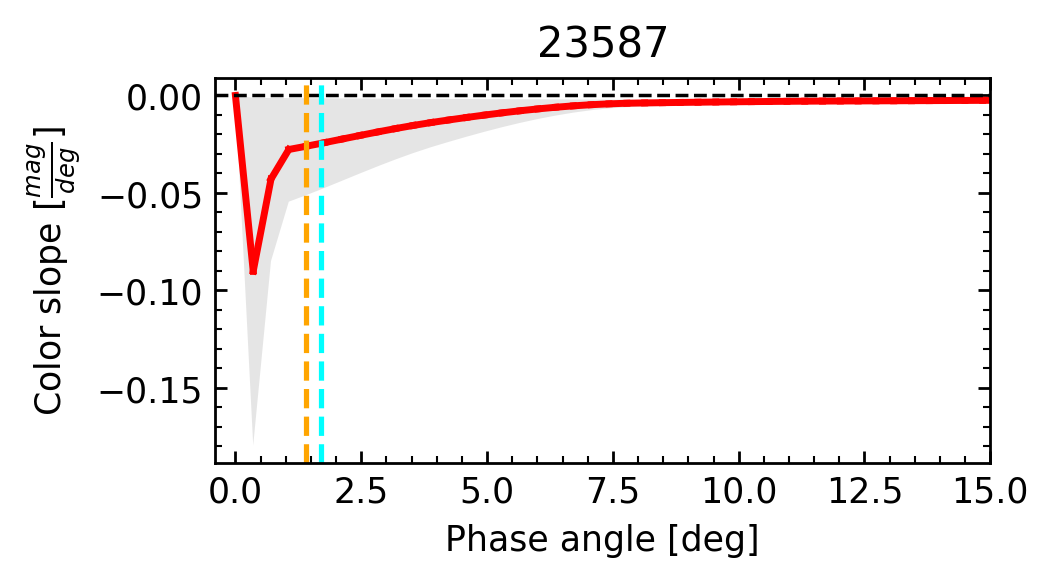}
\includegraphics[width=4cm]{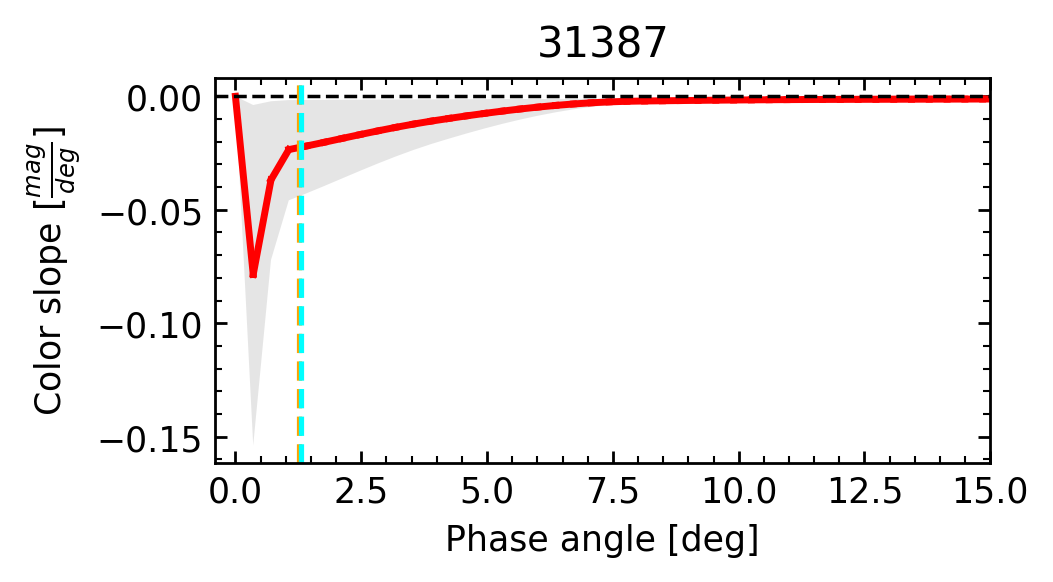}
\includegraphics[width=4cm]{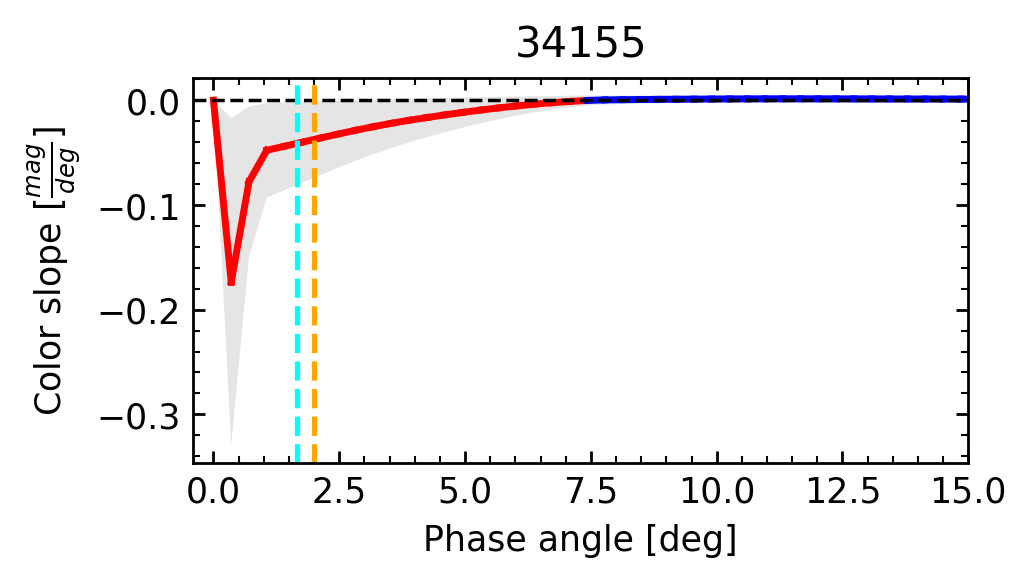}
\includegraphics[width=4cm]{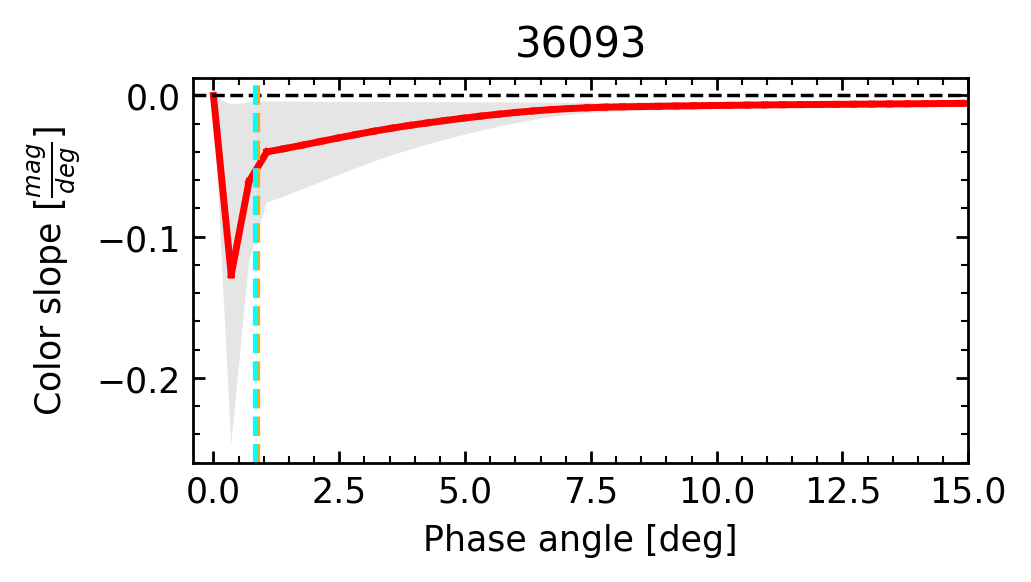}
\includegraphics[width=4cm]{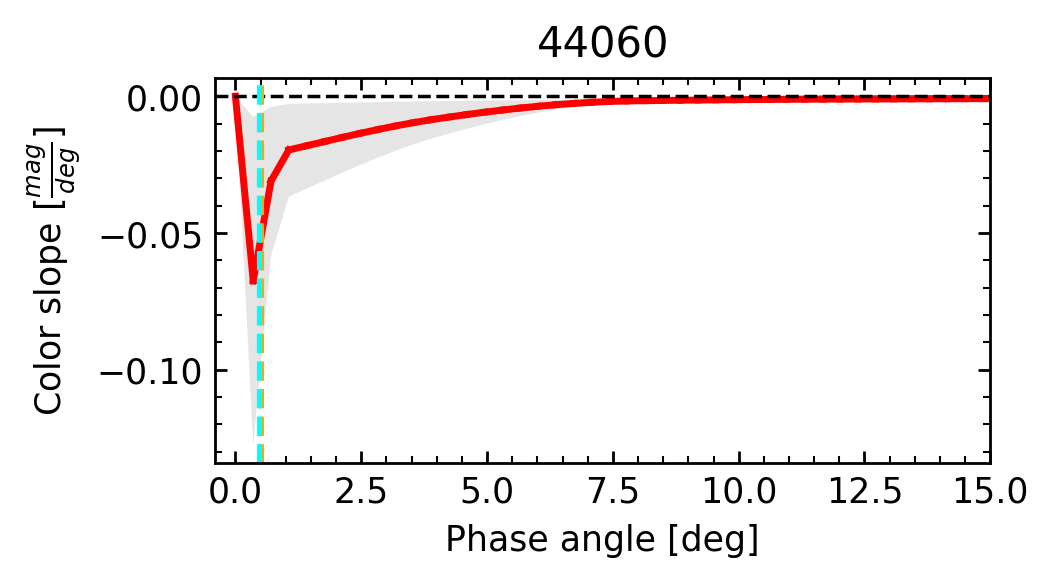}

\caption{(continued).}
\label{red2}
\end{figure}

\begin{figure}[ht]
\centering
\includegraphics[width=4cm]{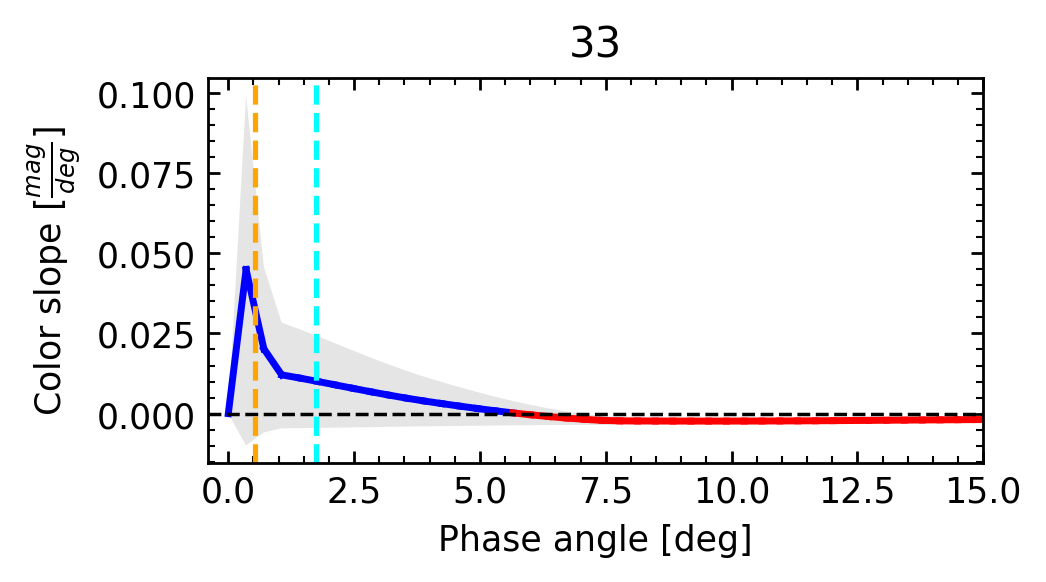}
\includegraphics[width=4cm]{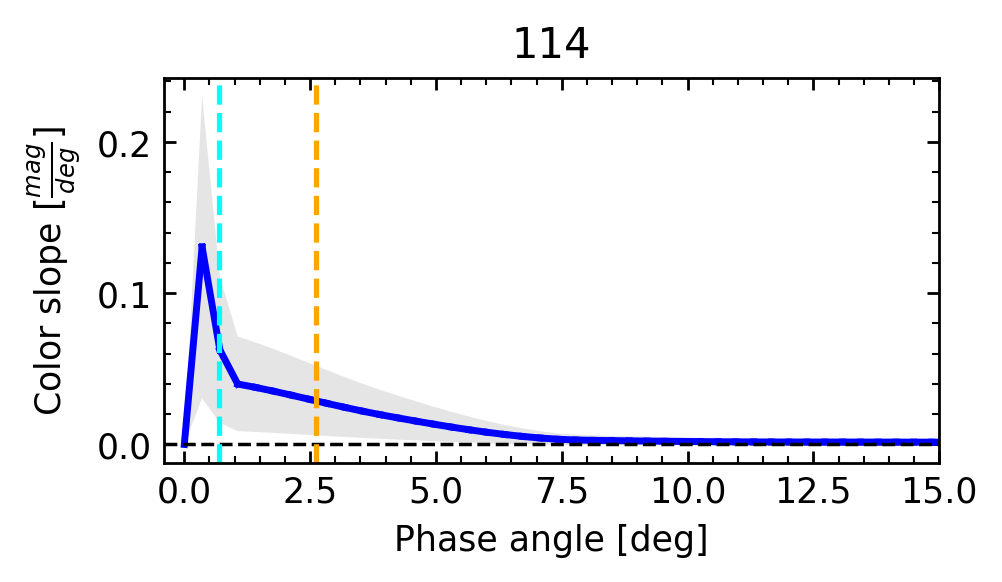}
\includegraphics[width=4cm]{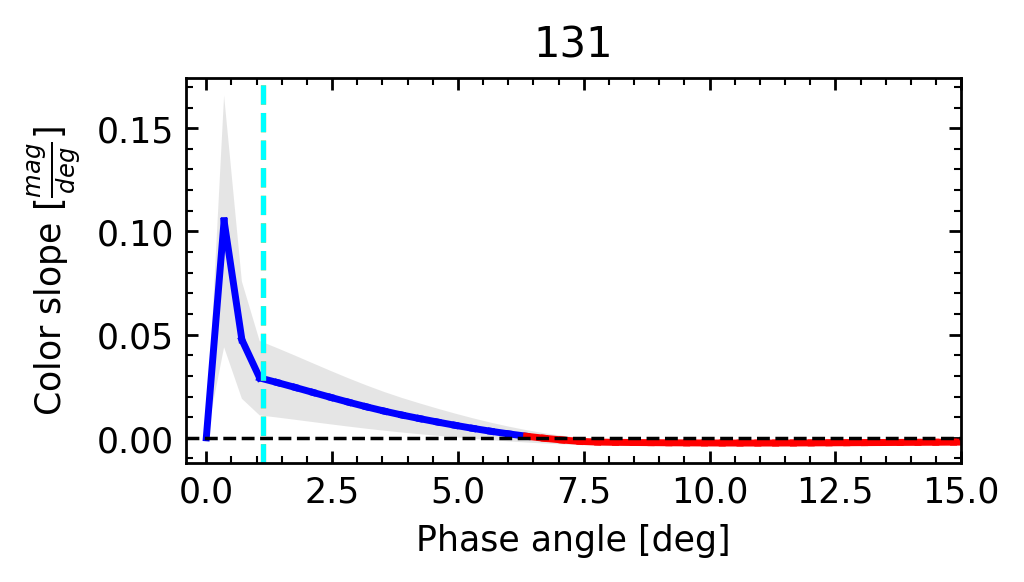}
\includegraphics[width=4cm]{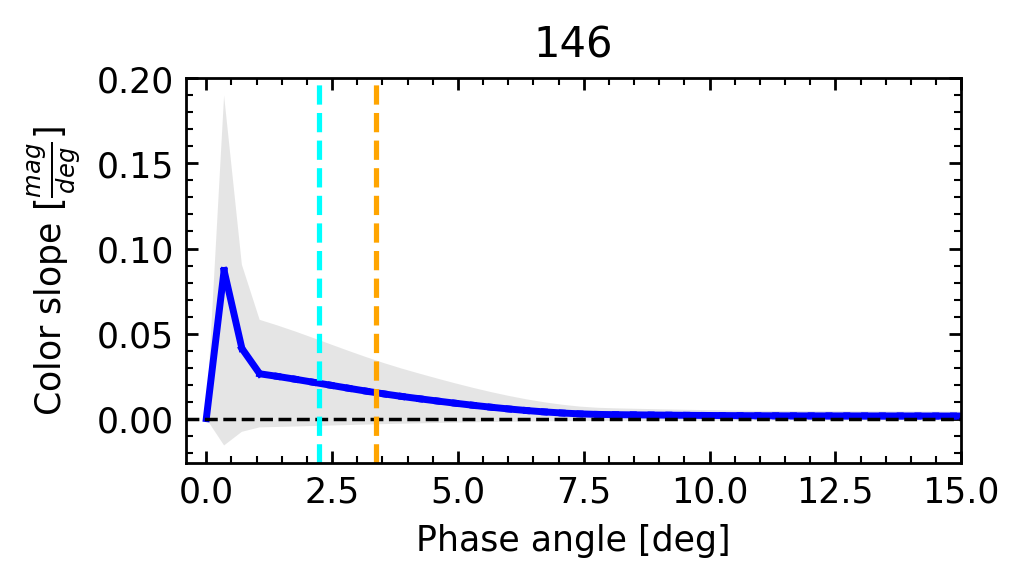}
\includegraphics[width=4cm]{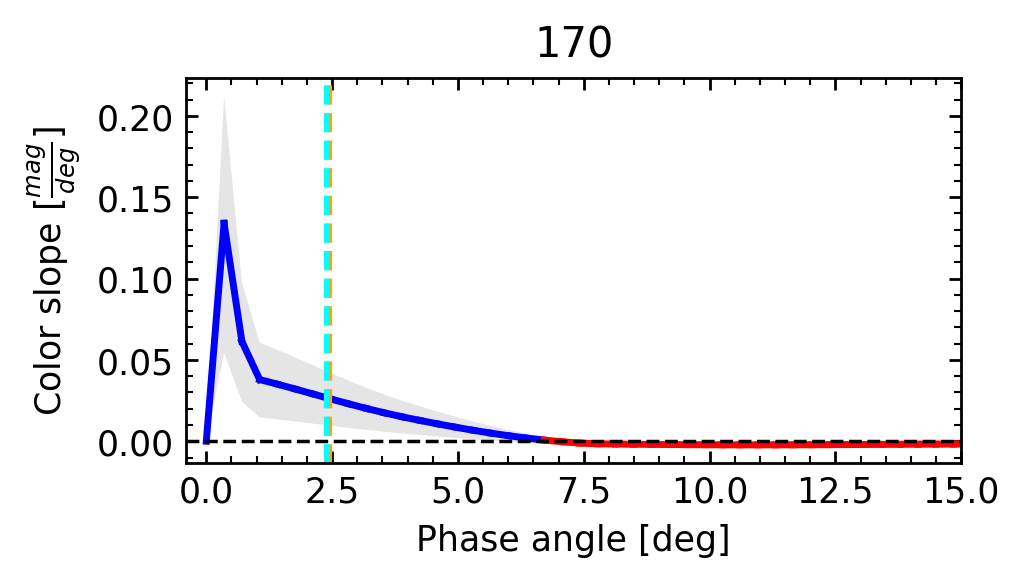}
\includegraphics[width=4cm]{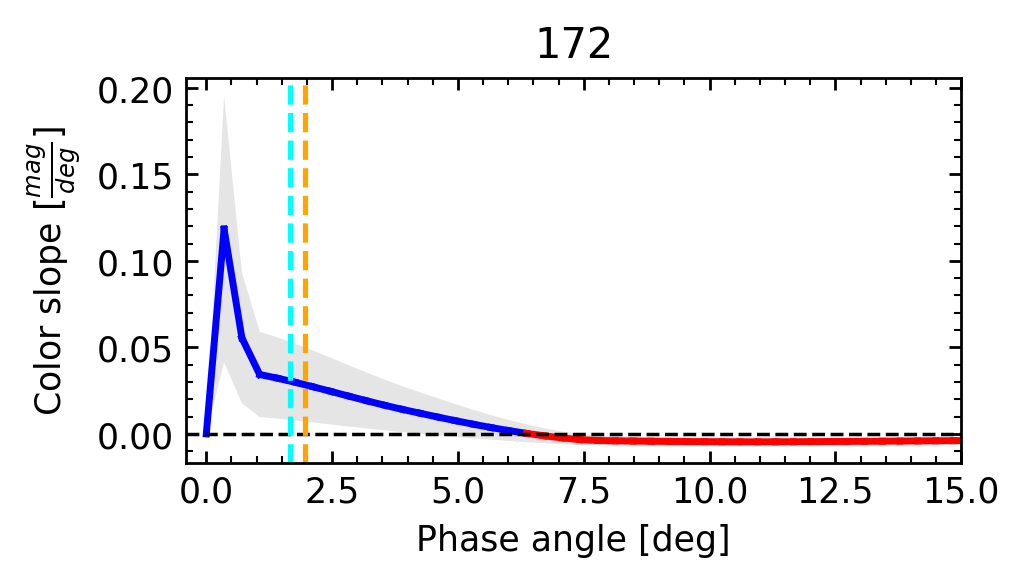}
\includegraphics[width=4cm]{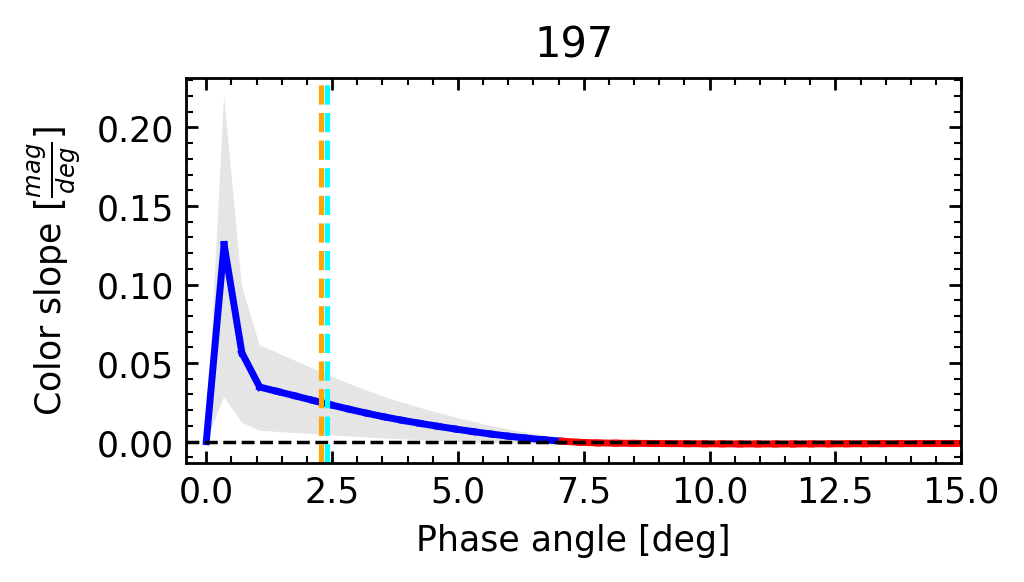}
\includegraphics[width=4cm]{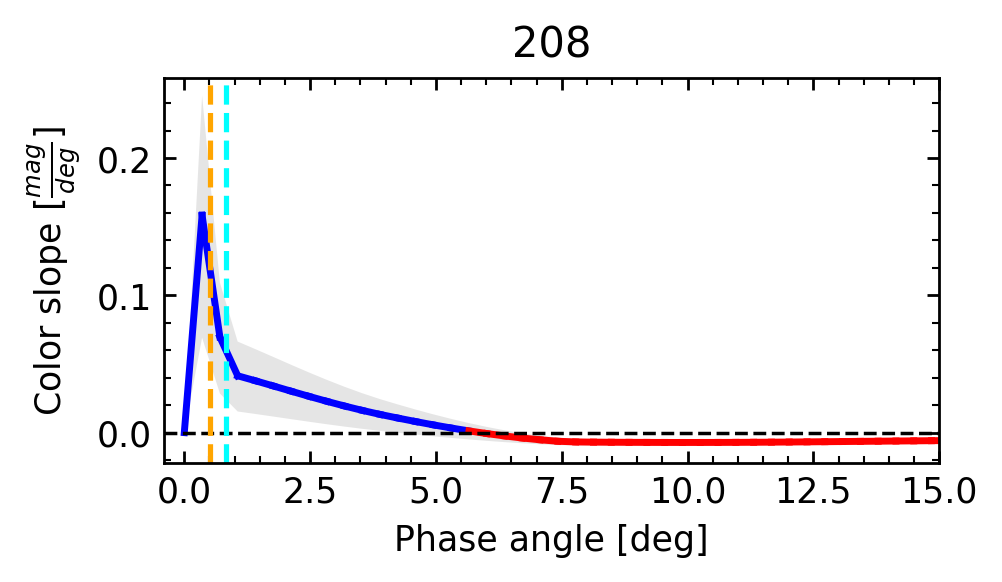}
\includegraphics[width=4cm]{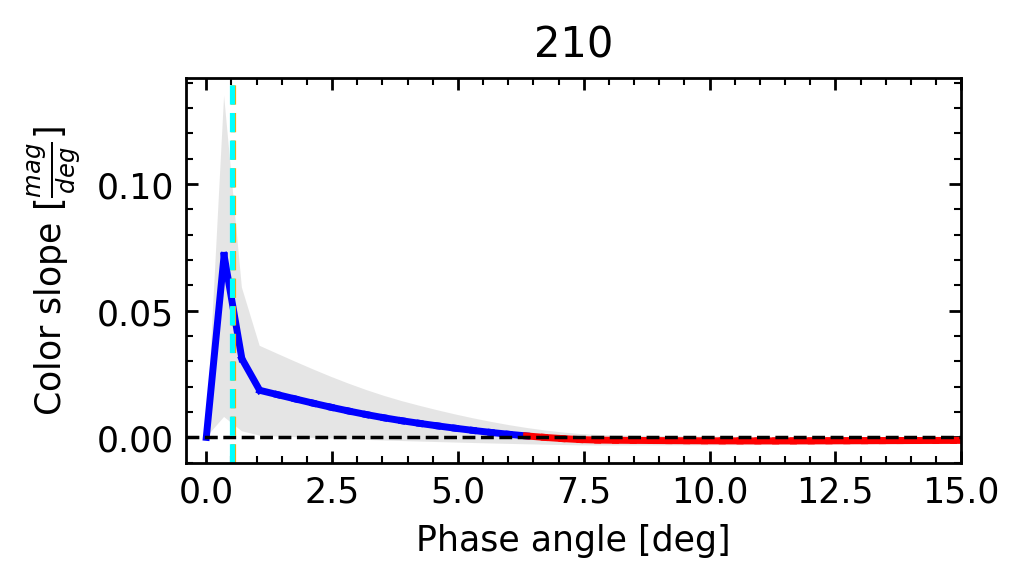}
\includegraphics[width=4cm]{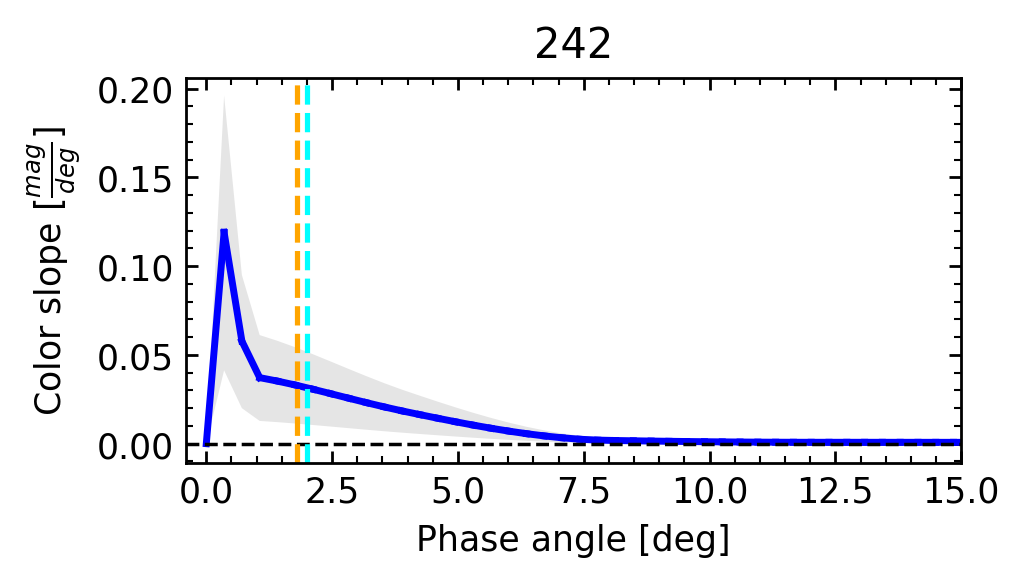}
\includegraphics[width=4cm]{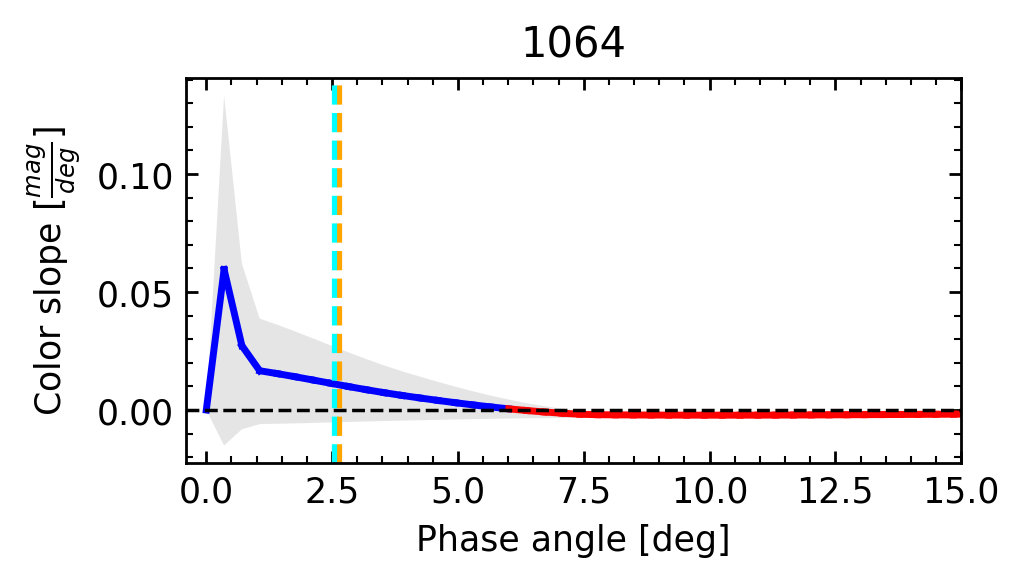}
\includegraphics[width=4cm]{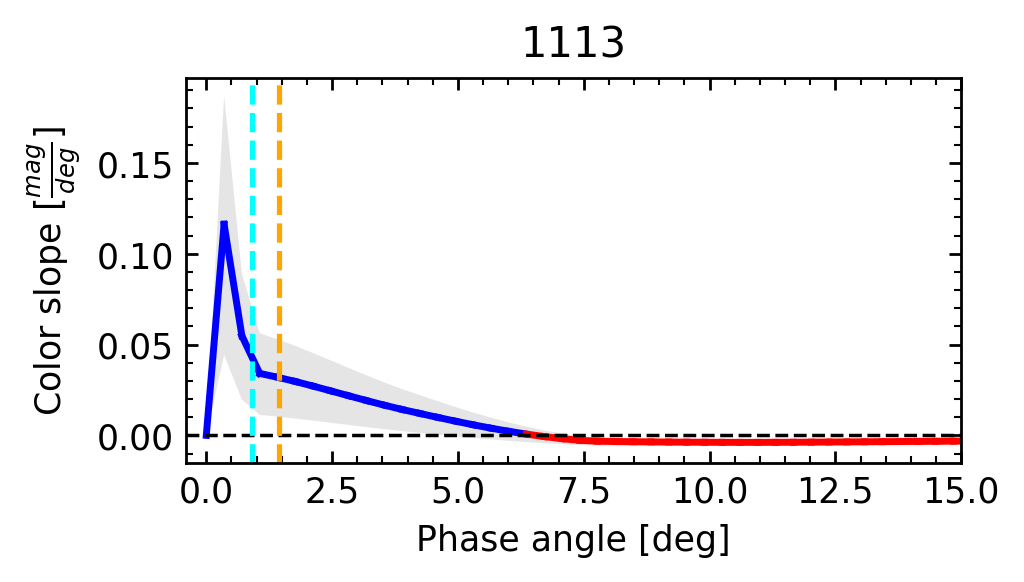}
\includegraphics[width=4cm]{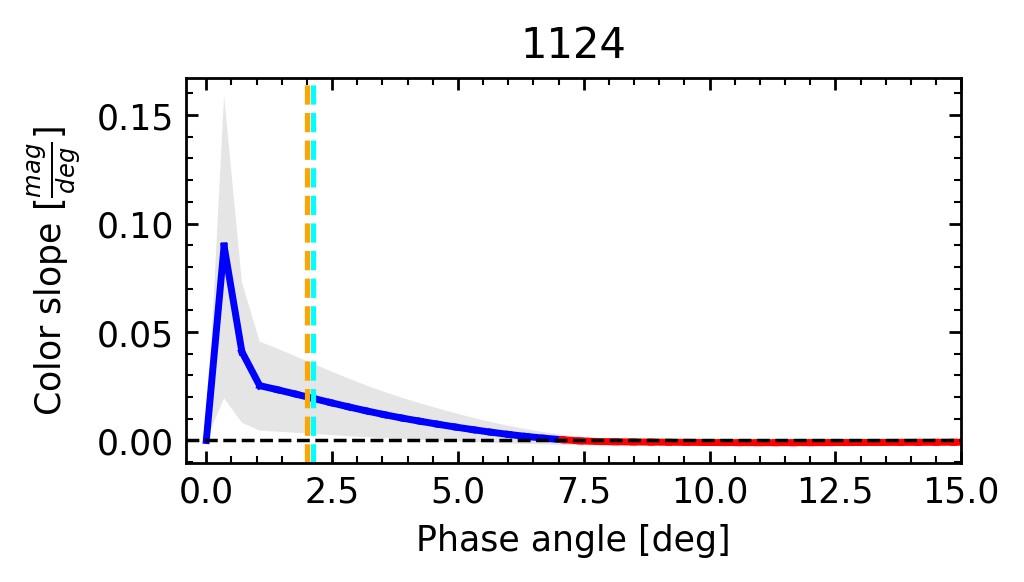}
\includegraphics[width=4cm]{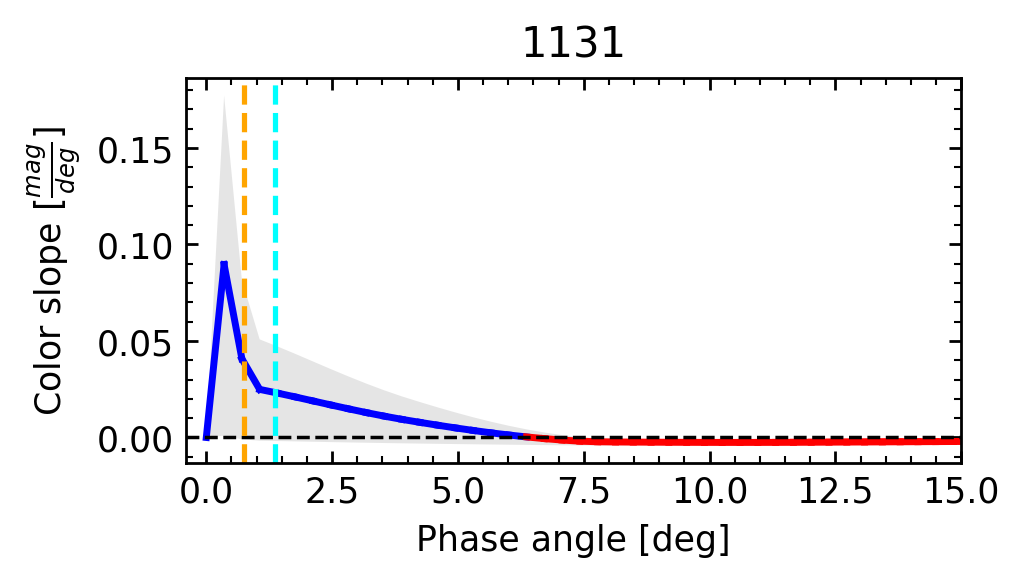}
\includegraphics[width=4cm]{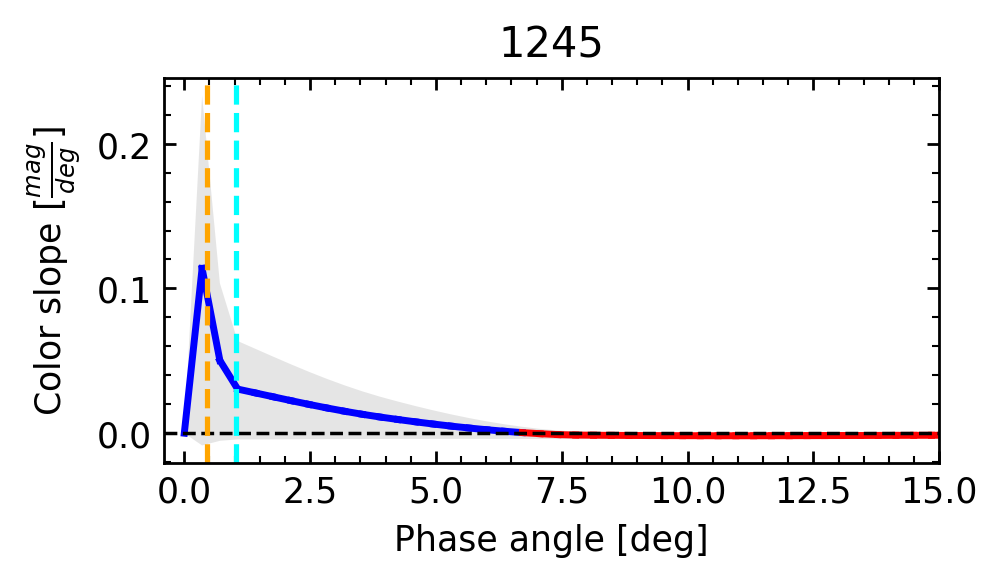}
\includegraphics[width=4cm]{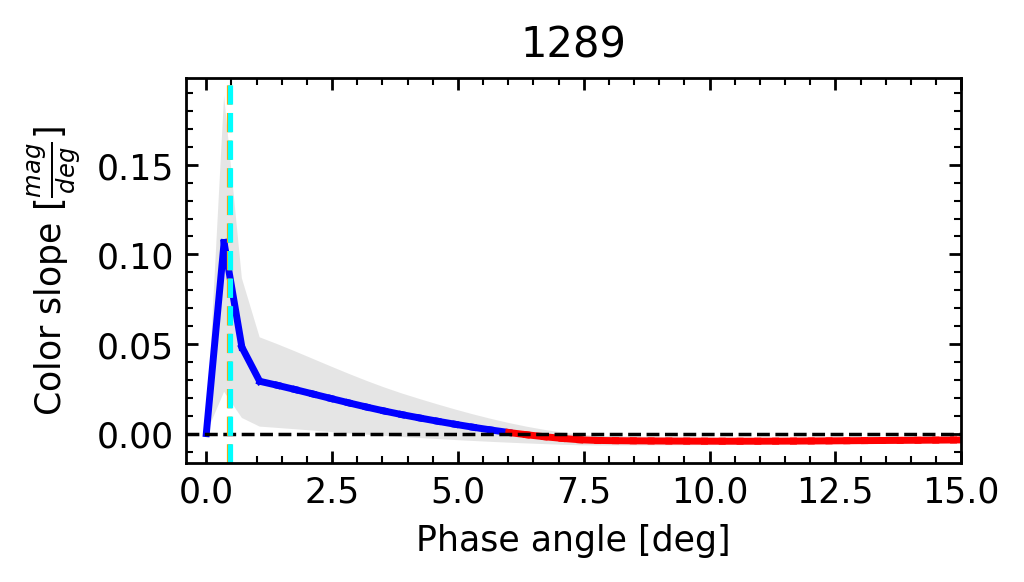}
\includegraphics[width=4cm]{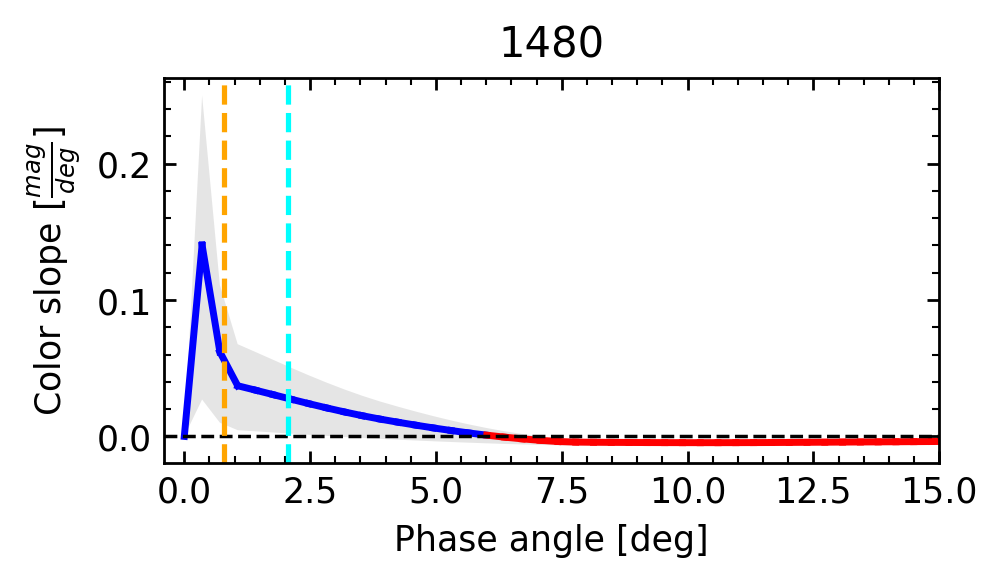}
\includegraphics[width=4cm]{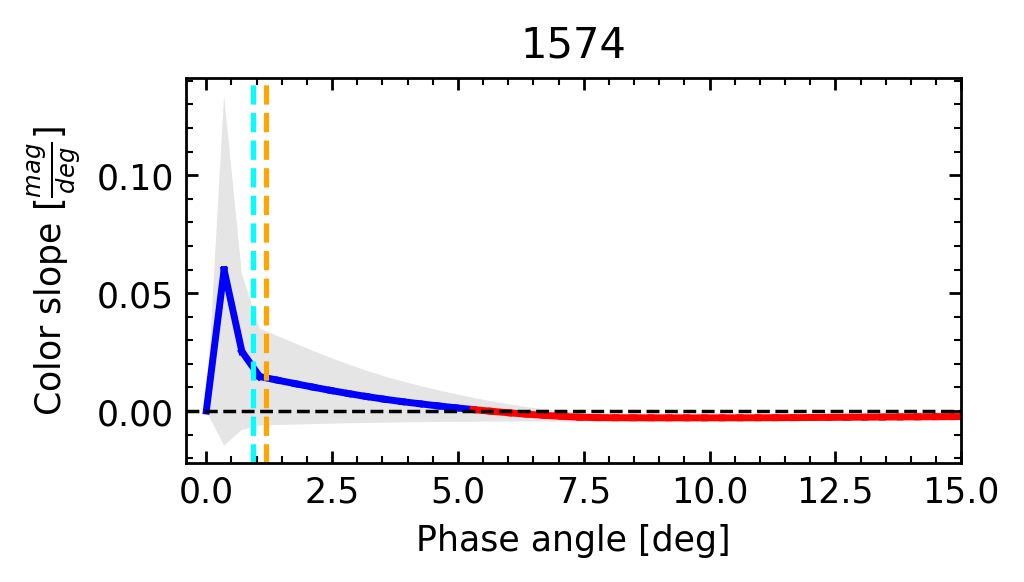}
\includegraphics[width=4cm]{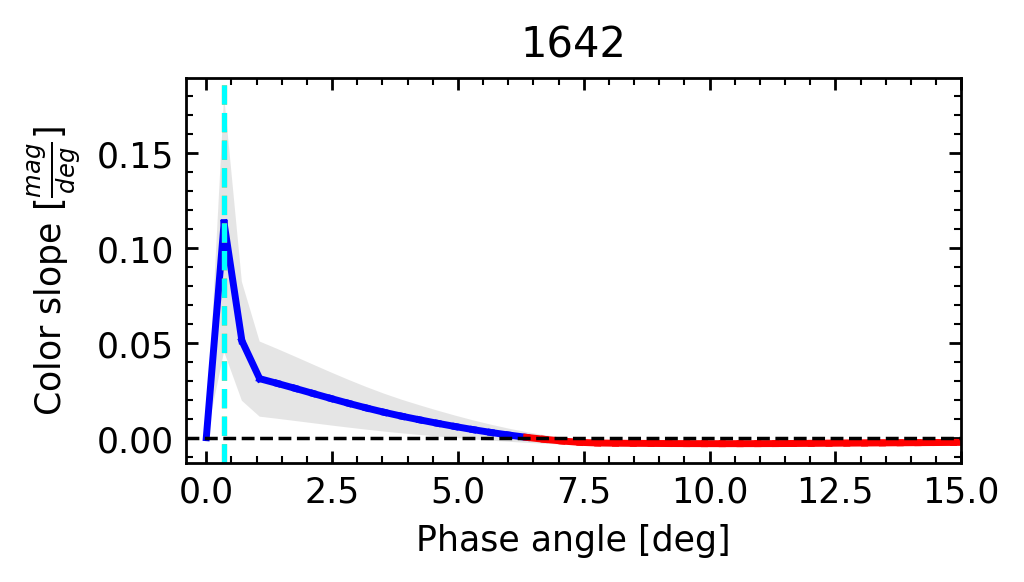}
\includegraphics[width=4cm]{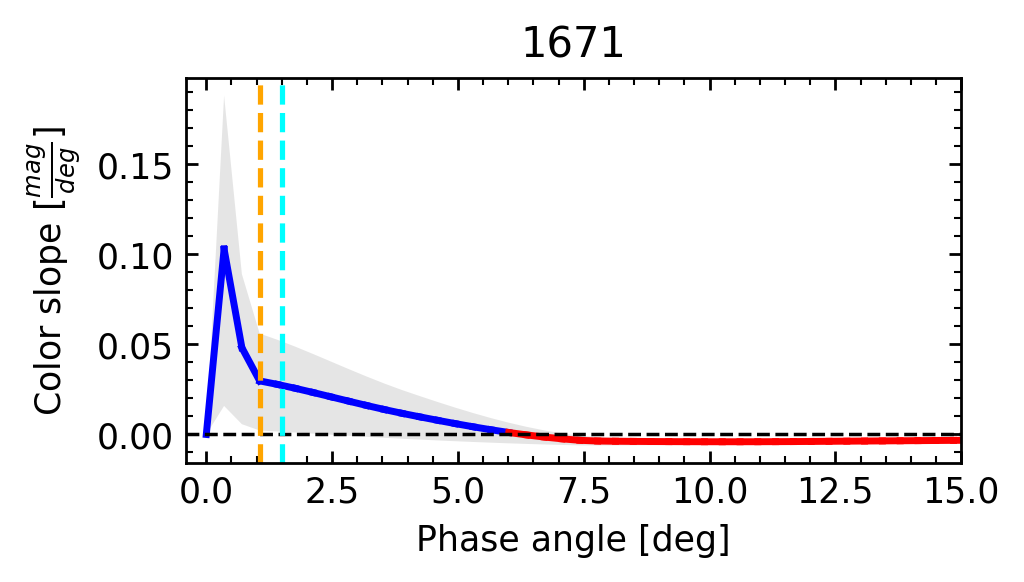}
\includegraphics[width=4cm]{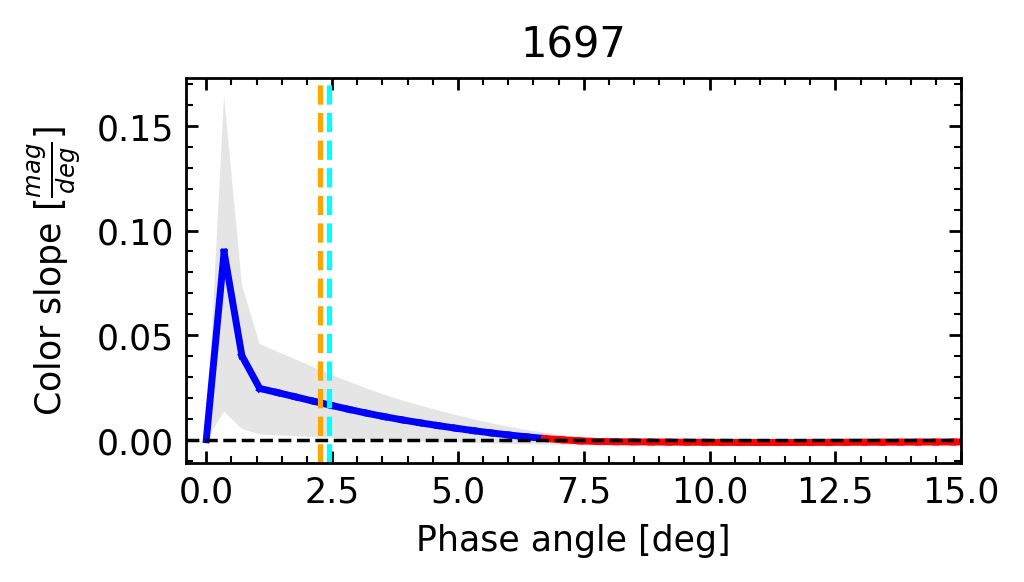}
\includegraphics[width=4cm]{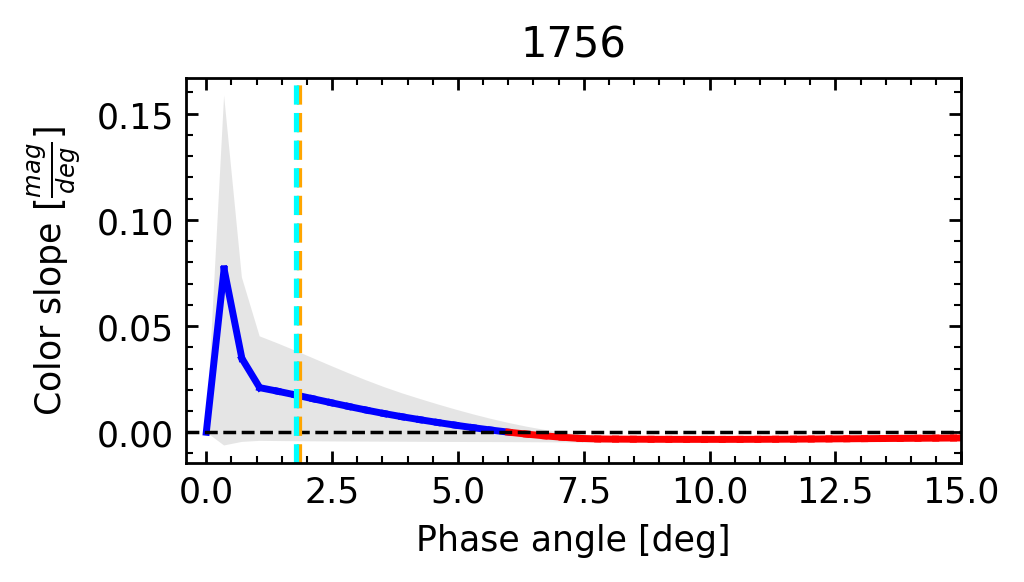}
\includegraphics[width=4cm]{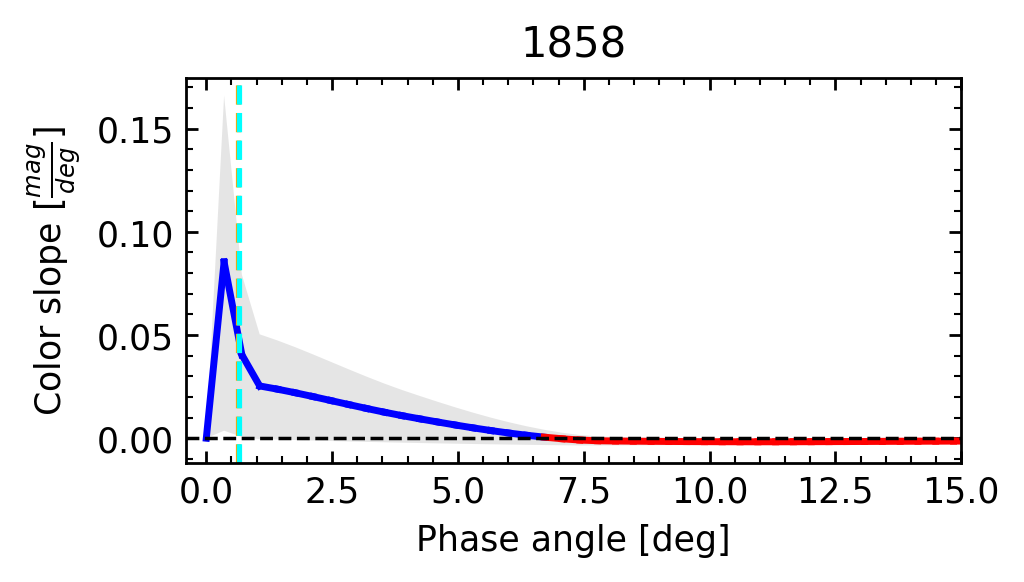}
\includegraphics[width=4cm]{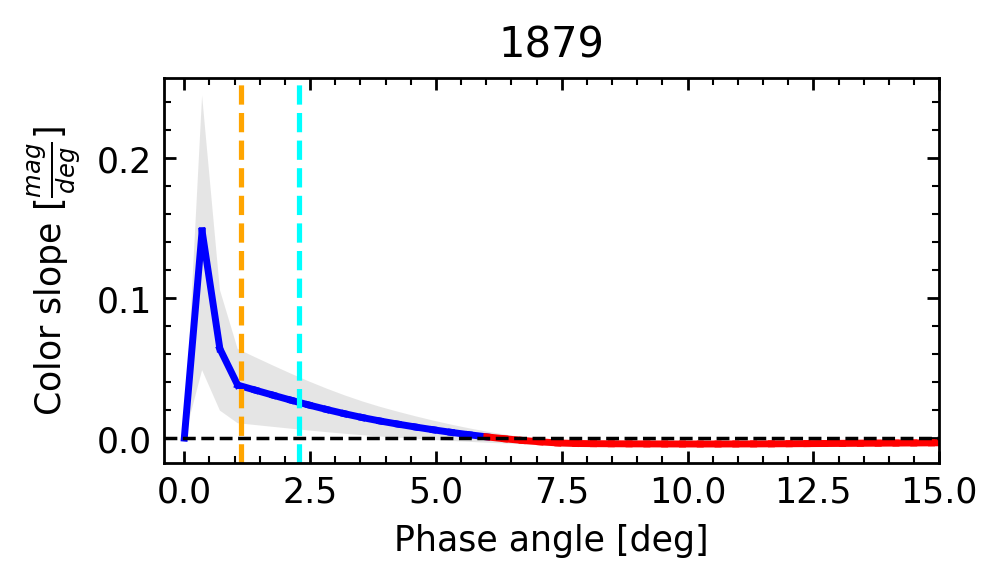}
\includegraphics[width=4cm]{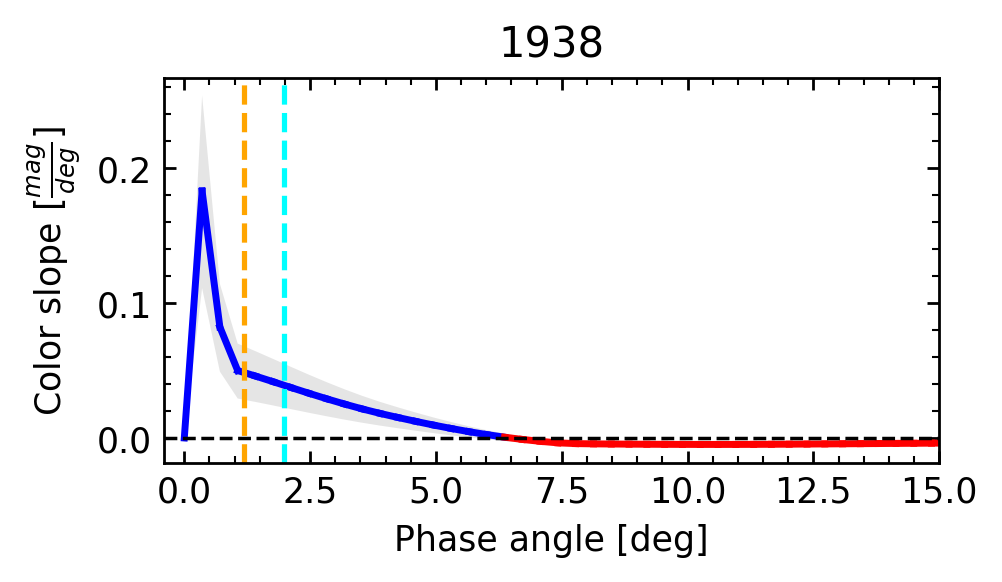}
\includegraphics[width=4cm]{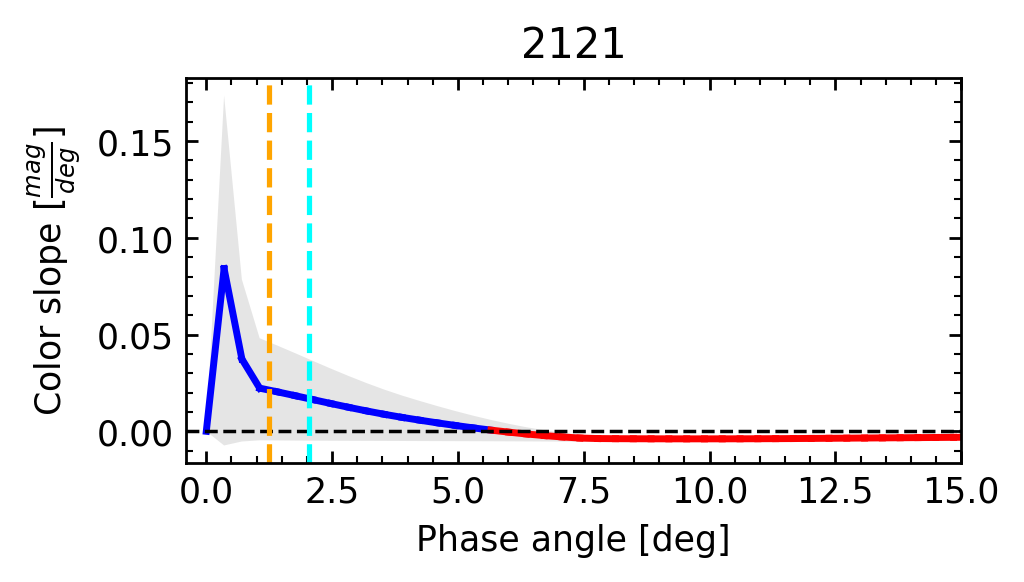}
\includegraphics[width=4cm]{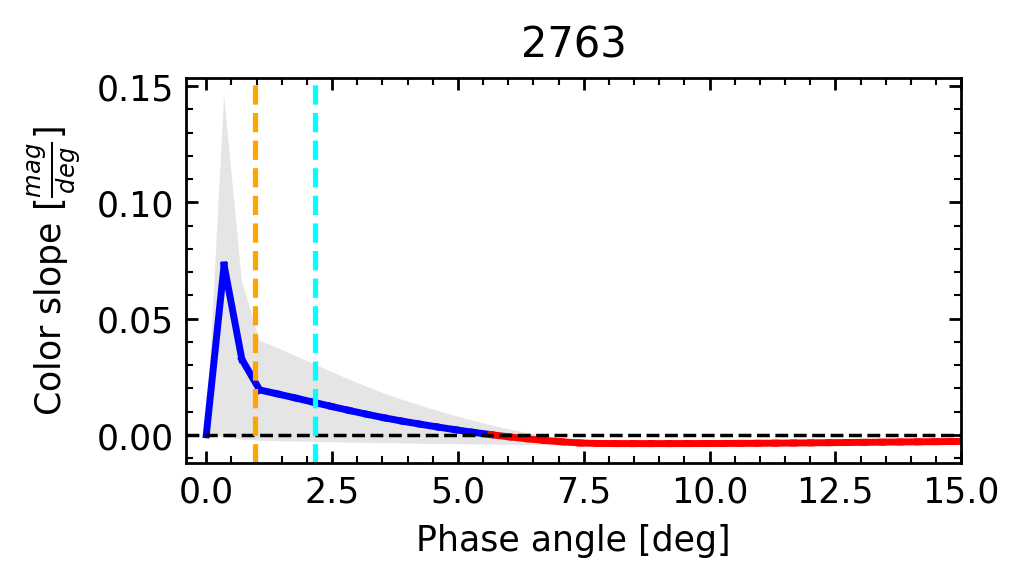}
\includegraphics[width=4cm]{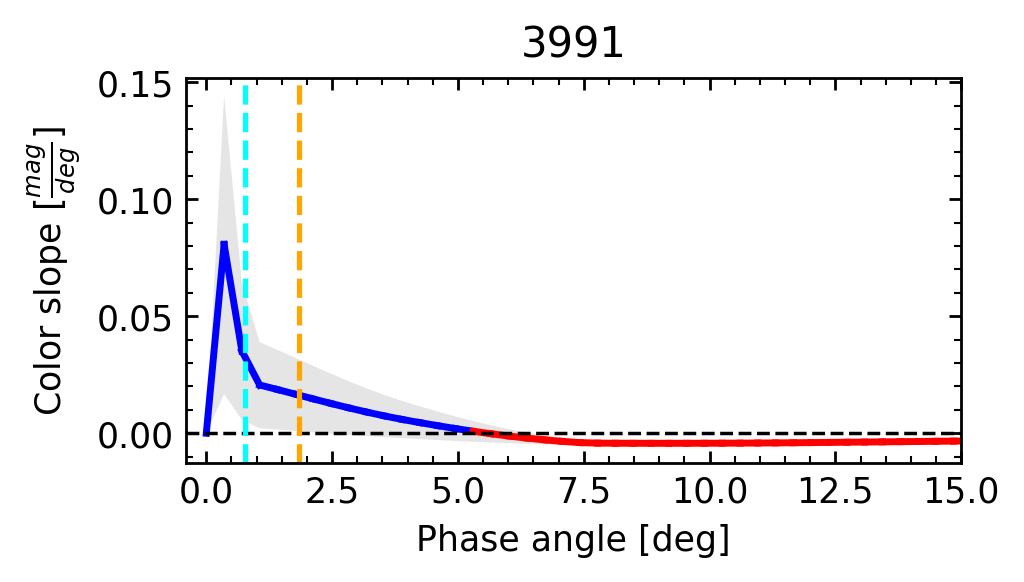}
\includegraphics[width=4cm]{color_slope_10144.png}
\includegraphics[width=4cm]{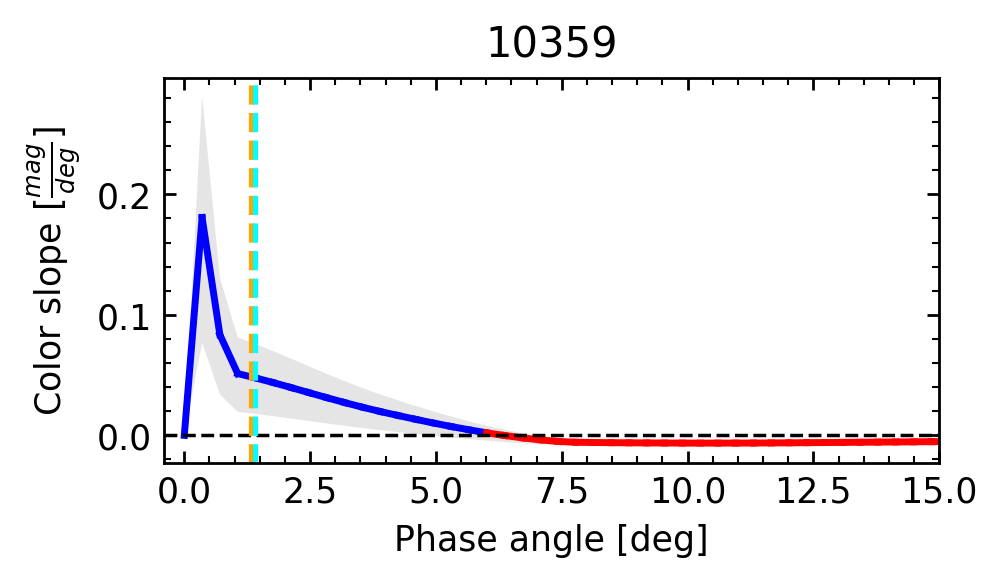}
\includegraphics[width=4cm]{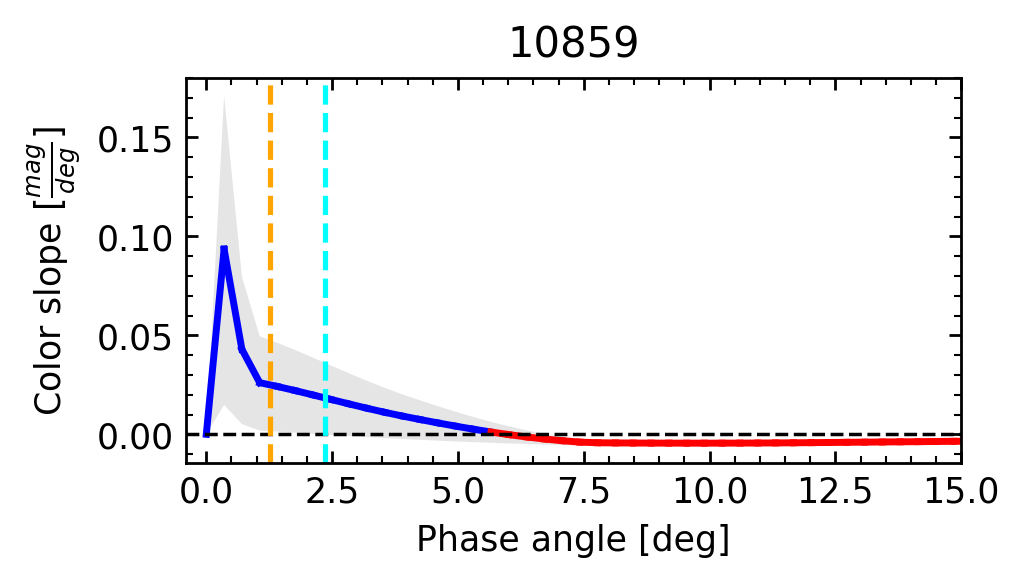}
\includegraphics[width=4cm]{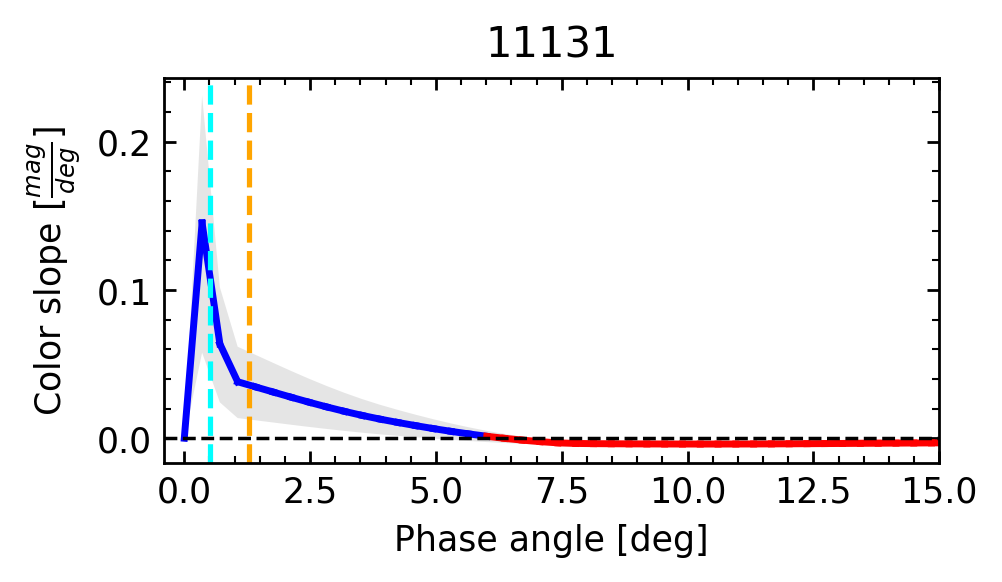}
\caption{Color slope, representing the difference between the first derivatives of the phase function with respect to the phase angle in the two filters for asteroids showing bluening effect. Vertical lines inidicates the minimum observed phase angles separately in both filters.}
\label{blue}
\end{figure}

\begin{figure}[ht]
\ContinuedFloat
\captionsetup{list=off,format=cont}
\centering
\includegraphics[width=4cm]{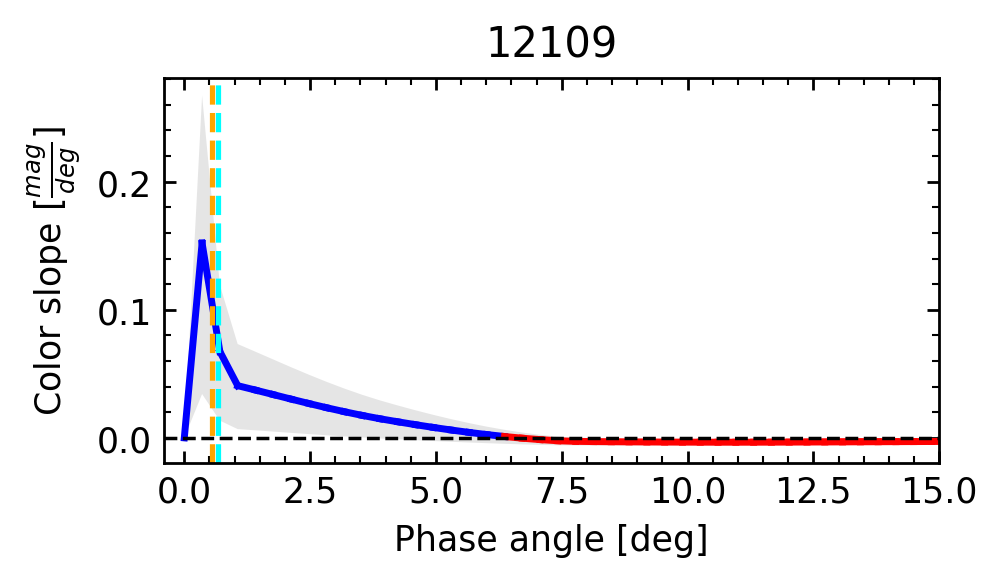}
\includegraphics[width=4cm]{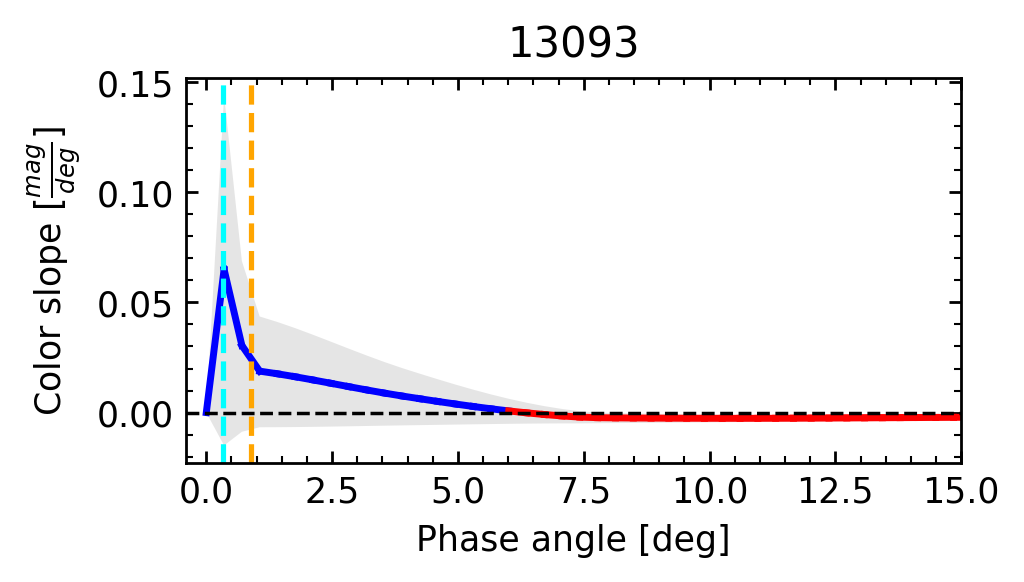}
\includegraphics[width=4cm]{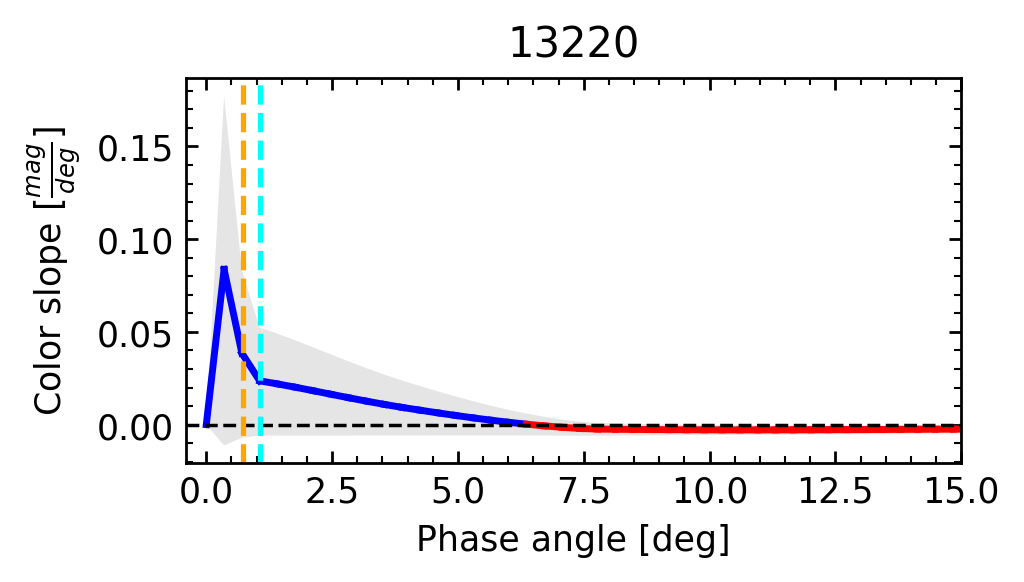}
\includegraphics[width=4cm]{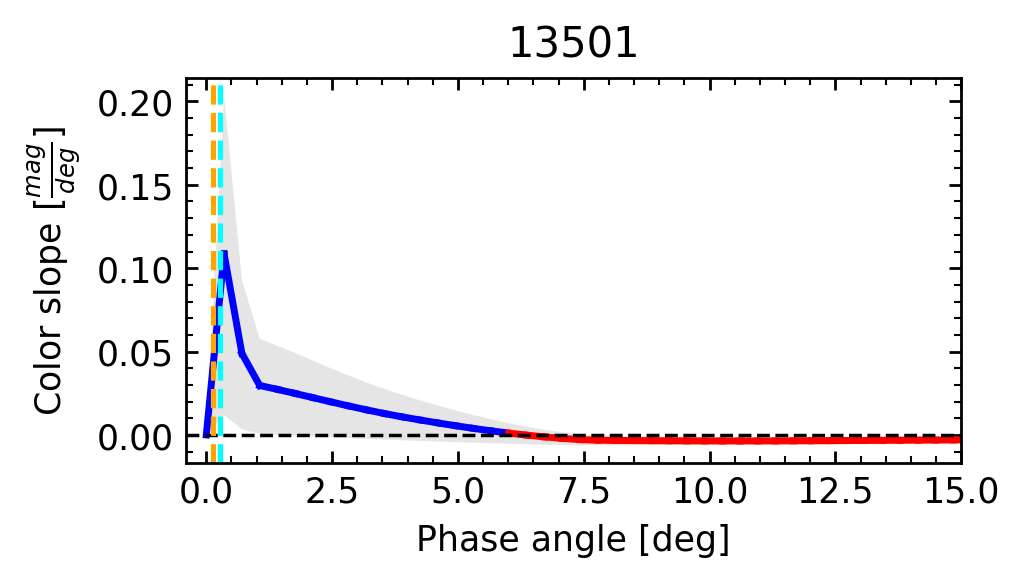}
\includegraphics[width=4cm]{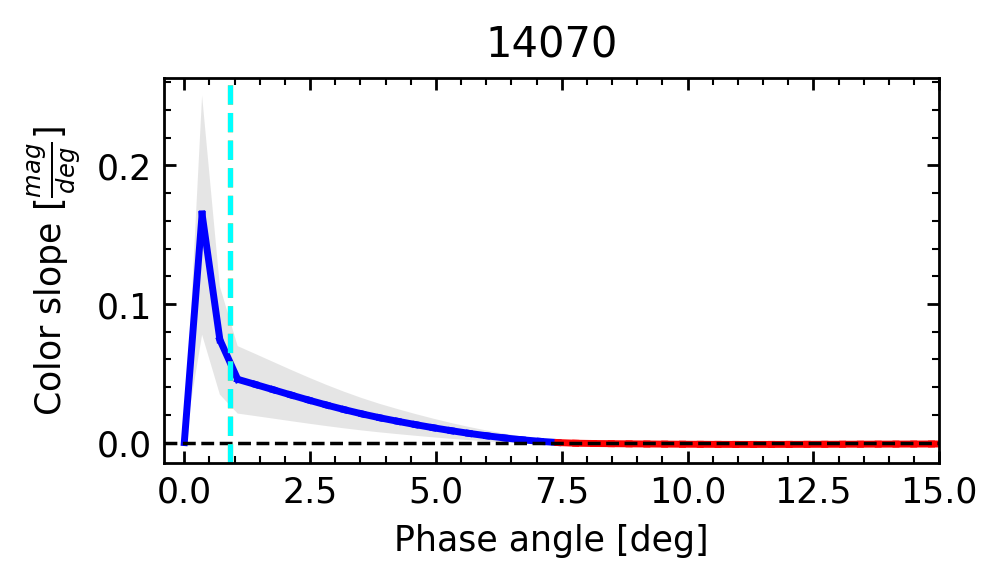}
\includegraphics[width=4cm]{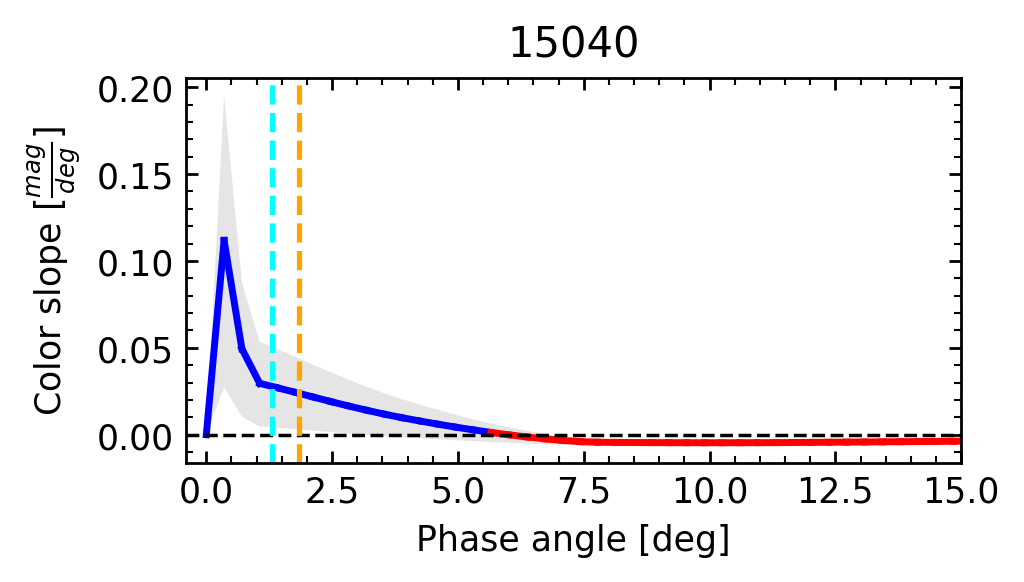}
\includegraphics[width=4cm]{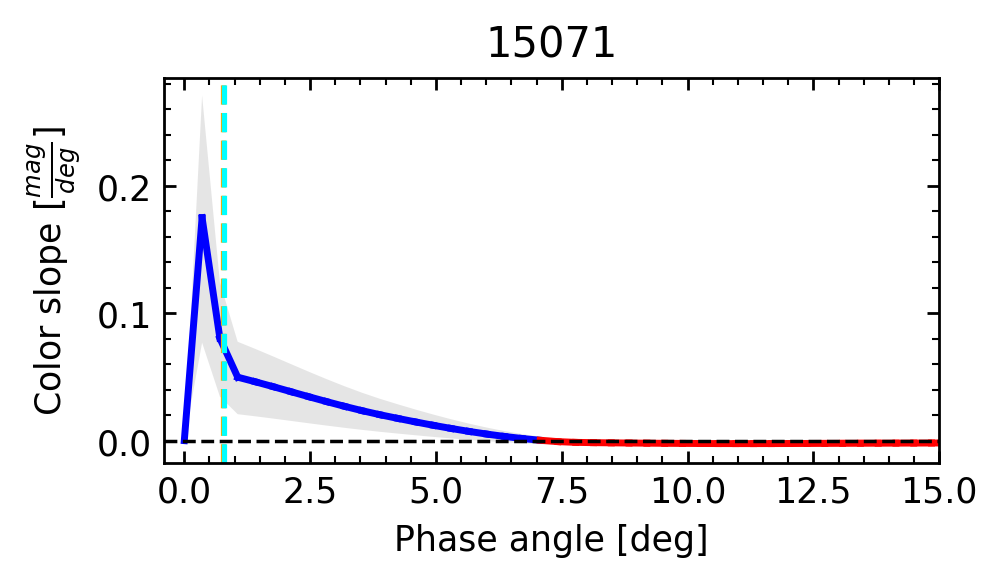}
\includegraphics[width=4cm]{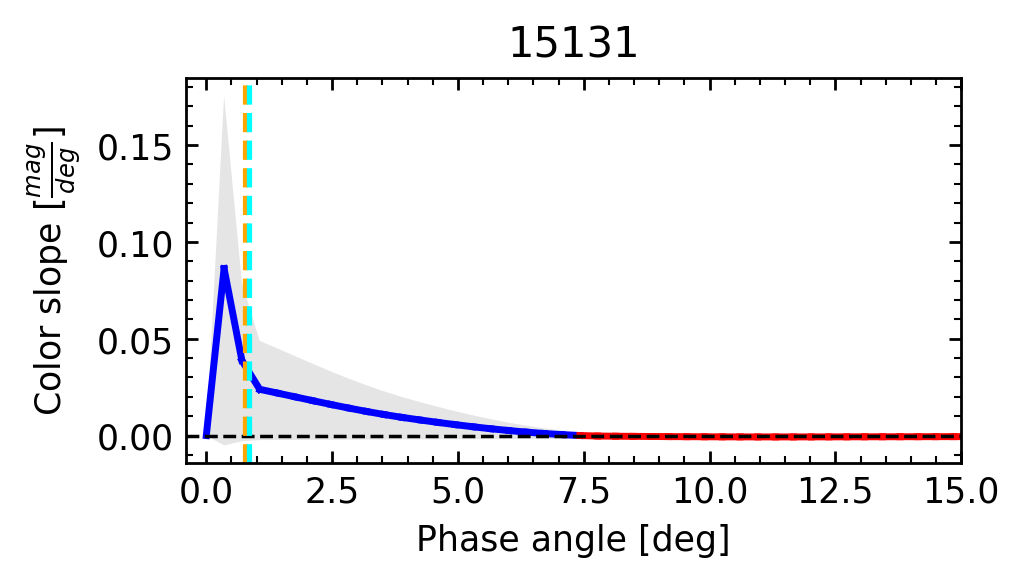}
\includegraphics[width=4cm]{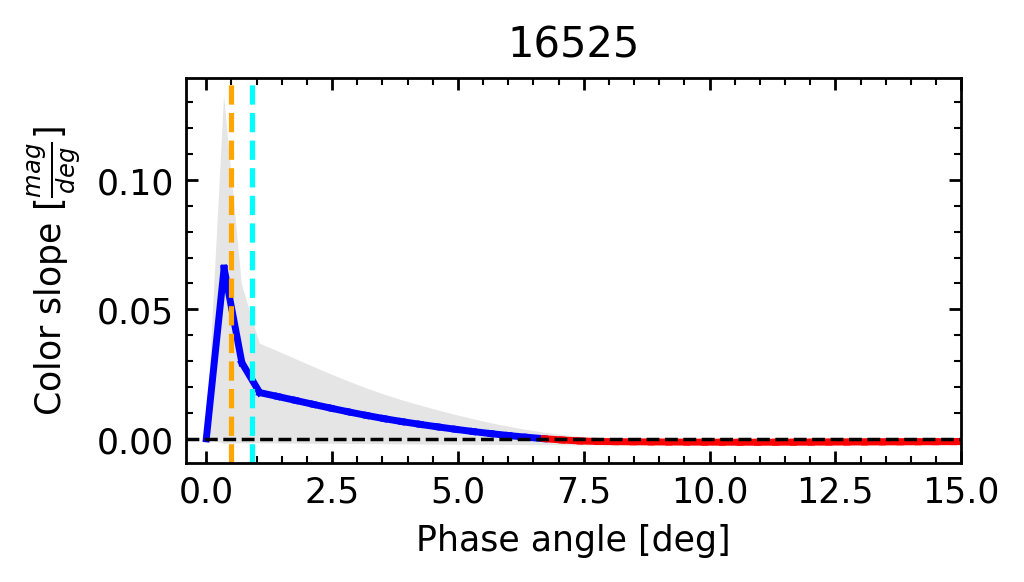}
\includegraphics[width=4cm]{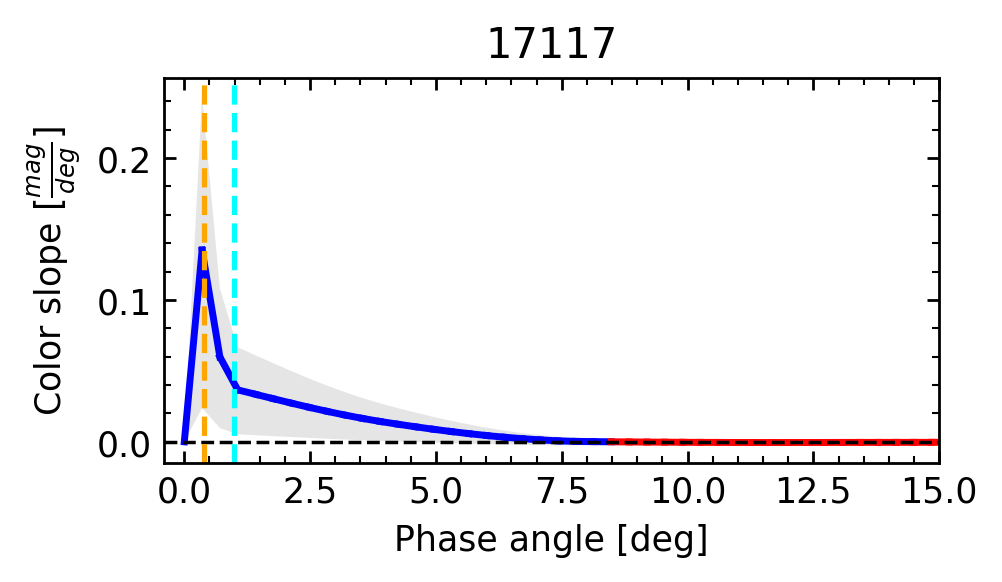}
\includegraphics[width=4cm]{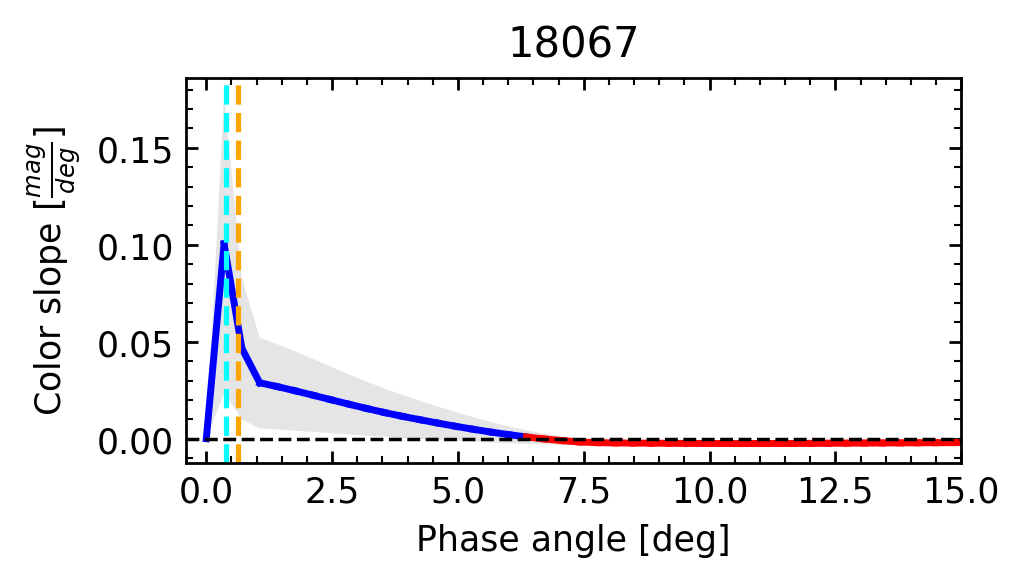}
\includegraphics[width=4cm]{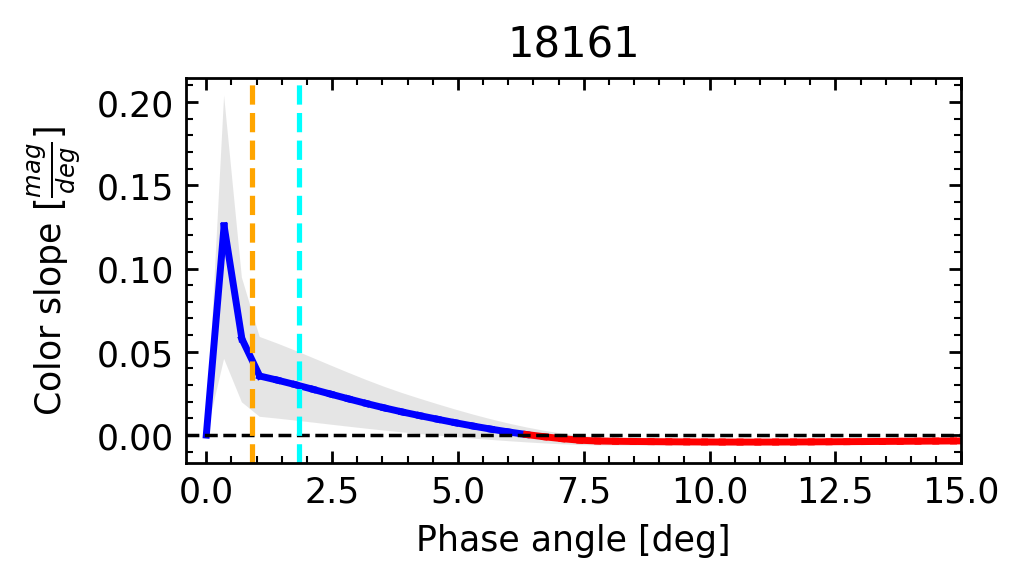}
\includegraphics[width=4cm]{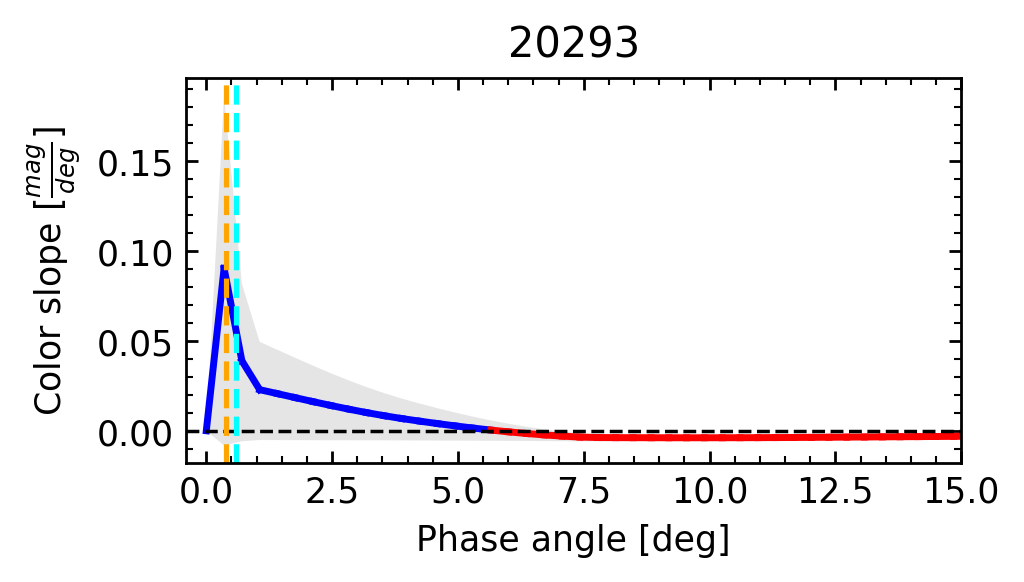}
\includegraphics[width=4cm]{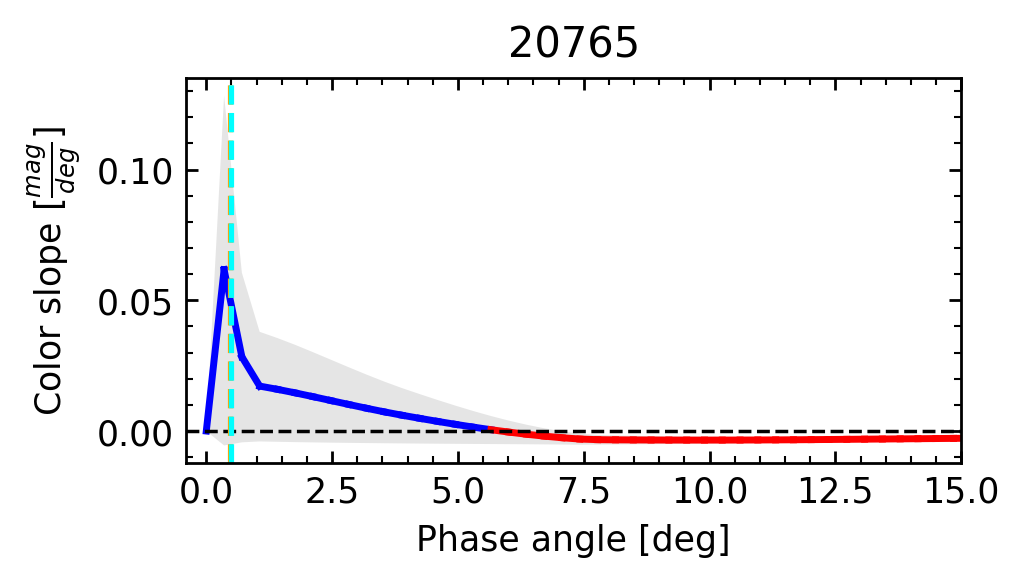}
\includegraphics[width=4cm]{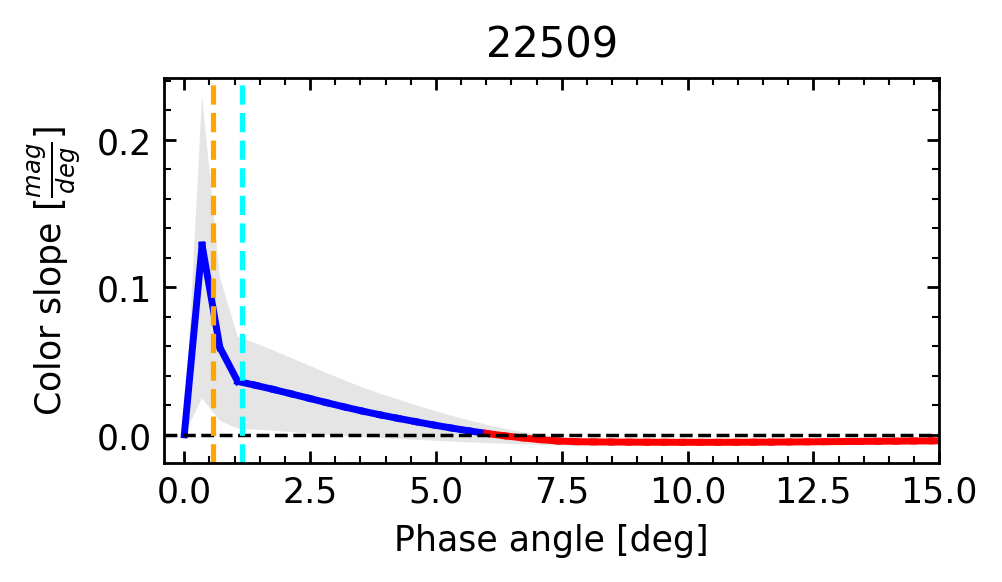}
\includegraphics[width=4cm]{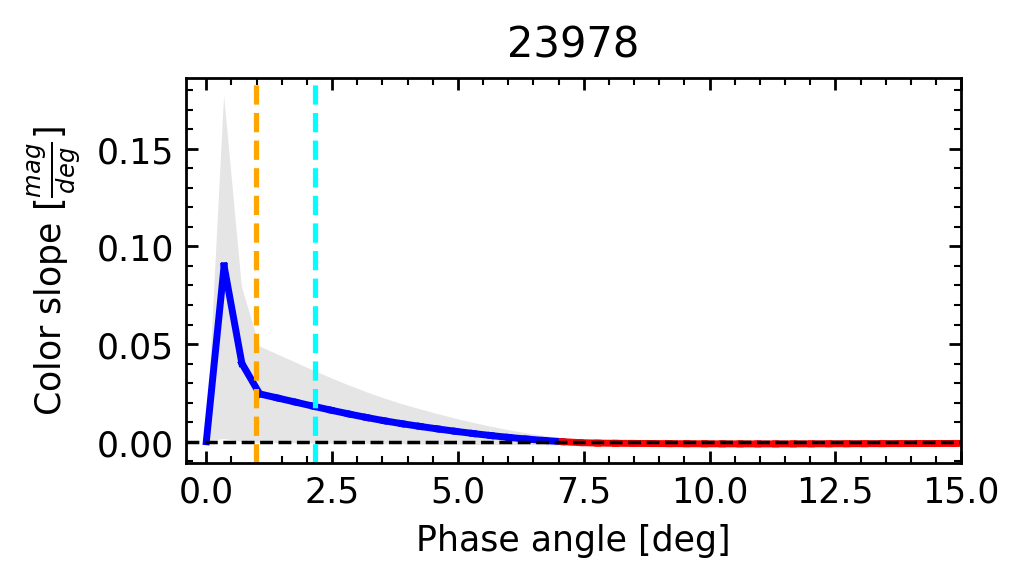}
\includegraphics[width=4cm]{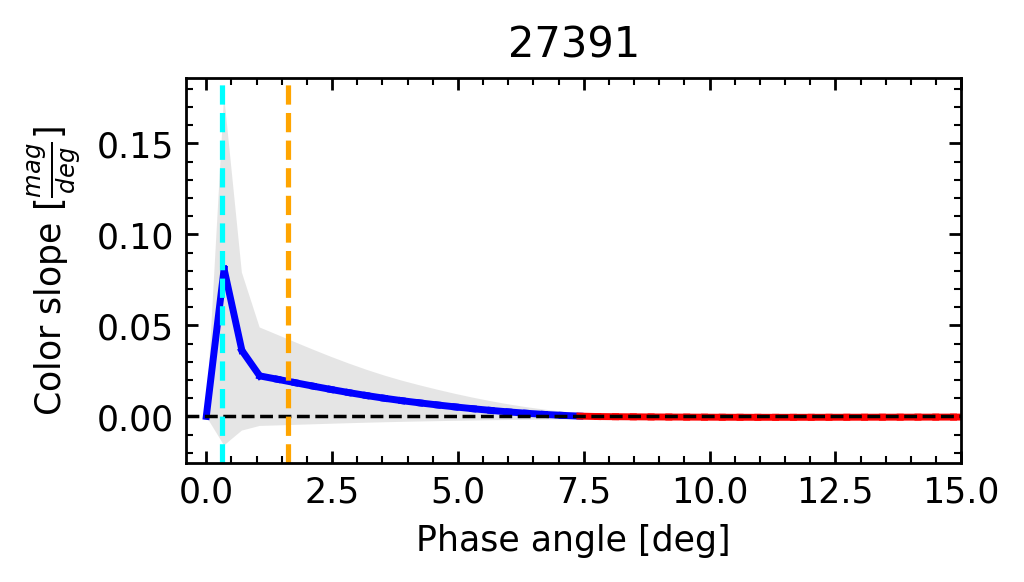}
\includegraphics[width=4cm]{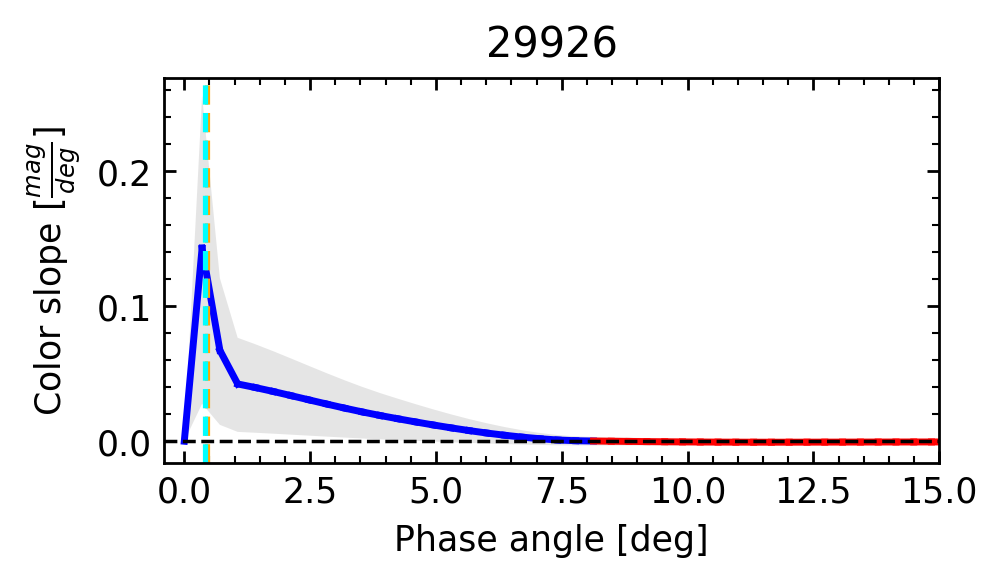}
\includegraphics[width=4cm]{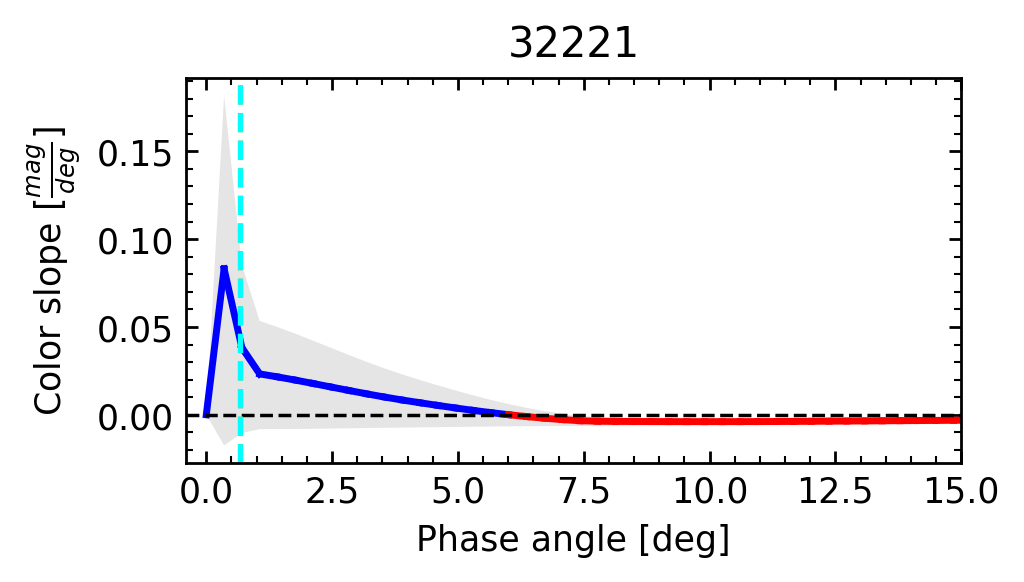}
\includegraphics[width=4cm]{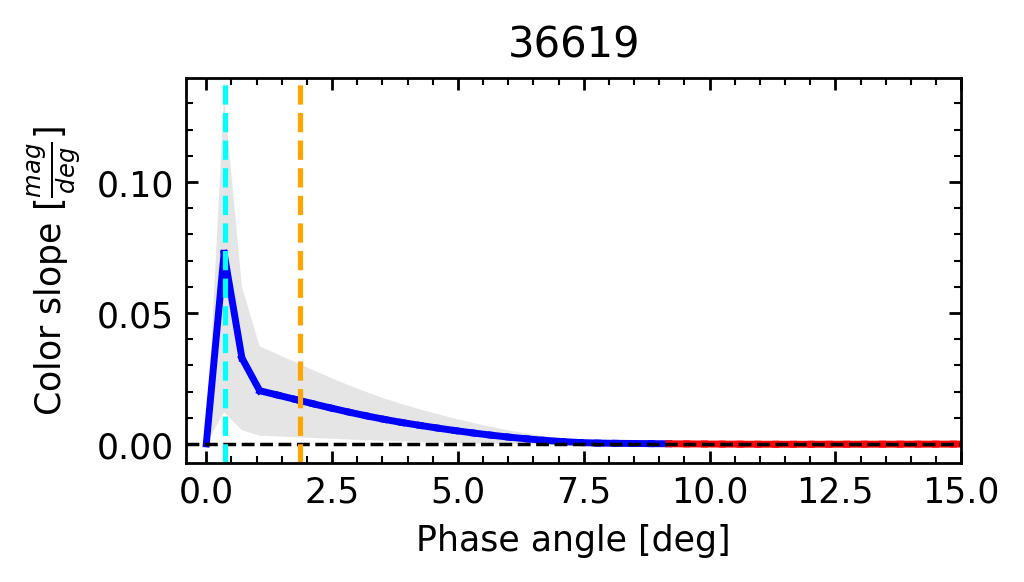}
\includegraphics[width=4cm]{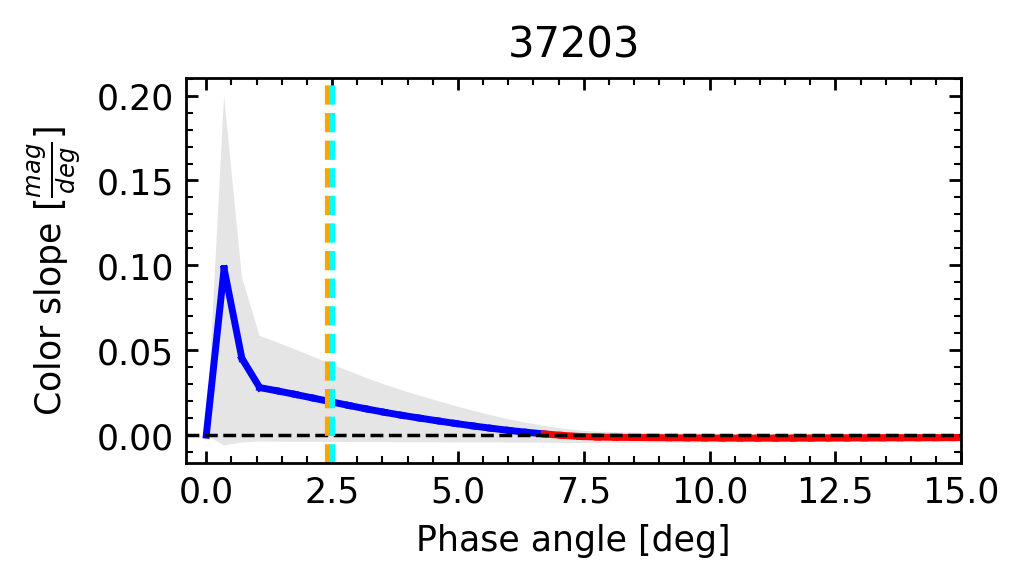}
\includegraphics[width=4cm]{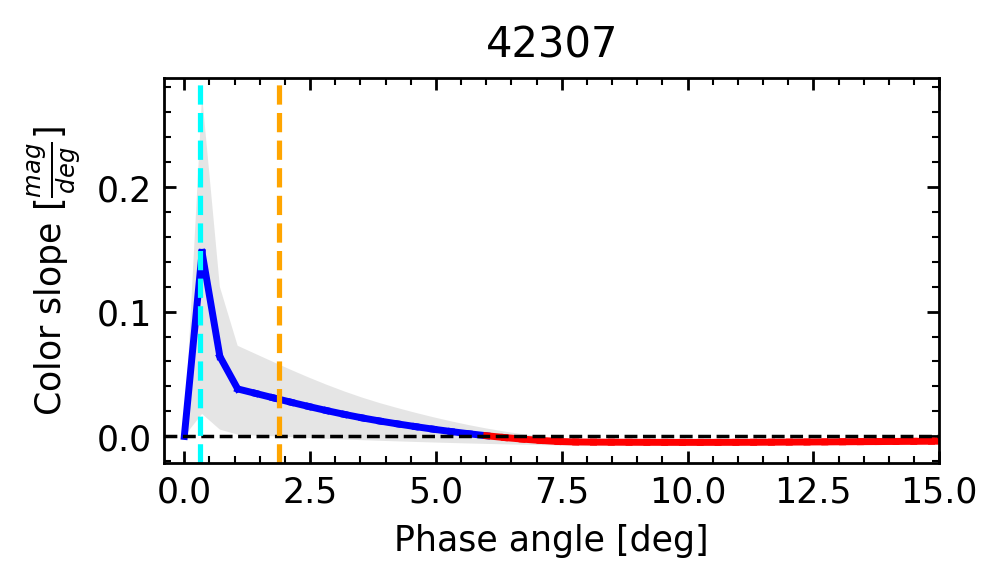}
\includegraphics[width=4cm]{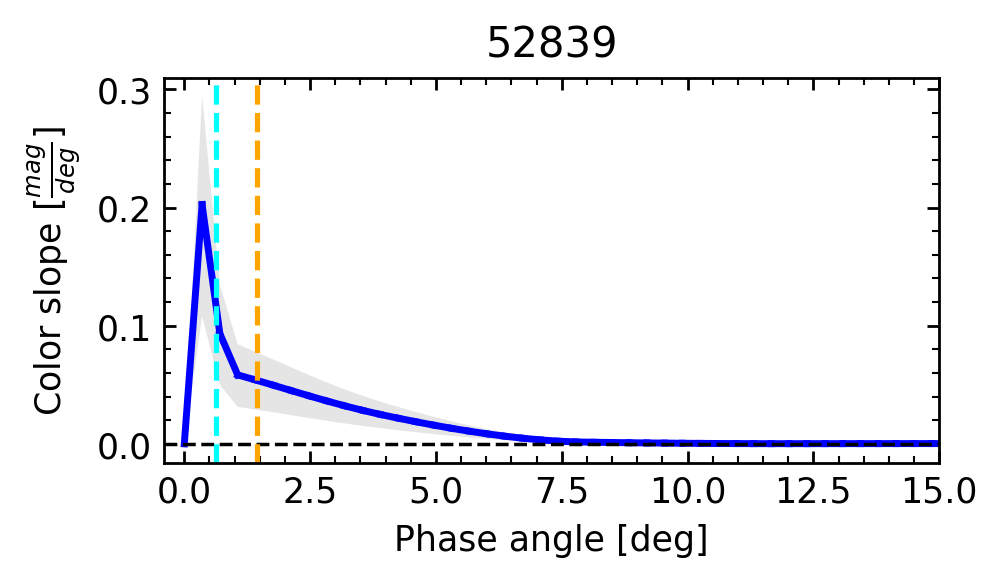}
\caption{(continued).}
\label{blue2}
\end{figure}

\end{document}